\documentclass[fleqn,usenatbib]{mnras}

\usepackage[T1]{fontenc}
\usepackage{ae,aecompl}
\usepackage{graphicx}	% Including figure files
\usepackage{amsmath}	% Advanced maths commands
\usepackage{amssymb}	% Extra maths symbols

\title{Retrieval analysis of 38 WFC3 transmission spectra and resolution of the normalisation degeneracy}

\author[Fisher \& Heng]{
Chloe Fisher$^{1,2}$\thanks{Email: chloe.fisher@csh.unibe.ch (CF)}
and Kevin Heng$^{1}$\thanks{E-mail: kevin.heng@csh.unibe.ch (KH)}
\\
$^{1}$University of Bern, Center for Space and Habitability, Gesellschaftsstrasse 6, CH-3012, Bern, Switzerland\\
$^{2}$University of Bern International 2021 Ph.D Fellowship}

\date{}

\pubyear{2018}

\begin{document}
\label{firstpage}
\pagerange{\pageref{firstpage}--\pageref{lastpage}}
\maketitle

\begin{abstract}
A comprehensive analysis of 38 previously published Wide Field Camera 3 (WFC3) transmission spectra is performed using a hierarchy of nested-sampling retrievals: with versus without clouds, grey versus non-grey clouds, isothermal versus non-isothermal transit chords and with water, hydrogen cyanide and/or ammonia.  We revisit the ``normalisation degeneracy": the relative abundances of molecules are degenerate at the order-of-magnitude level with the absolute normalisation of the transmission spectrum.  Using a suite of mock retrievals, we demonstrate that the normalisation degeneracy may be partially broken using WFC3 data alone, even in the absence of optical/visible data and without appealing to the presence of patchy clouds, although lower limits to the mixing ratios may be prior-dominated depending on the measurement uncertainties.  With James Webb Space Telescope-like spectral resolutions, the normalisation degeneracy may be completely broken from infrared spectra alone.  We find no trend in the retrieved water abundances across nearly two orders of magnitude in exoplanet mass and a factor of 5 in retrieved temperature (about 500--2500 K).  We further show that there is a general lack of strong Bayesian evidence to support interpretations of non-grey over grey clouds (only for WASP-69b and WASP-76b) and non-isothermal over isothermal atmospheres (no objects).  35 out of 38 WFC3 transmission spectra are well-fitted by an isothermal transit chord with grey clouds and water only, while 8 are adequately explained by flat lines.  Generally, the cloud composition is unconstrained.
\end{abstract}


\section{Introduction}
\label{sect:intro}

At the time of writing, we are in the transitional period between the Hubble and James Webb Space Telescopes (HST and JWST).  In the foreseeable future, WFC3 transmission spectra spanning 0.8--1.7 $\mu$m will be superceeded by NIRSpec data ranging from 0.6--5 $\mu$m and at enhanced spectral resolution.  It is therefore timely to perform a uniform theoretical analysis of a consolidated dataset of WFC3 transmission spectra, which is the over-arching motivation behind the current study.

\subsection{Observational motivation: a statistical study of cloudy atmospheres}

Following the work of \cite{iyer16}, \cite{fu17} recently conducted a statistical study of the transmission spectra of 34 exoplanets (mostly hot Jupiters) measured using WFC3 onboard HST, which were mostly gathered from \cite{tsi18}.  In order to isolate the spectral feature due to water\footnote{Technically, it is due to a collection of unresolved water lines.}, they quantified the strength of absorption between 1.3--1.65 $\mu$m, relative to the continuum, in terms of the number of pressure scale heights, which they represented by $A_H$.  Based on the finding that both $A_H$ and the equilibrium temperature ($T_{\rm eq}$) follow log-normal distributions, \cite{fu17} concluded that their sample of $A_H$ is affected by observational bias.  \cite{tsi18} defined an Atmospheric Detectability Index (ADI) to quantify the strength of detection of the water feature, but do not explicitly link the ADI to any trends in cloud properties.  They concluded that all of their WFC3 transmission spectra, except for WASP-69b, are consistent with the presence of a grey cloud deck.

Our intention is to build upon the \cite{fu17} and \cite{tsi18} studies by subjecting their WFC3 sample to a detailed atmospheric retrieval study and elucidating the presence of assumptions, limitations and trends.  It follows the principle that the same datasets should be analysed by different groups (using different codes and techniques) within the community, so as to check for the consistency and robustness of theoretical interpretations \citep{fortney16}.

\begin{figure}
\vspace{-0.1in}
\begin{center}
\includegraphics[width=\columnwidth]{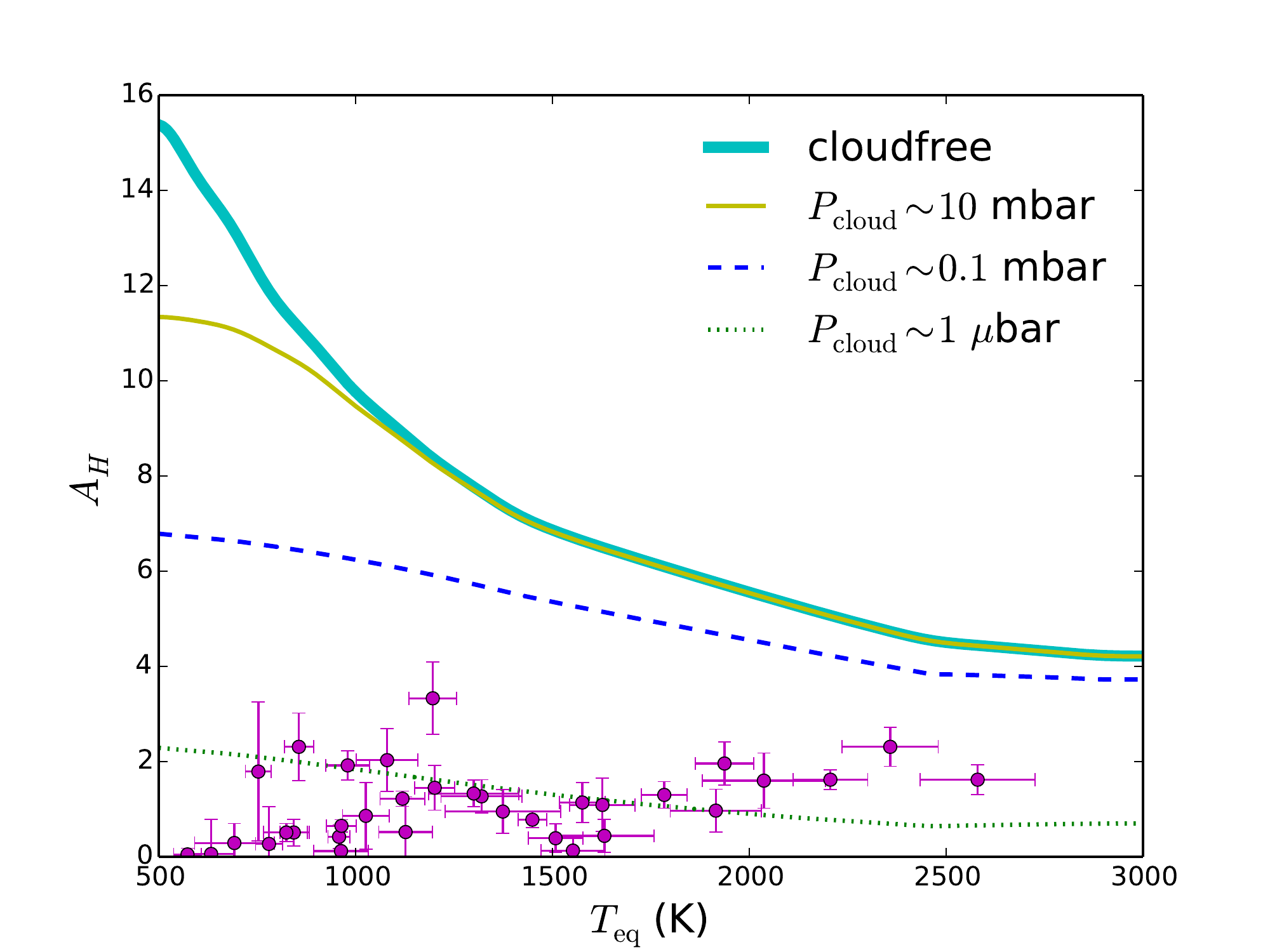}
\end{center}
%\vspace{-0.1in}
\caption{Strength of WFC3 water feature, $A_H$, in terms of pressure scale heights as a function of the equilibrium temperature.  Also shown are the theoretical predictions of $A_H$ for cloud-free and cloudy atmospheres. For the latter, the curves correspond to transit chords probing $P_{\rm cloud} \sim 1$ $\mu$bar, $\sim 0.1$ mbar and $\sim 10$ mbar if the opacity was solely due to grey clouds and the gravity is $\sim 10^3$ cm s$^{-2}$.  It is apparent that all of the 34 atmospheres are cloudy if only water is assumed to be present.}
\label{fig:ah}
\end{figure}

From a theoretical standpoint, $A_H$ is an elegant quantity to examine, because the difference in transit radii between the peak of the water feature and the continuum is simply
\begin{equation}
A_H = \ln{\left( \frac{\kappa_{\rm max}}{\kappa_{\rm min}} \right)},
\end{equation}
where $\kappa_{\rm max}$ and $\kappa_{\rm min}$ are the maximum and minimum values of the water opacity in the WFC3 range of wavelengths.  The preceding equation naturally derives from equation (\ref{eq:radius}), if the volume mixing ratio of water ($X_{\rm H_2O}$) is assumed to be uniform across altitude, and is free of the normalisation degeneracy (see next subsection).  Its simplicity allows us to do a first check on if the 34 objects in the sample gathered by \cite{fu17} have cloudy atmospheres.

In Figure \ref{fig:ah}, we show curves of $A_H$ for completely cloud-free atmospheres by assuming that the temperature (sampled by the water opacity) is the equilibrium temperature.  Also shown are curves of $A_H$ corresponding to cloudy atmospheres with constant opacities.  For example, an opacity of 1 cm$^2$ g$^{-1}$ corresponds to a transit chord probing a pressure $\sim 0.1$ mbar if only clouds (and not molecules) are present in the atmosphere.  By comparing these theoretical curves to the measured data points of \cite{fu17}, we tentatively conclude that all of the 34 transiting exoplanets in their sample have cloudy atmospheres.  It is one of the goals of the present study to examine if this conclusion is robust.  Assuming that the temperature is some fraction of the equilibrium temperature merely translates the theoretical curves along the horizontal axis (not shown).  

\begin{figure}
\vspace{-0.1in}
\begin{center}
\includegraphics[width=\columnwidth]{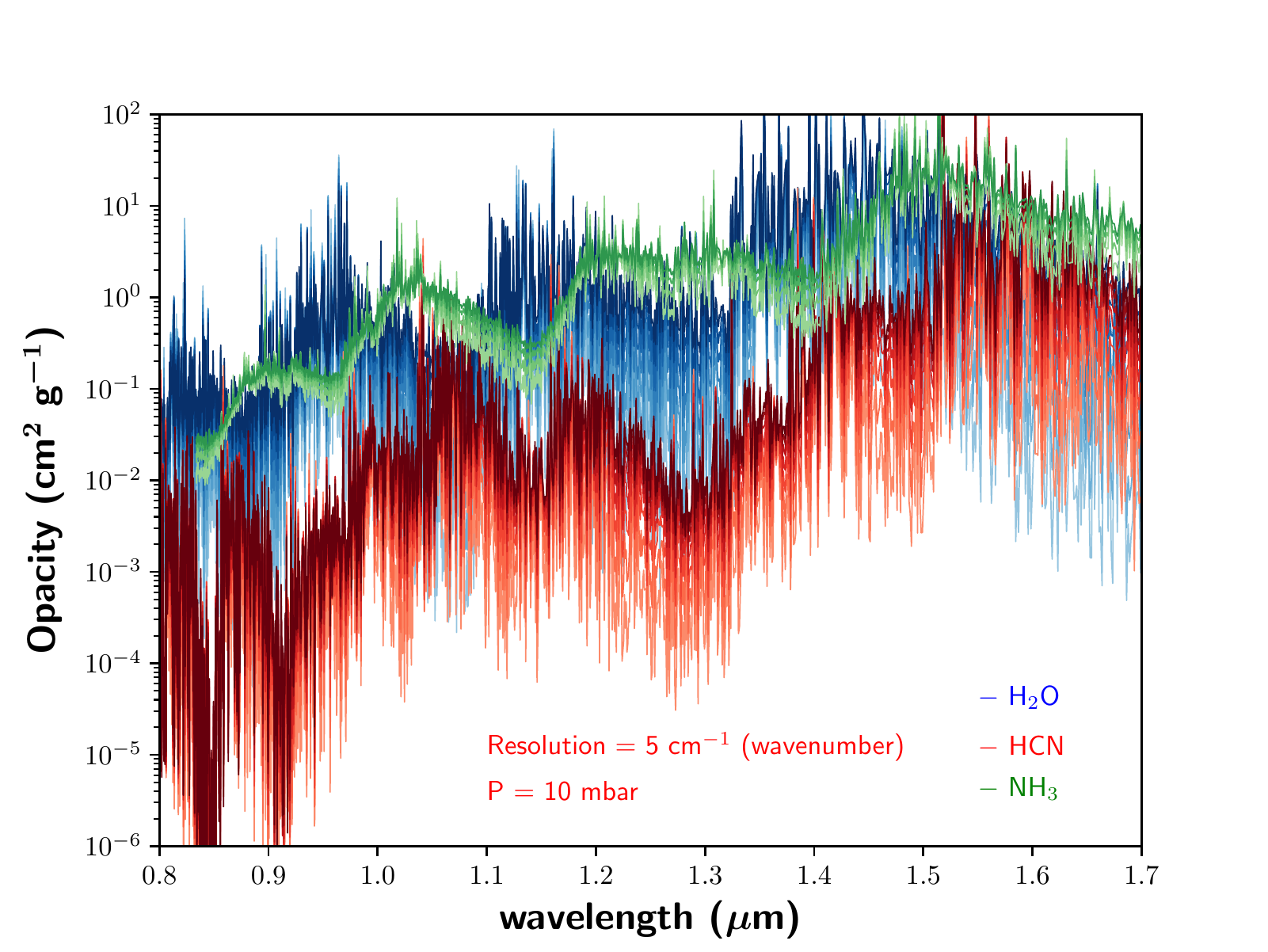}
\end{center}
%\vspace{-0.1in}
\caption{Opacities of water, hydrogen cyanide and ammonia as functions of wavelength.  The \texttt{ExoMol} spectroscopic line list was used as input for computing these opacities.  For water and hydrogen cyanide, we show a sequence of opacities from 900--2100 K.  Darker shades of the same colour correspond to higher temperatures.  For ammonia, the temperature sequence is terminated at 1500 K, because \texttt{ExoMol} does not provide any data for higher temperatures.  The spectral resolution is 5 cm$^{-1}$ and the pressure is fixed at $P=10$ mbar, because these values are what we use in our retrievals (see text for details). }
\label{fig:opacities}
\end{figure}

There is an additional, supporting argument for the atmospheres being cloudy.  By visual inspection of measured WFC3 transmission spectra, we noticed that the continuum blue-wards of the 1.4 $\mu$m water feature tends to be somewhat flat, in contrast to the opacity of water which tends to be rather structured at these wavelengths (Figure \ref{fig:opacities}).  This suggests that most, if not all, of the WFC3 transmission spectra measured so far are probing cloudy atmospheres---at least at the atmospheric limbs.  However, this argument becomes less clear if ammonia and hydrogen cyanide are present, as they may mimic these effects on the spectra.  

In the current study, one of our goals is to formalise this finding by performing atmospheric retrieval, within a nested-sampling framework (e.g., \citealt{skilling06,feroz08,feroz09,feroz13,bs13,wald15,lavie17,tsi18}), on each of the 34 objects in the \cite{fu17} sample.  We construct a hierarchy of models with increasing levels of sophistication: cloud-free model (2 parameters), cloudy model with constant/grey cloud opacity (3 parameters), cloudy model with non-grey opacity (6 parameters).  It is assumed that the main molecular absorber is water.  If hydrogen cyanide (HCN) and ammonia (NH$_3$) are added to the analysis \citep{mm17a}, then it adds two more free parameters for a maximum of 8 parameters for the isothermal model.  Our non-isothermal model adds another parameter.  For comparison, \cite{mm17a} employ a 16-parameter model based partly on the heritage from \cite{madhu09}. 

We use the computed Bayesian evidence \citep{trotta08} from the retrievals to select the best model given the quality of the data, and hence determine if the atmospheres are cloudy, if cloud properties may be meaningfully constrained, and if NH$_3$ and/or HCN are detected in a given WFC3 spectrum.  Unlike the approach adopted by \cite{mm17a}, we do not test for patchy clouds.  Rather, we test essentially for whether the cloud particles are small or large (compared to the wavelengths probed).

%The implication is that our proposed strategy for a self-bootstrapping procedure to break the $R_0$-$P_0$-$X_{\rm H_2O}$ degeneracy is plausible, since any model constructed to analyze any of these 34 WFC3 transmission spectra needs to necessarily include a treatment of clouds.  An additional goal is to perform cloud-free and cloudy retrievals within a nested-sampling framework, and quantify the degree of cloudiness in all of the WFC3 transmission spectra measured so far.  For the cloudy models, we will test both grey and non-grey models, which represent clouds with large and small particles, respectively.

\subsection{Theoretical motivation: the normalisation degeneracy}
\label{subsect:motivation}

Atmospheric retrievals of transmission spectrum typically specify a plane-parallel model atmosphere, assume azimuthal symmetry and then trace a transit chord through a set of atmospheric columns (each approximated by a plane-parallel atmosphere) \citep{madhu09,bs12,bs13,line13,wald15}.  This brute-force procedure for calculating the transmission spectrum was previously described by \cite{brown01} and \cite{h01}.  In the current study, our intention is to build a nested-sampling retrieval framework around a validated analytical model for computing the transmission spectrum that bypasses the need for a brute-force calculation.  Complementary to previous retrieval studies, we make a different set of investments, approximations and simplifications.

Building on the work of \cite{lec08}, \cite{ds13} and \cite{bs17}, \cite{hk17} demonstrated that an analytical expression for the isothermal transit chord of an atmosphere,
\begin{equation}
R = R_0 + H \left( \gamma + E_1 + \ln{\tau} \right),
\label{eq:radius}
\end{equation}
is accurate enough\footnote{Meaning that the errors incurred are smaller than the noise floor (about 50 parts per million) of HST and the expected noise floor of JWST.} to model WFC3 transmission spectra for atmospheres with temperatures $\sim 1000$ K or hotter, where we have
\begin{equation}
\tau \equiv \frac{\kappa P_0}{g} \sqrt{\frac{2\pi R_0}{H}}.
\end{equation}
The pressure scale height is given by $H$, the Euler-Mascheroni constant by $\gamma \approx 0.57721$ and the surface gravity by $g$.  The exponential integral of the first order is given by $E_1(\tau)$ \citep{abram,arfken}, which has the argument $\tau$.  For a WFC3 spectrum dominated by water, the opacity is $\kappa \propto X_{\rm H_2O}$, where $X_{\rm H_2O}$ is the volume mixing ratio of water.  Equation (\ref{eq:radius}) assumes that $R_0 < R$; if the layer of the atmosphere located at $R_0$ is opaque in the WFC3 bandpass ($\tau \gg 1$), then the $E_1$ term may be dropped.

\begin{figure}
%\vspace{-0.1in}
\begin{center}
\includegraphics[width=\columnwidth]{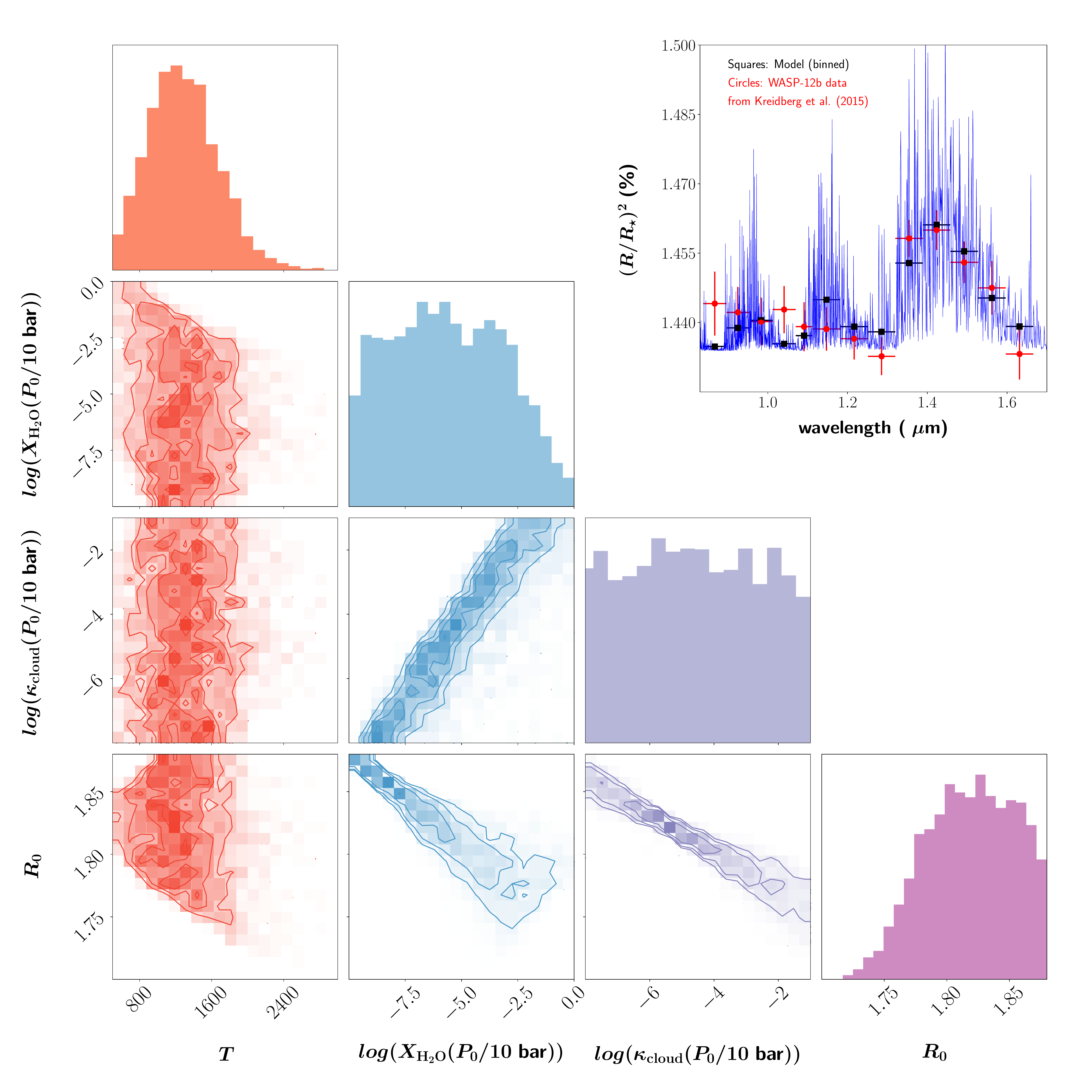}
\end{center}
%\vspace{-0.1in}
\caption{Posterior distributions of the water volume mixing ratio ($X_{\rm H_2O}$), temperature ($T$), grey/constant cloud opacity ($\kappa_{\rm cloud}$) and reference transit radius ($R_0$; uniform prior from $1.7$--$1.88 ~R_{\rm J}$) from a retrieval analysis of the WASP-12b transmission spectrum.  The degeneracies between $R_0$ and the other quantities is apparent; $R_0$ is bounded from below because $X_{\rm H_2O}$ is bounded from above by unity.  In this test, we have set $P_0=10$ bar but in our subsequent retrieval of the WASP-12b WFC3 transmission spectrum we will fit for $P_0$ (see text and Figure \ref{fig:veryhot}).   The measured data and best-fit model are shown in the top-right panel.  The physical units of $T$ and $\kappa_{\rm cloud}$ are K and cm$^2$ g$^{-1}$, respectively, while $R_0$ is given in units of Jupiter radii ($R_{\rm J}$).  This retrieval assumes a constant mean molecular mass and ignores the effect of collisional induced absorption, which we will explore later in the current study.}
\label{fig:demo}
\end{figure}

\begin{figure*}
%\vspace{-0.1in}
\begin{center}
\includegraphics[width=\columnwidth]{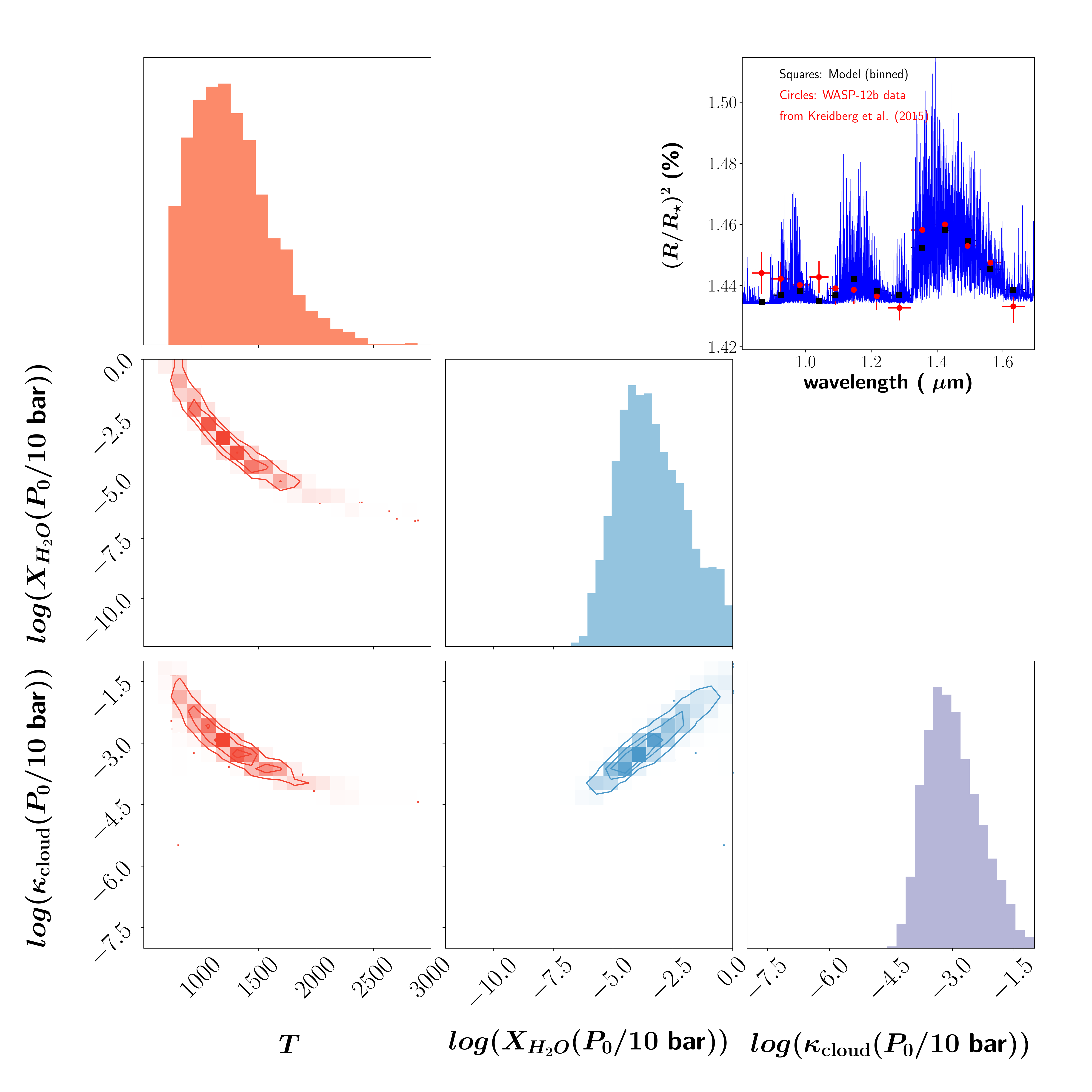}
\includegraphics[width=\columnwidth]{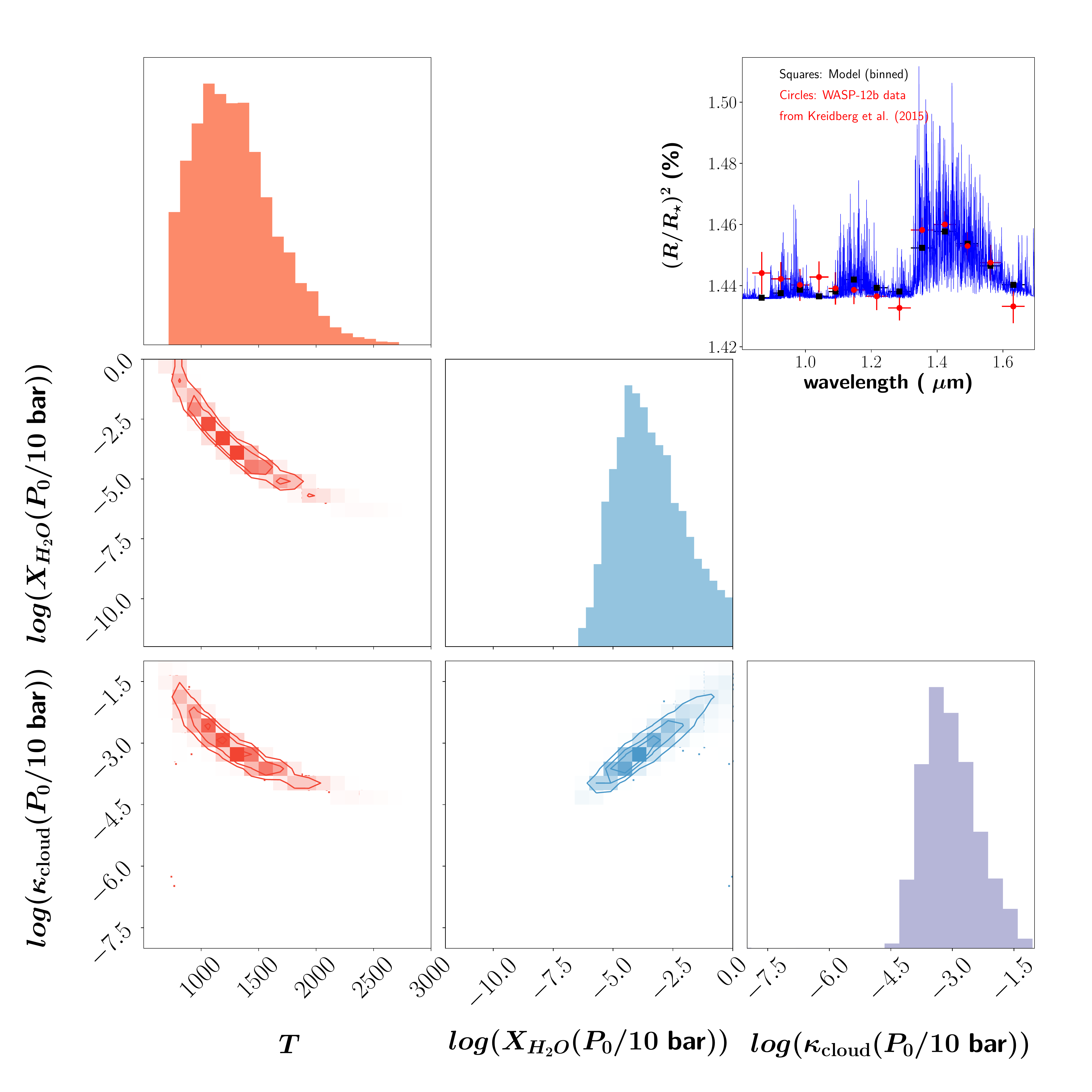}
\includegraphics[width=\columnwidth]{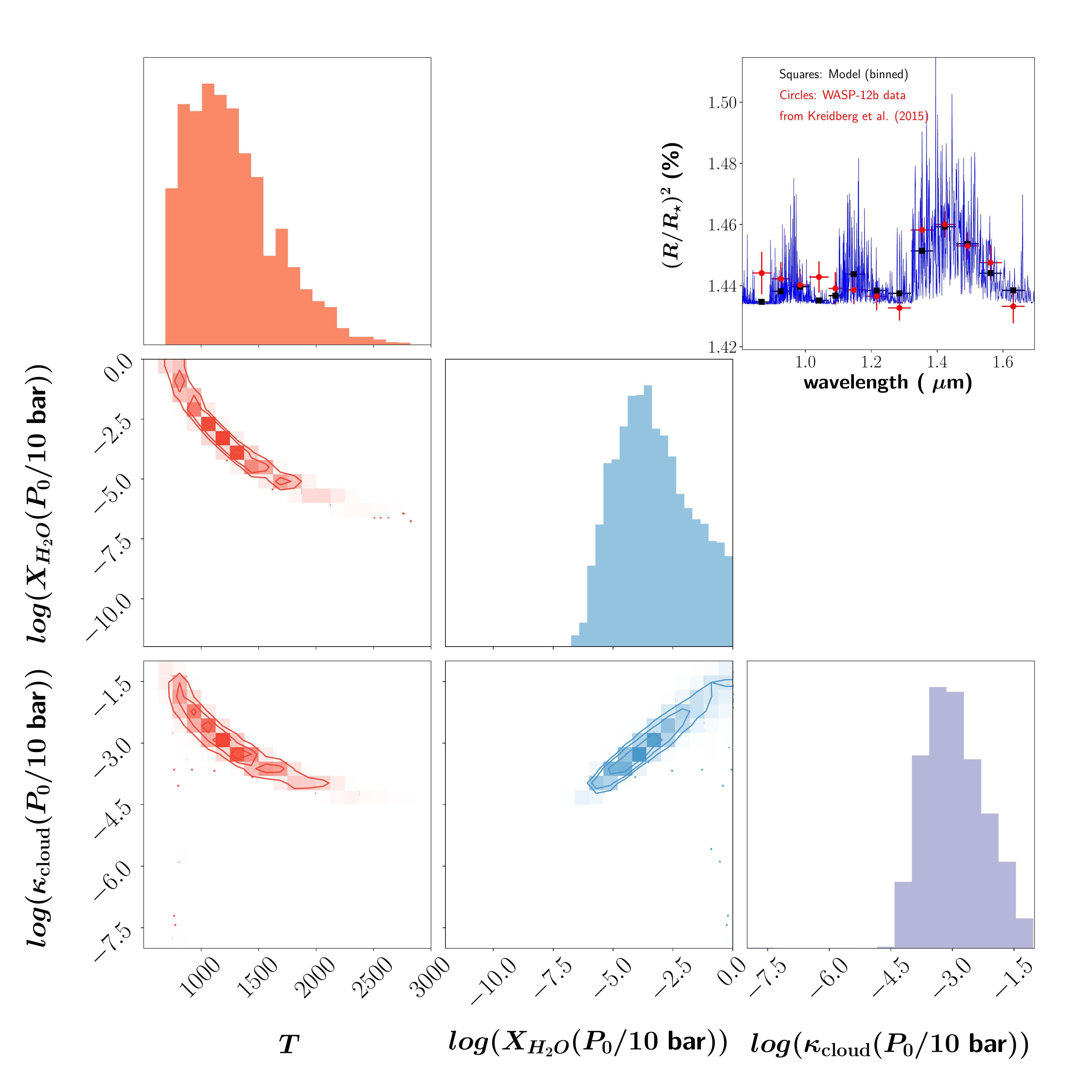}
\includegraphics[width=\columnwidth]{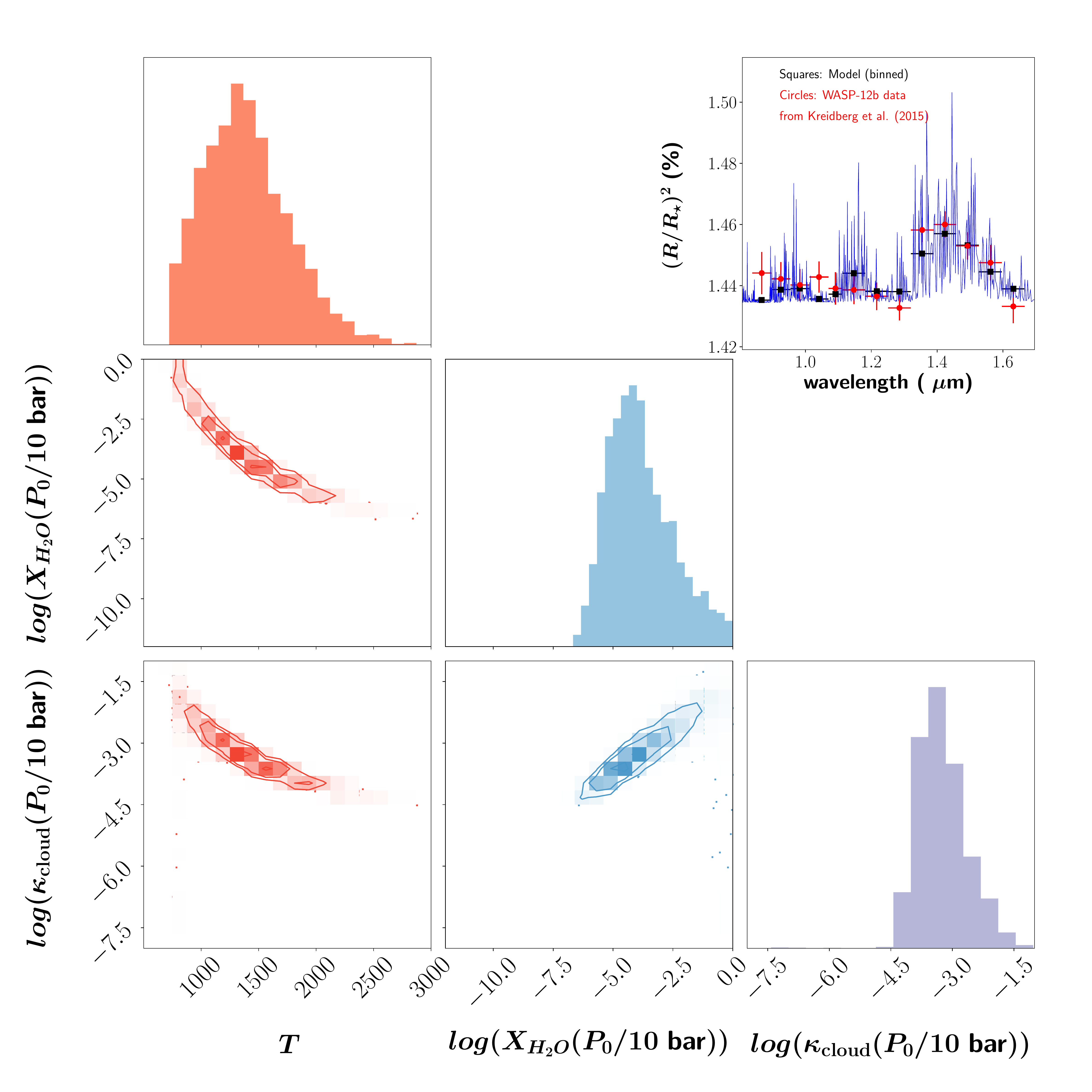}
\end{center}
\vspace{-0.1in}
\caption{Same as Figure \ref{fig:demo}, but using spectral resolutions of 1 cm$^{-1}$ (top left panel), 2 cm$^{-1}$ (top right panel), 5 cm$^{-1}$ (bottom left panel) and 10 cm$^{-1}$ (bottom right panel) for the \texttt{ExoMol} water opacity.}
\label{fig:resolution}
\end{figure*}

Equation (\ref{eq:radius}) straightforwardly shows that there exists a three-way degeneracy between the reference transit radius ($R_0$), reference pressure ($P_0$) and $X_{\rm H_2O}$, which was first noticed numerically\footnote{Our stand is that a numerical demonstration of an effect alone does not qualify as attaining full understanding of it, until its theoretical (analytical) formalism has been elucidated.} by \cite{bs12} and \cite{g14}.  The values of $R_0$ and $P_0$, as well as the relationship between them, are a priori unknown, because it is akin to having prior knowledge of the structure of the exoplanet.  It is apparent that a small change in $R_0$ causes a large variation in $X_{\rm H_2O}$.  Furthermore, it is $X_{\rm H_2O} P_0$, and not $X_{\rm H_2O}$ alone, that is being retrieved from the data.  It is worth emphasising that these obstacles do not exist in the forward problem, where one makes a specific set of assumptions (e.g., solar metallicity, chemical equilibrium) and computes the transmission spectrum, but they are front and center in the inverse problem.  \cite{hk17} pointed out these issues, but they did not examine them further within a Bayesian retrieval framework, which partially motivates the current study.

Figure \ref{fig:demo} shows a retrieval calculation performed using a new code (\texttt{HELIOS-T}) presented as part of the current study, which we constructed specifically to perform fast retrievals on transmission spectra at low spectral resolution.\footnote{At high spectral resolution, the fully resolved spectral lines may span many orders of magnitude in pressure between the line peaks and wings, thereby violating the isobaric assumption.}  (A detailed description of methodology will come later in \S\ref{sect:method}.)  It demonstrates that while the temperature may be robustly retrieved, there are order-of-magnitude degeneracies associated with the water mixing ratio and cloud opacity that arise from small variations of $R_0$ (in the third significant figure), as previously elucidated by \cite{hk17}.  In the present study, we wish to examine if $R_0$ or $P_0$ may be used as a fitting parameter to break the normalisation degeneracy.  We further examine if the normalisation degeneracy may be broken with WFC3 transmission spectra alone, or if JWST-like spectra is needed.

%It is worth emphasising that while we have treated $R_0$ as a fitting parameter for the purpose of illustrating the ``normalisation degeneracy", there should be a unique relationship between $R_0$ and $P_0$.  In other words, once $P_0$ is stated, $R_0$ should be known in theory and the two quantities cannot be treated as independent fitting parameters.  Part of the goal of the present study is to quantify these degeneracies further for the \cite{fu17} sample of WFC3 transmission spectra.  In the case of cloud-free atmospheres (in the optical range of wavelengths) $R_0(P_0)$ may be uniquely computed, as we will demonstrate explicitly for WASP-17b and WASP-31b.

%If the continuum is dominated by clouds, then the cloud opacity should directly yield a unique value for the pressure being probed.  This is conceptually equivalent to using the spectral slope in the optical range of wavelengths to determine a unique $R(P)$ relationship and break the $R_0$-$P_0$-$X_{\rm H_2O}$ degeneracy---a procedure that only holds if the spectral slope may be solely attributed to Rayleigh scattering by hydrogen molecules.  In the current study, we pursue this alternative route of using the blue continuum in the WFC3 bandpass to ``self-bootstrap" the retrieval.  But in order to proceed, we first need to convince ourselves that WFC3 transmission spectra probe, to date, cloud exoplanetary atmospheres.

\subsection{Layout of current study}

In \S\ref{sect:method}, we describe our theoretical methodology, including how we compute transit radii and opacities.  In \S\ref{sect:results}, we perform suites of tests, a detailed analysis of the 38 WFC3 transmission spectra in the \cite{tsi18} and \cite{dewit18} samples and elucidate trends among the retrieved quantities.  The implications of our results are discussed in \S\ref{sect:discussion}.  Table 1 lists our assumptions for the prior distributions of parameters.  Table 2 summarises our retrieval results.  Table 3 summarises some of the input parameters for the retrievals.

Our new nested-sampling retrieval code for transmission spectra, \texttt{HELIOS-T}, is part of our open-source suite of tools for analysing exoplanetary atmospheres known as the Exoclimes Simulation Platform (\texttt{www.exoclime.org} or \texttt{https://github.com/exoclime}).

\section{Methodology}
\label{sect:method}

\begin{table*}
\label{tab:priors}
\begin{center}
\caption{Assumed prior distributions for retrievals and values of physical constants used}
\begin{tabular}{lcccc}
\hline
\hline
Quantity & Symbol & Range & Assumption & Units \\
\hline
\hline
Temperature & $T$ & (100, 2900) & Uniform & K \\
Water mixing ratio & $X_{\rm H_2O} $ & $\left( 10^{-13}, 1 \right)$ & Log-uniform & -- \\
Hydrogen cyanide mixing ratio & $X_{\rm HCN}$ & $\left( 10^{-13}, 1 \right)$ & Log-uniform & -- \\
Ammonia mixing ratio & $X_{\rm NH_3}$ & $\left( 10^{-13}, 1 \right)$ & Log-uniform & -- \\
Grey cloud opacity & $\kappa_{\rm cloud}$ & $\left( 10^{-12}, 10^{2} \right)$ & Log-uniform & cm$^{2}$ g$^{-1}$ \\
%$P$ & $\left( 10^{-5}, 1 \right)$ & Log-uniform & bar \\
Opacity normalisation for non-grey cloud model & $\kappa_0$ & $\left( 10^{-10}, 10^{-1} \right)$ & Log-uniform & cm$^{2}$ g$^{-1}$ \\
Composition parameter in non-grey cloud model & $Q_0$ & $\left( 1, 100 \right)$ & Uniform & -- \\
Index in non-grey cloud model & $a$ & $\left( 3, 6 \right)$ & Uniform & -- \\
Monodisperse, spherical cloud particle radius & $r_{\rm c}$ & $\left( 10^{-7}, 10^{-1} \right)$ & Log-uniform & cm \\
Non-isothermal temperature profile parameter & $b$ & $\left( {-30}, {-1} \right)$ , $\left( 1, 30 \right)$ & Uniform & -- \\
Reference transit radius$^\dagger$ & $R_0$ & $\left(1.619,1.799 \right)$ & Uniform & $R_{\rm J}$ \\
Reference pressure & $P_0$ & $\left(10^{-1}, 10^3 \right)$ & Log-uniform & bar \\
\hline
Equatorial radius of Jupiter & $R_{\rm J}$ & $7.1492 \times 10^9$ & -- & cm \\
Mass of hydrogen atom & $m_{\rm H}$ & $m_{\rm amu}$ & -- & cm \\
Atomic mass unit & $m_{\rm amu}$ & $1.66053904 \times 10^{-24}$ & -- & g \\
Boltzmann constant & $k_{\rm B}$ & $1.38064852 \times 10^{-16}$ & -- & erg K$^{-1}$ \\
\hline
\hline
\end{tabular}\\
%\vspace{0.05in}
$\dagger$: Only used in the test retrievals of WASP-17b (\S\ref{subsect:wasp17_mocks}).
\end{center}
\end{table*}

\subsection{Transmission spectra}

As explained in \S\ref{sect:intro}, equation (\ref{eq:radius}) describes our forward model for transforming a given temperature, surface gravity, opacity function, reference transit radius and reference pressure into a transmission spectrum.  The accuracy of equation (\ref{eq:radius}) has previously been demonstrated by \cite{hk17} and we will not repeat the analysis and explanation here.  To test for non-isothermality, we use another formula derived by \cite{hk17},
\begin{equation}
R = R_0 + H \tau^{1/b} \left( \gamma + E_1 + \ln \tau \right)
\end{equation}
where the reference optical depth is now given by
\begin{equation}
\tau = \frac{\pi P_0 \kappa}{2g} \sqrt{\frac{2 R_0 \left \vert b \right \vert}{H}}.
\end{equation}
We again have $E_1 = E_1(\tau)$.  The dimensionless index $b$ is the ratio of the non-isothermal to the isothermal scale height.  If the values of $\vert b \vert$ are much larger than unity, then it means that the behavior is close to being isothermal.  Essentially, our simplified temperature-pressure profile is described by 2 parameters.

Our approach is complementary to other approaches in the literature that use more complicated prescriptions for temperature-pressure profiles.  For example, \cite{madhu09} and \cite{mm17a} use 9- and 7-parameter fitting functions, respectively.  Again, we make a different investment: we choose to simplify the temperature profile prescription in order to isolate the effects of the other parameters.  It allows us to more cleanly study degeneracies.

\subsection{Opacities}
\label{subsect:opacities}

3  Our H$_2$O, HCN and NH$_3$ opacities are taken from the \texttt{ExoMol} spectroscopic database \citep{barber06,y11,y13,barber14,yt14}.  In a single set of tests (see \S\ref{subsect:linelists}), we also use the \texttt{HITRAN} \citep{rothman87,rothman92,rothman98,rothman03,rothman05,rothman09,rothman13} and \texttt{HITEMP} \citep{rothman10} databases for water.  For a review of the spectroscopic databases, please see \cite{ty17}.  For the procedure on how to use the \texttt{ExoMol} inputs to compute opacities, we refer the reader to \cite{gh15}, Chapter 5 of \cite{heng17} and \cite{y18}.  Examples of opacities for all three molecules are given in Figure \ref{fig:opacities}.

The opacity function used in equation (\ref{eq:radius}) is given by
\begin{equation}
\begin{split}
\kappa =& \frac{X_{\rm H_2O} m_{\rm H_2O} \kappa_{\rm H_2O}}{m} + \frac{X_{\rm HCN} m_{\rm HCN} \kappa_{\rm HCN}}{m} \\
&+ \frac{X_{\rm NH_3} m_{\rm NH_3} \kappa_{\rm NH_3}}{m} + \kappa_{\rm cloud},
\end{split}
\end{equation}
where $m$ is the mean molecular mass, $m_{\rm H_2O}$ is the mass of the water molecule, $\kappa_{\rm H_2O}$ is the water opacity, $X_{\rm HCN}$ is the volume mixing ratio of hydrogen cyanide, $m_{\rm HCN}$ is the mass of the hydrogen cyanide molecule, $\kappa_{\rm HCN}$ is the hydrogen cyanide opacity, $X_{\rm NH_3}$ is the volume mixing ratio of ammonia, $m_{\rm NH_3}$ is the mass of the ammonia molecule, $\kappa_{\rm NH_3}$ is the ammonia opacity and $\kappa_{\rm cloud}$ is the cloud opacity.  

Denoting the atomic mass unit by $m_{\rm amu}$, the mean molecular weight ($\mu = m/m_{\rm amu}$) is given by
\begin{equation}
\mu = 2.4 X_{\rm H_2} +  \frac{X_{\rm H_2O} m_{\rm H_2O}}{m_{\rm amu}} + \frac{X_{\rm HCN} m_{\rm HCN}}{m_{\rm amu}} + \frac{X_{\rm NH_3} m_{\rm NH_3}}{m_{\rm amu}}.
\label{eq:mmm}
\end{equation}
The mixing ratio of molecular hydrogen is determined by demanding that all mixing ratios sum to unity,
\begin{equation}
1.1 X_{\rm H_2} + X_{\rm H_2O} + X_{\rm HCN} + X_{\rm NH_3} = 1,
\end{equation}
where we have assumed that the helium mixing ratio follows cosmic abundance ($X_{\rm He} = 0.1 X_{\rm H_2}$). 

The molecular opacities are sampled at 1 mbar for the first suite of tests (\S\ref{subsect:wasp12_tests}; to ensure continuity with \citealt{hk17}) and 10 mbar for our second suite of tests (\S\ref{subsect:wasp17_mocks}) and actual results (see \S\ref{subsect:all}).  The cloud mixing ratio is subsumed into $\kappa_{\rm cloud}$.  The opacity associated with collision-induced absorption (both H$_2$-H$_2$ and H$_2$-He) are taken from \cite{rothman13}.

An unresolved physics problem inherent in the computation of opacities concerns the effects of pressure broadening.  The spectral lines of \textit{isolated} atoms and molecules are described rather well by a Voigt profile.  As a population, collisions between them become important at high enough pressures, which modifies the shape of the far line wings of the profile.  It remains unknown exactly what ``far" means and how to compute these modified profiles.  In practice, various workers in the field have resorted to truncating the Voigt profiles at some ad hoc distance from line centre (see \citealt{gh15} and references therein).  For this study, we use a line-wing cutoff of 100 cm$^{-1}$.  Fortunately, since transmission spectra probe pressures that are tenuous enough such that pressure broadening has a negligible effect for $\sim 1000$ K atmospheres, this is not a limiting issue.

Another limitation is that, at the time of writing, the NH$_3$ opacities do not exist for temperatures above 1600 K \citep{y11}.  In the absence of these data, we set the opacity for NH$_3$ to be zero for temperatures above 1600 K.

\subsection{Cloud models}

We consider both grey and non-grey clouds.  For our grey cloud model, we assume a constant cloud opacity, which is physically equivalent to assuming that the cloud particles are much larger than the WFC3 wavelengths being probed.  Our non-grey cloud model uses the opacity of \citep{kh18},
\begin{equation}
\kappa_{\rm cloud} = \frac{\kappa_0}{Q_0 x^{-a} + x^{0.2}},
\end{equation}
where $x= 2 \pi r_c / \lambda$ is the dimensionless size parameter, $r_c$ is the particle radius and $\lambda$ is the wavelength.  In their study of 32 condensate species, \cite{kh18} showed that $Q_0 \approx 0.1$--65 is a proxy for cloud composition with larger values corresponding to more volatile species.  For example, water ice has $Q_0=64.98$ and olivine has $Q_0 \approx 10$.  The index $a$ ranges from 3 to 7; $a=4$ corresponds to Rayleigh scattering.  Our non-grey cloud model has 4 free parameters: $\kappa_0$, $Q_0$, $r_c$ and $a$.  The immediate implication of the preceding equation is that if the cloud is grey ($a \approx 0$), then the composition cannot be decisively constrained.

Conceptually, the treatment of \cite{lee13} and \cite{kh18} are identical in that they both allow smooth transitions between the Rayleigh and large-particle regimes.  However, \cite{lee13} assumed $a=4$, whereas \cite{kh18} calibrated $Q_0$ and $a$ against a larger library of species and composition.

\subsection{Data}

For 30 out of 38 objects, the WFC3 transmission spectra were obtained from \cite{tsi18} and provided in electronic form by the first author (A. Tsiaras 2018, private communication).  For WASP-17b, WASP-19b, GJ 1214b and HD97658b, the WFC3 transmission spectra were obtained from \cite{mandell13}, \cite{huitson13}, \cite{k14} and \cite{knutson14}, respectively.  The WFC3 transmission spectra of TRAPPIST-1d, e, f and g were taken from \cite{dewit18}.  The stellar radii and surface gravities for each object are listed in Table \ref{tab:parameters}.  Uncertainties in the stellar radii manifest themselves as uncertainties in the normalisation of the transmission spectra.

\section{Results}
\label{sect:results}

\subsection{Suite of tests on WASP-12b transmission spectrum}
\label{subsect:wasp12_tests}

\begin{figure}
\vspace{-0.1in}
\begin{center}
\includegraphics[width=\columnwidth]{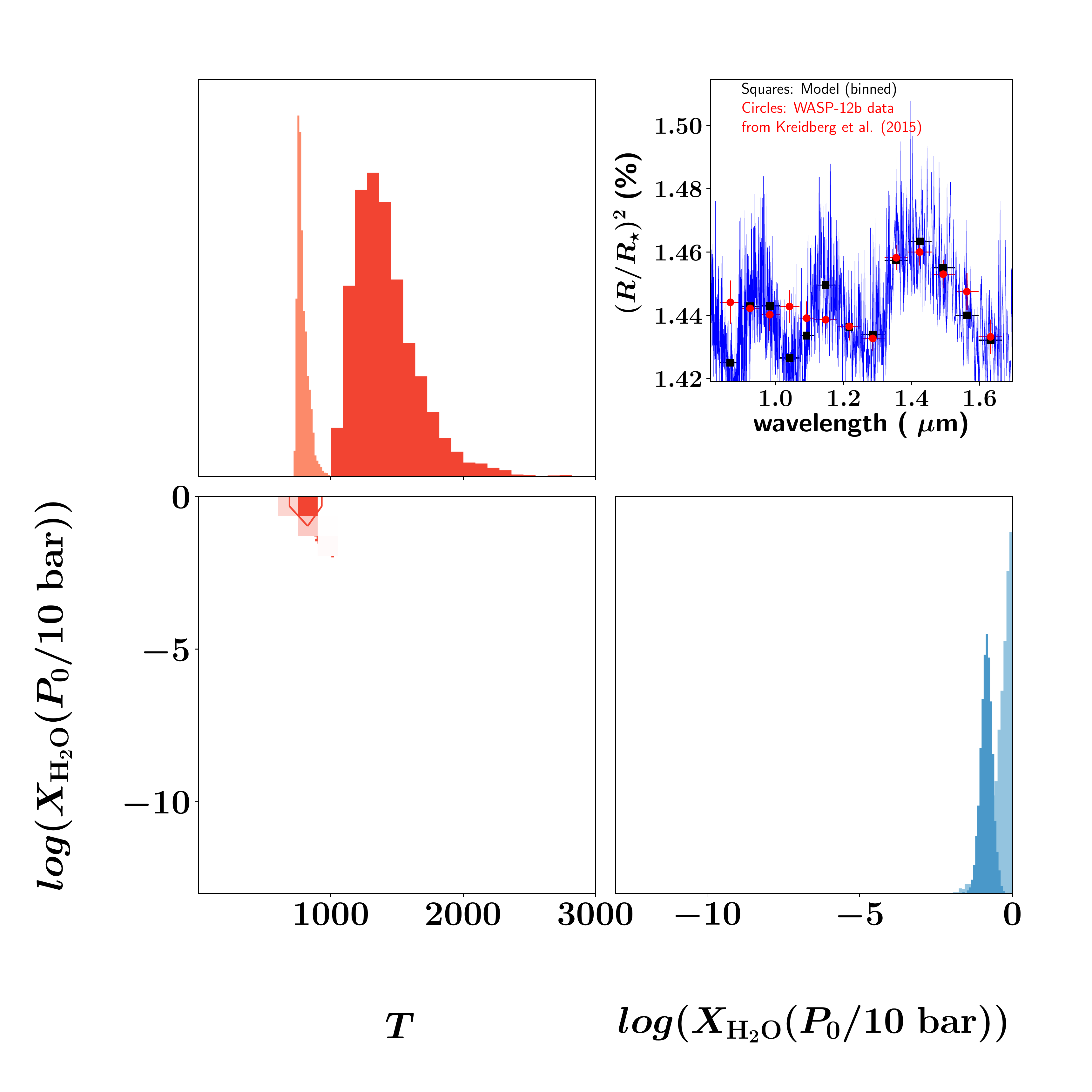}
\end{center}
%\vspace{-0.3in}
\caption{Same as Figure \ref{fig:demo}, but for cloud-free models in which we fix $m=2.4 m_{\rm H}$ ($\log{X_{\rm H_2O}}$ posterior bumps up against 0) versus a variable $m$ (posteriors distributions are in a darker shade) that takes into account water-rich atmospheres.}
\label{fig:cloud-free}
\end{figure}

\begin{figure*}
%\vspace{-0.1in}
\begin{center}
\includegraphics[width=\columnwidth]{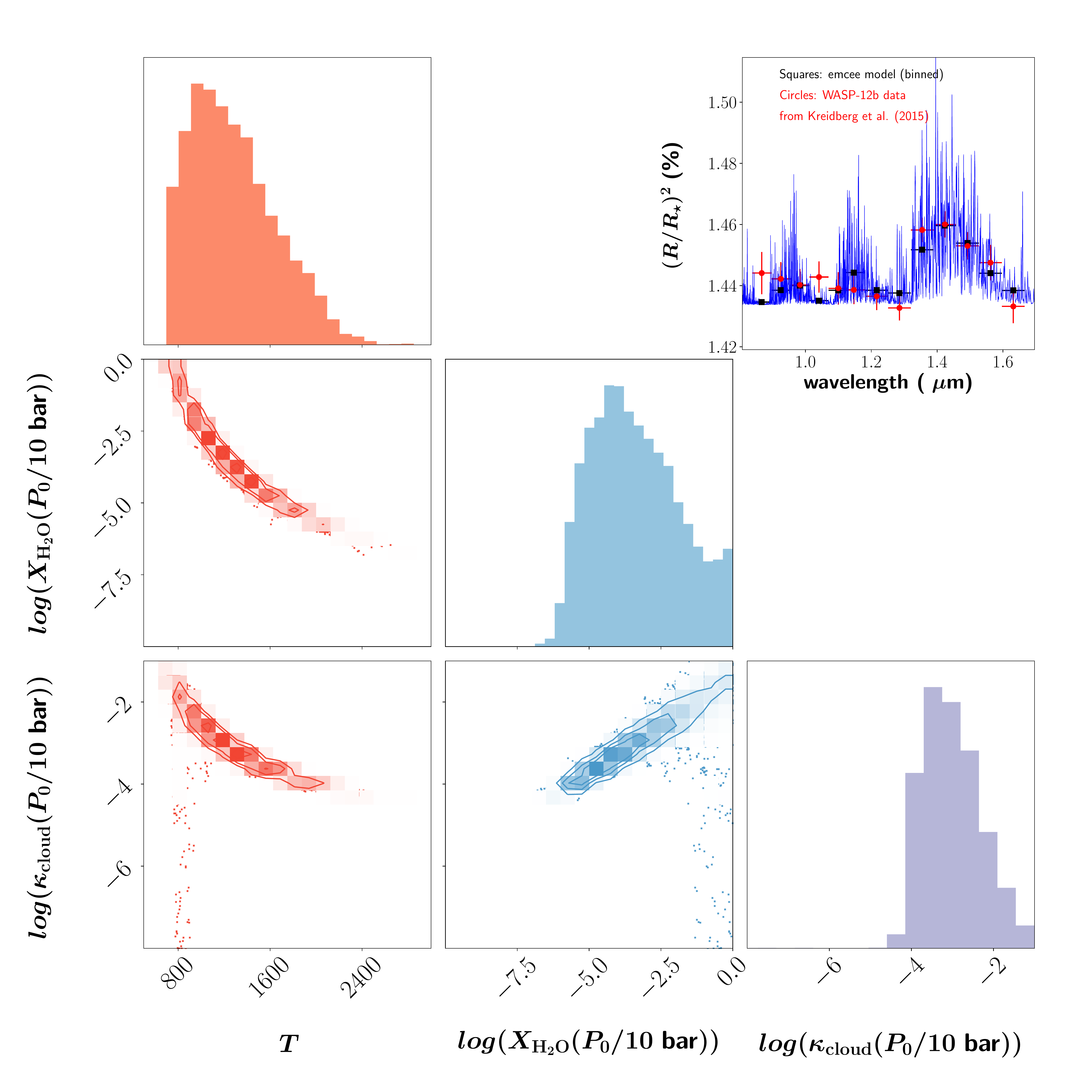}
\includegraphics[width=\columnwidth]{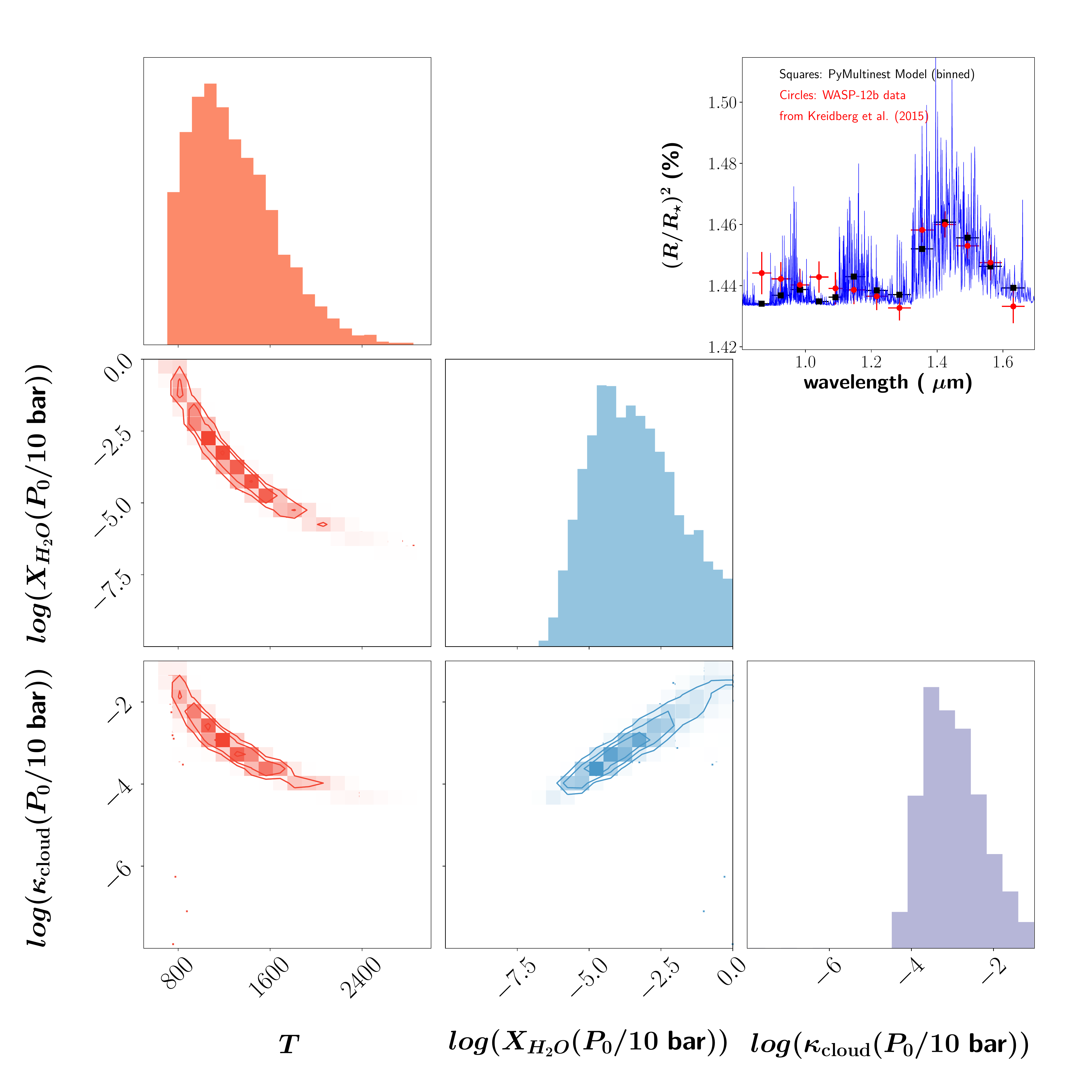}
\end{center}
%\vspace{-0.1in}
\caption{Same as Figure \ref{fig:demo}, but comparing a MCMC (left panel) versus nested-sampling (right panel) retrieval approach.}
\label{fig:mcmc_vs_ns}
\end{figure*}

To provide continuity between \cite{hk17} and the present study, we use the WFC3 transmission spectrum of WASP-12b (13 data points), measured by \cite{k15}, as our starting point for tests.  To cleanly isolate the effects studied, we begin with using a constant/grey cloud opacity.  Note that for these WASP-12b tests only (Figures \ref{fig:resolution} to \ref{fig:wasp12b_evidence}), the molecular opacities are sampled at 1 mbar, CIA is not included and $m$ is fixed at $2.4 m_{\rm H}$, where $m_{\rm H}$ is the mass of the hydrogen atom (which we take to be one atomic mass unit, $m_{\rm amu}$), unless otherwise stated.  In these tests only, we set $R_0 = 1.79 ~R_{\rm J}$ and $P_0=10$ bar.  These restrictions are lifted for the rest of the study.

\subsubsection{Spectral resolution of opacities}

In Figure \ref{fig:resolution}, we perform resolution tests associated with the sampling of the water opacity across wavenumber.  We show retrieval outcomes for spectral resolutions of 1, 2, 5 and 10 cm$^{-1}$.  For all of these values, the posterior distributions of $T$, $X_{\rm H_2O}$ and $\kappa_{\rm cloud}$ are somewhat similar.  Specifically, the retrieved temperatures are $1218^{+388}_{-297}$ K, $1252^{+393}_{-307}$ K, $1203^{+468}_{-323}$ K and $1363^{+403}_{-343}$ K, respectively.  The logarithm of the retrieved water volume mixing ratios are $-3.51^{+1.66}_{-1.26}$, $-3.61^{+1.68}_{-1.27}$, $-3.46^{+1.93}_{-1.49}$ and $-4.04^{+1.62}_{-1.14}$, respectively.  For the rest of the study, we will adopt a sampling resolution of 5 cm$^{-1}$.

\subsubsection{Cloudy versus cloud-free}

Another necessary check is to determine that cloudy models are necessary in the first place for WASP-12b.  In Figure \ref{fig:cloud-free}, we subject the WASP-12b WFC3 transmission spectrum to two cloud-free retrievals: the first has a fixed $m=2.4m_{\rm H}$, while the second has a variable $m$.  For the retrieval with a fixed $m$, the outcome is implausible as the water volume mixing ratio is $\sim 10\%$--$100\%$.  The retrieval with a variable $m$ produces more plausible posteriors, but even by visual inspection it is apparent that the cloud-free model struggles to match the somewhat flat spectral continuum blue-wards of the 1.4 $\mu$m water feature.  For the rest of the WFC3 transmission spectra, we will not show the posterior distributions associated with the cloud-free retrieval (unless it has the highest Bayesian evidence in the model hierarchy), but we will still include them in the overall analysis.

%Contrary to the claim of \cite{mm17a}, we do not find that the cloud-free model is a good estimator of the water abundance.  In fact, it biases the retrieved water abundance to unphysically high values.

\subsubsection{MCMC versus nested sampling}

The next logical test is to compare cloudy retrievals obtained using a Markov Chain Monte Carlo (MCMC) versus nested sampling approach.  For the former, we use the open-source \texttt{emcee} package \citep{fm13}.  For the latter, we use the open-source \texttt{PyMultiNest} package \citep{buchner14}.  Figure \ref{fig:mcmc_vs_ns} compares the outcome from this pair of retrievals.  It is reassuring that the posterior distributions of $T$, $X_{\rm H_2O}$ and $\kappa_{\rm cloud}$ are somewhat similar, although we note that the retrieval performed with MCMC produces higher values of the water volume mixing ratio in the tail of the distribution (towards $X_{\rm H_2O}=1$).  The reason to select the nested-sampling approach over MCMC is because it allows us to straightforwardly compute the Bayesian evidence associated with each model, which in turn allows us to formally apply Occam's Razor.

\subsubsection{Choice of spectroscopic databases: \texttt{HITRAN} versus \texttt{HITEMP} versus \texttt{ExoMol}}
\label{subsect:linelists}

\begin{figure*}
%\vspace{-0.1in}
\begin{center}
\includegraphics[width=\columnwidth]{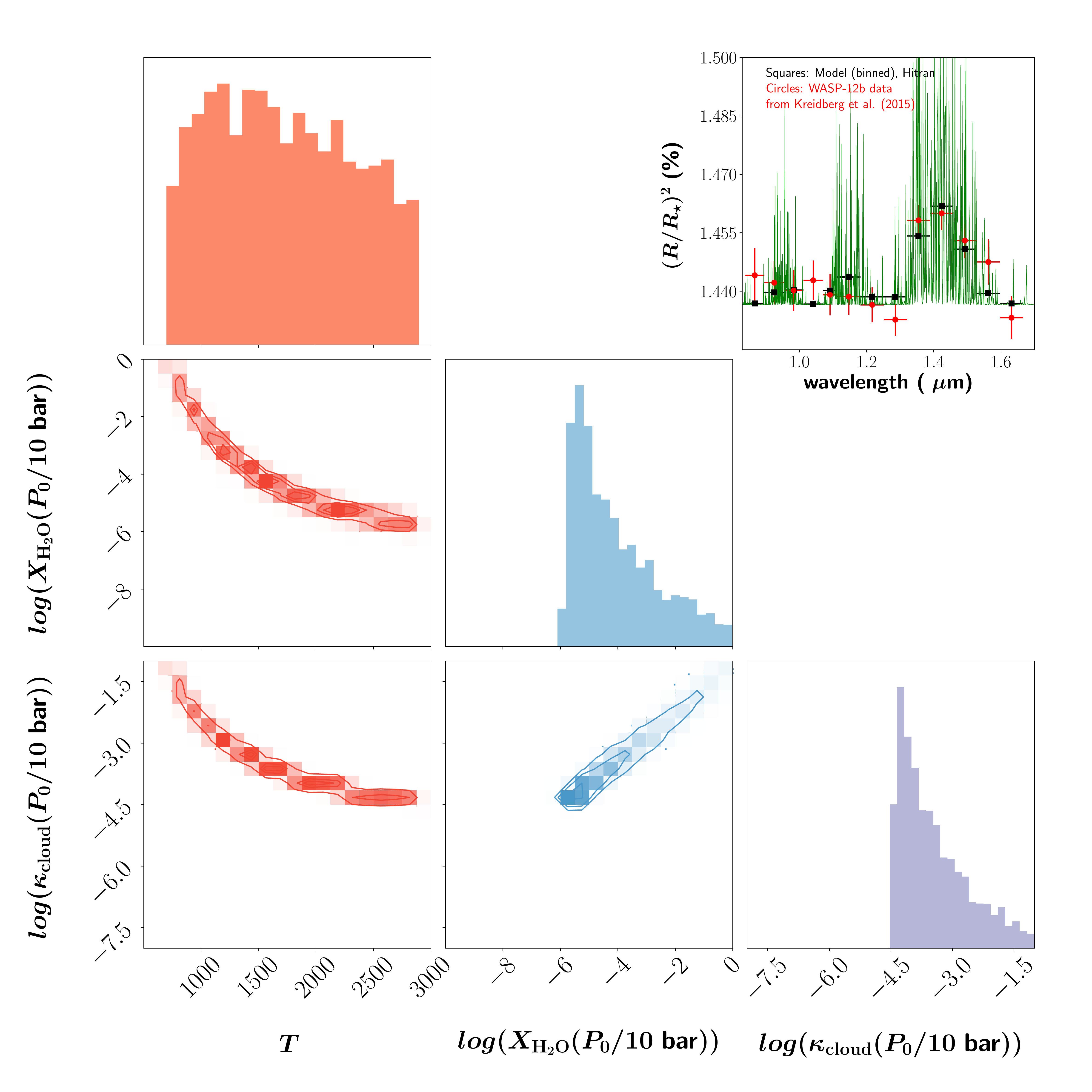}
\includegraphics[width=\columnwidth]{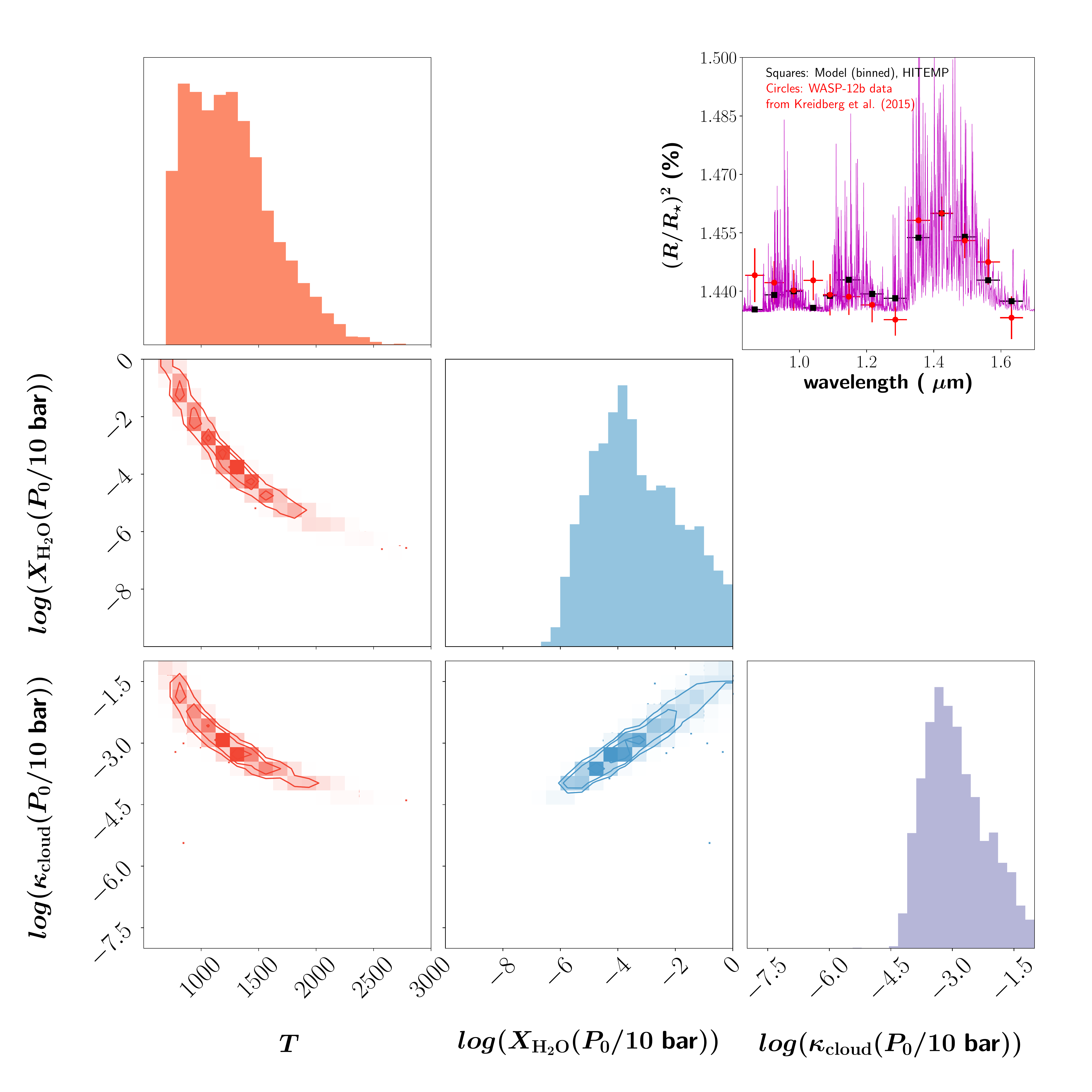}
\includegraphics[width=\columnwidth]{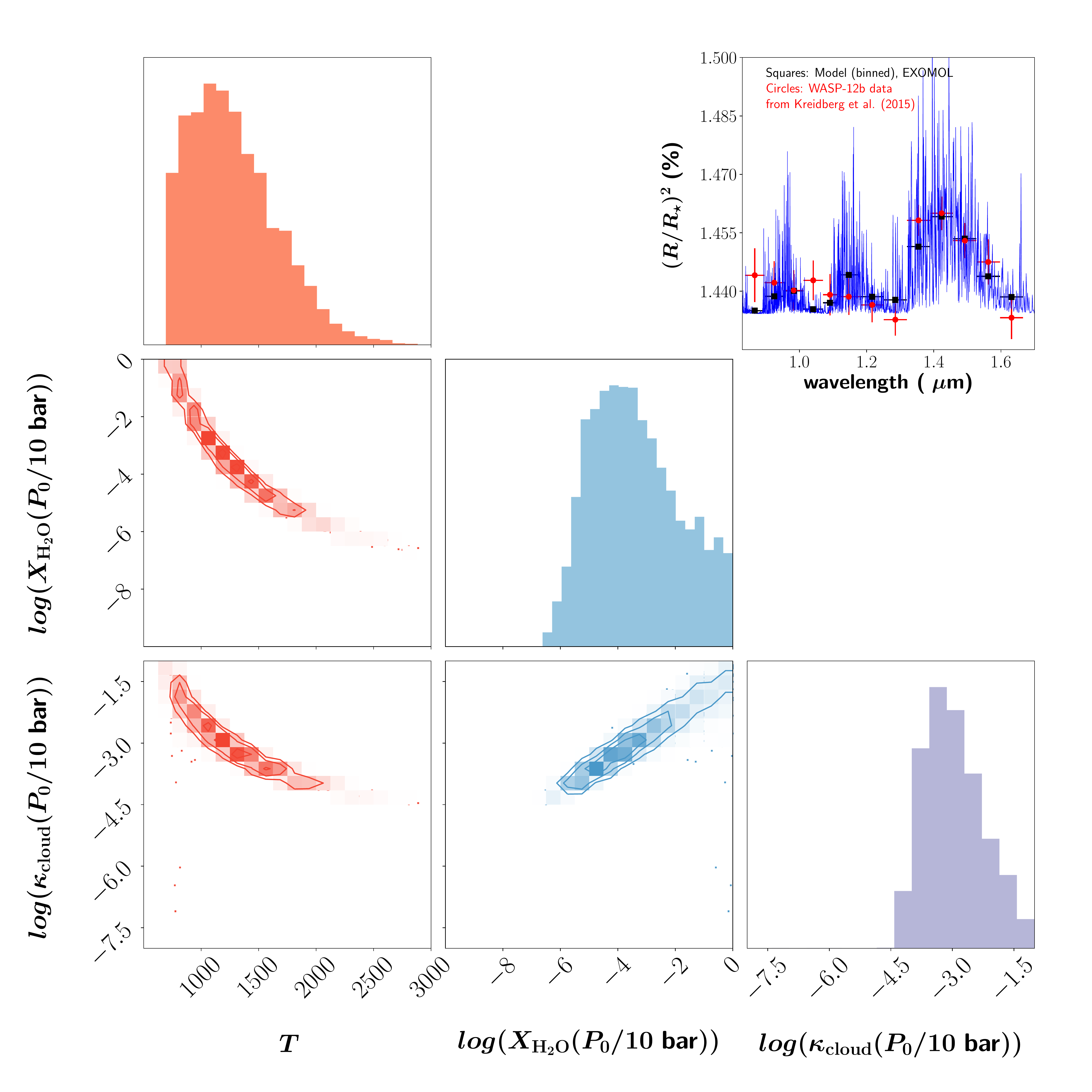}
\includegraphics[width=\columnwidth]{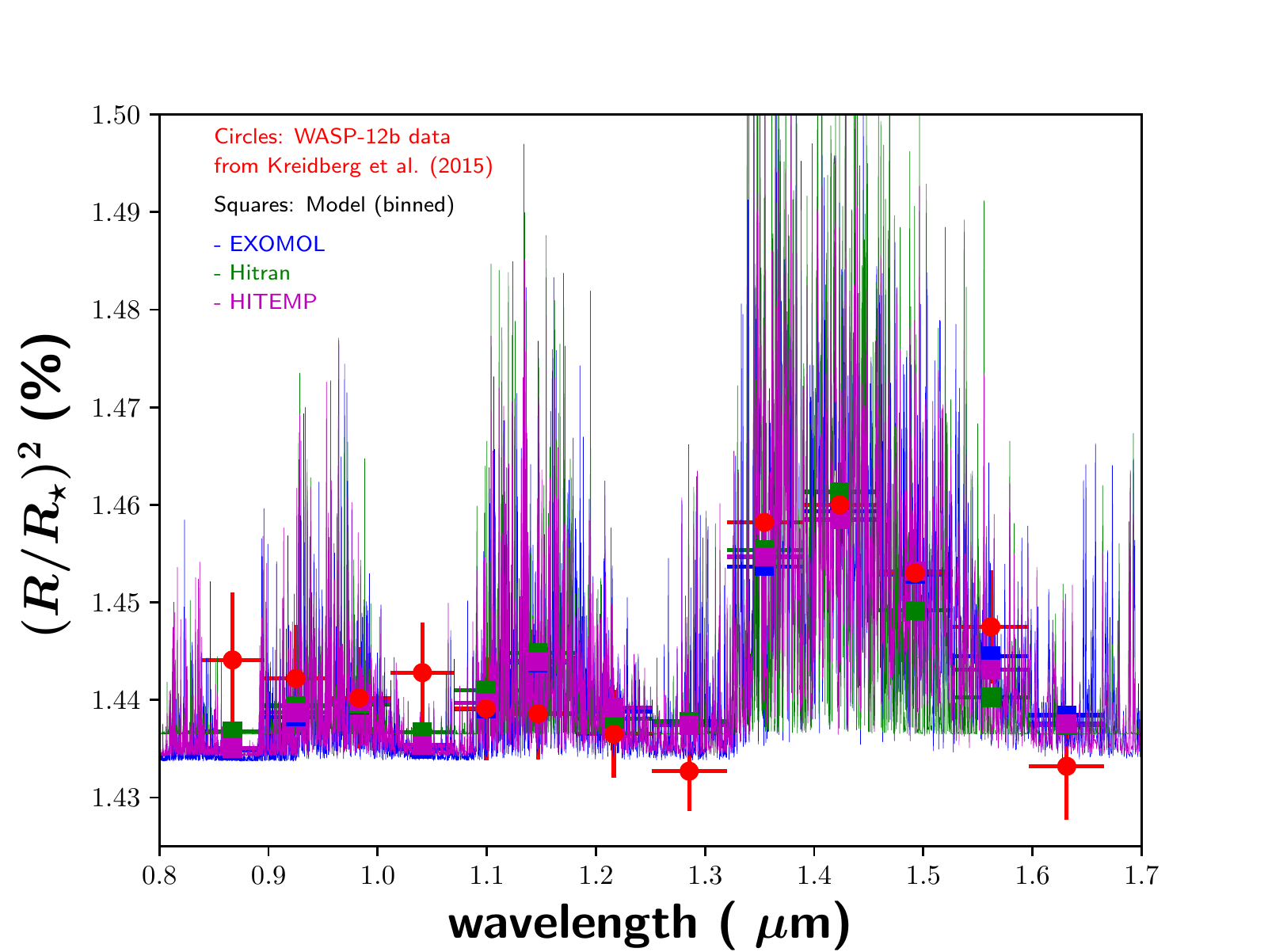}
\end{center}
%\vspace{-0.1in}
\caption{Same as Figure \ref{fig:demo}, but comparing the use of \texttt{HITRAN} (top left panel) versus \texttt{HITEMP} (top right panel) versus \texttt{ExoMol} (bottom left panel) spectroscopic line lists for water.  Additionally, the best-fit spectra are compared in the bottom right panel.}
\label{fig:line_lists}
\end{figure*}

Perhaps the most surprising outcome of our series of WASP-12b tests is shown in Figure \ref{fig:line_lists}, where we examine the retrieval outcomes using the \texttt{HITRAN}, \texttt{HITEMP} and \texttt{ExoMol} spectroscopic databases to construct the water opacity.  The main shortcoming with \texttt{HITRAN} is that it omits the weak lines of water that contribute prominently to the spectral continuum when $T \sim 1000$ K or hotter.  \texttt{HITEMP} addresses this issue somewhat, but it is widely accepted by the exoplanet community that \texttt{ExoMol} addresses this issue most completely to date (see \citealt{ty17} for a review).  With an equilibrium temperature in excess of 2500 K, WASP-12b is an ideal target for testing if discrepancies from retrievals arise from the use of different line lists.  Yet, Figure \ref{fig:line_lists} shows us that the choice of line list for the water opacity is irrelevant at the present spectral resolution and signal-to-noise attainable of the WFC transmission spectrum of WASP-12b.  It suggests that the retrievals performed on the other WFC3 transmission spectra are robust to the choice of spectroscopic line list.  Despite this finding, we persist in using the \texttt{ExoMol} line list in order to dispel any notion that our results lack robustness.

\subsubsection{Insensitivity to pressure broadening}

\begin{figure}
%\vspace{-0.1in}
\begin{center}
\includegraphics[width=\columnwidth]{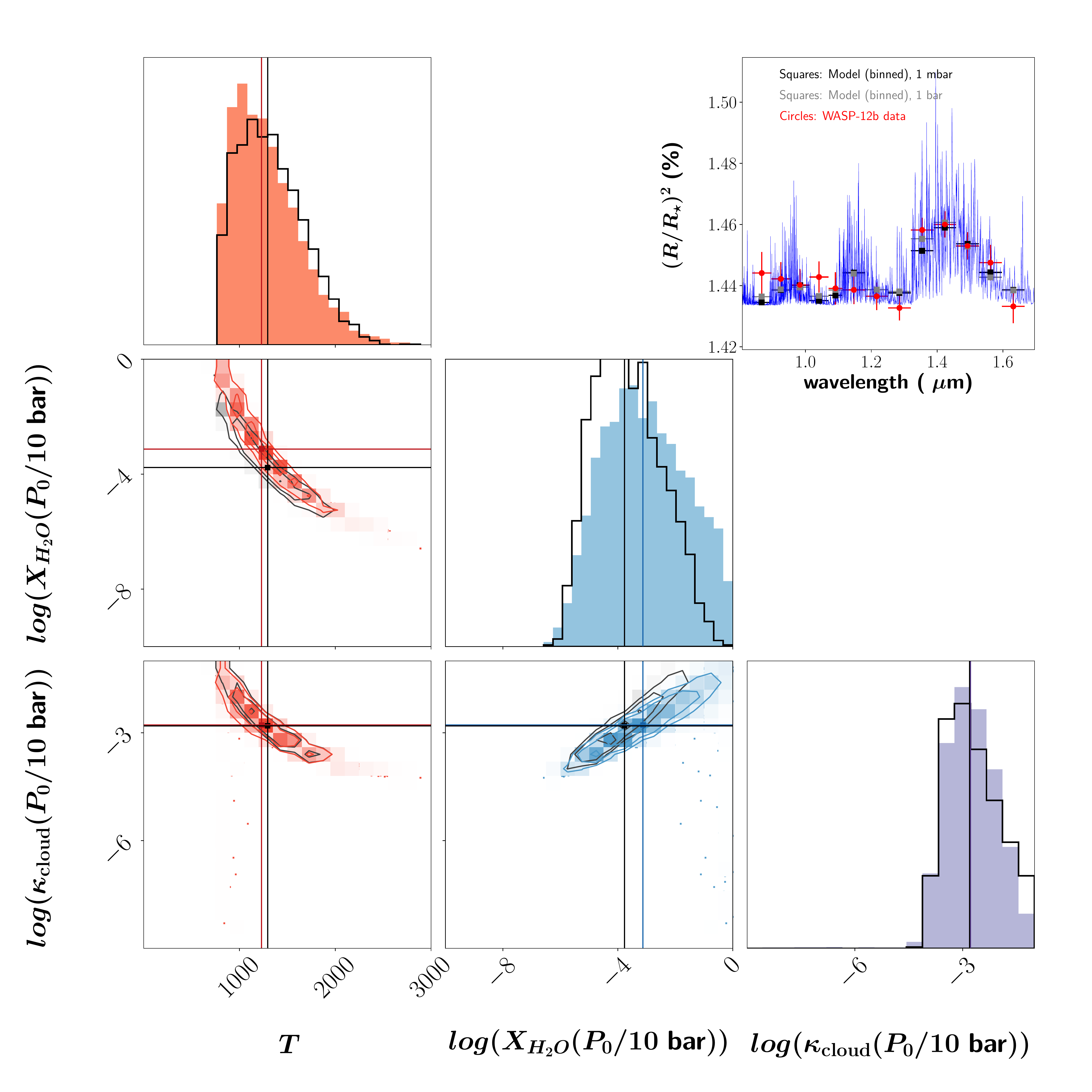}
\end{center}
%\vspace{-0.1in}
\caption{Same as Figure \ref{fig:demo}, but elucidating the effect of pressure broadening.  The posteriors are for $P=1$ mbar, while the posteriors associated with $P=1$ bar are overplotted as the solid curves. The vertical and horizontal lines represent the median values.} 
\label{fig:pressure_variations}
\end{figure}

Pressure broadening is an ill-defined source of uncertainty, because there is no first-principles theory to describe it (see discussion in \S\ref{subsect:opacities}).  Nevertheless, we quantify its effect as the final test in this WASP-12b suite.  Figure \ref{fig:pressure_variations} shows two retrievals with $P=1$ mbar versus 1 bar, which span the conceivable range of pressures probed by the WFC3 transmission spectrum.  The effects on the posterior distributions of the temperature, water mixing ratio and grey cloud opacity are minimal, even with a factor of 1000 difference in pressure between the pair of retrievals.

For the rest of the study, we will fix the pressure associated with pressure broadening at 10 mbar.  The reasoning is that departures from this value will result in minor errors to the retrieved posterior distributions, which are subsumed as errors in the grey cloud opacity.  Given that the exact functional form of pressure broadening cannot be specified from first principles, this is a reasonable approach, because it allows us to include pressure broadening in a more controlled way.

\subsubsection{Comparison of Bayesian evidence}

\begin{figure}
%\vspace{-0.1in}
\begin{center}
\includegraphics[width=\columnwidth]{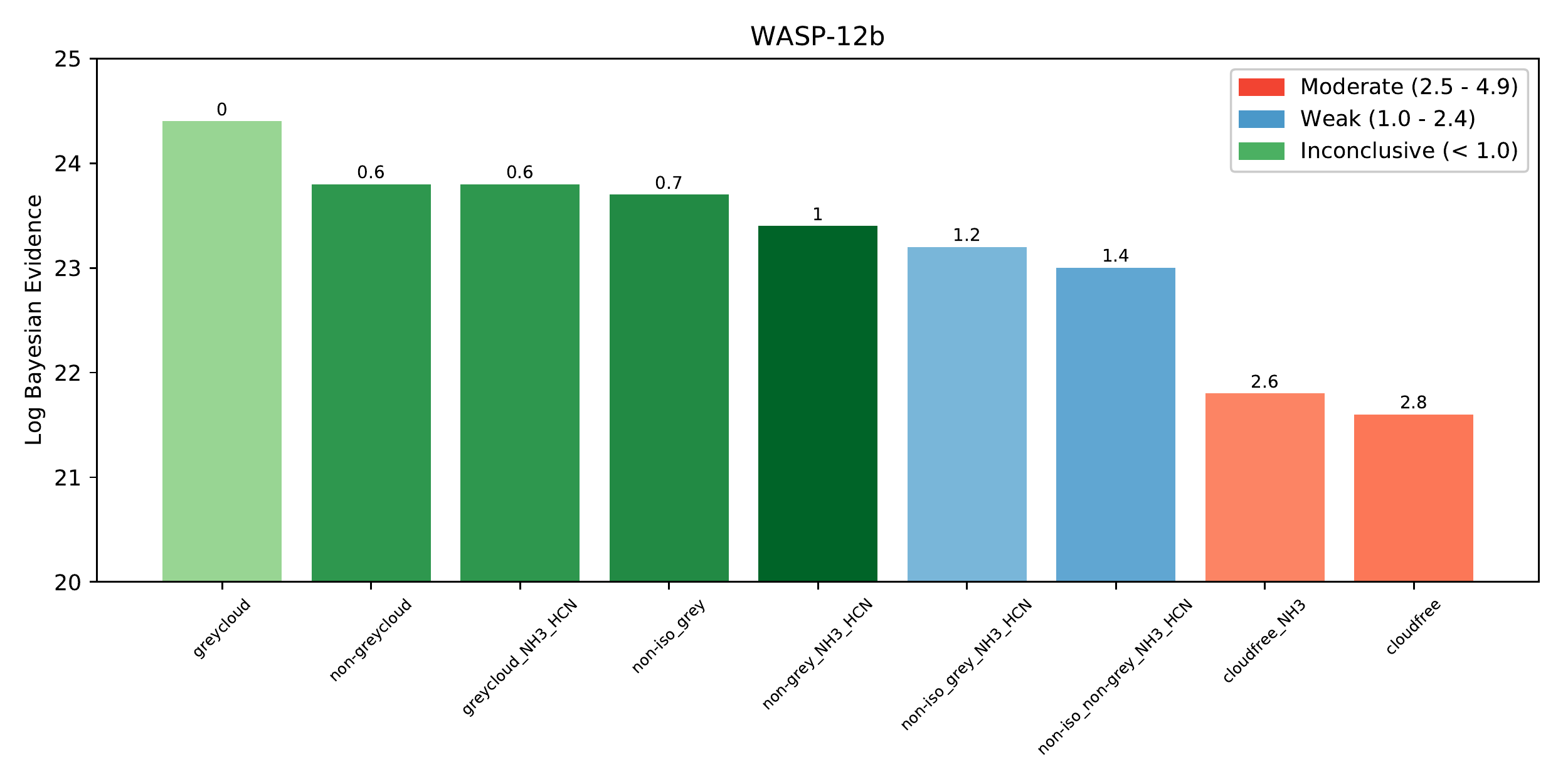}
\end{center}
%\vspace{-0.1in}
\caption{Logarithm of the Bayesian evidence and corresponding Bayes factor between each model compared to the best model (as indicated by the number on top of each bar).  The entry marked by ``0" is the best model, i.e., the model with the highest Bayesian evidence.  The legend lists the correspondence between the Bayes factor and the strength or weakness of the evidence in favour of a given model (compared to the best model).  For this set of Bayesian evidences only, we sample the opacities at 1 mbar, ignore CIA and use a fixed $m=2.4 m_{\rm H}$ (see text).}
\label{fig:wasp12b_evidence}
\end{figure}

Following these tests, we analyse the WFC3 transmission spectrum of WASP-12b using a hierarchy of models: with and without clouds, grey versus non-grey clouds, isothermal versus non-isothermal and with various permutations of the three molecules being present.  Figure \ref{fig:wasp12b_evidence} shows the Bayes factor for each model, which is the logarithm of the ratio of the Bayesian evidence of a given model compared to the best model.  The value of the Bayes factor may be interpreted as being weak, moderate or strong evidence for the best model in favour of a given model \citep{trotta08}.  It may also be used to infer that the comparison is inconclusive, i.e., there is no evidence to favour one model over the other, if the Bayes factor between them is less than unity.

A few conclusions may be drawn from Figure \ref{fig:wasp12b_evidence}.  First, cloud-free models are disfavoured.  Second, there is weak evidence for non-isothermal behaviour, non-grey clouds and the presence of HCN and/or NH$_3$, but overall the isothermal model with only water present and grey clouds is sufficient to fit the WFC3 transmission spectrum.  In other words, there is no evidence for more complicated models to be favoured.

Again, note that the molecular opacities are sampled at 1 mbar, CIA is not included, $m$ is fixed at $2.4 m_{\rm H}$ and we have fixed $R_0 = 1.79 ~R_{\rm J}$ and $P_0=10$ bar.  These assumptions will be lifted for WASP-12b in Figure \ref{fig:veryhot}.

\subsection{Breaking the normalisation degeneracy for cloud-free objects}
\label{subsect:break}

\subsubsection{Deriving $R_0(P_0)$: case study of WASP-17b}

\cite{heng16} previously concluded that the atmospheres of WASP-17b and WASP-31b are cloud-free based on optical transmission spectra recorded by STIS \citep{sing16}.  This conclusion was based on the reasoning that the sodium and potassium lines may serve as diagnostics for cloudiness.  The peaks of these resonant lines are hardly affected by clouds, but the line wings are, which makes the distance between the line peak and wing highly sensitive to the degree of cloudiness.  If the optical transit chord is cloud-free, then we may associate the measured optical spectral slope with Rayleigh scattering by hydrogen molecules (H$_2$), which yields a direct measurement of the pressure scale height \citep{lec08,heng16},
\begin{equation}
H = - \frac{1}{4} \frac{\partial R}{\partial \left( \ln{\lambda} \right)},
\end{equation}
where $\lambda$ is the wavelength.  Such an approach is possible only because we have $\kappa = X_{\rm H_2} m_{\rm H_2} \kappa_{\rm H_2}/m$, $X_{\rm H_2} \approx 1$ and $\kappa_{\rm H_2}$ is known from first principles.  If the optical transit chord is cloudy, then $\kappa = X_{\rm cloud} m_{\rm cloud} \kappa_{\rm cloud}/m$.  The cloud volume mixing ratio ($X_{\rm cloud}$), composition of the cloud particles (and hence their mass, $m_{\rm cloud}$) and opacity ($\kappa_{\rm cloud}$) are now generally unknown and cannot be uniquely retrieved from either the optical or WFC3 transmission spectra. 

%\cite{heng16} was unable to discern if WASP-19b is cloudy or if sodium and potassium are absent in its atmosphere, due to the non-detection of these lines.  Later, we will show that WASP-17b, WASP-19b, WASP-31b and XO-1b are consistent with being cloud-free, i.e., the WFC3 transmission spectra may be explained by a family of models (all with Bayes factors that are less than unity compared to the best model), which include the isothermal, cloud-free model with water only.  A literature search reveals that XO-1b does not appear to have published optical transmission spectra at the time of writing.

%HAT-P-26b has an optical transmission spectrum that is either associated with clouds or at least does not robustly indicate the presence of Rayleigh scattering by hydrogen molecules \citep{wakeford17}.  WASP-52b appears to have a somewhat flat optical transmission spectrum \citep{louden17}.

We use WASP-17b as a working example, for which \cite{heng16} previously estimated $H = 1896$ km using two data points from \cite{sing16} and $R_\star = 1.583 ~R_\odot$ \citep{southworth12}. In the current study, we fit a line to the optical spectral slope (comprising 15 data points) and derive $H=1950$ km (not shown).

In a hydrogen-dominated atmosphere, the opacity associated with Rayleigh scattering alone is $\kappa = \sigma_{\rm H_2}/m$.  The cross section for Rayleigh scattering by hydrogen molecules is \citep{su05},
\begin{equation}
\sigma_{\rm H_2} = \frac{24 \pi^3}{n_{\rm ref}^2 \lambda^4} \left( \frac{n_r^2 - 1}{n_r^2+2} \right)^2,
\end{equation}
where $n_{\rm ref} = 2.68678 \times 10^{19}$ cm$^{-3}$ and the real part of the index of refraction is \citep{cox}
\begin{equation}
n_r = 1.358 \times 10^{-4} \left[ 1 + 7.52 \times 10^{-3} ~\left(\frac{\lambda}{1 ~\mu\mbox{m}} \right)^{-2} \right] + 1.
\end{equation}
If the optical spectral slope is associated with H$_2$ Rayleigh scattering alone, then hydrostatic equilibrium allows us to derive a unique solution for $P_0$,
\begin{equation}
P_0 =  \frac{0.56 m g}{\sigma_{\rm H_2}} \sqrt{\frac{H}{2 \pi R_0}} ~\exp{\left( \frac{R-R_0}{H}\right)},
\label{eq:P0}
\end{equation}
based on equation (\ref{eq:radius}) and assuming that $R_0$ is associated with the part of the atmosphere that is opaque to both optical and infrared radiation.

For WASP-17b, we take $R=1.890 ~R_{\rm J}$ at $\lambda = 0.405$ $\mu$m from the measurements of \cite{sing16}.  We then select a reference radius that is three orders of magnitude in pressure greater than that probed by WFC3,
\begin{equation}
R_0 = \bar{R}_{\rm WFC3} - 6.908H,
\label{eq:R0}
\end{equation}
where $\bar{R}_{\rm WFC3}$ is the average value of the transit radius in the measured WFC3 bandpass.  The preceding expression assumes hydrostatic equilibrium.  For WASP-17b, we have $\bar{R}_{\rm WFC3} = 1.897 ~R_{\rm J}$ and $R_0 = 1.709 ~R_{\rm J}$.  Using the measured value of $R$ and equation (\ref{eq:P0}), we estimate that $P_0 = 8$ bar.  This means that the pressure probed in the WFC3 bandpass is, on average, about 8 mbar.  We note that the pressure probed at $\lambda = 0.405$ $\mu$m is about 10 mbar.

We do the same analysis for WASP-31b.  We use $R_\star = 1.252 ~R_\odot$ \citep{anderson11} and derived $H = 1619$ km.  \cite{heng16} previously derived $H=1390$ km based on using $R_\star = 1.12 ~R_\odot$ \citep{anderson11}.  We estimate $R_0 = 1.379 ~R_{\rm J}$ and $P_0=26$ bar, based on $R= 1.547 ~R_{\rm J}$ at $\lambda=0.4032$ $\mu$m.  This means that the WFC3 bandpass and the optical data point correspond to about 26 mbar and 15 mbar, respectively.  

These estimates are broadly consistent with our approach of assuming 10 mbar for the molecular opacities.

\subsubsection{Mock retrievals of WASP-17b: breaking the normalisation degeneracy}
\label{subsect:wasp17_mocks}

Using the derived $R_0=1.709 ~R_{\rm J}$ and $P_0=8$ bar, we perform suites of mock retrievals to study if the normalisation degeneracy may be broken.  A uniform prior distribution of 1.619 to $1.799~R_{\rm J}$ is set for $R_0$, while a log-uniform prior distribution of 0.1 to 1000 bar is set for $P_0$.

First, we create high-resolution mock transmission spectra with 100 data points that are representative of what will be possible with JWST.  The uncertainty on each data point is assumed to be 10 parts per million (ppm).  We explore pairs of retrievals in which $R_0$ is held fixed and $P_0$ is a fitting parameter, and vice versa.  Second, we create a hierarchy of mock spectra to gain understanding into the retrieval outcomes: three molecules with grey clouds, water only with grey clouds and water only (cloud-free).  All volume mixing ratios are set to $10^{-3}$ for illustration, with a grey-cloud opacity of $10^{-2}$cm$^2$g$^{-1}$. 

\begin{figure*}
\vspace{-0.1in}
\begin{center}
\includegraphics[width=0.9\columnwidth]{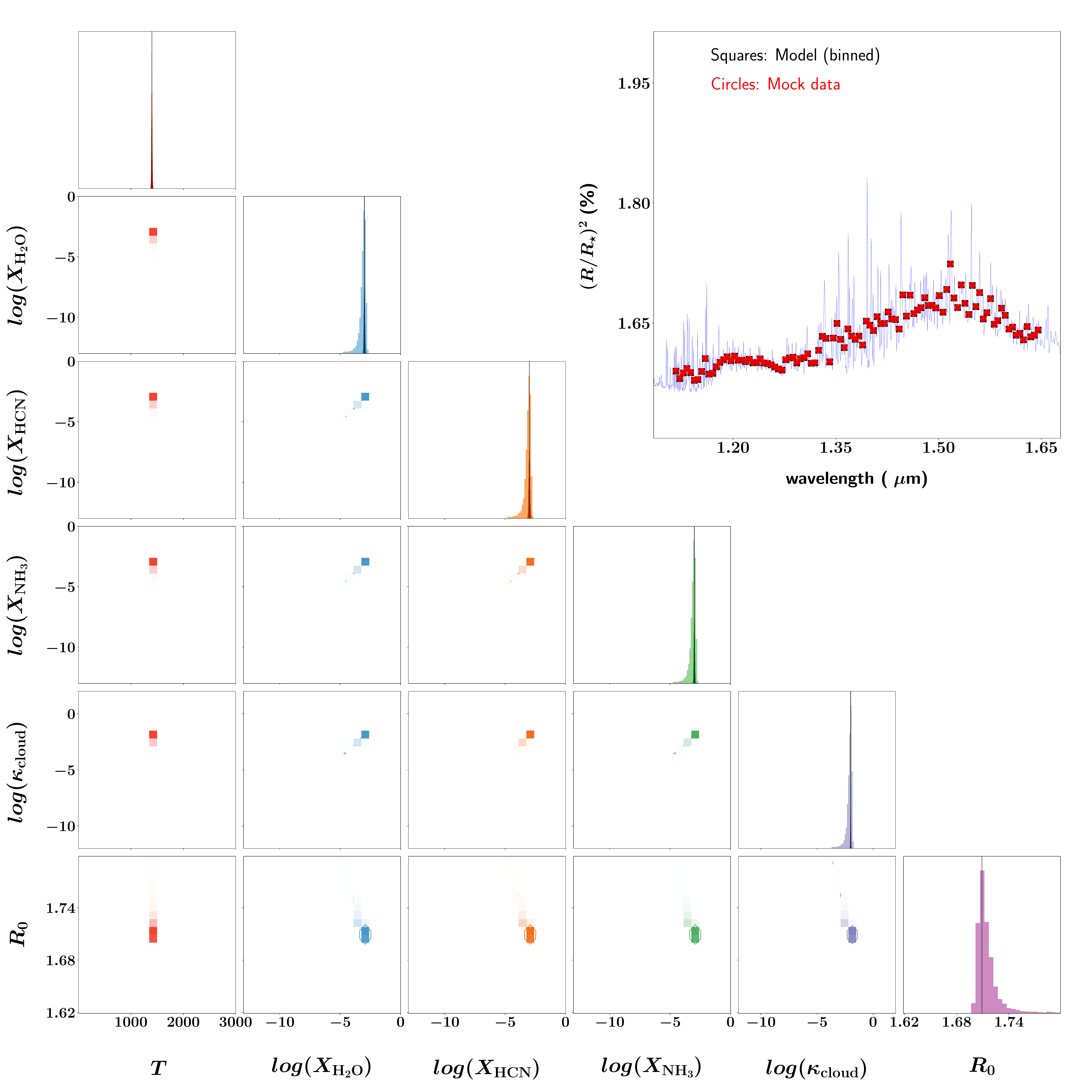}
\includegraphics[width=0.9\columnwidth]{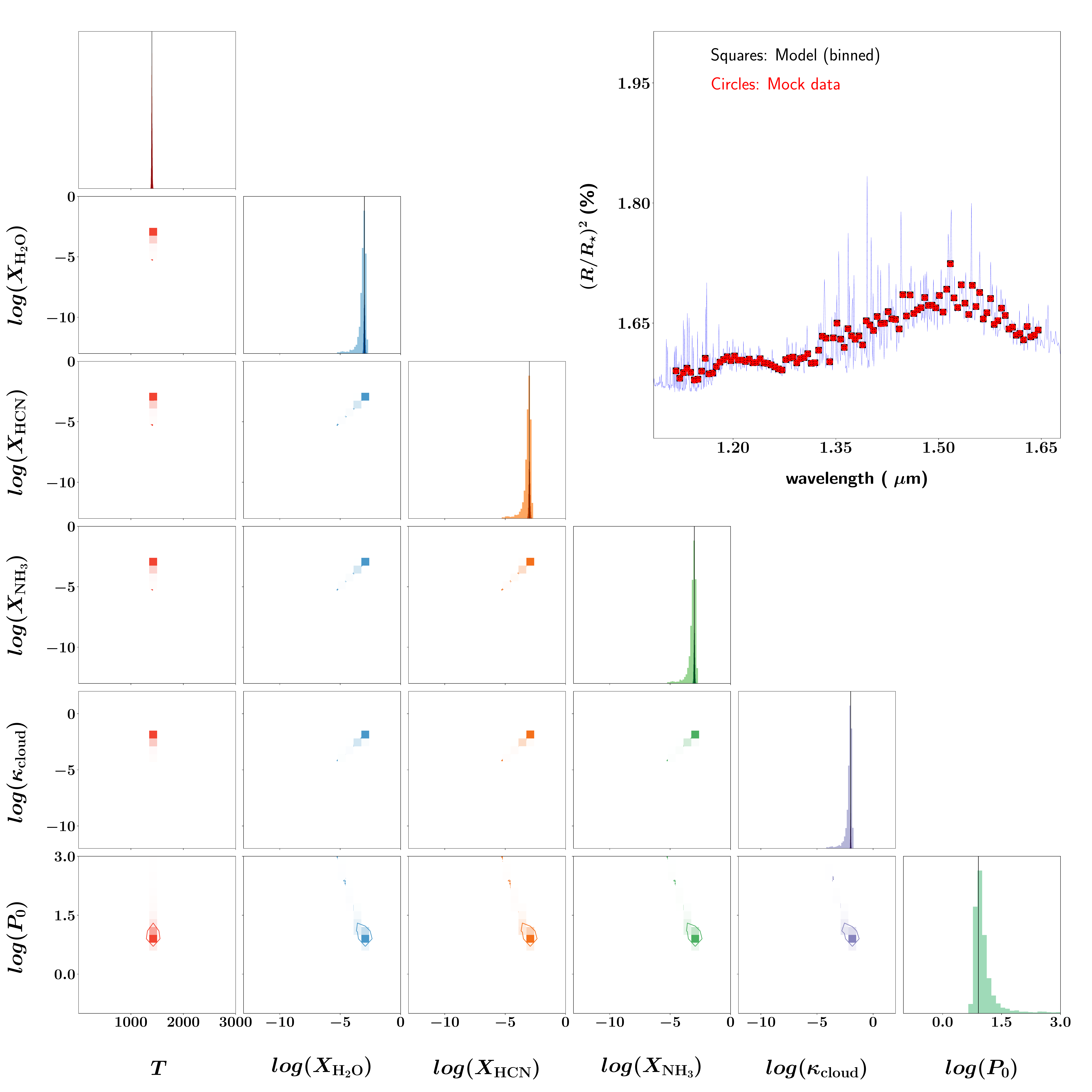}
\includegraphics[width=0.9\columnwidth]{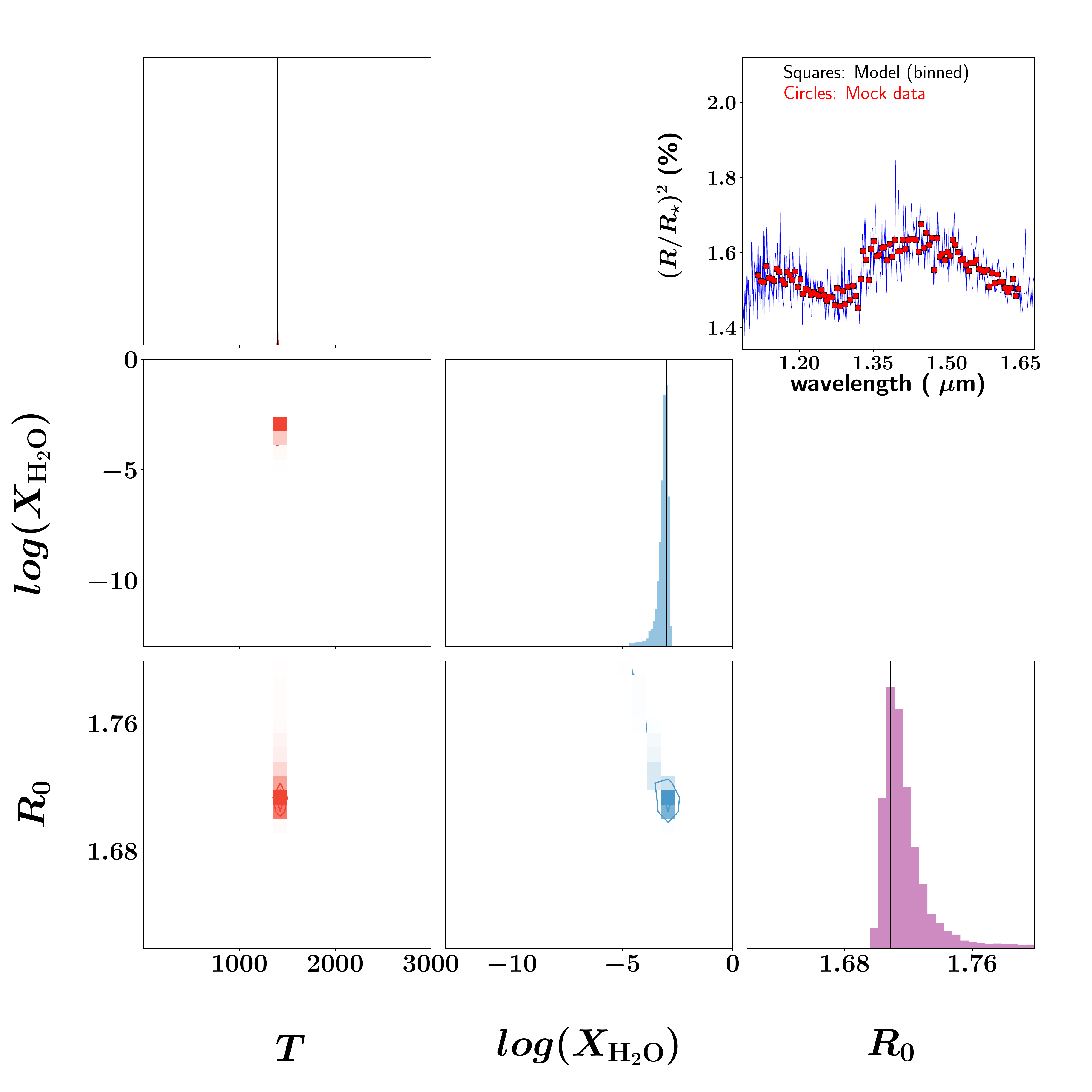}
\includegraphics[width=0.9\columnwidth]{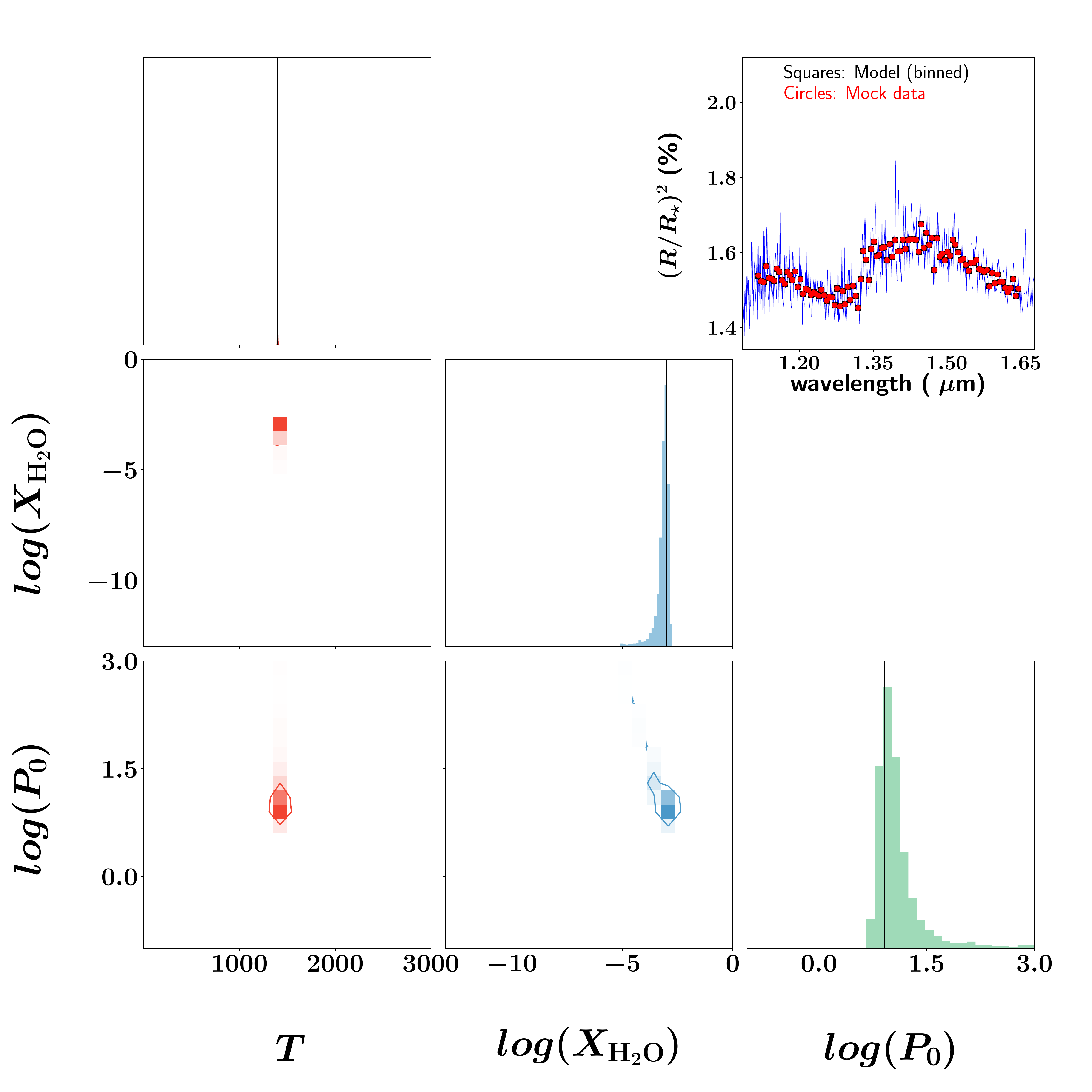}
\includegraphics[width=0.9\columnwidth]{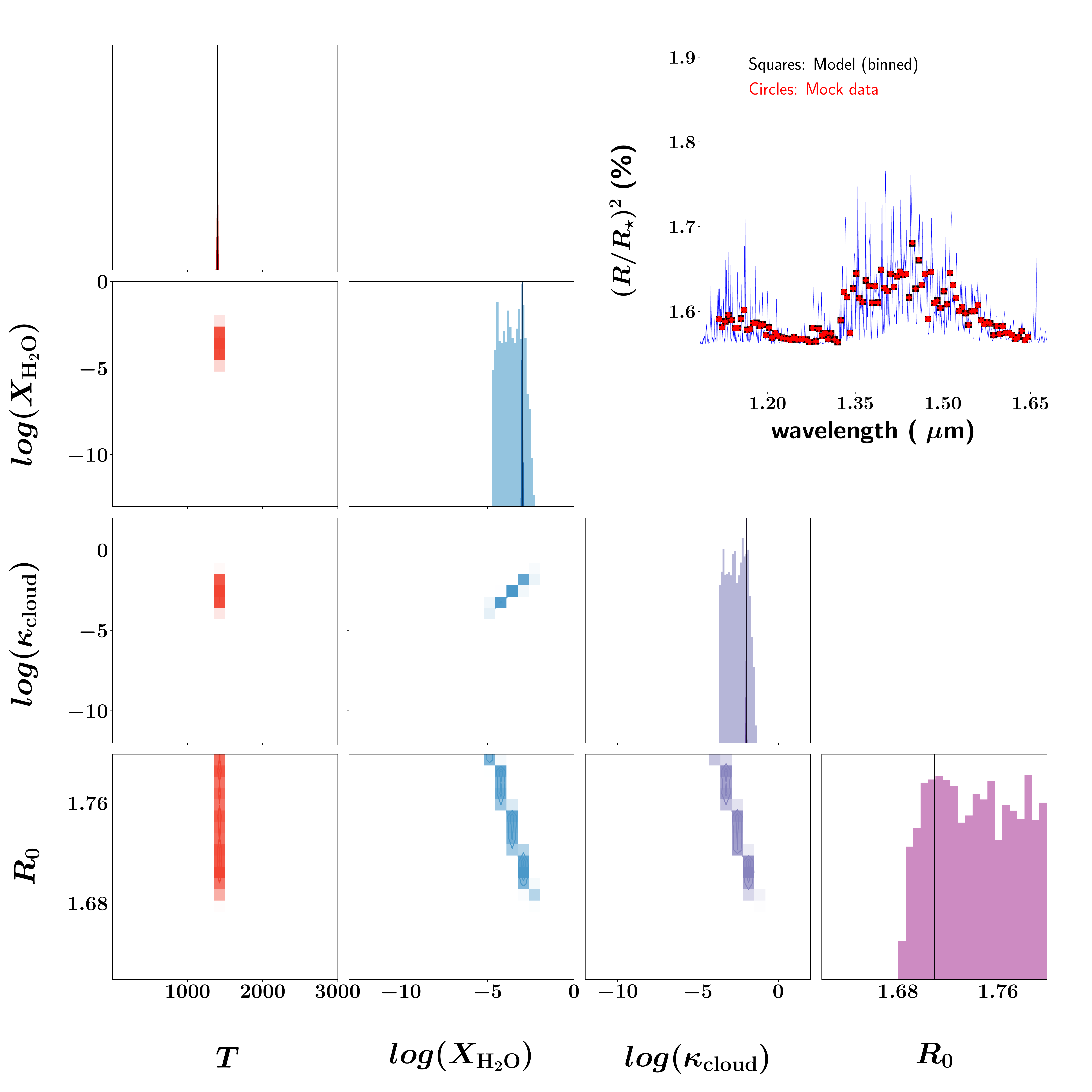}
\includegraphics[width=0.9\columnwidth]{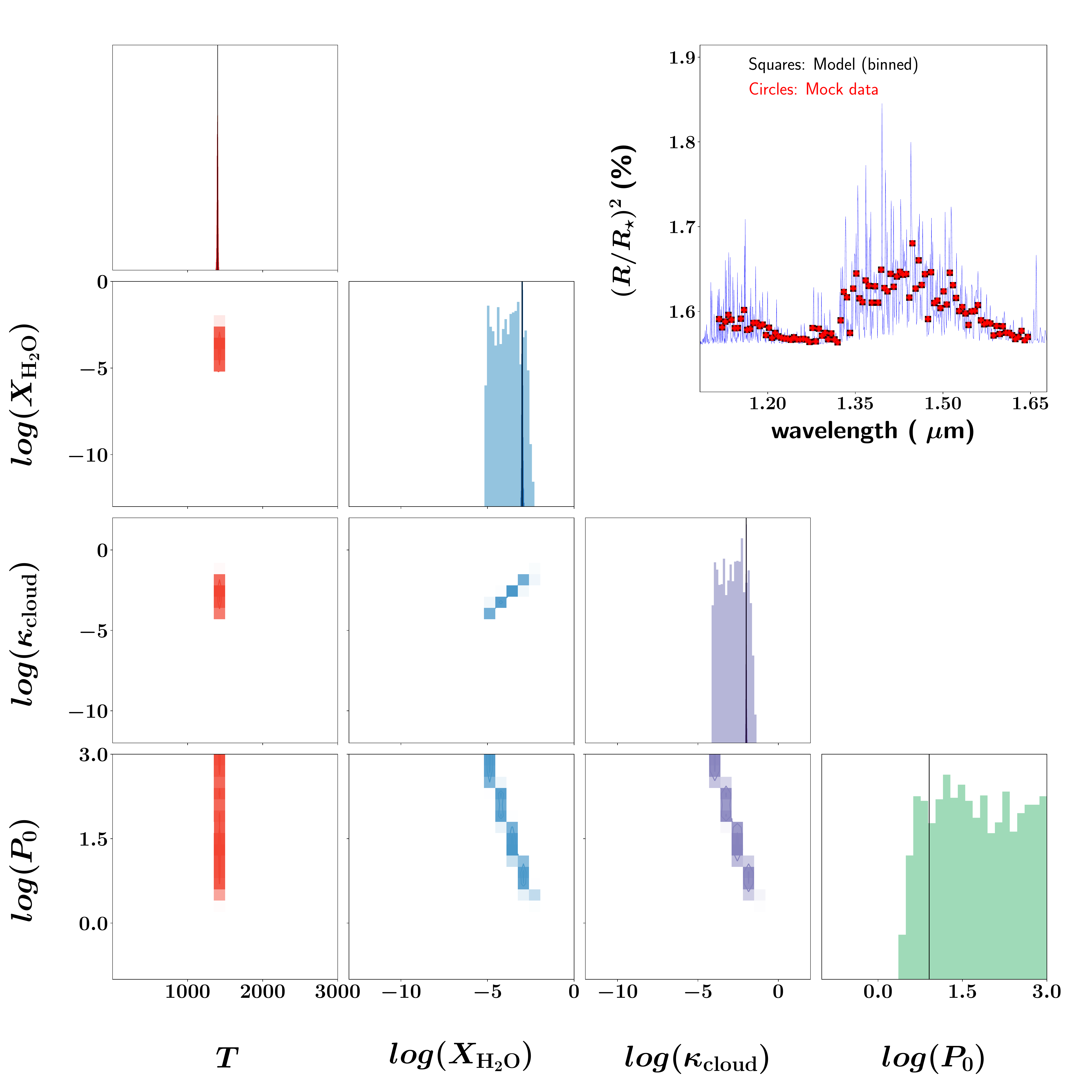}
\end{center}
\vspace{-0.1in}
\caption{High-resolution (JWST-like) mock retrievals for WASP-17b using $R_0=1.709 ~R_{\rm J}$ and $P_0=8$ bar.  The left column of retrievals hold $P_0$ fixed at 8 bar and fit for $R_0$, while the right column holds $R_0$ fixed at $1.709 ~R_{\rm J}$ and fit for $P_0$.  The top, middle and bottom rows are for three molecules with grey clouds, water only (cloud-free) and water only with grey clouds, respectively.  All mock retrievals assume isothermal atmospheres and uncertainties of 10 ppm.  Vertical lines indicate the true (input) values of the parameters.}
\label{fig:wasp17b_mock}
\end{figure*}

Figure \ref{fig:wasp17b_mock} shows the outcomes of 6 retrievals on high-resolution mock spectra.  Unexpectedly, the peaks of the narrow posterior distributions of all 6 parameters, including $R_0$ or $P_0$, land exactly on the true values.  The pair of cloud-free retrievals with water only also manages to locate the correct solution.  In fact, the posterior distribution on the temperature is essentially a narrow spike with no width.  This is straightforward to understand, because the temperature controls the ``stretch factor" in the spectrum and a unique solution is obtained by correctly fitting for the difference between the peaks and troughs of the spectrum.  By contrast, $R_0$ or $P_0$ serves as a ``translation factor", which shifts the spectrum up or down in transit radius or depth without altering its shape.  Further insight is obtained by examining a pair of retrievals with water only but with grey clouds present.  The presence of grey clouds provides an extra degree of freedom in the system in the form of a constant spectral continuum.  Grey clouds mute spectral features, which may be compensated by an increase in the volume mixing ratio of water, which is clearly seen in Figure \ref{fig:wasp17b_mock}.  Note that the normalisation degeneracy is simultaneously present, as increases in $X_{\rm H_2O}$ and $\kappa_{\rm cloud}$ are negated by decreases in $P_0$ or $R_0$.  The lower bound on the water mixing ratio in this pair of retrievals is artificial and is set by the chosen upper limit of our prior on $R_0$ or $P_0$.  This pair of cloudy retrievals with water only allows us to understand that the degeneracy may be broken, even in the presence of clouds, if multiple molecules are present to provide additional information on the shape of the spectrum.

\begin{figure*}
\vspace{-0.1in}
\begin{center}
\includegraphics[width=0.9\columnwidth]{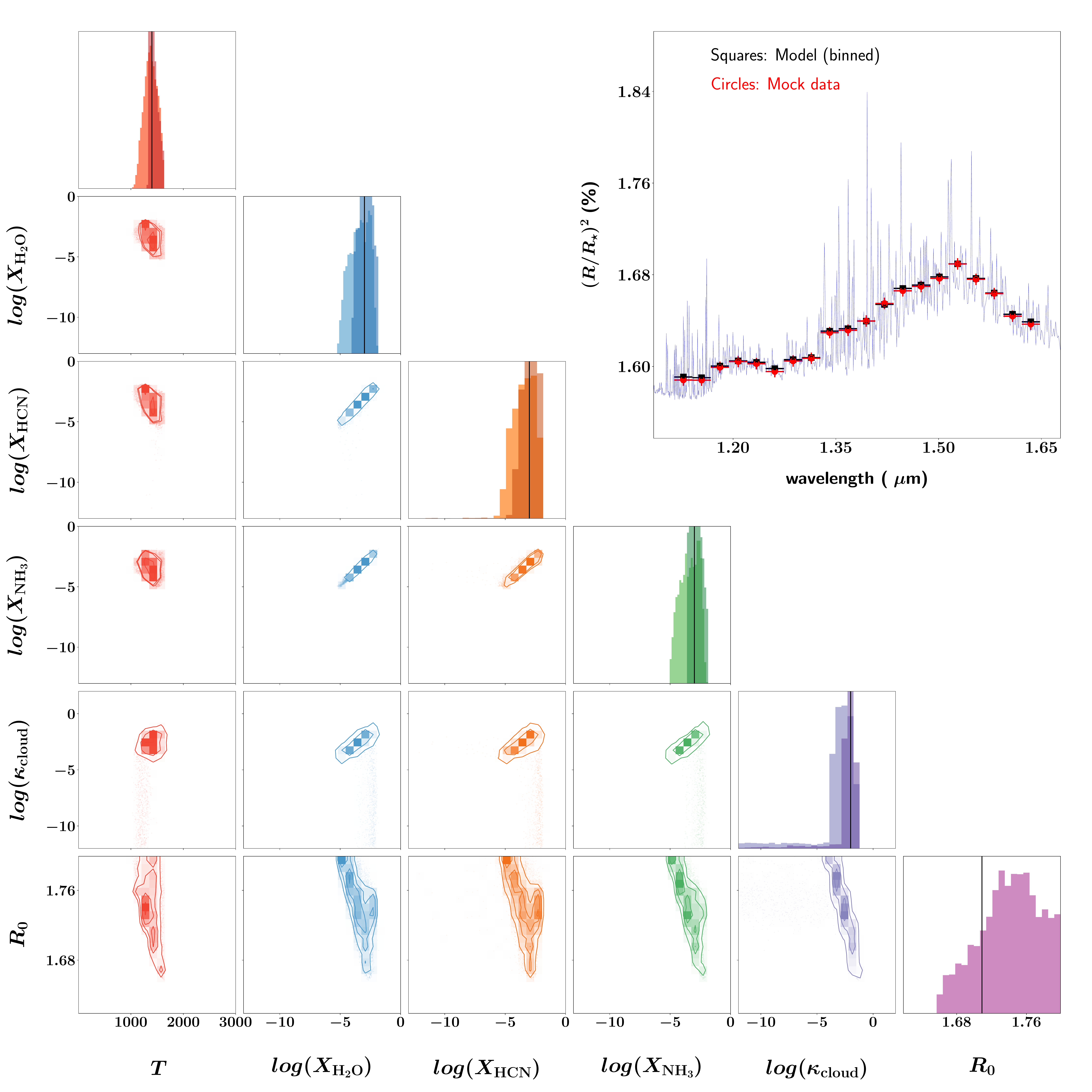}
\includegraphics[width=0.9\columnwidth]{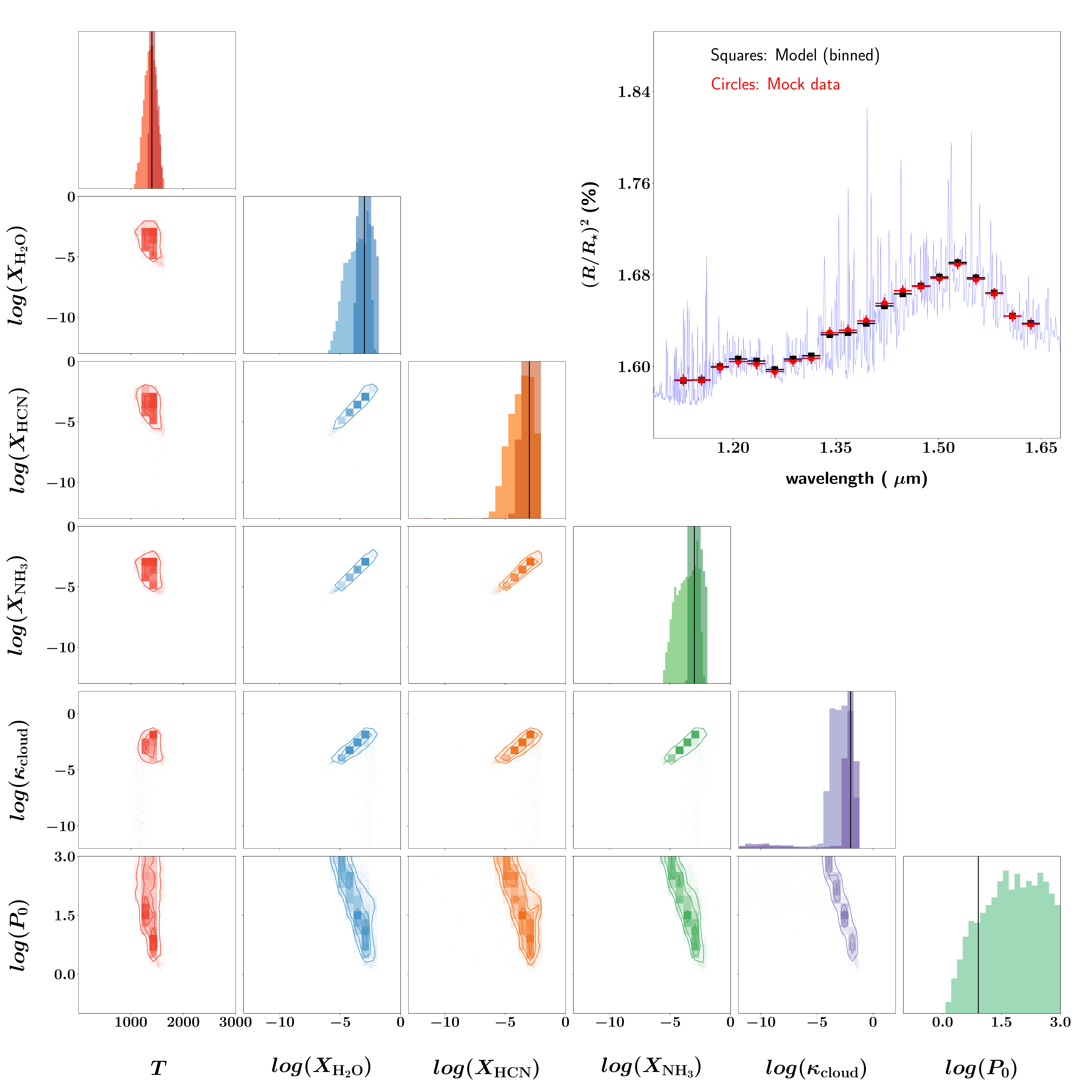}
\includegraphics[width=0.9\columnwidth]{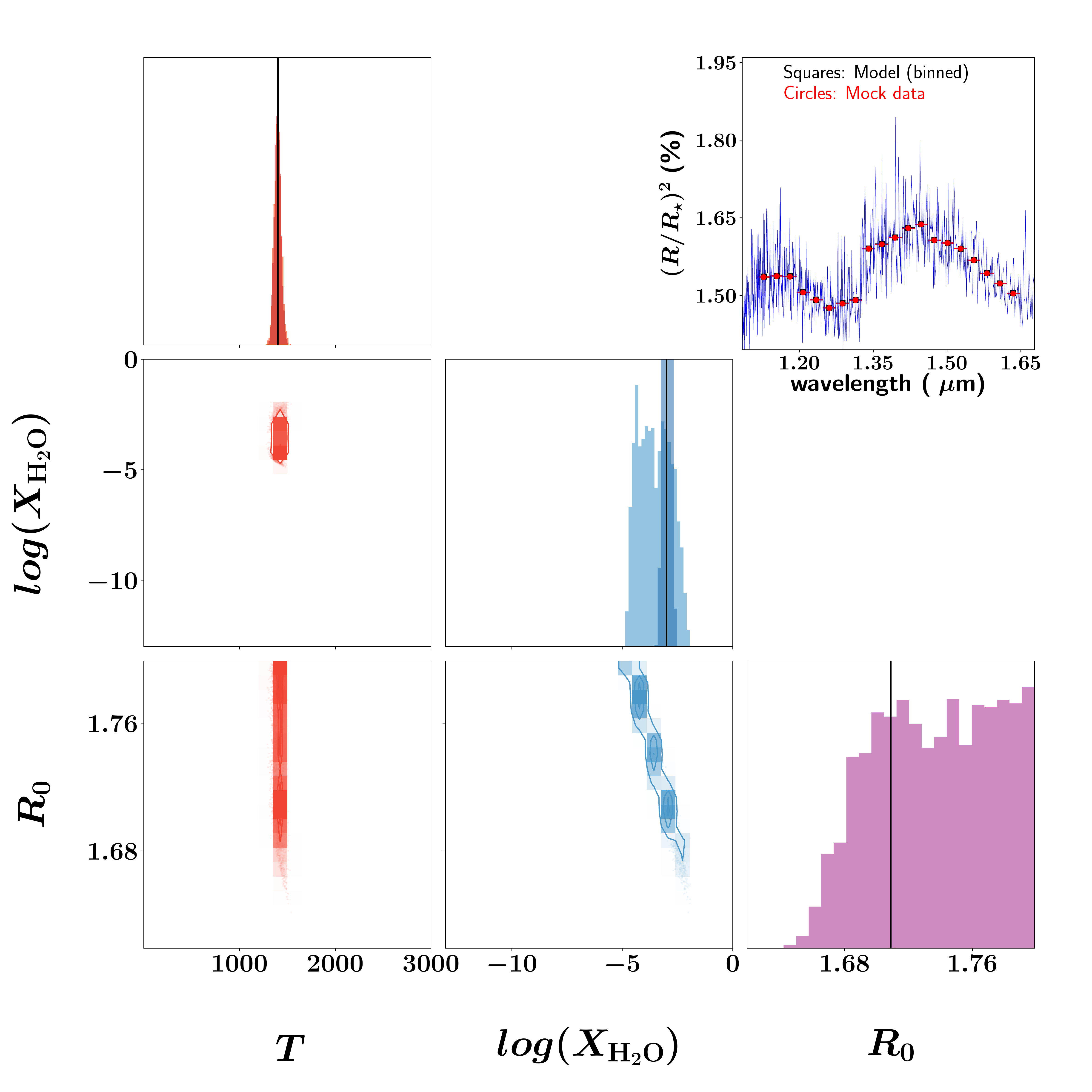}
\includegraphics[width=0.9\columnwidth]{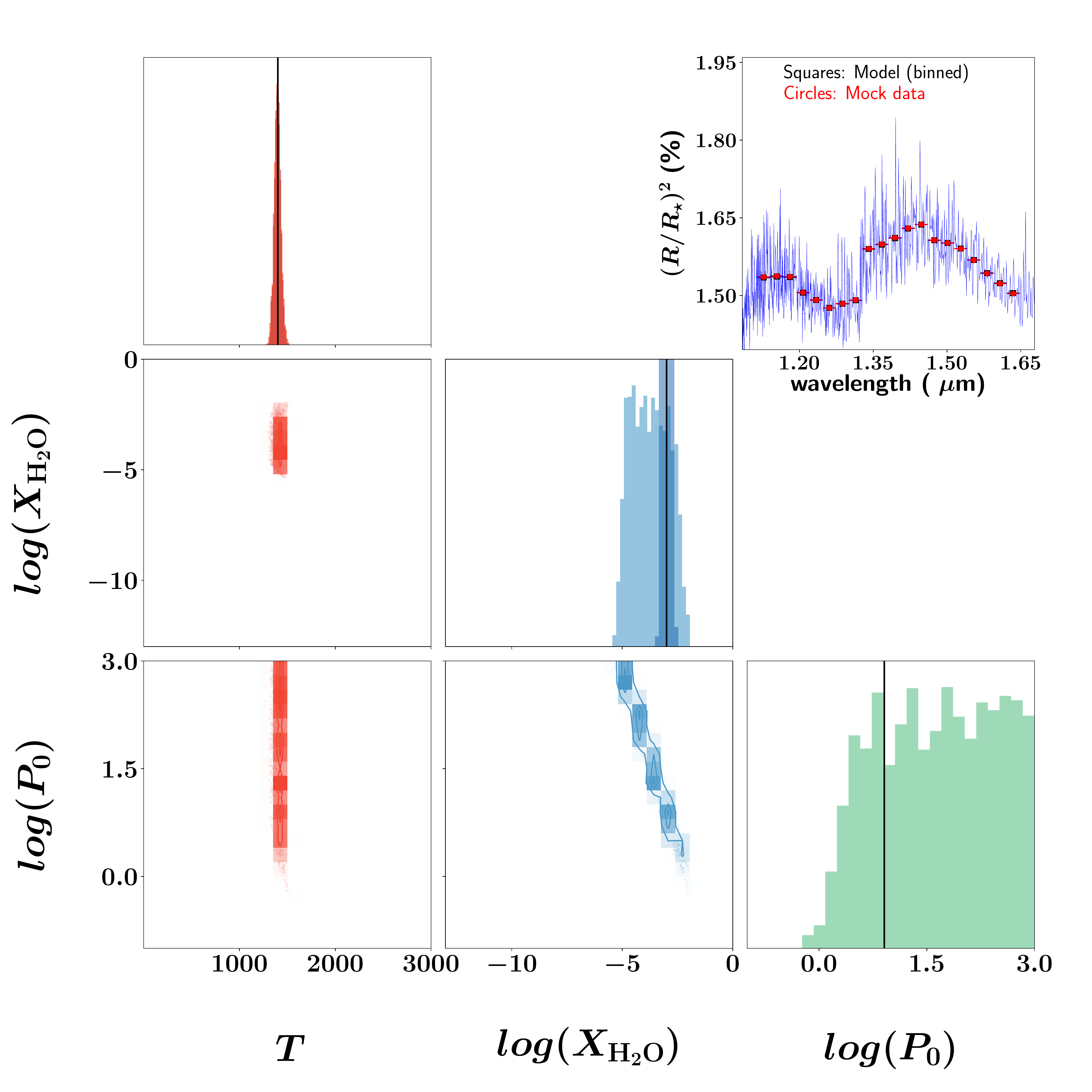}
\includegraphics[width=0.9\columnwidth]{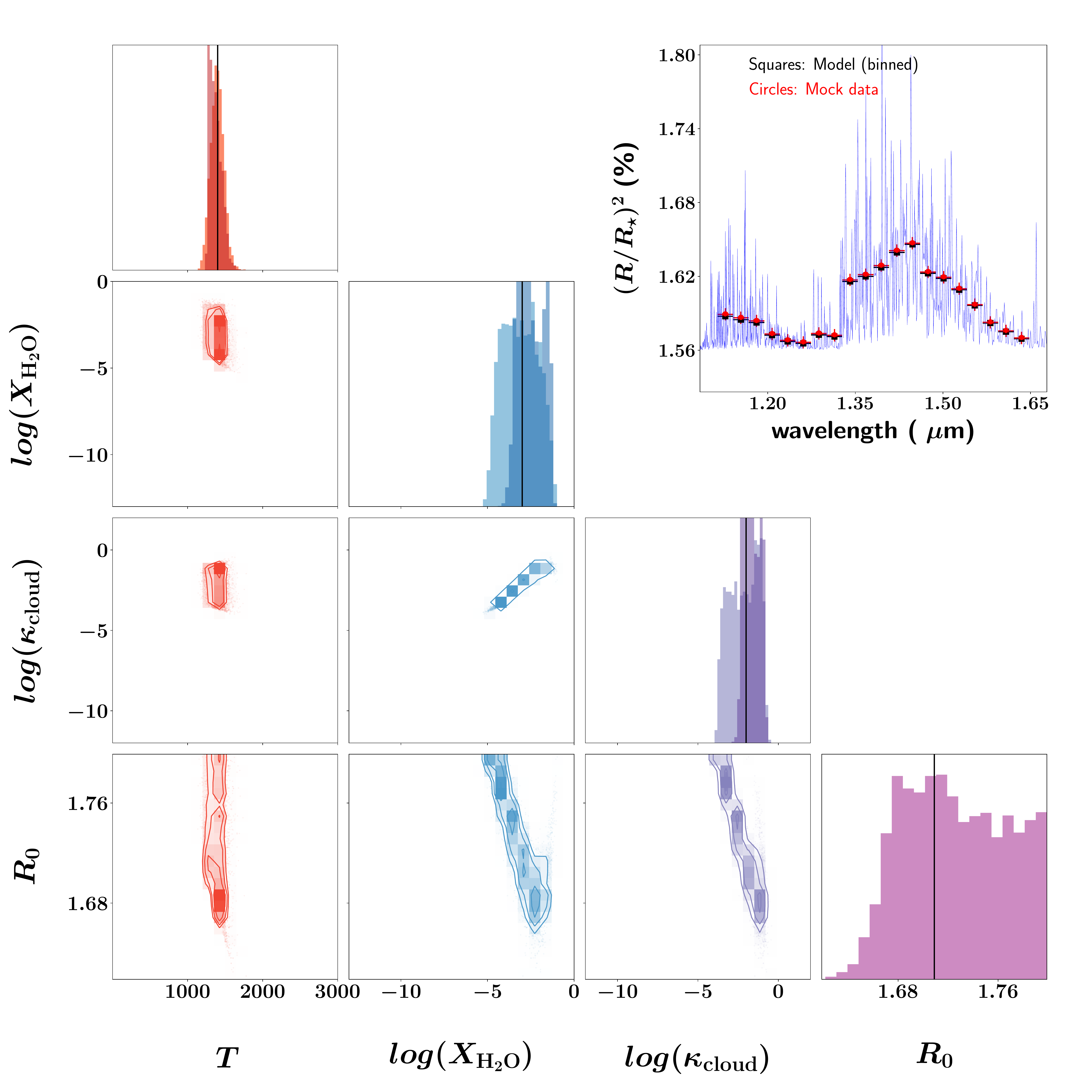}
\includegraphics[width=0.9\columnwidth]{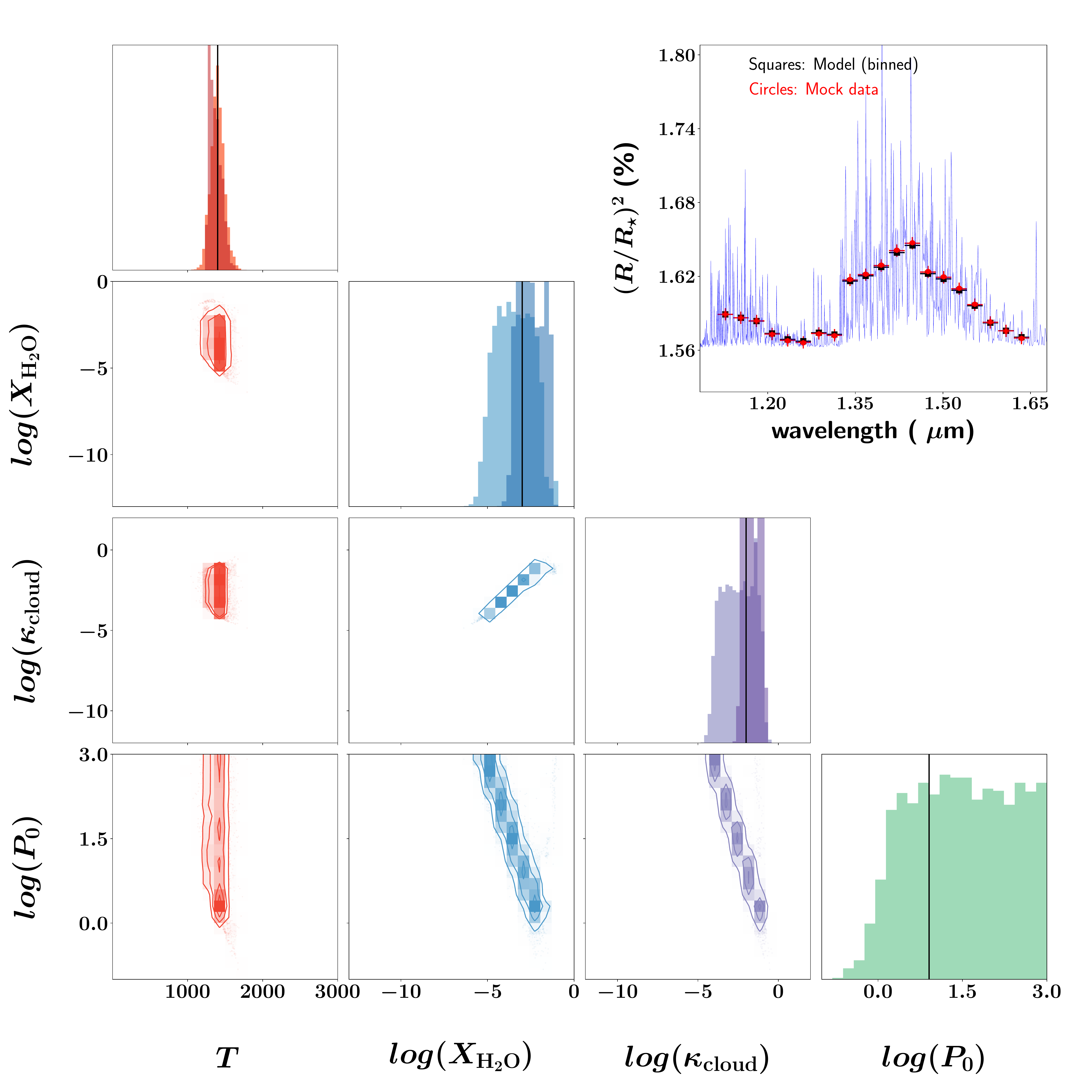}
\end{center}
\vspace{-0.1in}
\caption{Same as Figure \ref{fig:wasp17b_mock}, but for low-resolution (WFC3-like) spectra.  Additionally, the darker posterior distribution in each panel corresponds to an additional retrieval in which $R_0$ (left column) or $P_0$ (right column) is held fixed at its true value ($1.709 ~R_{\rm J}$ or 8 bar).  The uncertainties on each mock data point is assumed to be 50 ppm.}
\label{fig:wasp17b_mock2}
\end{figure*}

Figure \ref{fig:wasp17b_mock2} shows the same suite of retrievals but for a low-resolution, WFC3-like spectrum with 20 data points.  The uncertainty on each data point is assumed to be 50 ppm.  For each of the 6 retrievals, we perform an additional retrieval in which $R_0$ or $P_0$ is held fixed at its true value ($1.709 ~R_{\rm J}$ or 8 bar).  The lessons learnt and insights gleaned from the high-resolution retrievals carry over to the low-resolution ones.  Tight constraints are obtained on the temperature.  For the volume mixing ratios of the molecules, constraints are obtained at the order-of-magnitude level that encompass the true values, but it is important to note that the lower bounds are artefacts of assuming an upper limit for the prior of $R_0$ or $P_0$.  Unlike in the high-resolution regime, the low-resolution retrievals do not provide tight constraints on either $R_0$ or $P_0$.

The key lesson learnt is that, for meaningful retrieval outcomes to be obtained, we have to assume a reasonable range of prior values for $R_0$ or $P_0$.  Since we find it easier to have an intuition about $P_0$, we will set the range of 0.1--1000 bar as the prior on $P_0$.  It then becomes important to set a value of $R_0$ that corresponds to this range of $P_0$ values (see \S\ref{subsect:R0_values}).  

\begin{figure}
%\vspace{-0.1in}
\begin{center}
\includegraphics[width=\columnwidth]{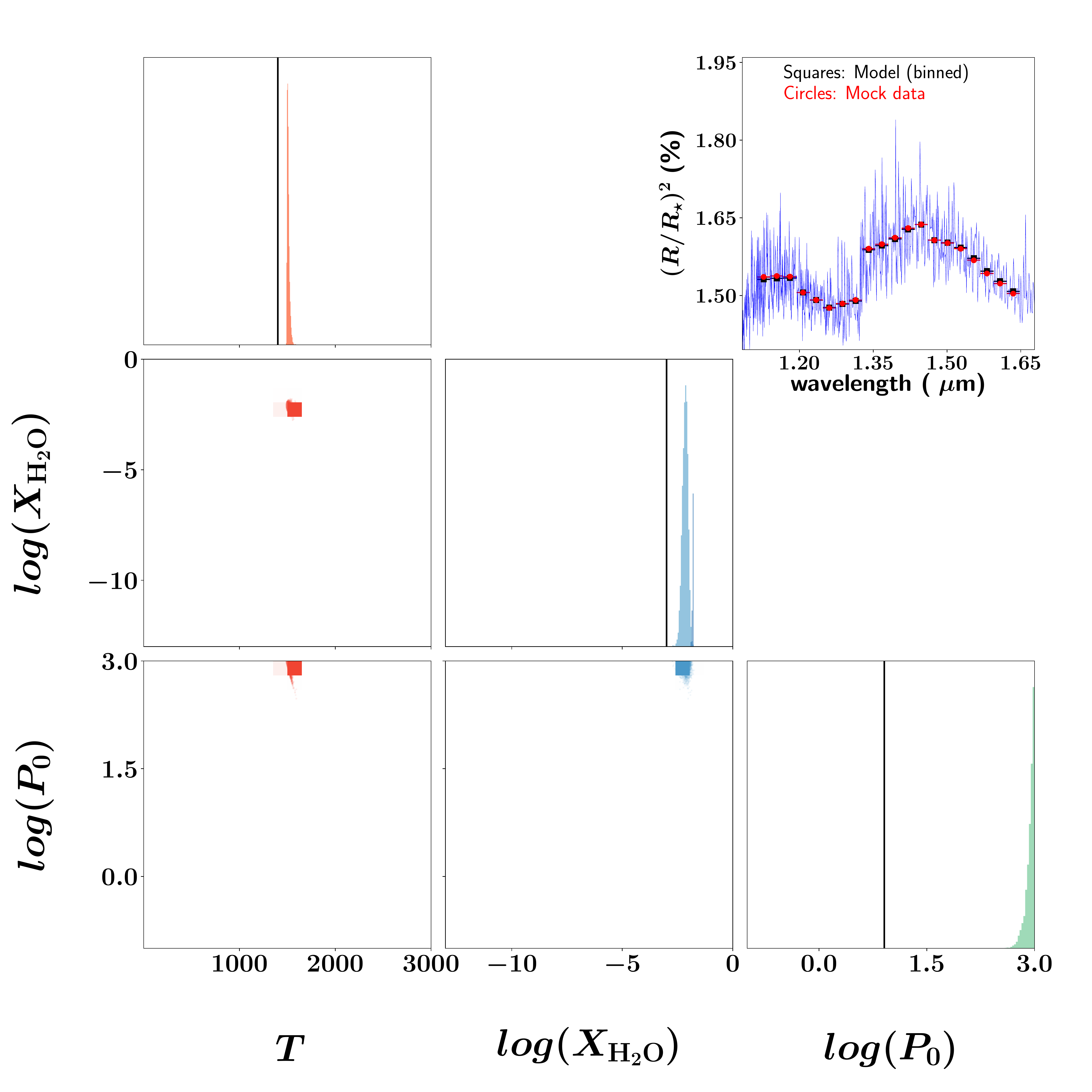}
\end{center}
%\vspace{-0.1in}
\caption{Low-resolution (WFC3-like) mock retrieval of WASP-17b for an isothermal, cloud-free atmosphere with water only.  The mock spectrum is created using $R_0=1.709 ~R_{\rm J}$ and $P_0=8$ bar.  The retrieval is performed with a fixed value of $R_0$ that is reduced by 10\% to $1.5381~R_{\rm J}$.  The corresponding value of $P_0$ now lies outside of its assumed prior range (0.1--1000 bar).  It is an illustration of how a bad assumption on $R_0$ can lead to an erroneous retrieval outcome.  We emphasise that there is no unique value of $R_0$ one can assume, but it is related to $P_0$ via hydrostatic equilibrium.  Retrievals with different $R_0$-$P_0$ pairs should yield the same outcome as long as the prior range of values of $P_0$ is set correctly.}
\label{fig:wrong}
\end{figure}

To illustrate this point, we perform an additional mock retrieval in which the value of $R_0$ is reduced and the corresponding $P_0$ value falls outside of the 0.1--1000 bar range.  Figure \ref{fig:wrong} shows that the posterior distribution for $P_0$ bumps up against the upper boundary of the prior distribution, which results in errors in the retrieved values of temperature and water mixing ratio.

\subsubsection{Catalogue of $R_0$ values for other objects}
\label{subsect:R0_values}

For the other 36 objects in our sample, we first assume the WFC3 bandpass to probe a pressure of 10 mbar.  We then use equation (\ref{eq:R0}) to estimate the value of $R_0$ that corresponds to 10 bar (Table \ref{tab:parameters}).  The pressure scale height is estimated using $H = k_{\rm B} T_{\rm eq} / m g$, where $k_{\rm B}$ is the Boltzmann constant and $T_{\rm eq}$ is the equilibrium temperature (as was done by \citealt{heng16}).  These $R_0$ values are then used as input in our retrievals.

We emphasize that while the value of $R_0$ is fixed to the tabulated value for each object, our retrievals ultimately use $P_0$ as a fitting parameter as justified by our tests in \S\ref{subsect:wasp17_mocks}.  The reason to use these values is to have $R_0$ be in approximately the range of values corresponding to 0.1--1000 bar, such that the retrieval will converge meaningfully.

%We use such a rough estimate of the $R_0(P_0)$ relationship for the other objects in our sample to prevent spurious model trends caused by significantly over- or under-estimating the value of $P_0$.  For example, if $R_0$ is fixed to some value but $P_0$ is higher than its true value, then the retrieval would compensate by producing a low temperature and/or low volume mixing ratios.  We note that the estimated value of $R_0 = 1.748 ~R_{\rm J}$ for WASP-12b is about 3.3 pressure scale heights smaller than the $R_0 = 1.79 ~R_{\rm J}$ value assumed by \cite{k15}, which implies that $R_0 = 1.79 ~R_{\rm J}$ corresponds to $\sim 1$ bar (rather than 10 bar).  This in turn implies that the volume mixing ratios reported by \cite{k15} should be reduced by about an order of magnitude.

\subsubsection{Collision-induced absorption}

\begin{figure}
%\vspace{-0.1in}
\begin{center}
\includegraphics[width=\columnwidth]{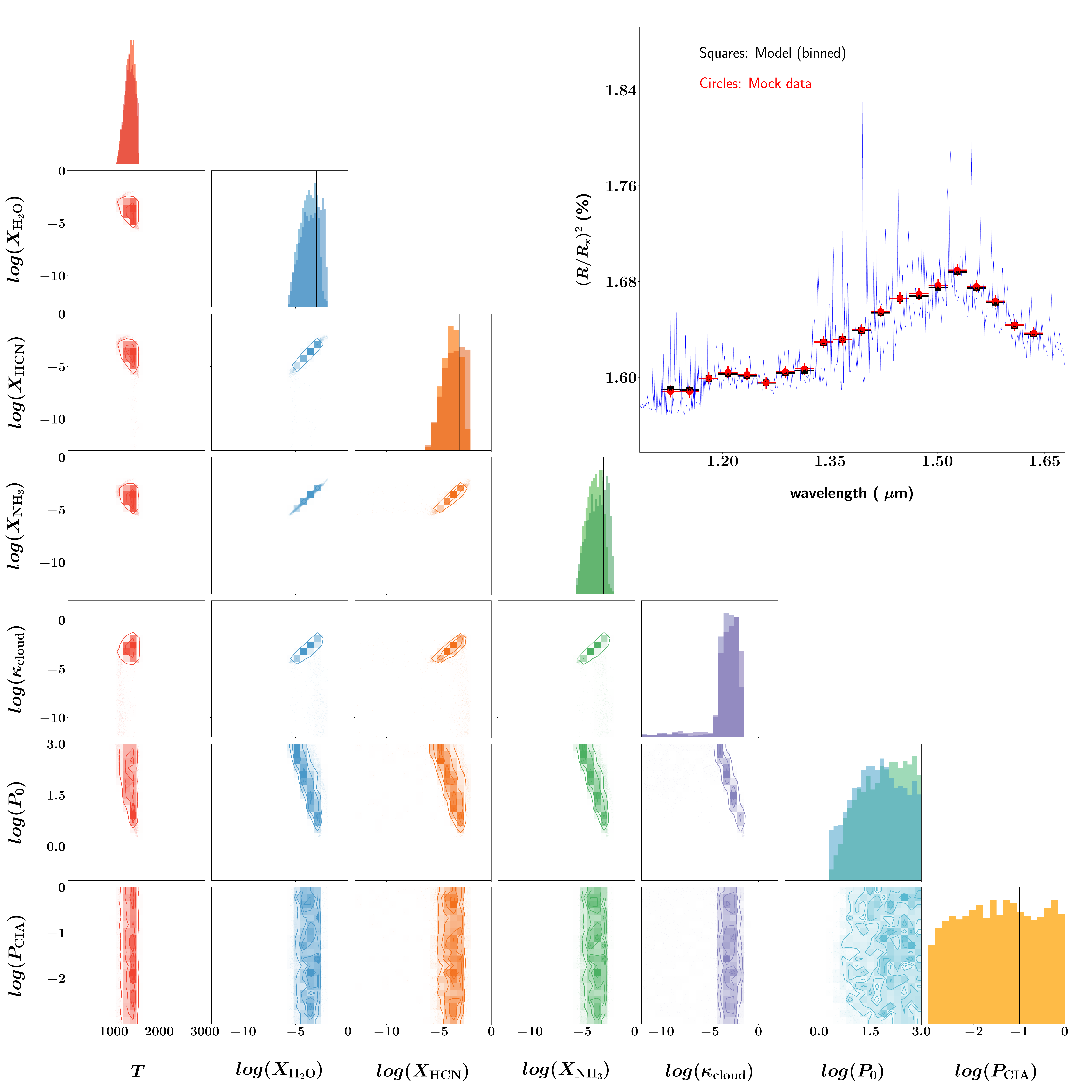}
\end{center}
%\vspace{-0.1in}
\caption{Low-resolution (WFC3-like) mock retrieval of WASP-17b for an isothermal atmosphere with grey clouds and all three molecules present.  The pressure associated with CIA is a fitting parameter of the retrieval; its true value is $P_{\rm CIA}=0.1$ bar.  The darker posterior distribution in each panel corresponds to an additional retrieval in which $P_{\rm CIA}$ is held fixed at its true value.  Vertical lines indicate the true (input) values of the parameters.}
\label{fig:cia}
\end{figure}

As a final test on mock WASP-17b spectra, we consider an isothermal model atmosphere with all three molecules present, grey clouds and CIA.  We set the pressure associated with CIA at 0.1 bar, but allow the retrieval to treat this pressure as a fitting parameter ($P_{\rm CIA}$).  Figure \ref{fig:cia} shows that the retrieval outcome is insensitive to the retrieved value of $P_{\rm CIA}$.  Similar to our treatment of pressure broadening, we set the pressure associated with CIA to be 0.1 bar for the rest of the study with the reasoning that any deviations from this value may be visualised as errors that are subsumed into the grey cloud opacity.  Figure 10 of \cite{tsi18} shows that CIA contributes a roughly flat continuum to the WFC3 spectrum.

\subsubsection{Retrieval analysis of WASP-17b WFC3 transmission spectrum}
\label{subsect:wasp17b}

\begin{figure}
%\vspace{-0.1in}
\begin{center}
\includegraphics[width=\columnwidth]{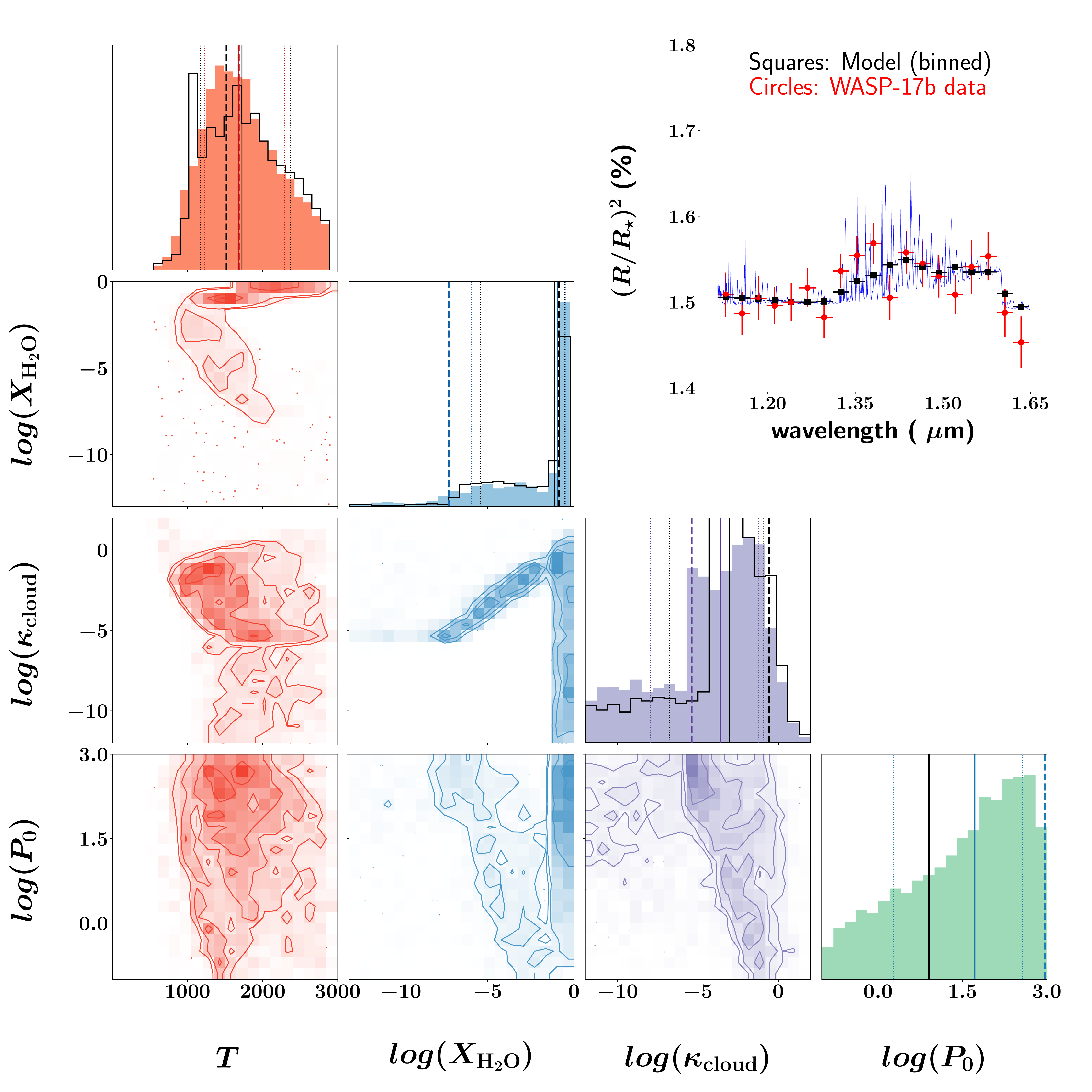}
\includegraphics[width=\columnwidth]{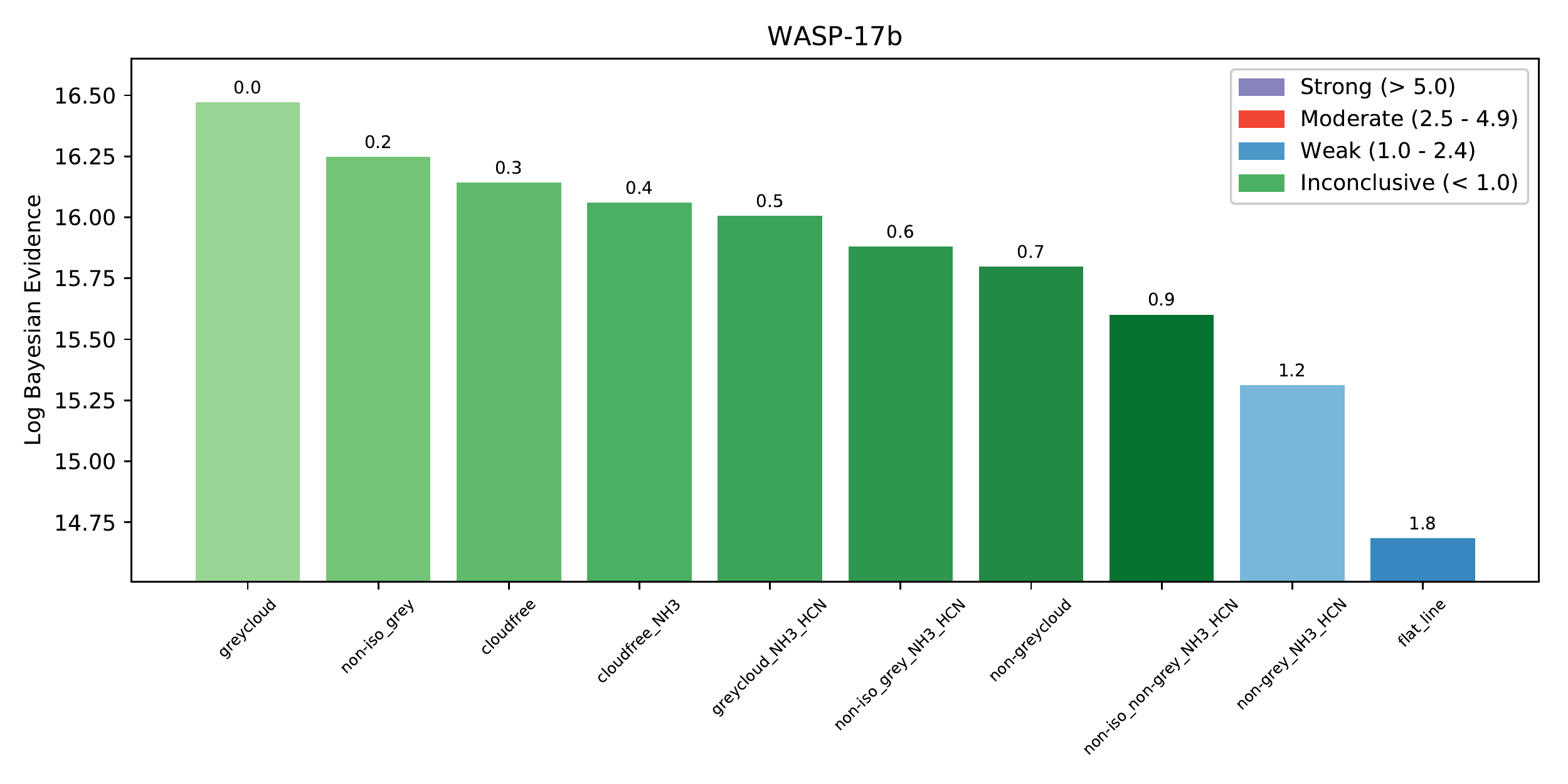}
\end{center}
\vspace{-0.1in}
\caption{Full atmospheric retrieval analysis of WASP-17b.  Top panel: Posterior distributions of parameters for the isothermal model with water only and grey clouds, which has the highest Bayes factor among the model hierarchy.  The vertical solid line is the median value of each posterior, while the vertical dotted lines are the 1-$\sigma$ uncertainties.  The vertical dashed line is the best-fit value of each posterior.  Also shown is a second retrieval where the $R_0$-$P_0$ relationship is determined by the values derived using optical data (see text).  Bottom panel: Logarithm of the Bayesian evidence and corresponding Bayes factor between each model compared to the best model.}
\label{fig:wasp17b_real}
\end{figure}

\begin{figure}
%\vspace{-0.1in}
\begin{center}
\includegraphics[width=\columnwidth]{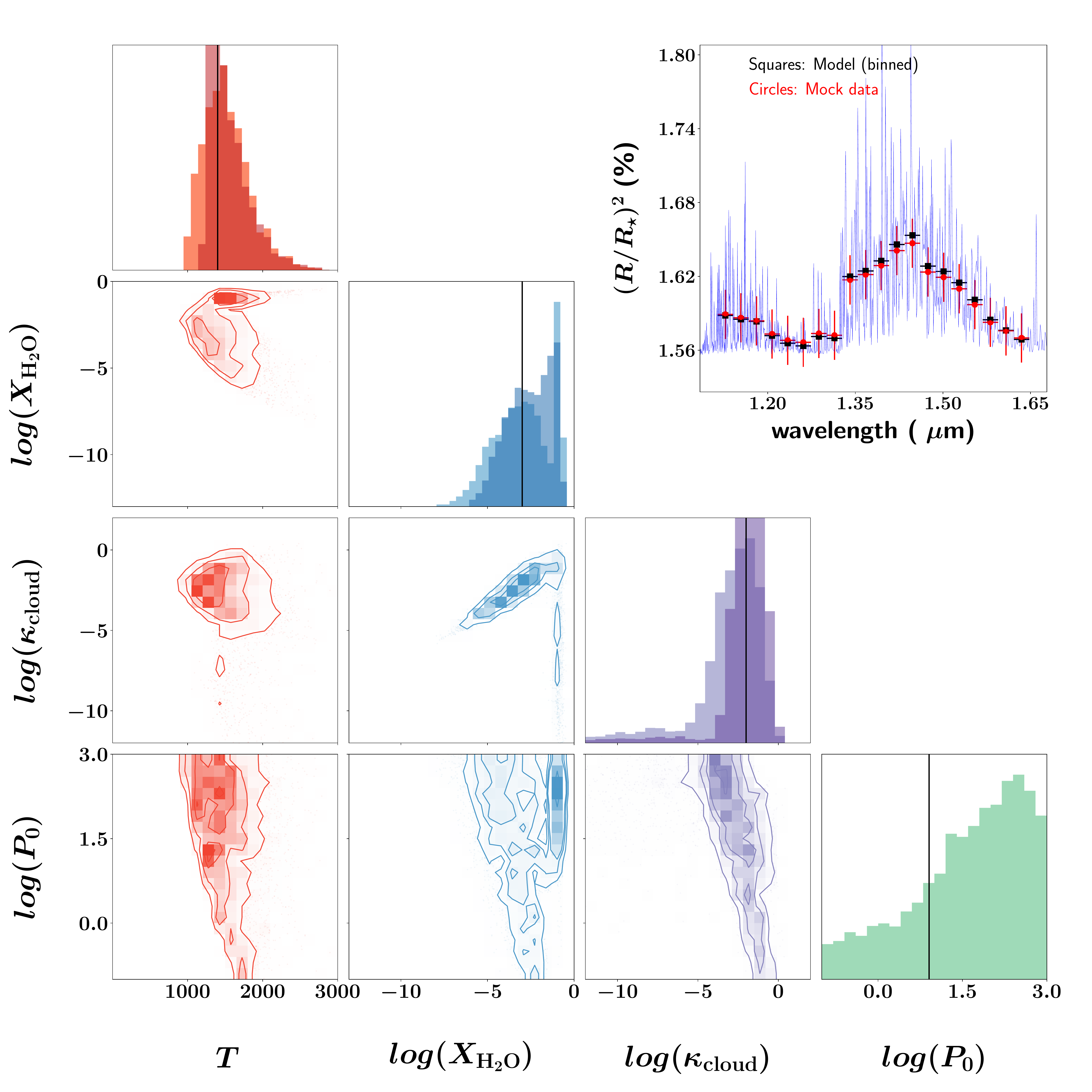}
\end{center}
%\vspace{-0.1in}
\caption{Additional low-resolution (WFC3-like) mock retrieval analysis of WASP-17b, but with larger uncertainties on each data point (200 ppm instead of 50 ppm).  The darker posterior distribution in each panel corresponds to a second retrieval in which $R_0$ is held fixed at its true value ($1.709 ~R_{\rm J}$).  This pair of retrievals should be compared to the lower right panel of Figure \ref{fig:wasp17b_mock2}.}
\label{fig:wasp17b_mock_larger_errors}
\end{figure}

Following our suite of tests, we now perform a full retrieval analysis on the WFC3 transmission spectrum of WASP-17b using a hierarchy of models.  Additionally, we attempt to fit the spectrum with a flat line (one parameter only) and compute its corresponding Bayesian evidence.  We see that there is weak evidence against the flat-line fit, but several models are consistent with the data (Figure \ref{fig:wasp17b_real}).  The isothermal model atmosphere with water only and grey clouds has the highest Bayesian evidence, which motivates us to display the posterior distributions of parameters associated with it in Figure \ref{fig:wasp17b_real}.  Alongside this retrieval, we perform a second retrieval where $R_0=1.709 ~R_{\rm J}$ and $P_0=8$ bar as derived using the optical spectral slope.  The posterior distribution for $P_0$ is only loosely constrained.  The median value of $P_0$ is a factor of about 6 larger than its true value (8 bar); its best-fit value hits the upper boundary of the prior distribution at 1000 bar.  

Yet, despite this inaccuracy in retrieving $P_0$, the posterior distributions of the pair of retrievals agree well.  This is somewhat surprising, because in our mock, low-resolution retrievals of WASP-17b we discovered that the volume mixing ratio of water is prior-dominated on its lower bound (and corresponds to the upper limit set on the prior of $P_0$).  To investigate this issue further, we ran an additional mock retrieval where the uncertainty on each data point is 200 ppm, instead of 50 ppm.  Figure \ref{fig:wasp17b_mock_larger_errors} shows that the pair of retrievals now have posterior distributions that are more similar to each other, which implies that the retrieval with variable $P_0$ is no longer as prior-dominated because there is now a larger parameter space of possibilities available to fit the mock spectrum.  However, the retrieval outcomes are still better (the posterior distributions are narrower) when the uncertainties are smaller.  The lesson learnt is that the lower bounds to volume mixing ratios retrieved from WFC3 transmission spectra may (or may not) be prior-dominated, depending on the measurement uncertainties.

\subsubsection{Retrieval analysis of WASP-31b WFC3 transmission spectrum}
\label{subsect:wasp31b}

\begin{figure}
%\vspace{-0.1in}
\begin{center}
\includegraphics[width=\columnwidth]{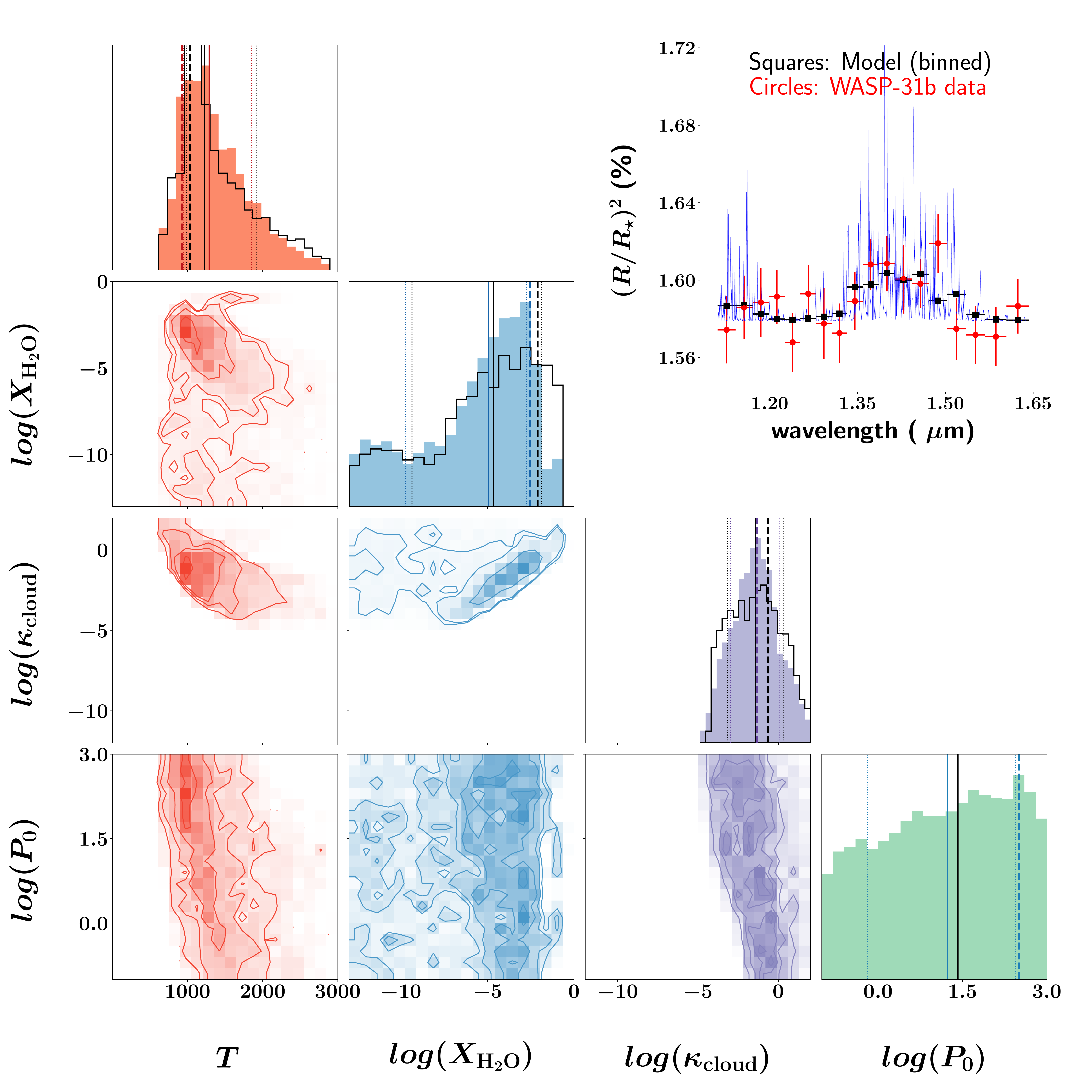}
\includegraphics[width=\columnwidth]{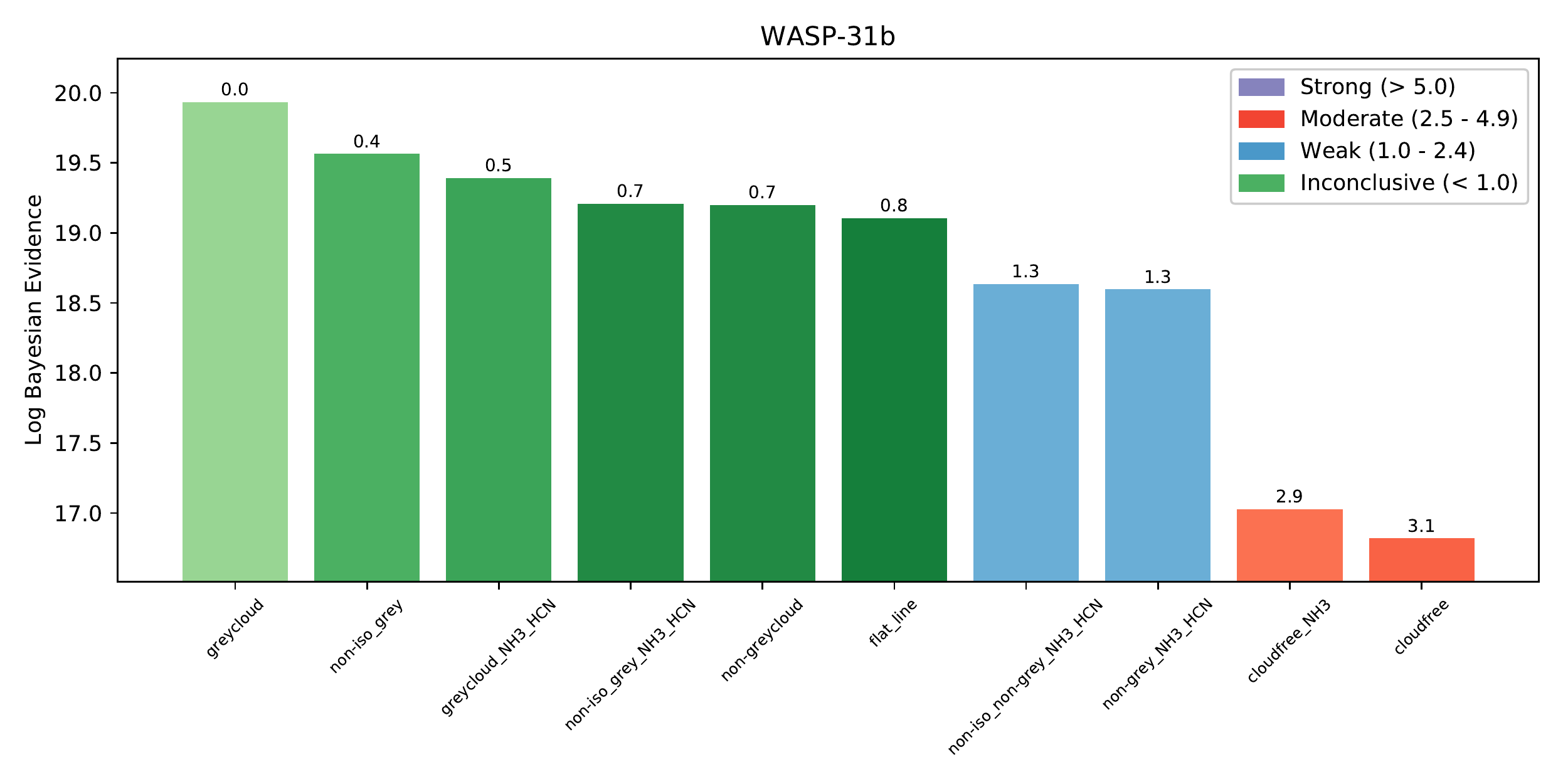}
\end{center}
\vspace{-0.1in}
\caption{Same as Figure \ref{fig:wasp17b_real}, but for WASP-31b.}
\label{fig:wasp31b_real}
\end{figure}

Since WASP-31b is the other object in our sample where we can robustly derive $R_0$ and $P_0$ from the optical spectral slope, we subject it to the same retrieval analysis we performed for WASP-17b.  In Figure \ref{fig:wasp31b_real}, we again subject the WFC3 transmission spectrum to a hierarchy of retrievals.  Again, the isothermal model with water only and grey clouds has the highest Bayesian evidence.  Two key differences are that the flat-line fit is not ruled out and that there is moderate evidence against cloud-free models.  As before, we perform a second retrieval with $R_0=1.379 ~R_{\rm J}$ and $P_0=26$ bar.   The median value of $P_0$ is about 16 bar and the best-fit value of $P_0$ almost hits the prior boundary at 594 bar, but despite these outcomes the posterior distributions of parameters from the pair of retrievals agree surprisingly well.

Our general conclusions from studying WASP-17b and WASP-31b are that $P_0$ can be robustly used as a fitting parameter as long as one's guess for $R_0$ corresponds to the range of prior values set on $P_0$.  Even if $P_0$ is not tightly constrained, the posterior distributions of the other parameters are, despite the low spectral resolution of the WFC3 transmission spectra.

%However, our retrieval of WASP-31b indicates that the atmosphere is cloudy in the near-infrared range of wavelengths probed by WFC3, even though it appears to be cloud-free in the optical.  Such a configuration is possible if the transit chord corresponding to the optical sits above the chord probed by WFC3.  Future infrared spectra of this benchmark object by JWST will shed light on this issue.

\subsection{Comparison of retrieval models for the other 36 WFC3 transmission spectra}
\label{subsect:all}

Following our suites of tests in \S\ref{subsect:wasp12_tests} and \S\ref{subsect:wasp17_mocks}, as well as our retrieval analyses of WASP-17b (\S\ref{subsect:wasp17b}) and WASP-31b (\S\ref{subsect:wasp31b}), we now apply our retrieval technique to the other 36 WFC3 transmission spectra in our sample.  For each object, we use the value of $R_0$ listed in Table \ref{tab:parameters} and allow $P_0$ to be a fitting parameter (with a log-uniform prior between 0.1 and 1000 bar).  

The results are shown in Table \ref{tab:results}, where the parameter values shown are the median and 1$\sigma$ uncertainties from the best model (highest Bayesian evidence).  Additionally, we ask several questions of the outcome.  An atmosphere is deemed to be cloudy if all of the cloud-free models have Bayes factors of unity or more.  Cloud-free atmospheres have \textit{only} cloud-free models with Bayes factors of less than unity.  If the models with Bayes factors of less than unity are a mixture of cloudy and cloud-free, then we tag the object with ``Maybe".  For non-grey clouds, our criterion is stricter: it refers only to objects where \textit{only} non-grey cloud models have Bayes factors of less than unity.

If the flat-line fit has a Bayes factor of less than unity, then we deem the retrieval to be inconclusive.  In these cases, we do not report any retrieved properties of the WFC3 transit chord.

%A general conclusion is that most of the objects in this sample have isothermal WFC3 transit chords with grey clouds.  Only XO-1b and WASP-19b are consistent with having cloud-free WFC3 transit chords.

\subsubsection{Ammonia may mimic cloudiness}

\begin{figure*}
%\vspace{-0.1in}
\begin{center}
\includegraphics[width=0.65\columnwidth]{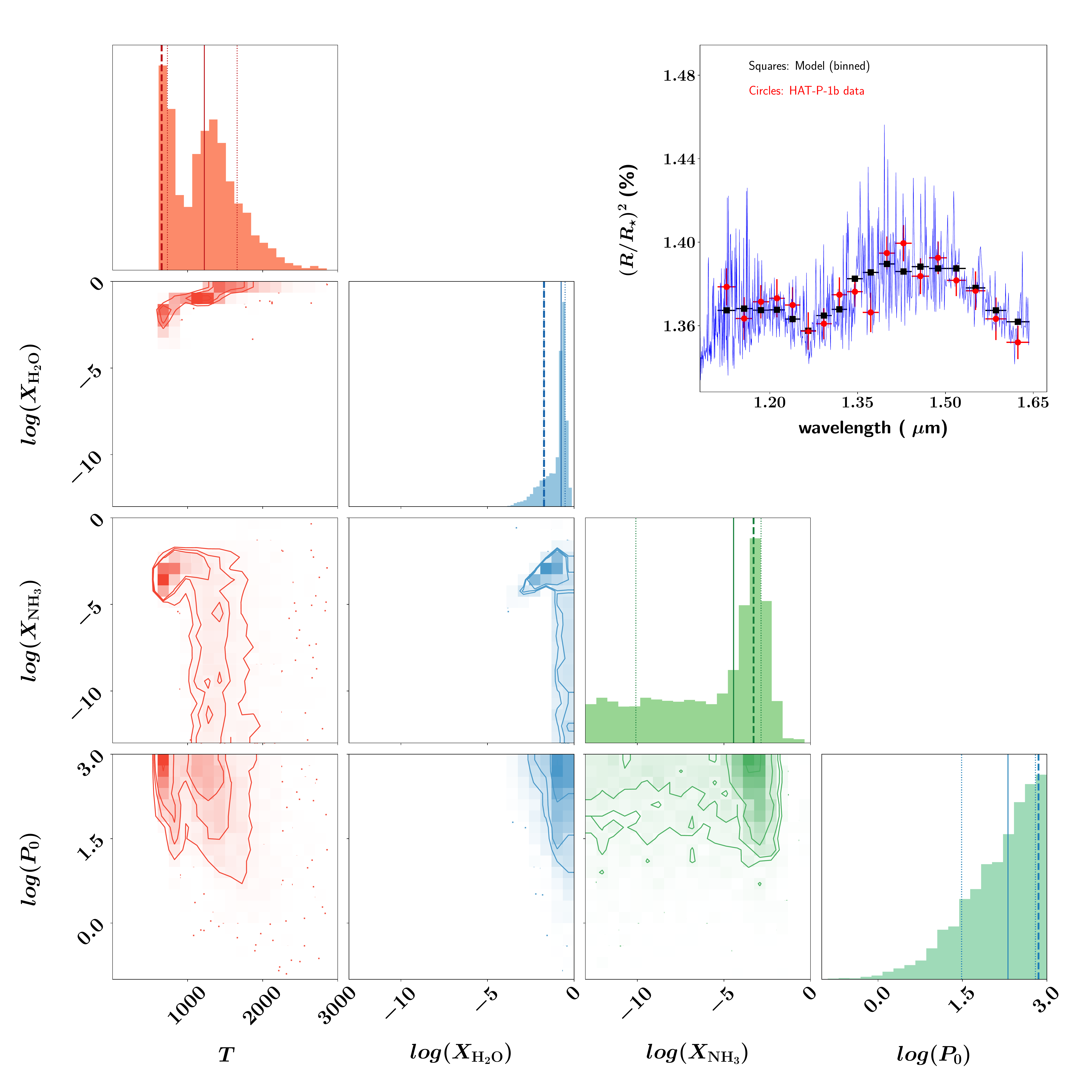}
\hspace{0.1in}
\includegraphics[width=1.2\columnwidth]{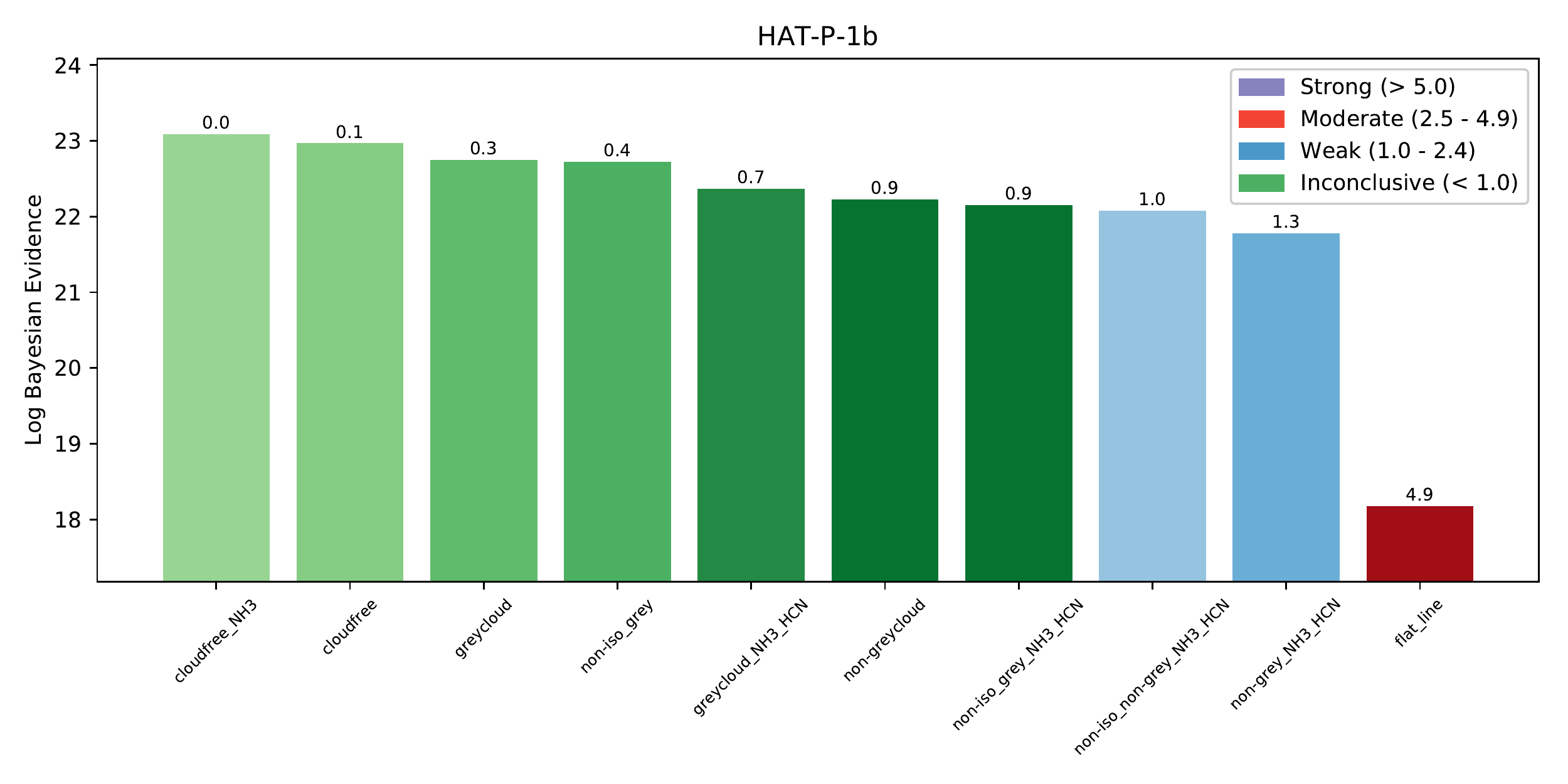}
\includegraphics[width=0.65\columnwidth]{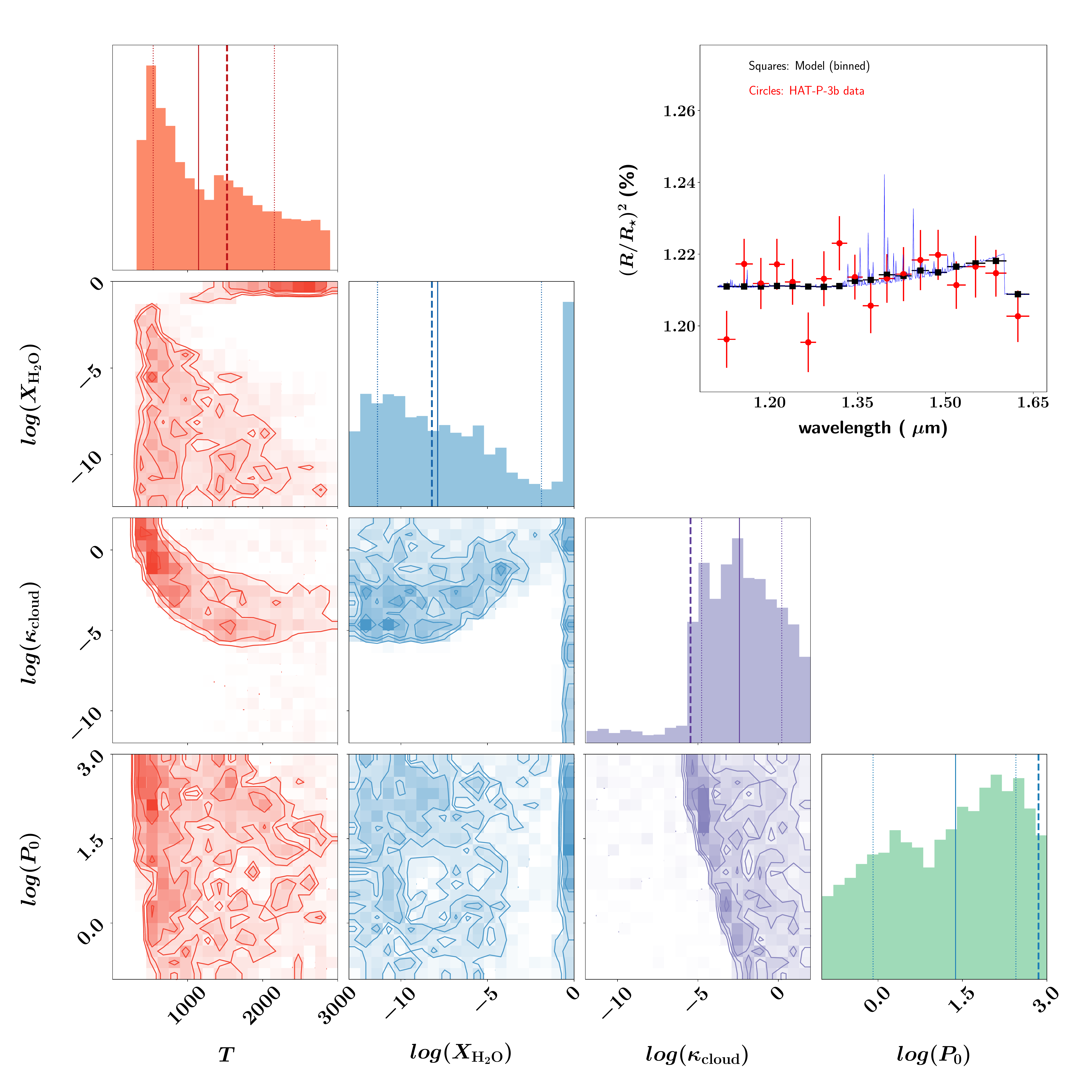}
\hspace{0.1in}
\includegraphics[width=1.2\columnwidth]{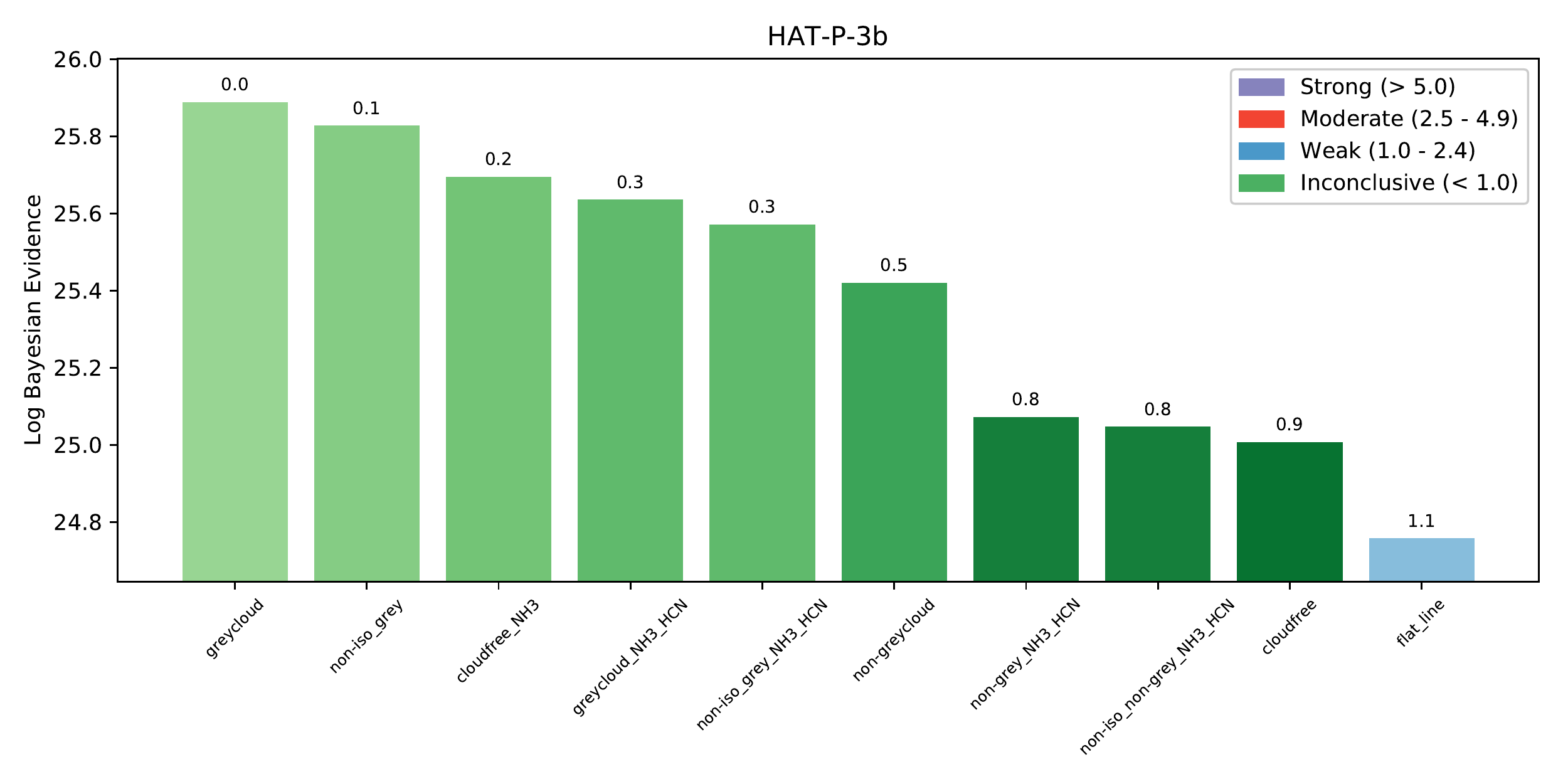}
\includegraphics[width=0.65\columnwidth]{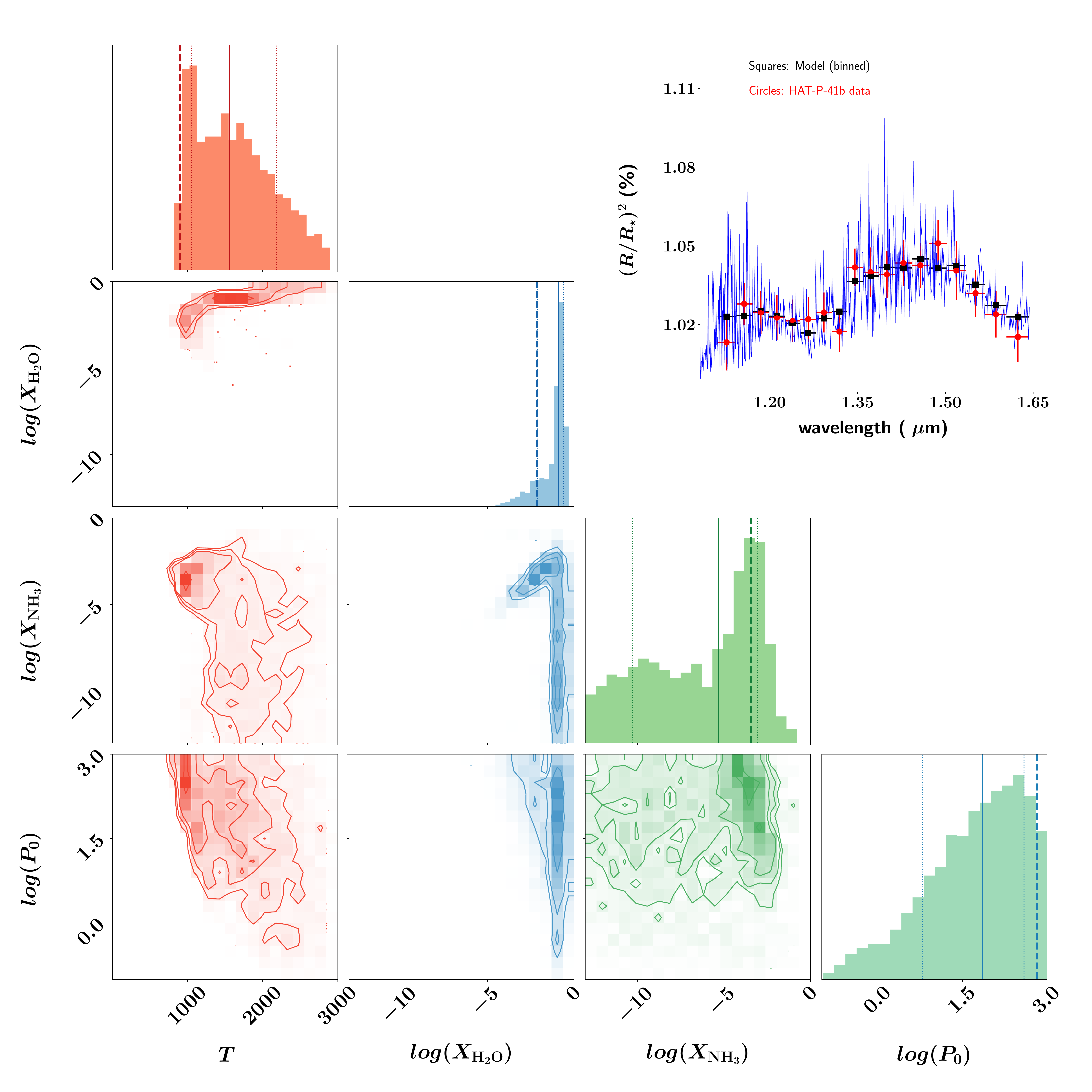}
\hspace{0.1in}
\includegraphics[width=1.2\columnwidth]{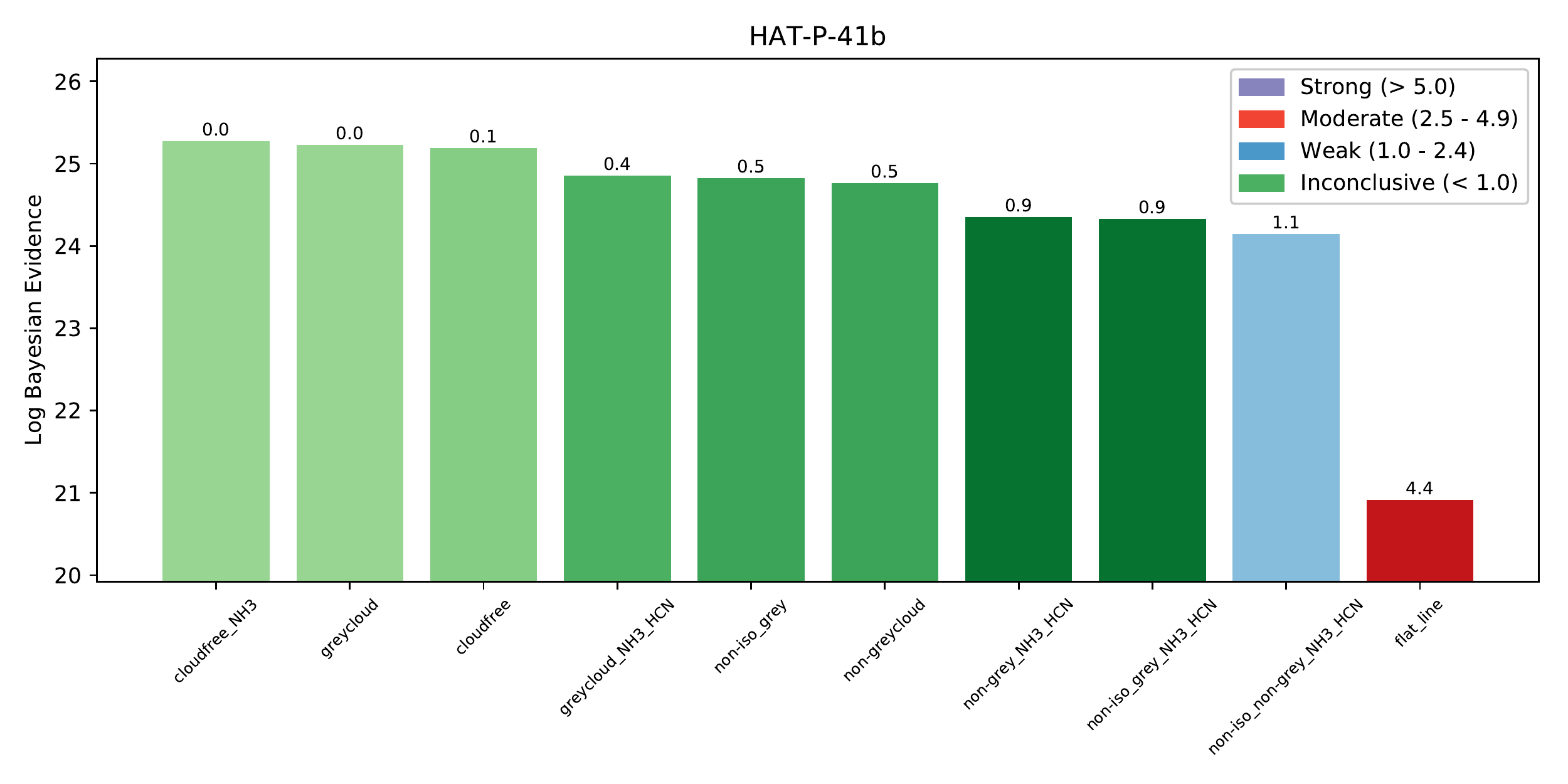}
\includegraphics[width=0.65\columnwidth]{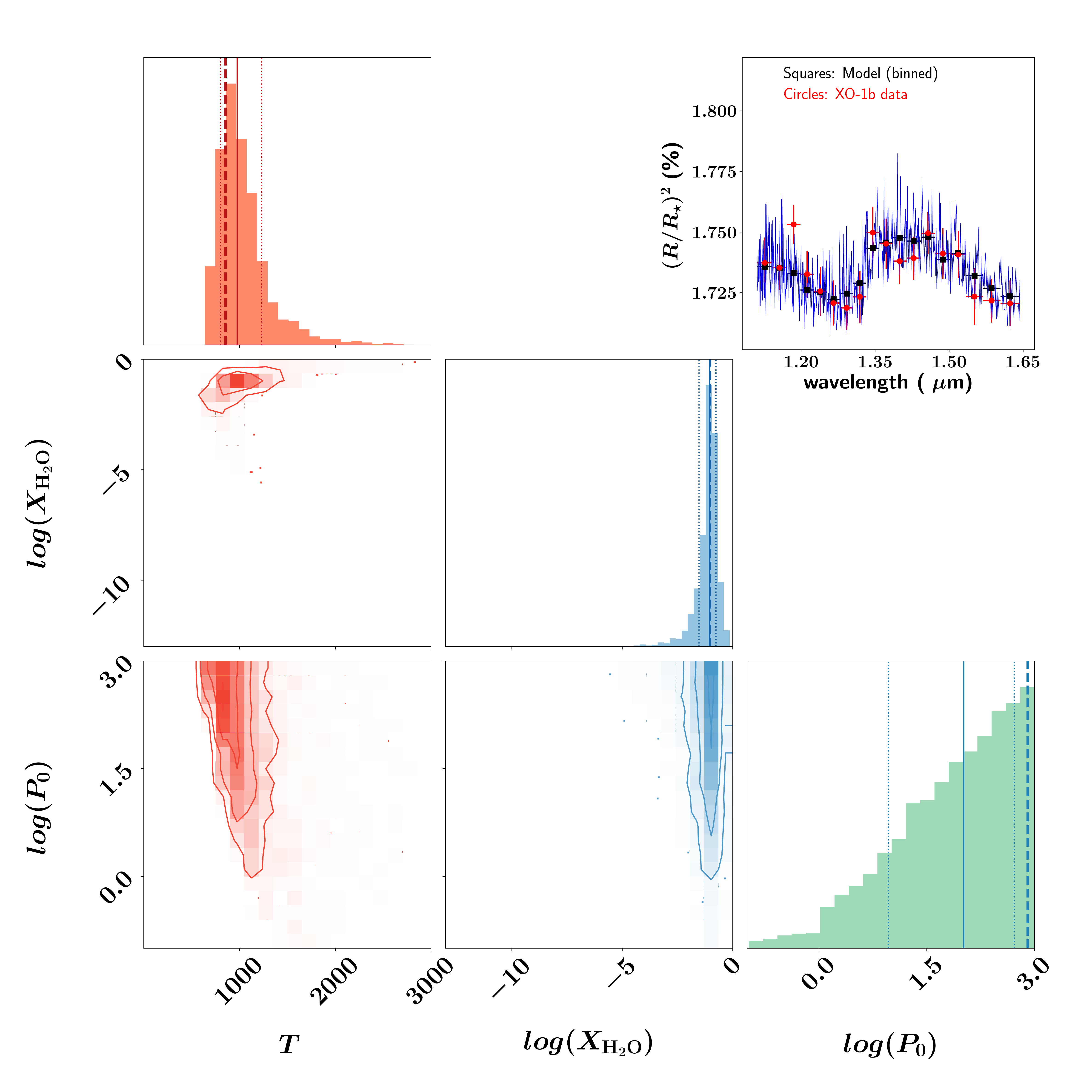}
\hspace{0.1in}
\includegraphics[width=1.2\columnwidth]{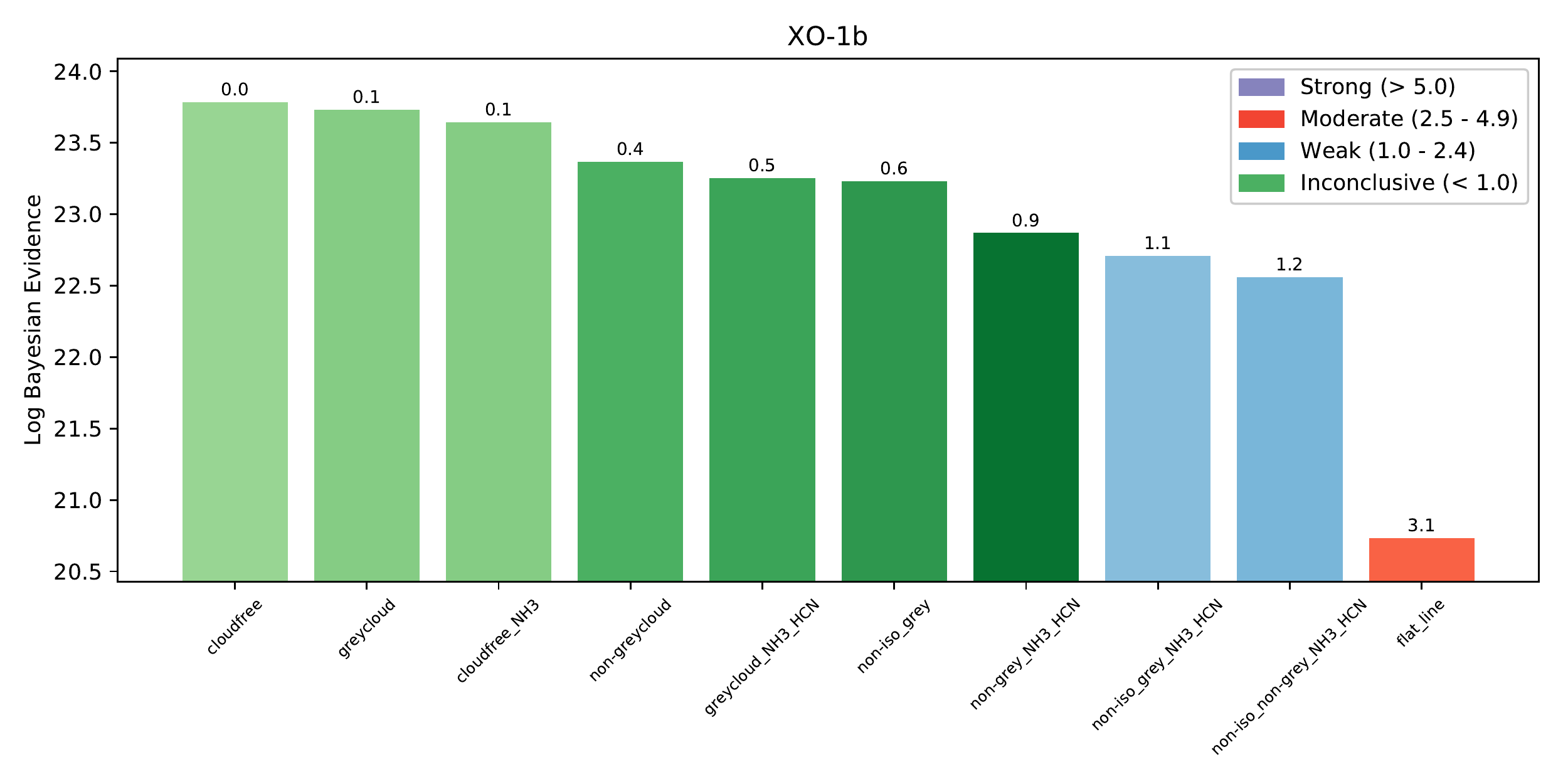}
\end{center}
\vspace{-0.1in}
\caption{Left column: Posterior distributions and synthetic spectrum for the best model (as selected by the Bayesian evidence).  Right column: Comparison of Bayesian evidence for objects for which it is not possible to distinguish between cloudy atmospheres containing water only versus cloud-free atmospheres with both water and ammonia present.   The solid, dotted and dashed vertical lines represent the median value, the 1-$\sigma$ uncertainties associated with the median and the best-fit value of each posterior distribution, respectively.  XO-1b is one of two objects with the highest Bayesian evidence for the cloudfree, isothermal model with water only (excluding the TRAPPIST-1 exoplanets).}
\label{fig:cloud-free_vs_nh3}
\end{figure*}

By visual inspection of the atmospheric opacities (Figure \ref{fig:opacities}), we had suspected that it would be possible for ammonia to mimic the flattening of the spectral continuum blue-wards of the 1.4 $\mu$m water feature.  Note that the one-parameter flat-line fits are disfavoured.  Figure \ref{fig:cloud-free_vs_nh3} shows 4 examples of objects (HAT-P-1b, HAT-P-3b, HAT-P-41b and XO-1b) where the Bayes factor between the model with grey clouds and water only versus the cloud-free model with water and ammonia is below unity, indicating that there is a lack of Bayesian evidence to favour one model over the other \citep{trotta08}.  This interpretation holds also for WASP-17b (Figure \ref{fig:wasp17b_real}), WASP-19b (Figure \ref{fig:veryhot}), HAT-P-38b and HD149026b (Figure \ref{fig:saturns}) and HAT-P-11b (Figure \ref{fig:neptunes}).  

With WFC3 transmission spectra, the cautionary tale is that cloudiness may be mimicked by the presence of ammonia and this occurs for 9 out of 38 objects in our sample.

\subsubsection{Prototypical hot Jupiters}

\begin{figure*}
%\vspace{-0.2in}
\begin{center}
\includegraphics[width=0.65\columnwidth]{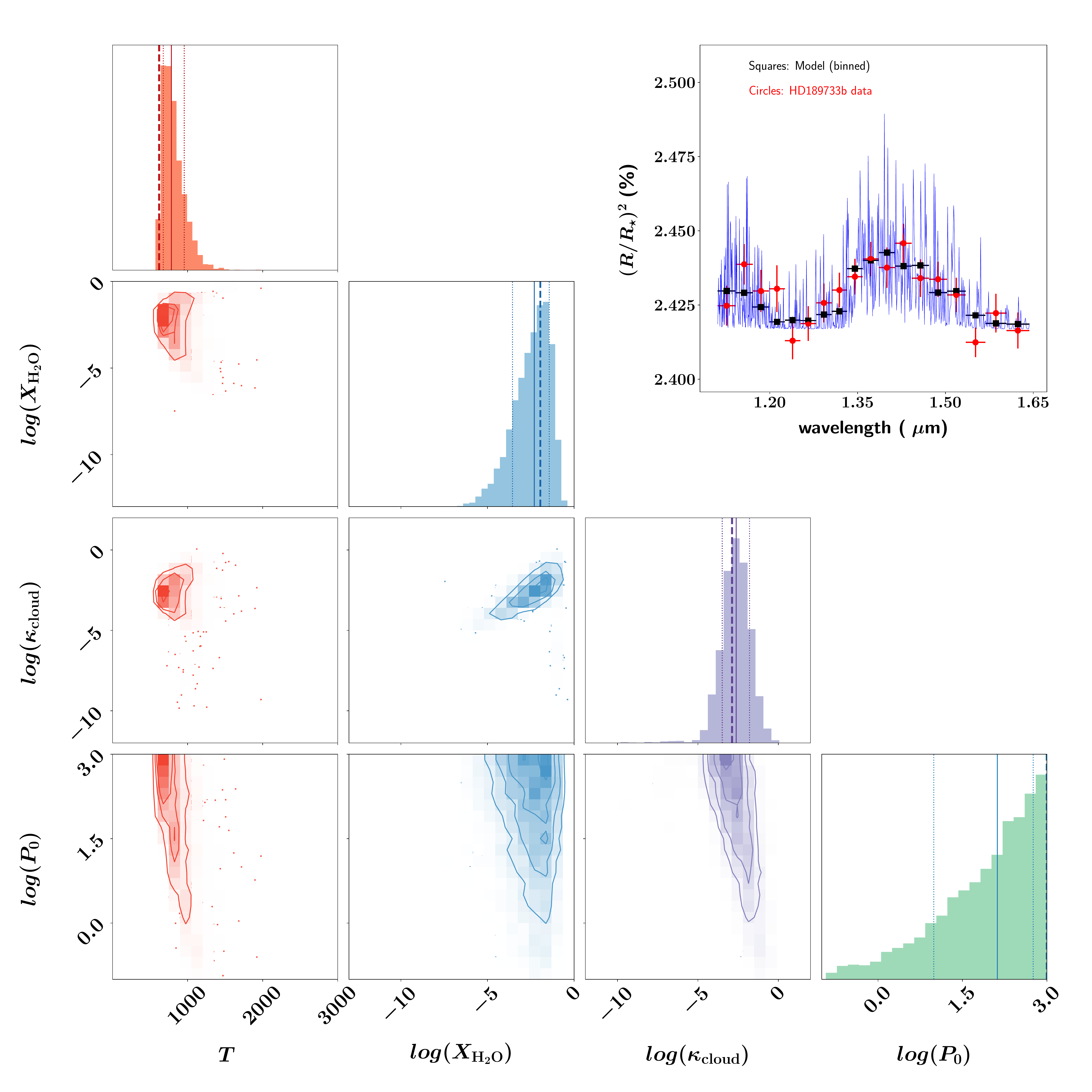}
\hspace{0.1in}
\includegraphics[width=1.2\columnwidth]{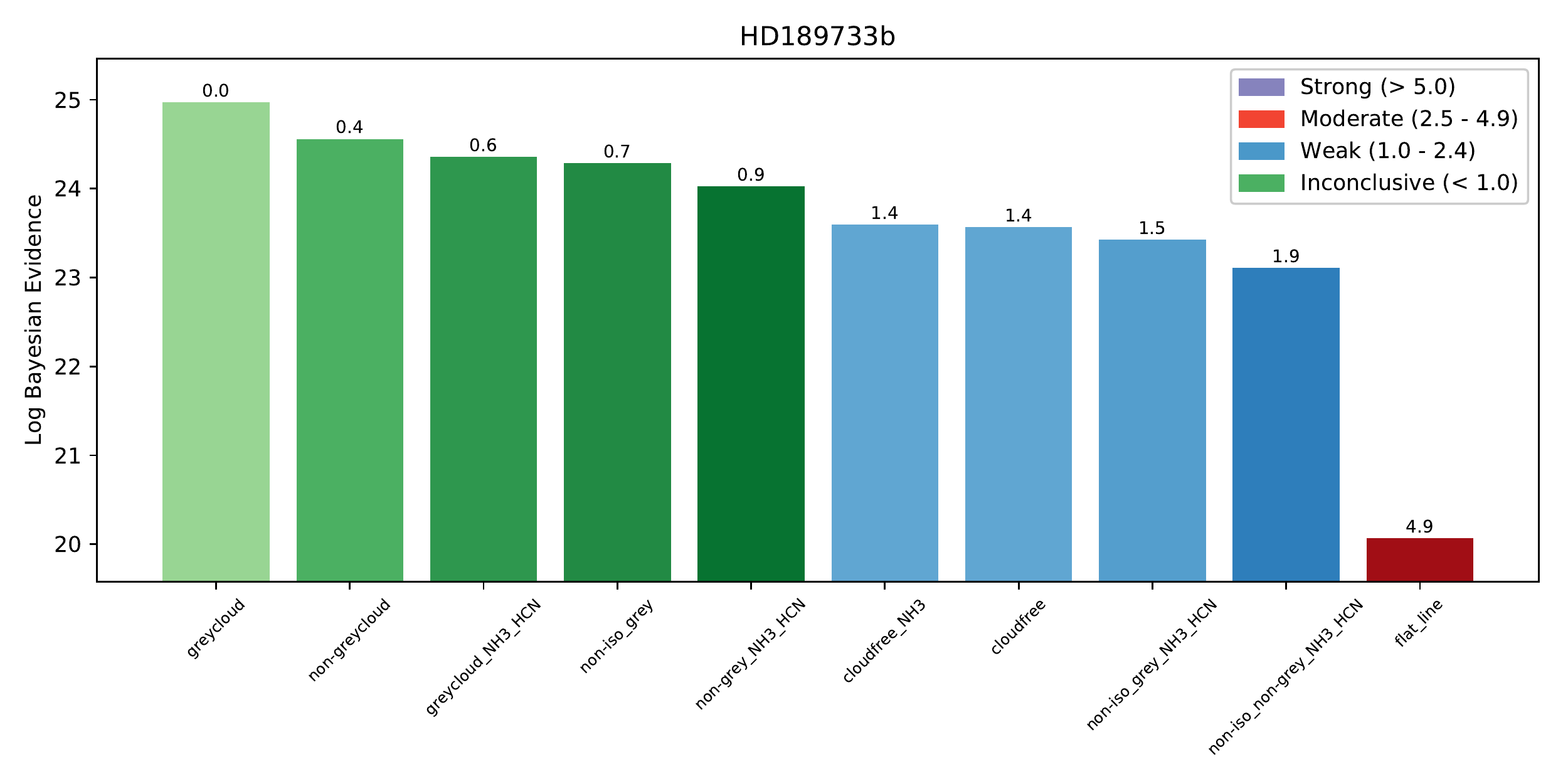}
\includegraphics[width=0.65\columnwidth]{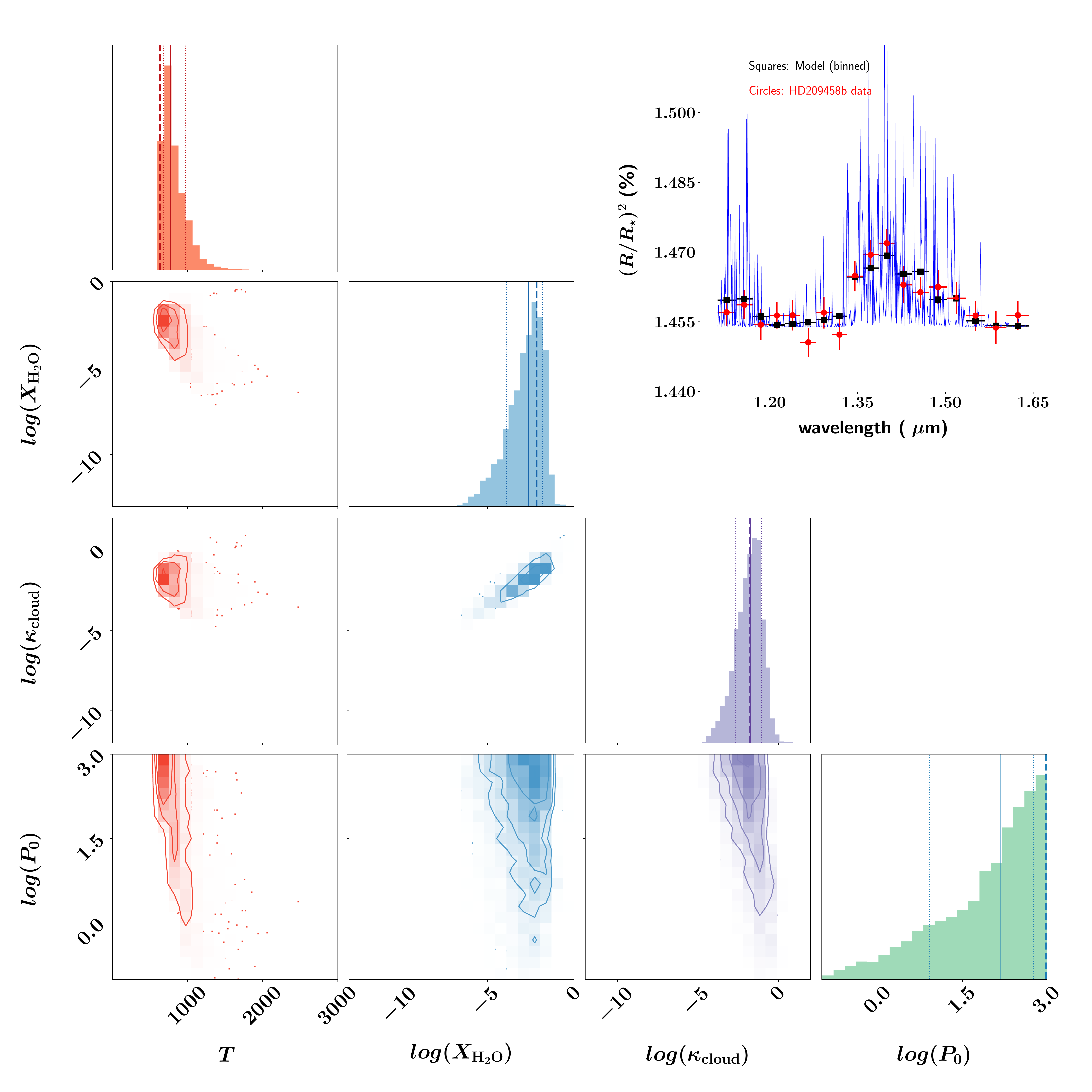}
\hspace{0.1in}
\includegraphics[width=1.2\columnwidth]{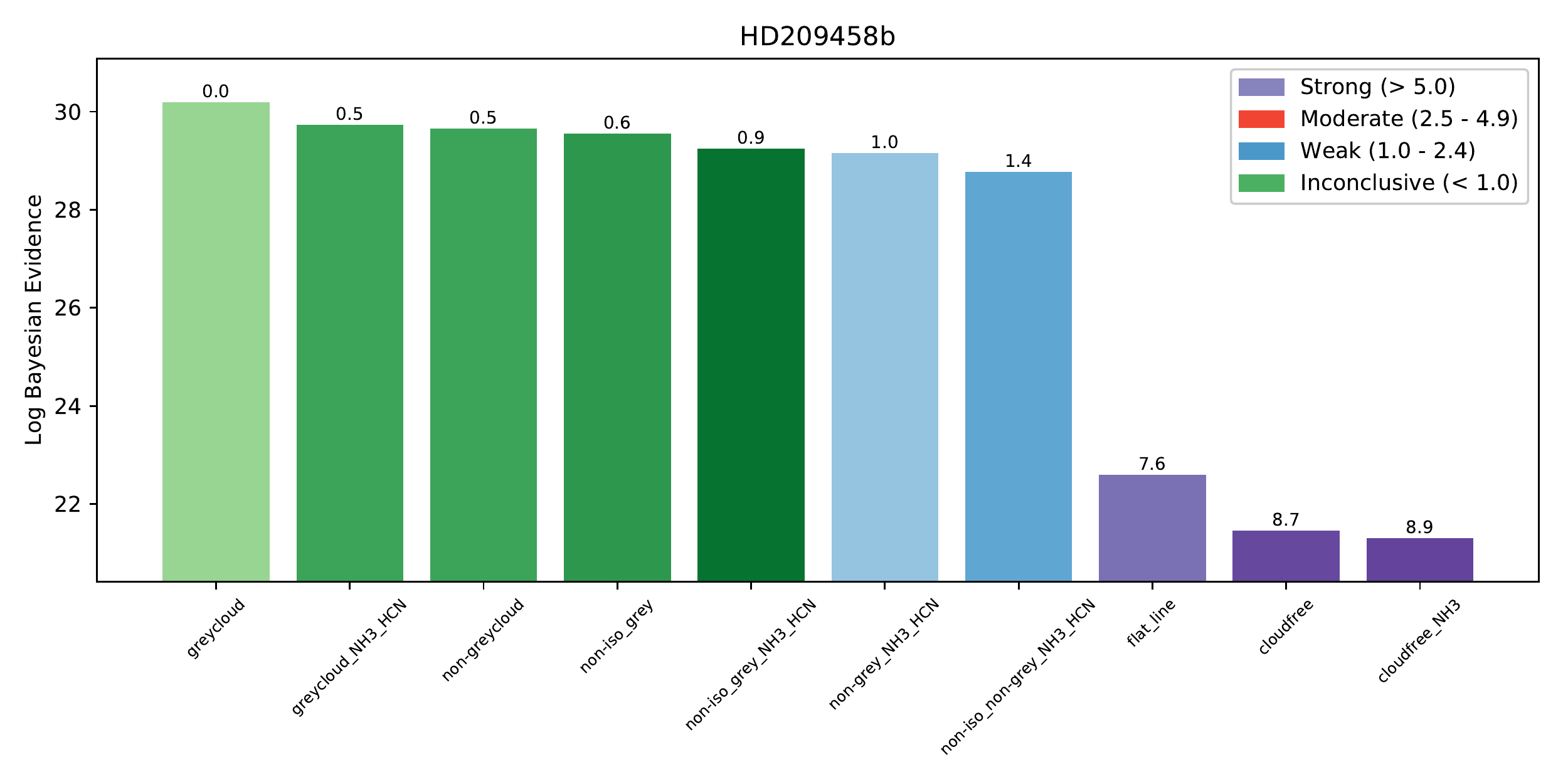}
\end{center}
\vspace{-0.1in}
\caption{Same as Figure \ref{fig:cloud-free_vs_nh3}, but for the prototypical hot Jupiters HD 189733b and HD 209458b.}
\label{fig:prototypes}
\end{figure*}

HD 189733b and HD 209458b are among the most studied hot Jupiters so far.  In Figure \ref{fig:prototypes}, we see that the WFC3 data definitively rules out cloud-free WFC3 transit chords for HD 209458b, and weakly rules them out for HD 189733b.  The simplest cloudy model, which is that of an isothermal atmosphere with grey clouds and water only, explains the WFC3 data well for both prototypical hot Jupiters. For HD 209458b, our retrieved temperature of $\approx 800$ K is roughly consistent with \cite{mm17a}, but our retrieved water abundance of $\log(X_{\rm H_2O}) = -2.65^{+0.81}_{-1.24}$ is more than two orders of magnitude higher than their retrieved value of $-5.24^{+0.36}_{-0.27}$.  It is unclear how to compare these values, because it is unclear how \cite{mm17a} have broken the normalisation degeneracy.  Unlike \cite{mm17a}, we find a lack of evidence for the detection of either NH$_3$ or HCN.  For example, the isothermal model with grey clouds and water only versus that with all three molecules have a Bayes factor of 0.5, indicating that one cannot favour one model over the other \citep{trotta08}.  For HD 189733b, we compare our results with those of \cite{madhu14} in \S\ref{subsect:subsolar}.

\subsubsection{Early Release Science (ERS) objects}

WASP-39b and WASP-43b are among the ERS objects proposed for JWST \citep{batalha17}.  WASP-63b is an ERS object for HST \citep{kil17}.  Additionally, WASP-43b is one of the few hot Jupiters to have multi-wavelength phase curves from HST, due to its sub-day orbit that circumvents the thermal breathing obstacle with HST \citep{stevenson14}.  None of the three objects are cloud-free in the WFC3 bandpass, and the simplest cloudy model fits the WFC3 data well.  There is no definitive evidence for the detection of either HCN or NH$_3$.  For WASP-63b, this is consistent with the analysis of \cite{kil17}.  For WASP-43b, our retrieved $\log(X_{\rm H_2O})=-2.89^{+1.13}_{-3.07}$ is broadly consistent with the $-3.6^{+0.8}_{-0.9}$ value reported by \cite{k14b}, although it should be noted that \cite{k14b} included carbon dioxide, carbon monoxide and methane in their analysis, while we excluded these molecules and included ammonia and hydrogen cyanide instead.  Interestingly, \cite{k14b} reported a logarithm of the ``reference pressure" of $-2.4^{+0.6}_{-0.4}$ (pressure in bar), which is broadly consistent with the pressure of 10 mbar that we assume the WFC3 bandpass to probe.  It is unclear how to compare the reference pressures between the two studies.

\subsubsection{Very hot Jupiters}

In our sample, 4 objects have equilibrium temperatures exceeding 2000 K: WASP-12b, WASP-19b, WASP-76b and WASP-121b.  For WASP-12b, the WFC3 transmission spectrum may be explained by models with HCN and NH$_3$ and also models with only water (i.e., these models all fall within Bayes factors of less than unity), which implies that we are unable to offer any estimate on the carbon-to-oxygen ratio, unlike in \cite{k15}. Our retrieved $\log(X_{\rm H_2O}) = -3.02^{+1.09}_{-1.36}$ is broadly consistent with the $\sim 10^{-4}$--$10^{-2}$ value reported by \cite{k15}.  In the case of WASP-19b, a cloud-free model with water only is a viable explanation---a rare occurrence in our sample.  WASP-76b is an interesting object in that several scenarios are strongly ruled out: cloud-free with either water only or water and ammonia, the simplest cloudy model, etc. In fact, it seems to show strong evidence for any model featuring a non-grey cloud. 

\subsubsection{Other hot Jupiters}

Figures \ref{fig:others_1} and \ref{fig:others_2} show the retrieval outcomes for 7 other hot Jupiters.  In all cases, cloud-free models are either unlikely or ruled out.  All of these 7 objects have WFC3 transmission spectra that may be explained by model atmospheres with grey clouds, meaning that non-grey clouds are not necessary to explain the data.  WASP-101b is the only object where HCN is detected at significant levels, while only upper limits are obtained on the abundances of H$_2$O and NH$_3$.

\subsubsection{Saturns}

Figures \ref{fig:saturns} and \ref{fig:saturns_2} show the retrieval outcomes for 6 Saturn-mass (0.2--$0.4 M_{\rm J}$) exoplanets.  WASP-39b, an ERS object, also belongs to this category.  With the exception of WASP-69b, the WFC3 transmission spectra are explained by the simplest cloudy model. WASP-69b requires non-grey clouds along its transit chord to explain the WFC3 data. For HAT-P-18b, HAT-P-38b and HD 149026b, the isothermal cloud-free model with water only provides a viable explanation for the data; several other models also have Bayes factor of less than unity.

\subsubsection{Neptunes}

There is strong evidence against a cloud-free interpretation of the somewhat flat WFC3 transmission spectra of the exo-Neptunes GJ 436b and GJ 3470b (Figure \ref{fig:neptunes}).  For GJ 436b, this is consistent with the findings of \cite{knutson14b}.  In fact, the WFC3 transmission spectrum of GJ 436b can simply be fit by a one-parameter flat line, rendering it impossible to report atmospheric properties in a meaningful sense.  HAT-P-26b does not have a flat transmission spectrum and cloud-free interpretations are strongly ruled out (Bayes factor exceeding 5.0).  \cite{wakeford17} previously analysed the transmission spectrum of HAT-P-26b, which includes STIS, WFC3 and Spitzer data spanning 0.5--5 $\mu$m, using a suite of models incorporating carbon monoxide, carbon dioxide, methane and water.  Using the Bayesian information criterion, they disfavoured cloud-free models.  Our WFC3-only analysis is consistent with the conclusion of \cite{wakeford17}.  The best model, in terms of the Bayesian evidence, is the simplest cloudy one: an isothermal atmosphere with grey clouds and water only, but a variety of cloud models have Bayes factors below unity compared to this best model.

\subsubsection{Super Earths}

Besides being a super Earth, GJ 1214b is the prototypical example of a flat transmission spectrum \citep{k14}.  The retrieval outcome in Figure \ref{fig:searths} corroborates this view and it is unsurprisingly that a one-parameter flat-line fit suffices.  In our analysis, HD 97658b is inconclusively favoured by a cloud-free model with water and NH$_3$, though the quantities of ammonia required to match the data may be implausibly high.  More data is needed to corroborate or refute this finding.  

\subsubsection{TRAPPIST-1 exoplanets}

\cite{dewit18} previously measured somewhat flat WFC3 transmission spectra for TRAPPIST-1d, e, f and g.  We note an ongoing debate concerning the robustness of these measured WFC3 transmission spectra, as it has been argued that the shapes of the spectral bandheads may have been contaminated by starspots and faculae from TRAPPIST-1 \citep{ducrot18,morris18,rackham18}.  Nevertheless, we will analyze these spectra as given.  \cite{dewit18} ruled out cloud-free, H$_2$-dominated atmospheres for TRAPPIST-1d, e and f, but not for g.  We wish to corroborate or refute this conclusion and also to go slightly further, by considering both Earth-like ($m=29 ~m_{\rm H}$) or H$_2$-dominated (variable $m$ as defined in equation [\ref{eq:mmm}]) atmospheres in two separate suites of retrievals.  

For Earth-like atmospheres, the WFC3 spectra are explained by the majority of the models in our hierarchy.  With the exception of TRAPPIST-1d, there is weak evidence against the WFC3 transmission spectra being explained by a flat line.  This is unsurprising (compared to the retrievals with H$_2$-dominated atmospheres), because for a nitrogen-dominated atmosphere the scale height is an order of magnitude smaller than for the H$_2$-dominated atmosphere, which implies that even small departures from a flat line require spectral features spanning several scale heights to explain the data.  Overall, when Earth-like atmospheres are assumed, the retrieval analyses are inconclusive.  

When H$_2$-dominated atmospheres are assumed, we rule out cloud-free atmospheres with water only for TRAPPIST-1d, e and f. For all four exoplanets, the WFC3 transmission spectrum is adequately explained by a one-parameter flat-line fit, which implies that atmospheric properties cannot be meaningfully retrieved.

We do not consider arguments based on the evolution of the exoplanet or atmospheric escape, as they are out of the scope of the present study.  Our inclusion of the TRAPPIST-1 exoplanets is for completeness and they will not be included in our analysis of the trends associated with the water volume mixing ratios in \S\ref{subsect:trends}.  However, when compiling population statistics, we will include the outcomes only from the retrievals of the TRAPPIST-1 exoplanets assuming Earth-like atmospheres.

\subsection{Trends}
\label{subsect:trends}

All of the techniques developed and tests performed in this study culminate in a singular result: to examine if there are trends in the retrieved atmospheric properties.  In particular, we wish to examine if $X_{\rm H_2O}$ correlates with the equilibrium temperature ($T_{\rm eq}$), retrieved temperature ($T$) or mass of the exoplanet ($M$).  The equilibrium temperature is a proxy for the strength of insolation or stellar irradiation.  Previous studies have plotted the ``metallicity" versus the exoplanet mass and claimed a correlation between the two quantities \citep{k14b,wakeford17,wakeford18,arc18,mansfield18,nikolov18}.
 
In Figure \ref{fig:trends}, we find little to no evidence for a correlation between $X_{\rm H_2O}$ and $M$, $T_{\rm eq}$ or $T$.  If the abundance of water is assumed to be a direct proxy for the elemental abundance of oxygen (see \S\ref{subsect:metallicity}), then this outcome runs contrary to previous claims.  There is a lack of correlation between $\kappa_{\rm cloud}$ and $T_{\rm eq}$, which has two implications.  First, it suggests that our inferred $X_{\rm H_2O}$ values are not biased by the degree of cloudiness (or haziness) in these atmospheres.  Second, the majority of atmospheric transit chords probed by WFC3 appear to have $\kappa_{\rm cloud} \sim 10^{-2}$ cm$^2$ g$^{-1}$ (corresponding to $\sim 10$ mbar), regardless of the surface gravity or strength of insolation.  The lack of evidence for non-grey clouds implies that the particle radii are $r_c \gtrsim 0.1$ $\mu$m.  Overall, these outcomes may be interpreted as the transit chords being affected by haze.\footnote{We adopt the planetary science definition of ``cloud" versus ``haze": the former is formed thermochemically, while the latter is formed photochemically.}  The ratio of the retrieved to the equilibrium temperatures ($T/T_{\rm eq}$) appears to have a lower limit of about 0.5.

It is unclear how to relate our results to claimed correlations between the bulk metallicity of exoplanets and their masses based on the analysis of mass-radius relations \citep{mf11,thorngren16}.

\section{Discussion}
\label{sect:discussion}

\subsection{Comparison to a previous retrieval study}

It is natural to compare our study to \cite{tsi18}, since the WFC3 transmission spectra of 30 objects from our sample are taken from it.  Furthermore, some of the modelling choices made by \cite{tsi18} are the same as ours: isothermal transit chord, nested sampling.  Our cloud models differ, because \cite{tsi18} use the formulation of \cite{lee13}, which also allows for a smooth transition between the Rayleigh and large-particle regimes but predates \cite{kh18}, and also assume a cloud-top boundary (which we do not).  Furthermore, \cite{tsi18} include methane, carbon monoxide, carbon dioxide, titanium oxide (TiO) and vanadium oxide (VO) in their retrievals in addition to water and ammonia; they do not include hydrogen cyanide.  By contrast, we only include water, ammonia and hydrogen cyanide in our model hierarchy.  Inevitably, these choices lead to differences in some of the retrieval outcomes.

Table 3 of \cite{tsi18} lists their retrieved water volume mixing ratios.  For GJ 436b, HAT-P-12b, WASP-29b, WASP-31b, WASP-67b and WASP-80b, we do not report any retrieved atmospheric properties, unlike for \cite{tsi18}, as the one-parameter flat-line fit is among the models with Bayes factors of less than unity.  For GJ 3470b, HAT-P-1b, HAT-P-3b, HAT-P-11b, HAT-P-17b, HAT-P-18b, HAT-P-26b, HAT-P-32b, HAT-P-38b, HD 149026b, HD 189733b, HD 209458b, WASP-12b, HAT-P-41b, WASP-43b, WASP-52b, WASP-63b, WASP-69b, WASP-74b, WASP-101b, WASP-121b and XO-1b, our retrieved water mixing ratios are broadly consistent with those of \cite{tsi18}.  For WASP-39b and WASP-76b, our retrieved water mixing ratios differ at the order-of-magnitude level compared to \cite{tsi18}.  Interestingly, these two objects also have the highest values of the Atmospheric Detectability Index (ADI) in the \cite{tsi18} sample of 30 objects.

Of particular interest is WASP-76b, which is one of two objects in our sample that requires a non-grey cloud to fit the data.  \cite{tsi18} reported that their retrieval favours a cloudfree interpretation, because the non-flat spectral continuum blueward of the 1.4-$\mu$m water feature may be fitted by the spectral features of TiO and VO.  \cite{tsi18} remark that their retrieved $\log{X_{\rm TiO}} \sim -2.5$ is ``likely unphysical".  Our retrieval yields $\log{X_{\rm H_2O}}=-5.3 \pm 0.61$, which is inconsistent with the $\log{X_{\rm H_2O}}=-2.7 \pm 1.07$ reported by \cite{tsi18}.  The WFC3 transmission spectrum of WASP-76b demonstrates that a wider wavelength range is required to resolve the degeneracy associated with these modelling choices, which will be provided by JWST spectra.

It is unclear why our retrieval outcome for WASP-39b differs from that of \cite{tsi18}, because they did not publish the full set of posterior distributions for this object, unlike for WASP-76b in their Figure 11.  For example, it is unclear if the high value of the ADI for WASP-39b translates into a cloud-free interpretation (which is the case for WASP-76b).

\subsection{Is there evidence for non-grey clouds?  Is cloud composition constrained?}

Cloud models of varying sophistication have been employed in retrieval models.  Our approach is somewhat different in that we include in our hierarchy of retrievals both grey and non-grey cloud models, as well as a one-parameter flat line.  For 8 out of 38 objects, the WFC3 transmission spectrum is explained by a flat line.  For 35 out of 38 objects, an isothermal grey cloud model with water only is sufficient to explain the data.  Only WASP-69b and WASP-76b have WFC3 transmission spectra that require an explanation by model atmospheres with non-grey clouds along the transit chord. Otherwise, there is generally no evidence for non-grey clouds being present in the sample of 38 objects.  

Since the cloud composition may only be inferred for non-grey clouds, this implies that the composition is generally unconstrained, which is consistent with the conclusion drawn by \cite{tsi18}.  Even for WASP-69b and WASP-76b, the parameter $Q_0$ is largely unconstrained because it spans the entire range of values set by the prior.

Given the retrieval outcomes, our approach to not consider patchy clouds \citep{lp16} is justified.  We have also shown that the normalisation degeneracy may be broken without appealing to the more complicated patchy cloud model, which was invoked by \cite{mm17a} to break the degeneracy.

\subsection{Is there evidence for non-isothermal transit chords?}

For all 38 objects in our sample, we find a lack of strong Bayesian evidence to support non-isothermal transit chords probed by WFC3.

\subsection{How prevalent is HCN or NH$_3$?}

Based on the best model selected by the Bayesian evidence, we find that only 6 objects have tentative evidence for the detection of ammonia: HAT-P-1b, HAT-P-17b, HAT-P-38b, HAT-P-41b, WASP-101b and HD 97658b. However, the retrieved value for HD 97658b is $\log(X_{\rm NH_3})=-0.48^{+0.19}_{-0.23}$, which may be unphysically high. This is unsurprising as our model contains no chemistry, so there is nothing to prevent unphysical values being retrieved. HAT-P-17b and WASP-101b also have tentative detections of hydrogen cyanide.

\subsection{Subsolar water abundances in hot Jupiters?}
\label{subsect:subsolar}

\cite{madhu14} previously analysed the WFC3 transmission spectra of HD 189733b, HD 209458b and WASP-12b using cloud-free retrieval models and found $\log(X_{\rm H_2O}) =-5.20^{+1.68}_{-0.18}$, $-5.27^{+0.65}_{-0.16}$ and $-5.35^{+1.85}_{-1.99}$, respectively.  They concluded that the water abundances from these hot Jupiters are subsolar by about 1--2 orders of magnitude.  By contrast, our retrievals find $\log(X_{\rm H_2O})$ values that are several orders of magnitude higher: $-2.3^{+0.87}_{-1.26}$, $-2.65^{+0.81}_{-1.24}$ and $-3.02^{+1.09}_{-1.36}$, respectively.  We estimate that $\log(X_{\rm H_2O}) \approx -3.2$ assuming chemical equilibrium, solar abundance and a pressure of 10 mbar, which suggests that our retrieved water abundances are super- rather than sub-solar as claimed by \cite{madhu14}.  The discrepancy arises from the retrievals of \cite{madhu14} being cloud-free, while we have included a cloud model that smoothly transitions between the Rayleigh and large-particle regimes.  It is consistent with the fact that cloud opacity diminishes the strength of the water feature, which may be negated by increasing $X_{\rm H_2O}$.  

\subsection{What does the ``metallicity" mean when interpreting spectra of exoplanetary atmospheres?}
\label{subsect:metallicity}

Several published studies have plotted the ``metallicity" (in ``solar" units) versus the mass of the exoplanet with entries from the Solar System gas and ice giants overplotted \citep{k14b,wakeford17,wakeford18,arc18,mansfield18,nikolov18}.  As already elucidated by \cite{heng18}, there are several caveats to these plots.  First, the ``metallicity" is predominantly O/H in these studies.  Second, the ``mixing ratio of water at solar abundance" is a temperature- and pressure-dependent statement.  Given a fixed value of O/H, the mixing ratio of water still depends on temperature and pressure.  In other words, it is a \textit{function} and not a number.  Third, the conversion factor between the water mixing ratio and O/H is not always unity and depends on the elemental abundances (O/H, C/H, etc), carbon-to-oxygen ratio, temperature, pressure, photochemistry, atmospheric mixing, condensation, etc.  For all of these reasons, we have chosen to present our retrieved water abundances as they are in Figure \ref{fig:trends}, rather than convert them to O/H.

%The assessment that WASP-17b and WASP-31b are cloud-free in the optical was based on low-resolution STIS data obtained by \cite{sing16}.  It will be insightful to examine the optical spectra of these objects using resolution $\sim 10^5$ data from ground-based spectrographs, as has been done for HD 189733b \citep{wyttenbach15}, especially with the upcoming ESPRESSO spectrograph that is capable of measuring radial velocities $\sim 10$ cm s$^{-1}$.  Using ultra-high-resolution optical spectra to break the normalisation degeneracy would enable JWST spectra of these objects to be decisively interpreted, and would usher in the era of high-precision atmospheric chemistry using transmission spectroscopy.

\vspace{0.1in}

{\scriptsize We acknowledge partial financial support from the Center for Space and Habitability (CSH), the PlanetS National Center of Competence in Research (NCCR), the Swiss National Science Foundation, a European Research Council (ERC) Consolidator Grant and the Swiss-based MERAC Foundation.  C.F. acknowledges partial financial support from a University of Bern International 2021 Ph.D Fellowship.  We thank Simon Grimm, Daniel Kitzmann, Maria Oreshenko, Shami Tsai and Matej Malik for constructive conversations.}

\begin{figure*}
\begin{center}
\includegraphics[width=0.65\columnwidth]{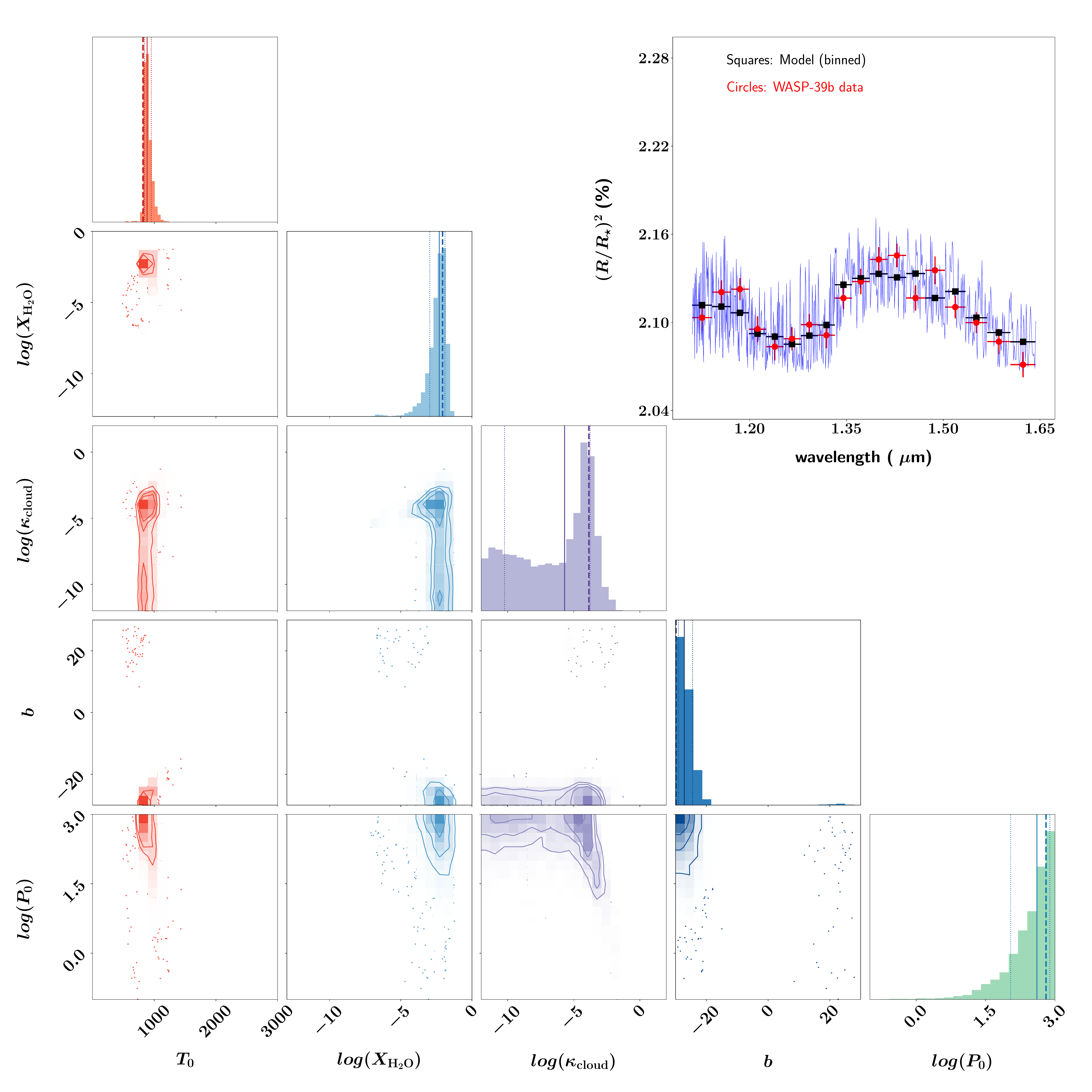}
\hspace{0.1in}
\includegraphics[width=1.2\columnwidth]{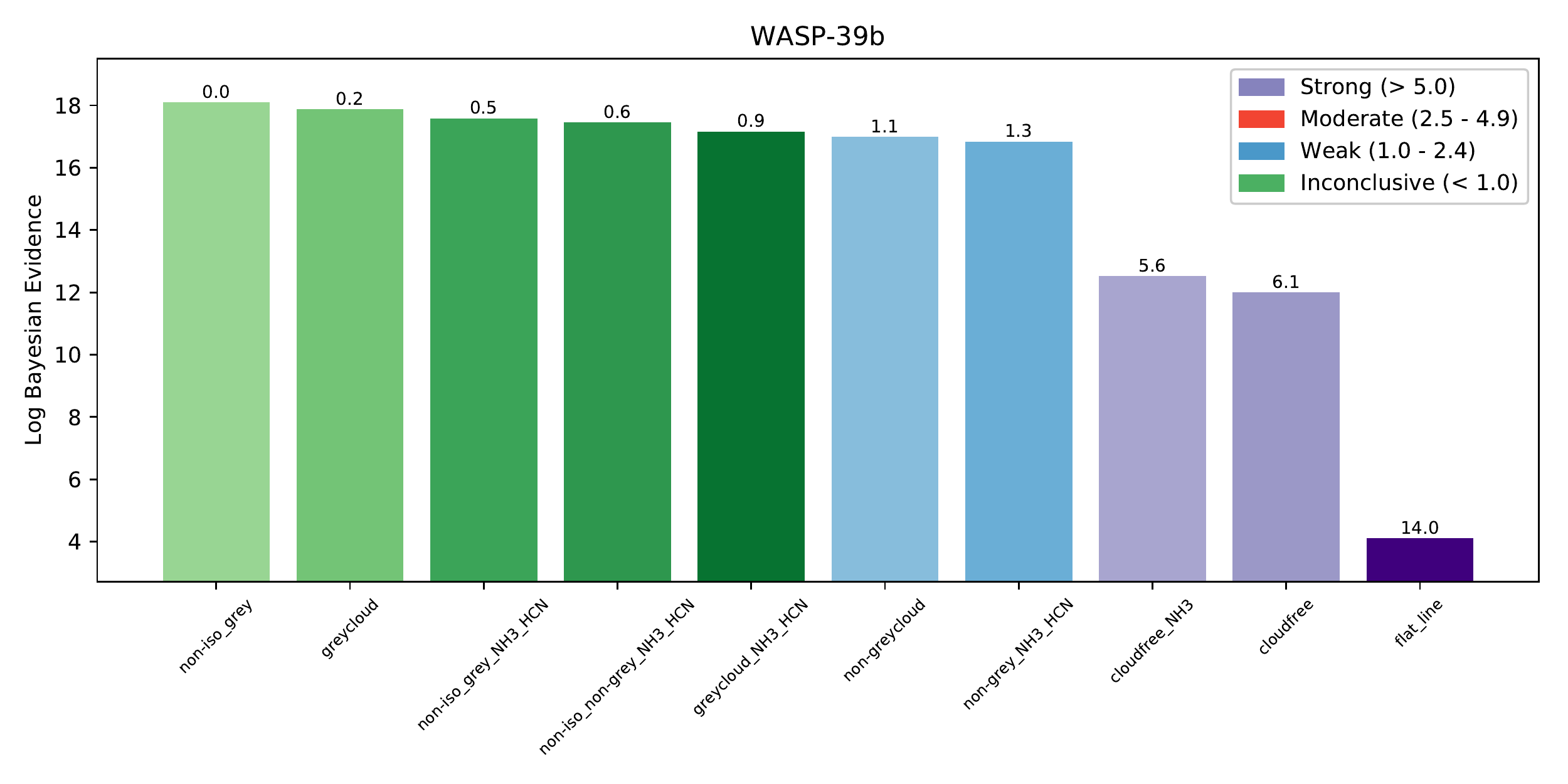}
\includegraphics[width=0.65\columnwidth]{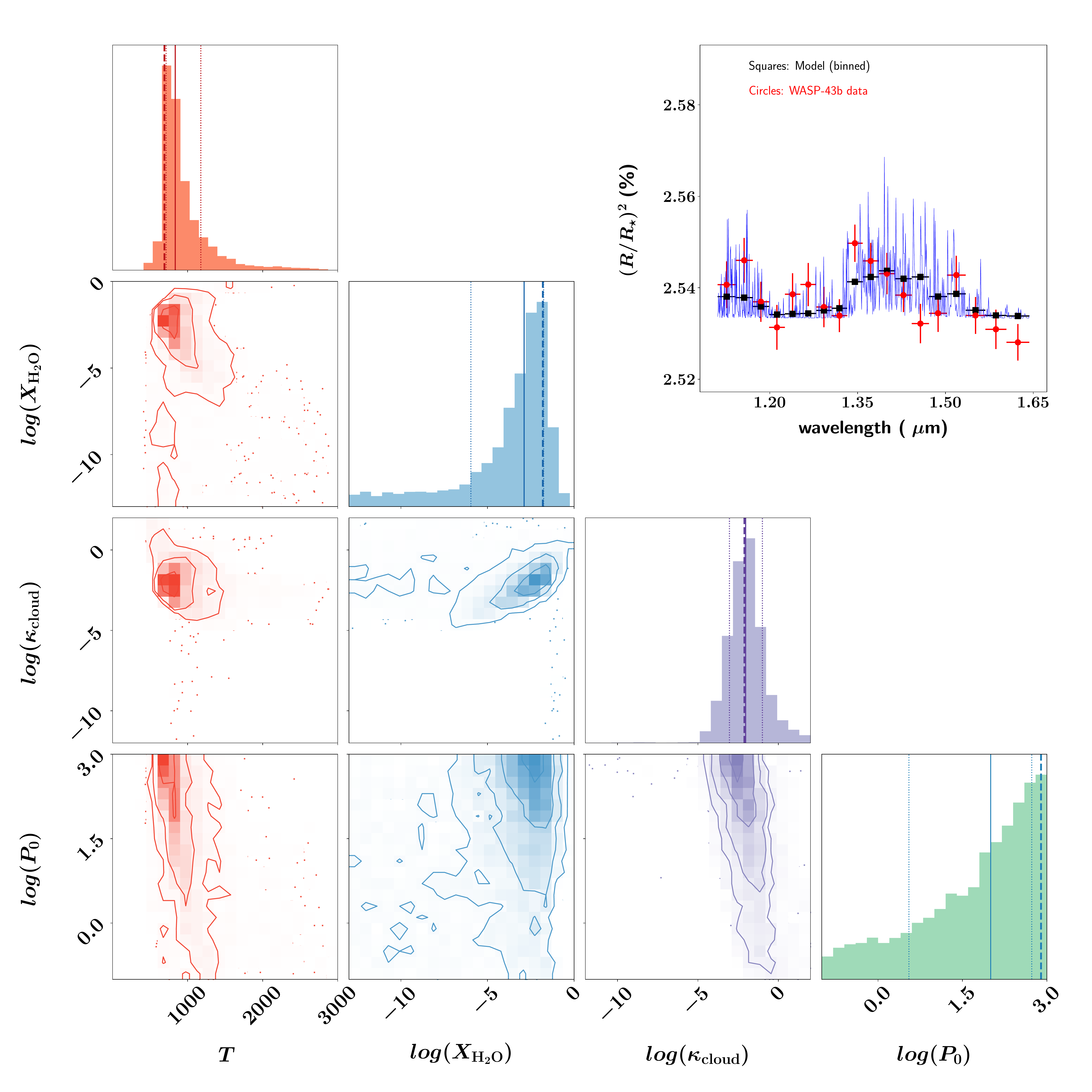}
\hspace{0.1in}
\includegraphics[width=1.2\columnwidth]{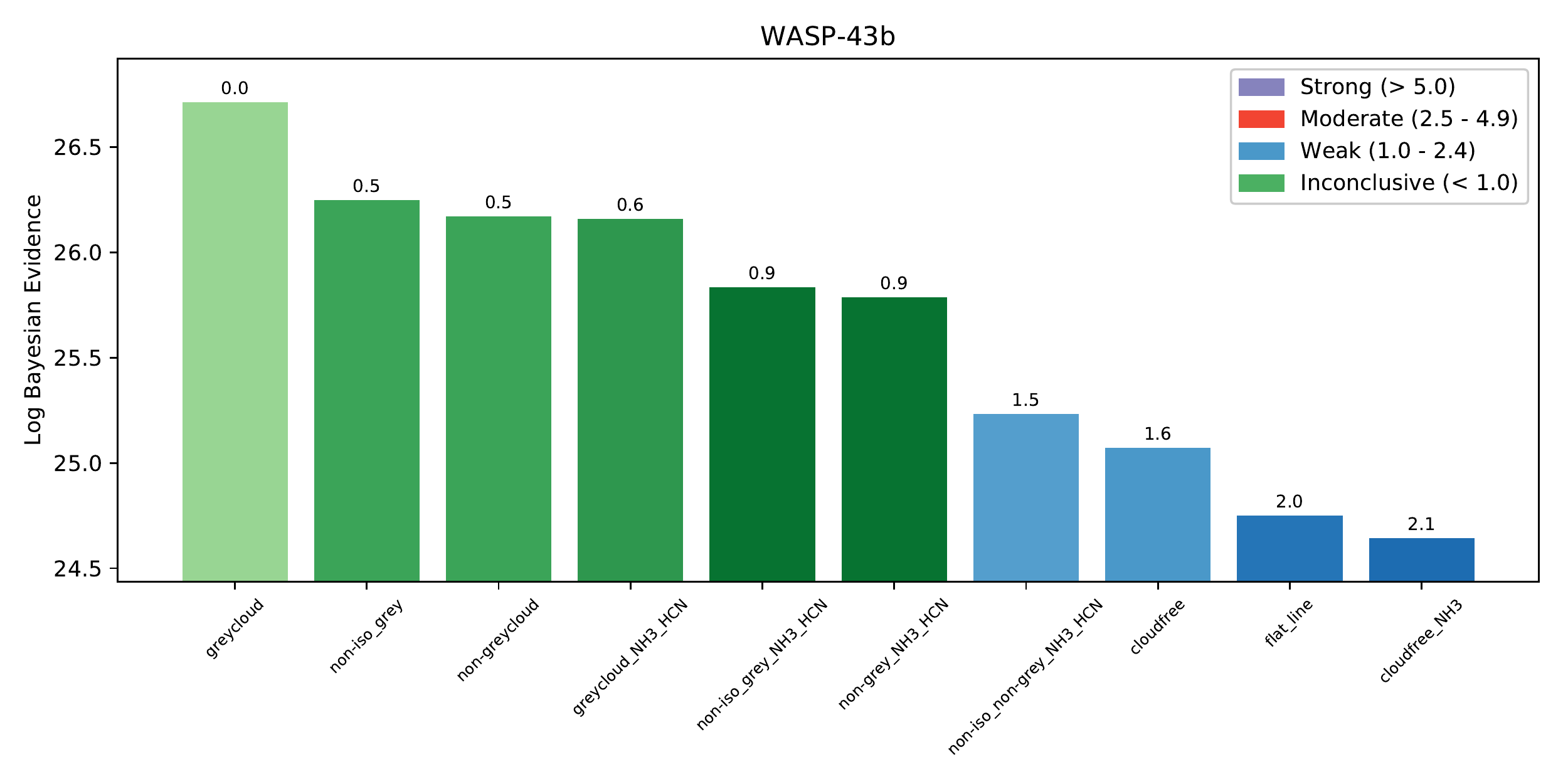}
\includegraphics[width=0.65\columnwidth]{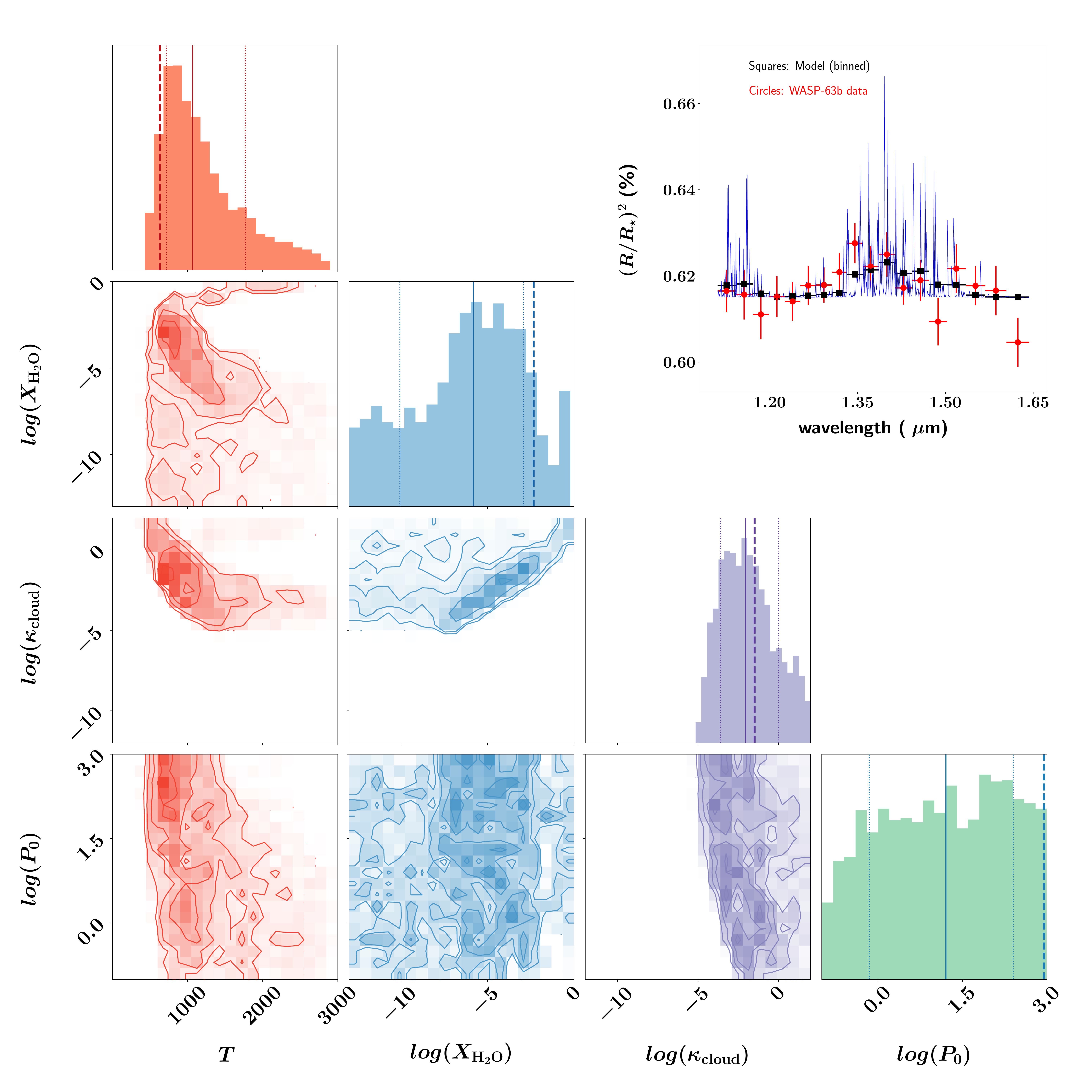}
\hspace{0.1in}
\includegraphics[width=1.2\columnwidth]{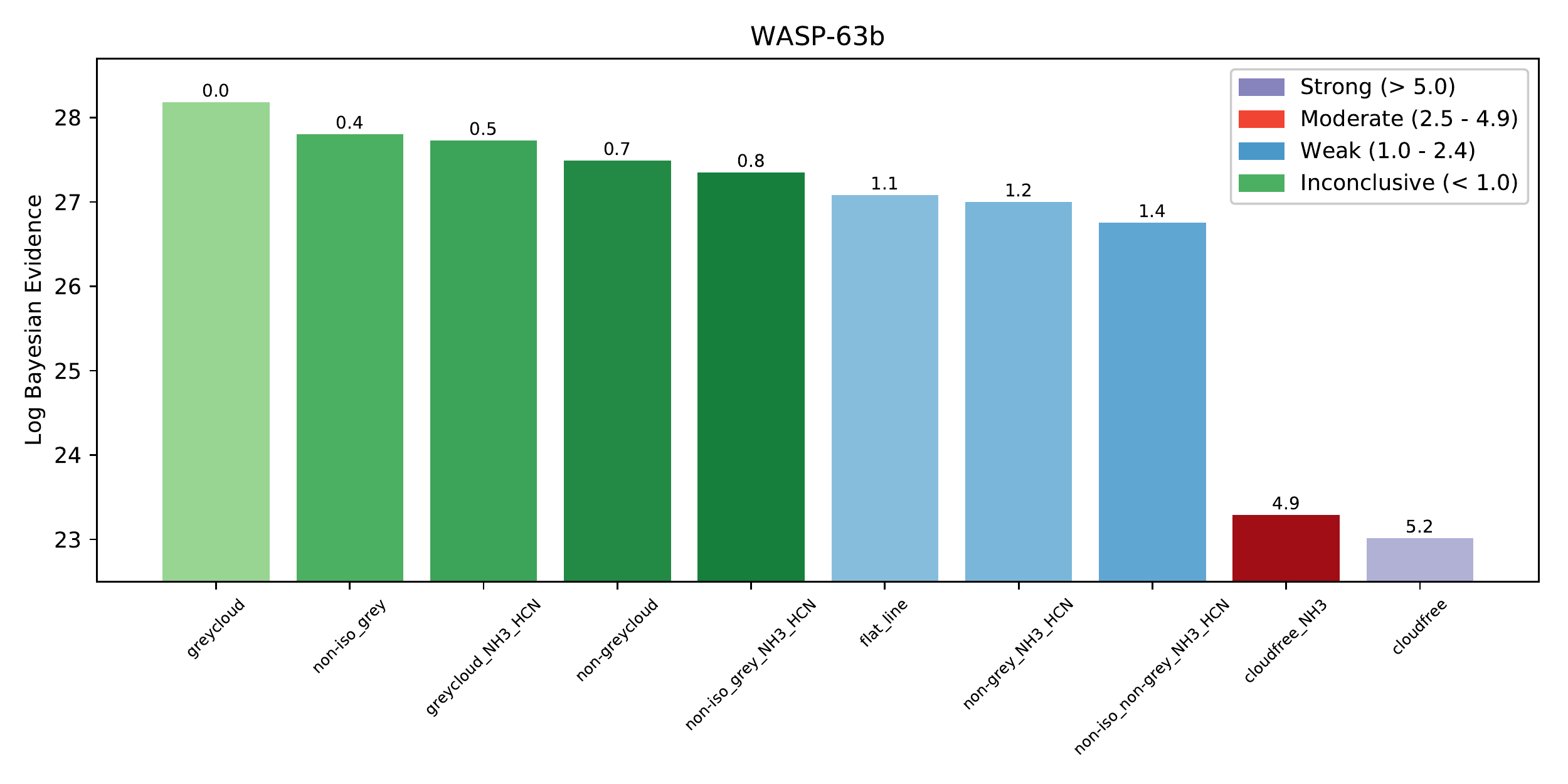}
\end{center}
\vspace{-0.1in}
\caption{Same as Figure \ref{fig:cloud-free_vs_nh3}, but for Early Release Science (ERS) objects: WASP-39b and WASP-43b (for JWST) and WASP-63b (for HST).}
\label{fig:ers}
\end{figure*}

\begin{figure*}
\begin{center}
\includegraphics[width=0.65\columnwidth]{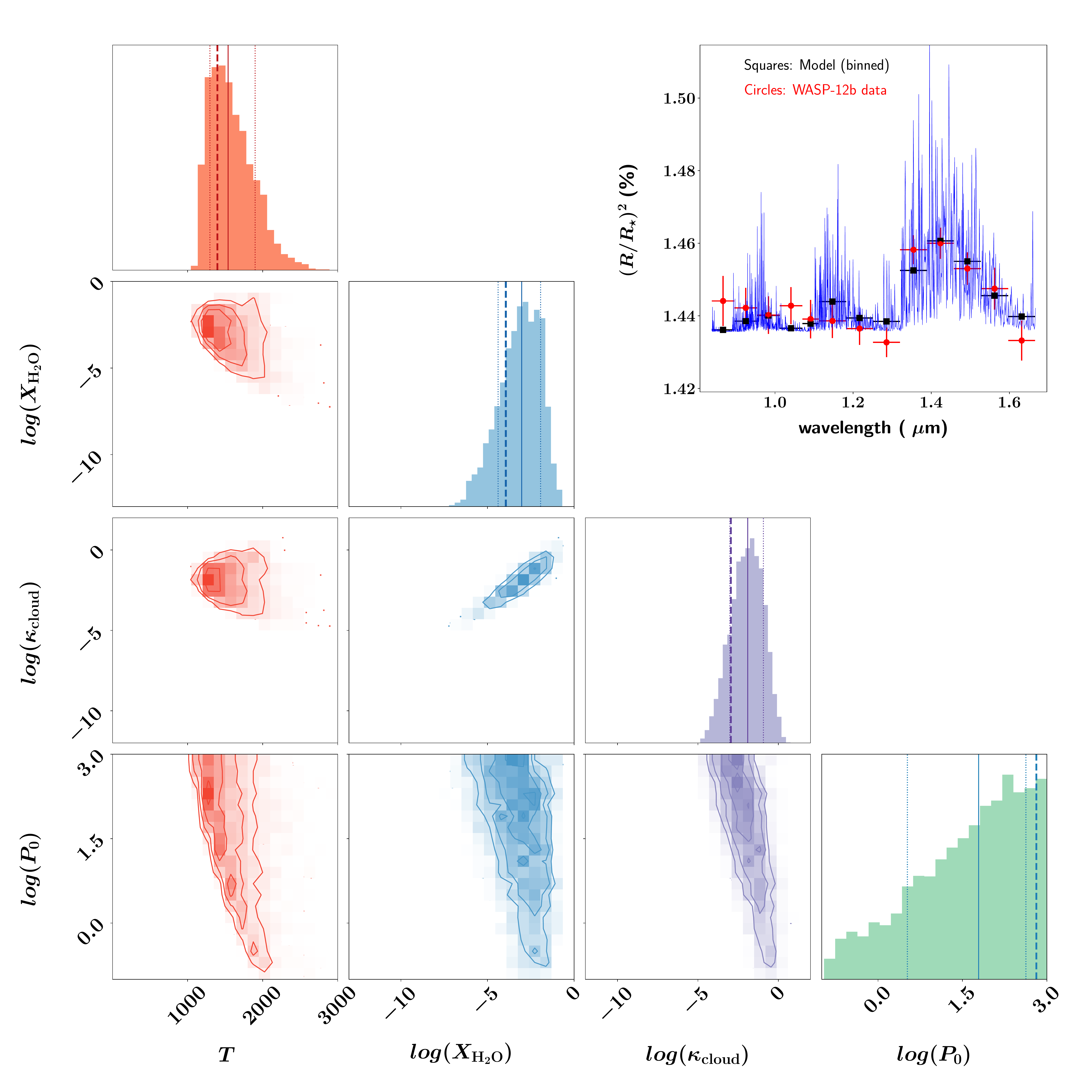}
\hspace{0.1in}
\includegraphics[width=1.2\columnwidth]{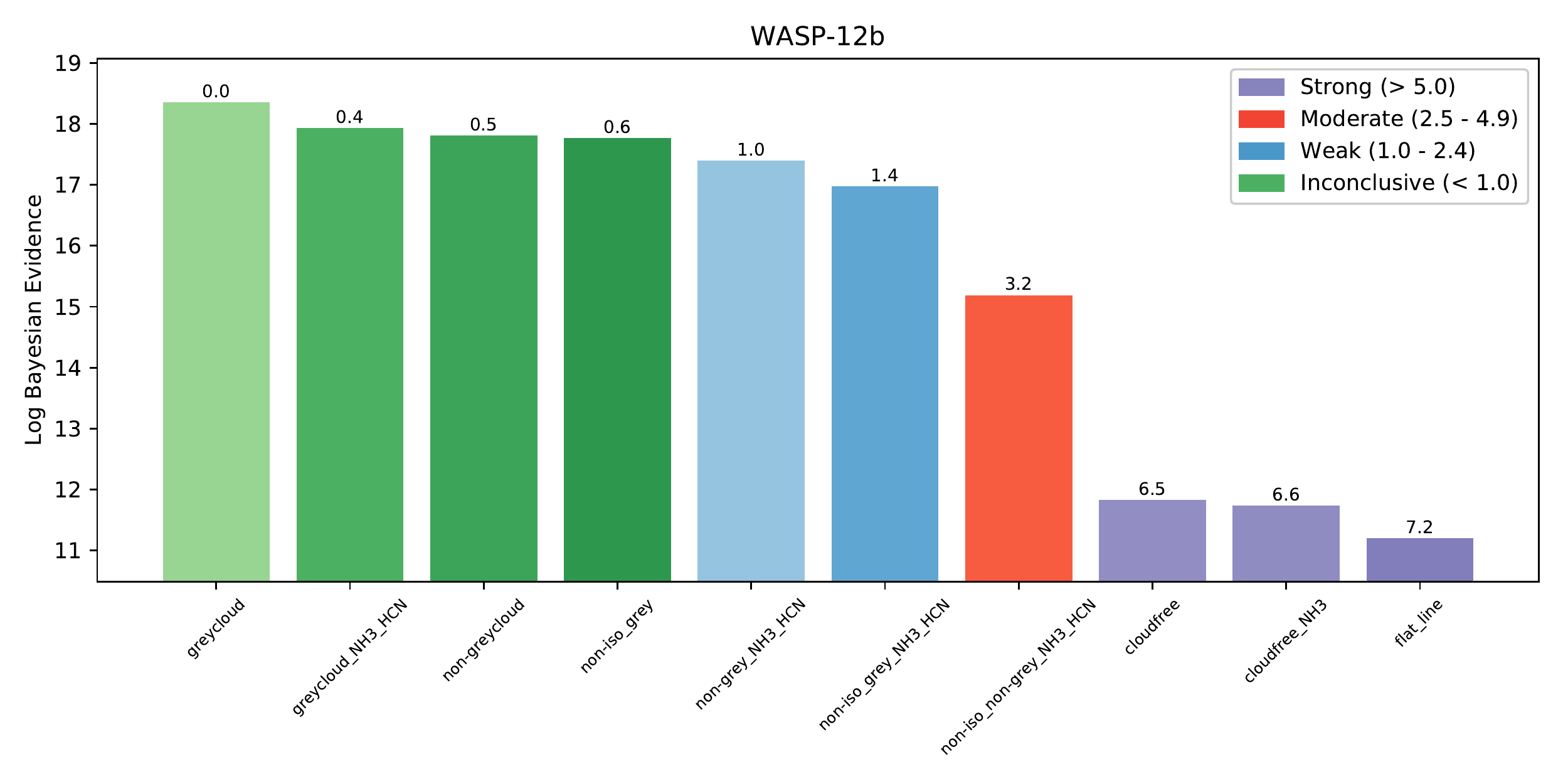}
\includegraphics[width=0.65\columnwidth]{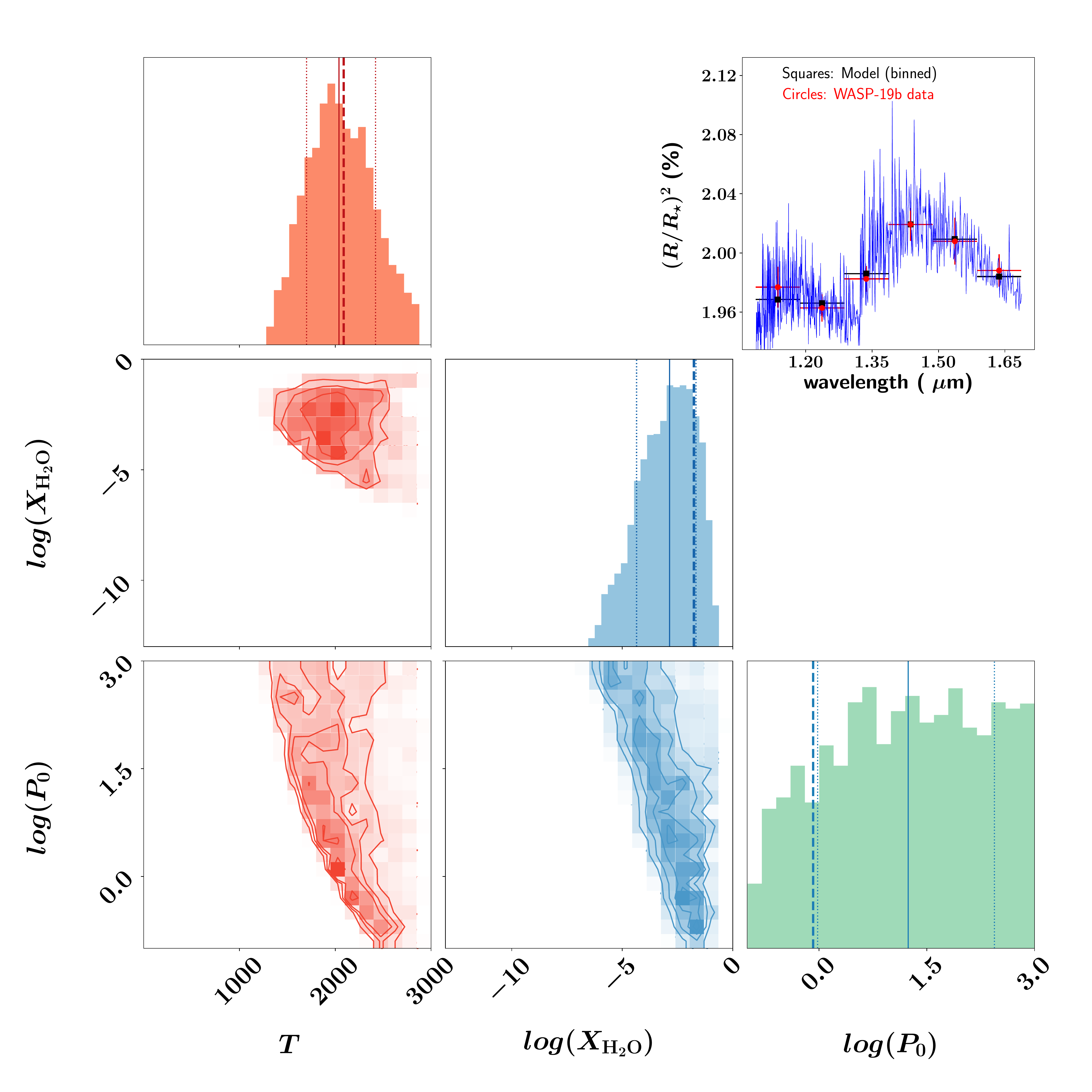}
\hspace{0.1in}
\includegraphics[width=1.2\columnwidth]{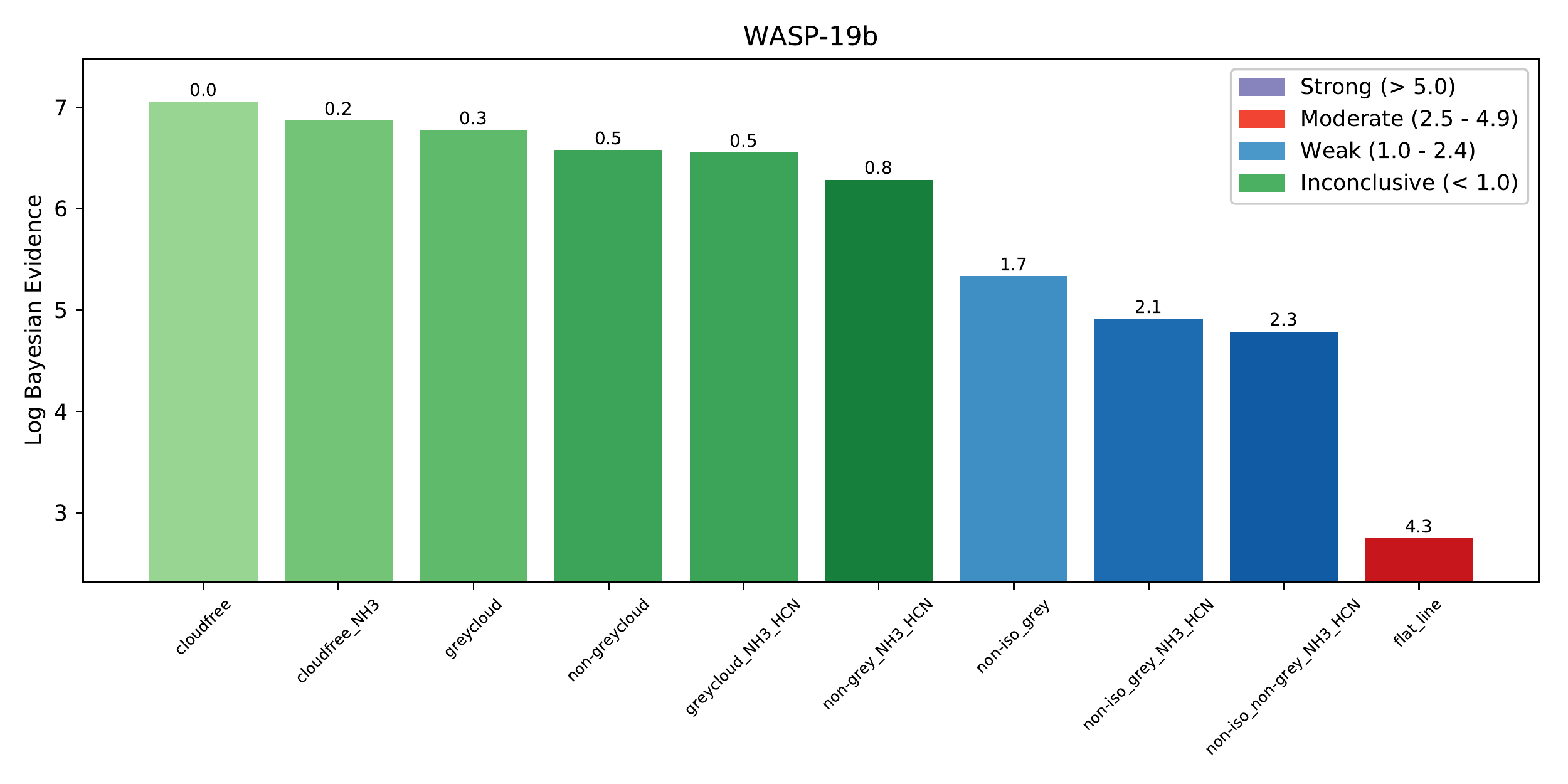}
\includegraphics[width=0.65\columnwidth]{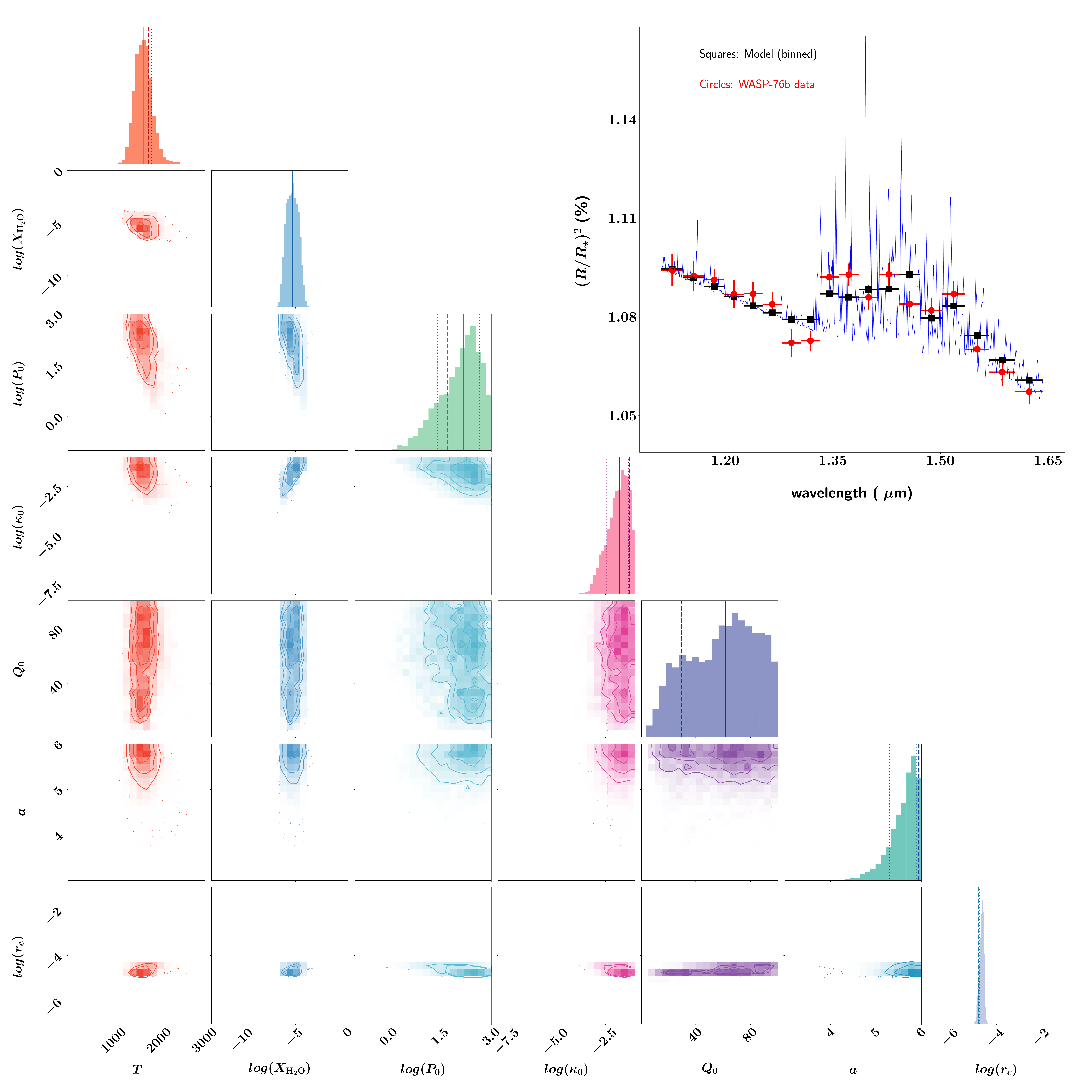}
\hspace{0.1in}
\includegraphics[width=1.2\columnwidth]{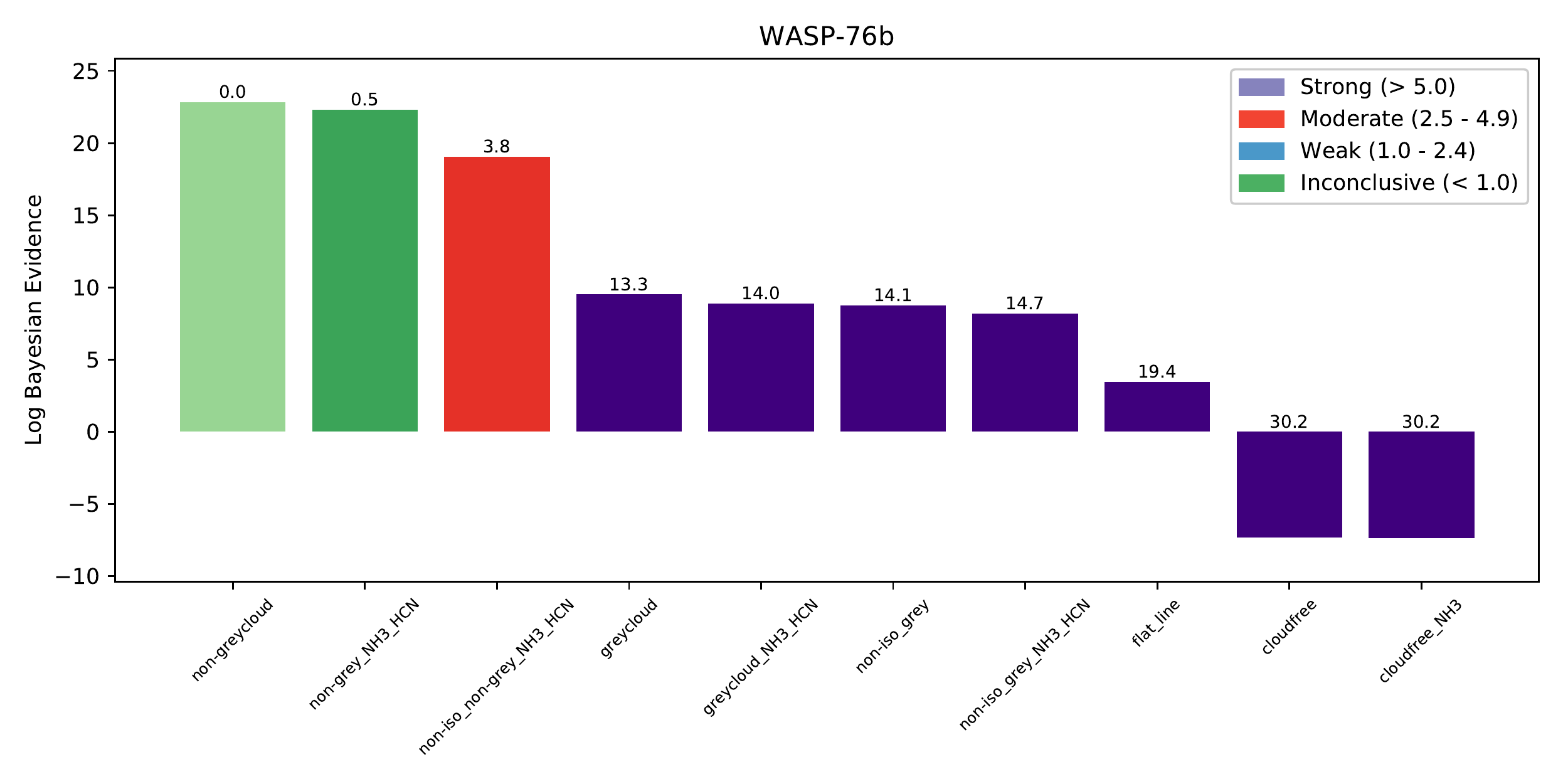}
\includegraphics[width=0.65\columnwidth]{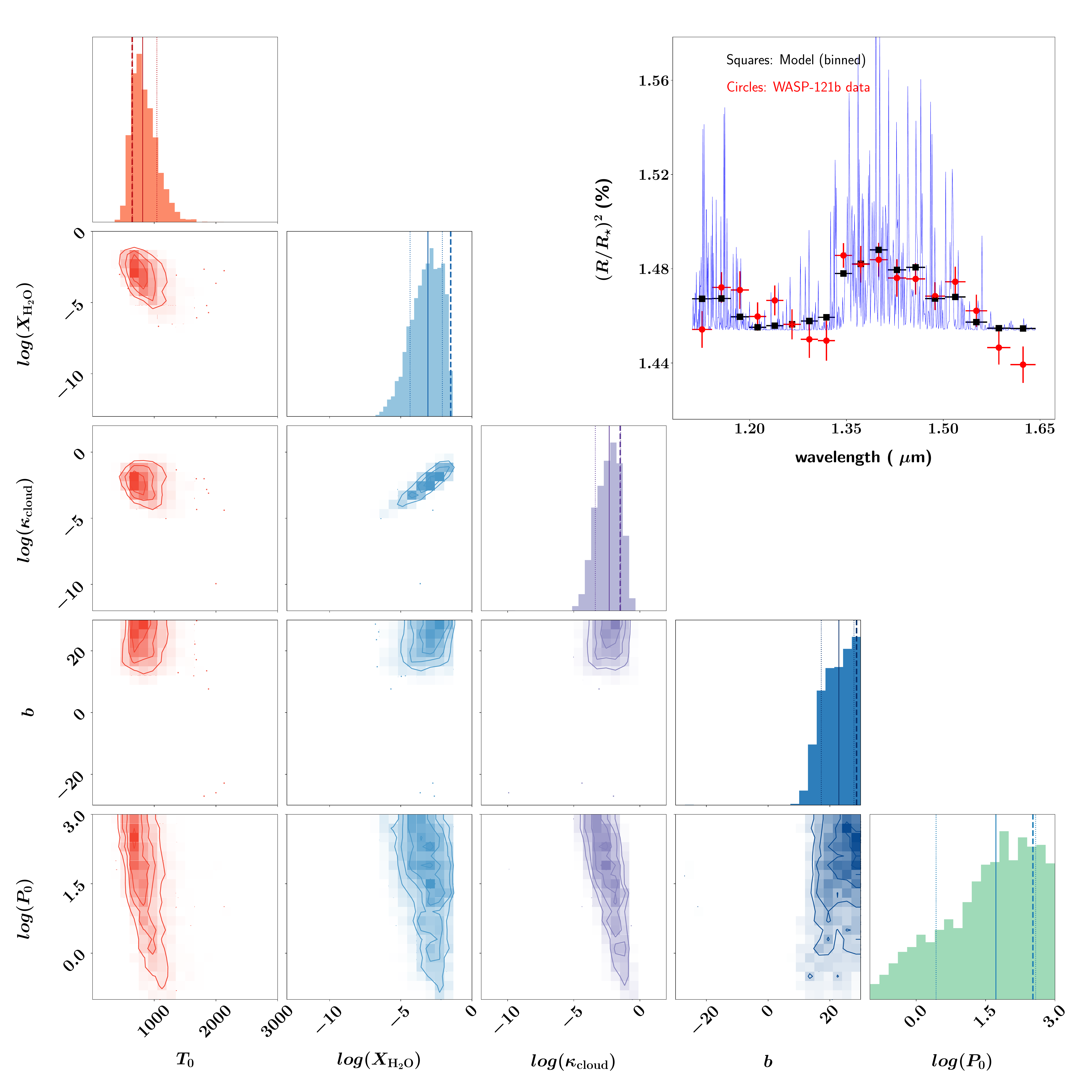}
\hspace{0.1in}
\includegraphics[width=1.2\columnwidth]{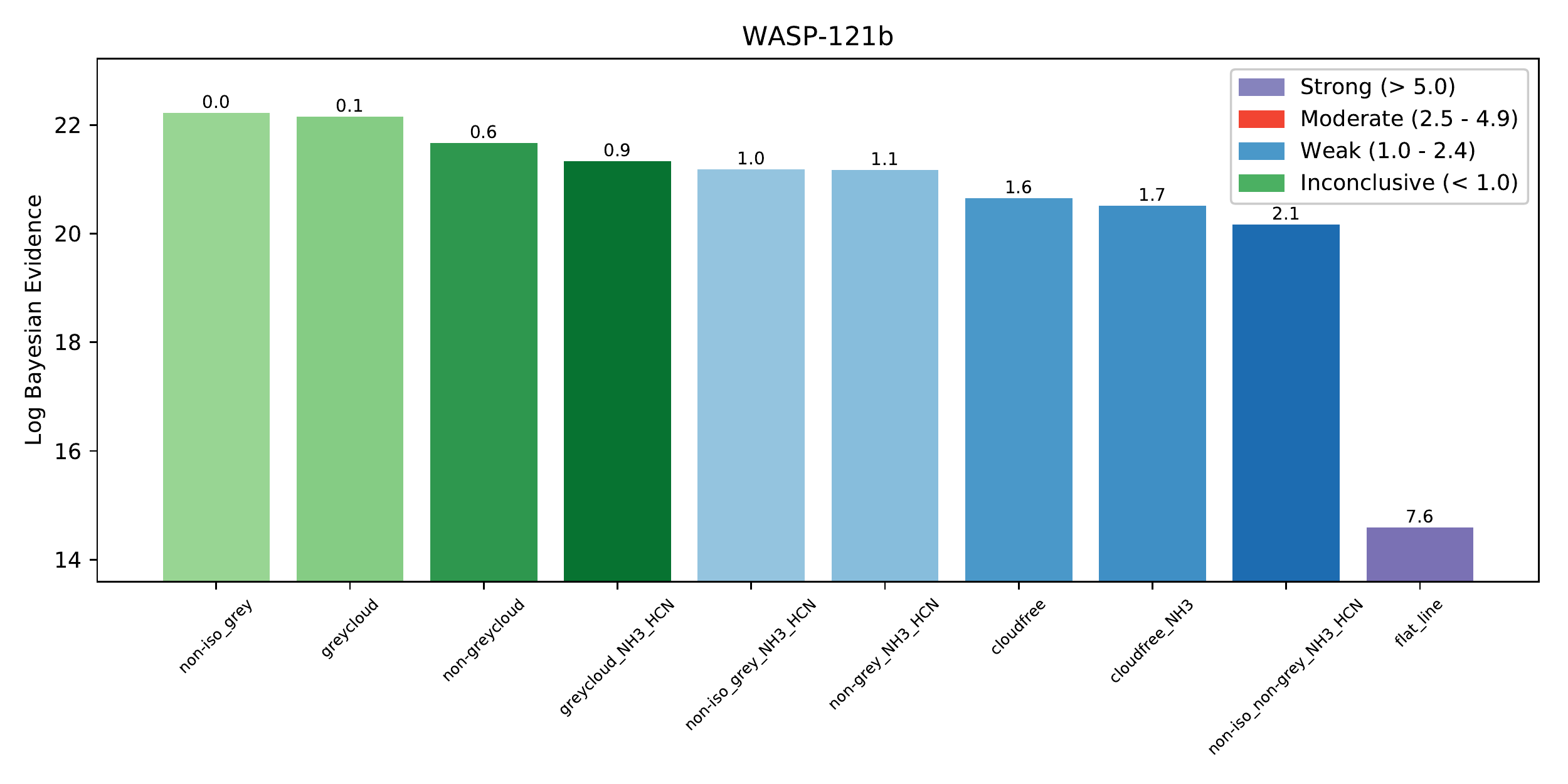}
\end{center}
\vspace{-0.1in}
\caption{Same as Figure \ref{fig:cloud-free_vs_nh3}, but for very hot Jupiters ($T_{\rm eq}>2000$ K): WASP-12b, WASP-19b, WASP-76b and WASP-121b.  WASP-19b is one of two objects with the highest Bayesian evidence for the cloudfree, isothermal model with water only (excluding the TRAPPIST-1 exoplanets).  WASP-76b is one of two objects where non-grey clouds are needed to fit the data.}
\label{fig:veryhot}
\end{figure*}

\begin{figure*}
\begin{center}
\includegraphics[width=0.65\columnwidth]{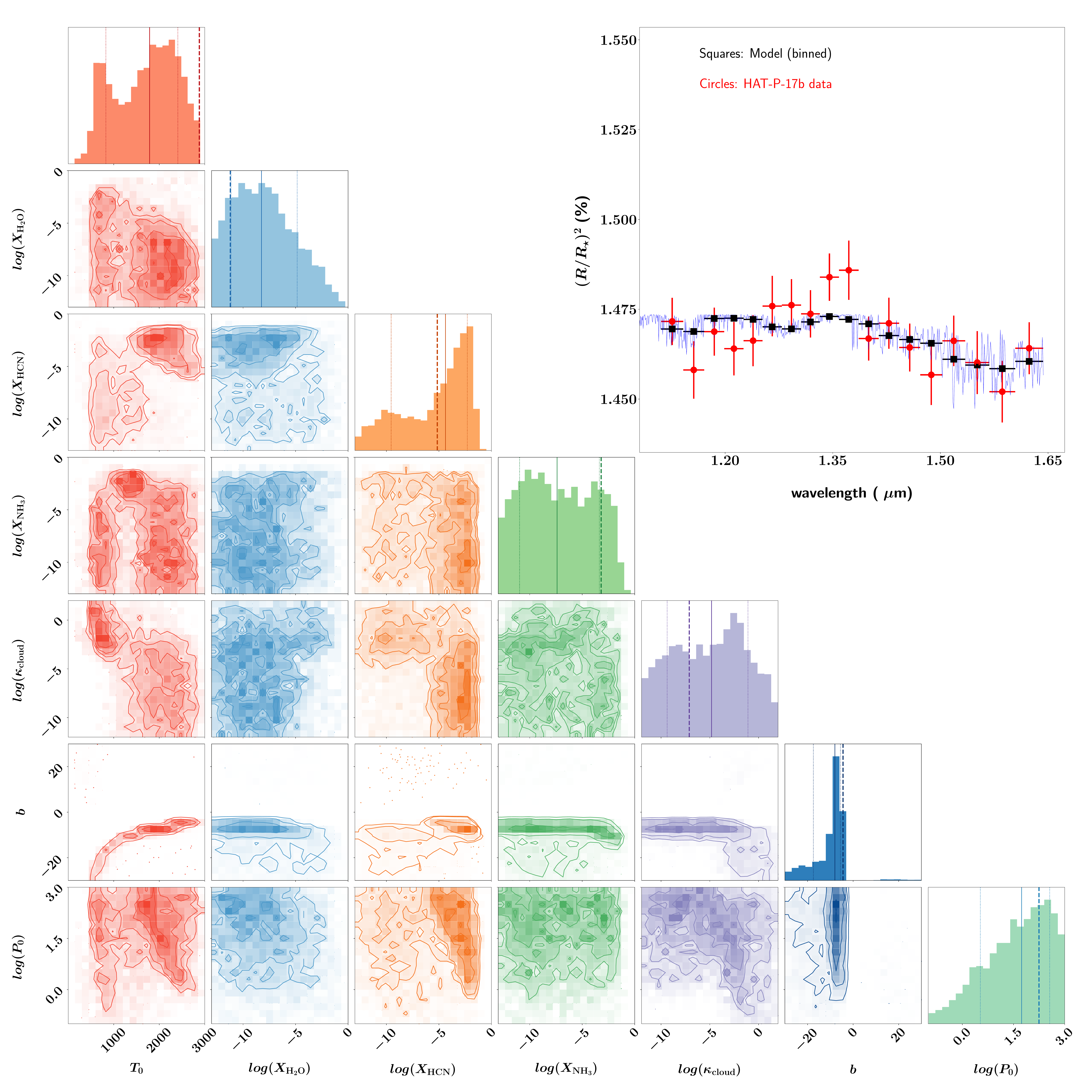}
\hspace{0.1in}
\includegraphics[width=1.2\columnwidth]{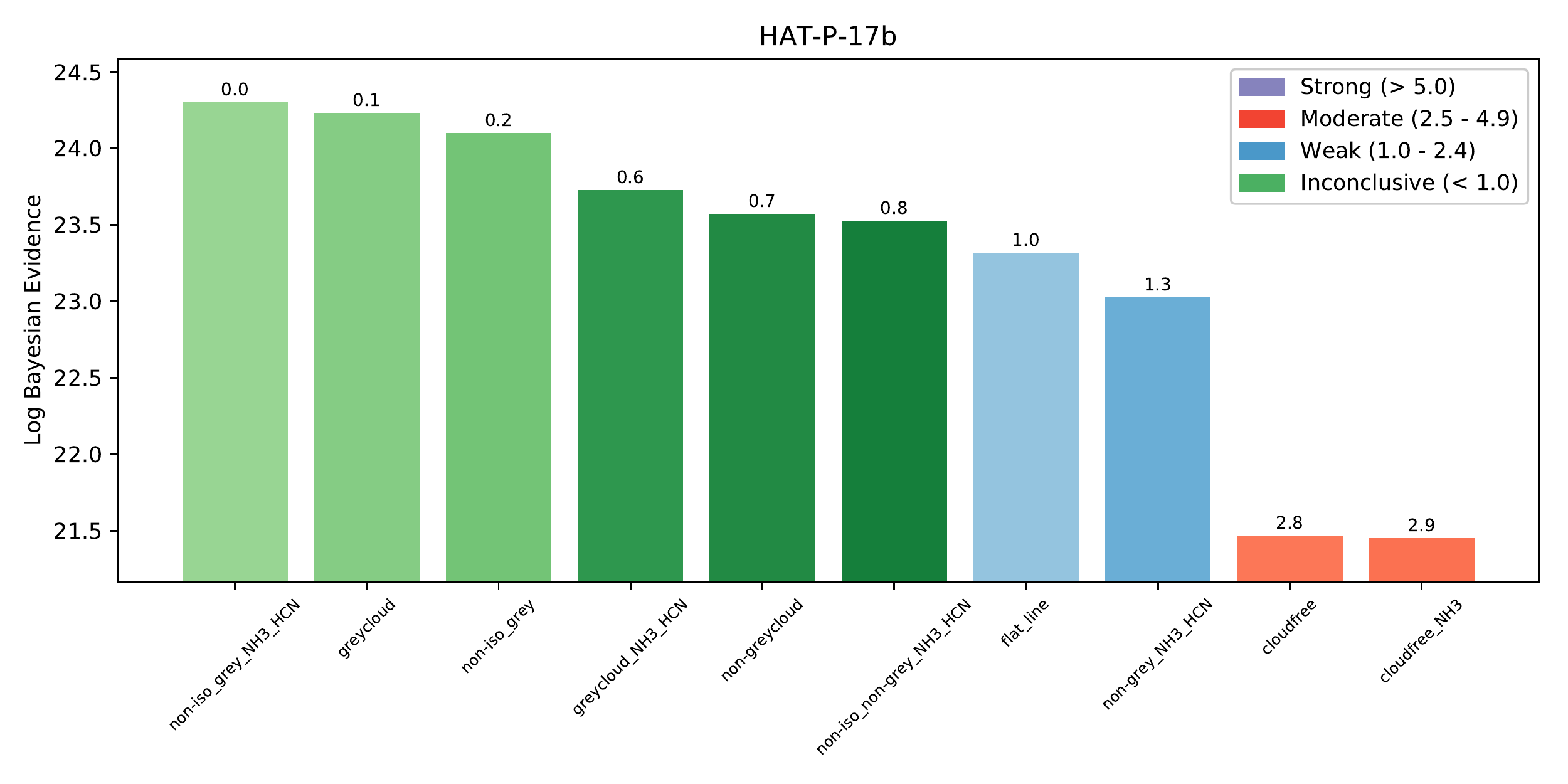}
\includegraphics[width=0.65\columnwidth]{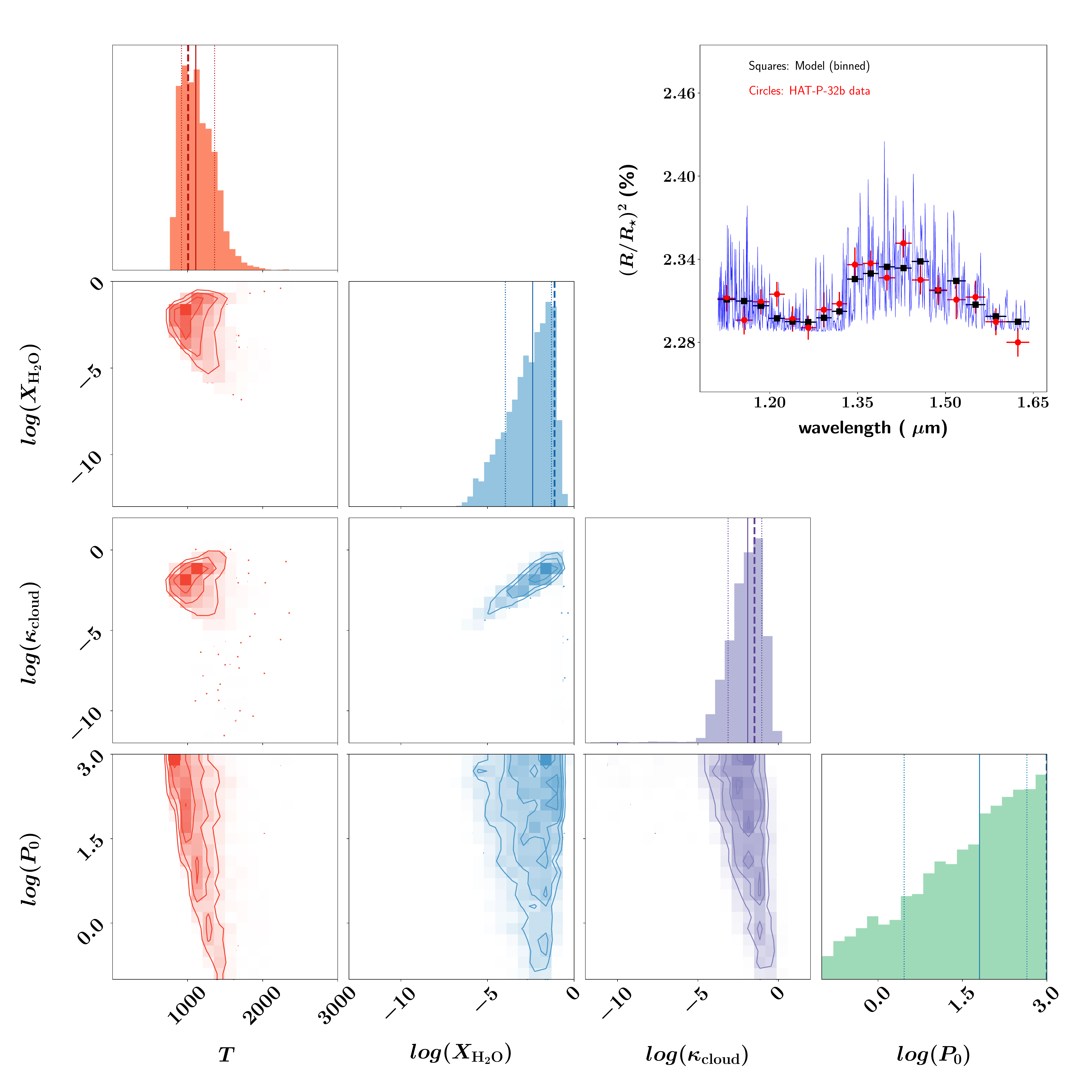}
\hspace{0.1in}
\includegraphics[width=1.2\columnwidth]{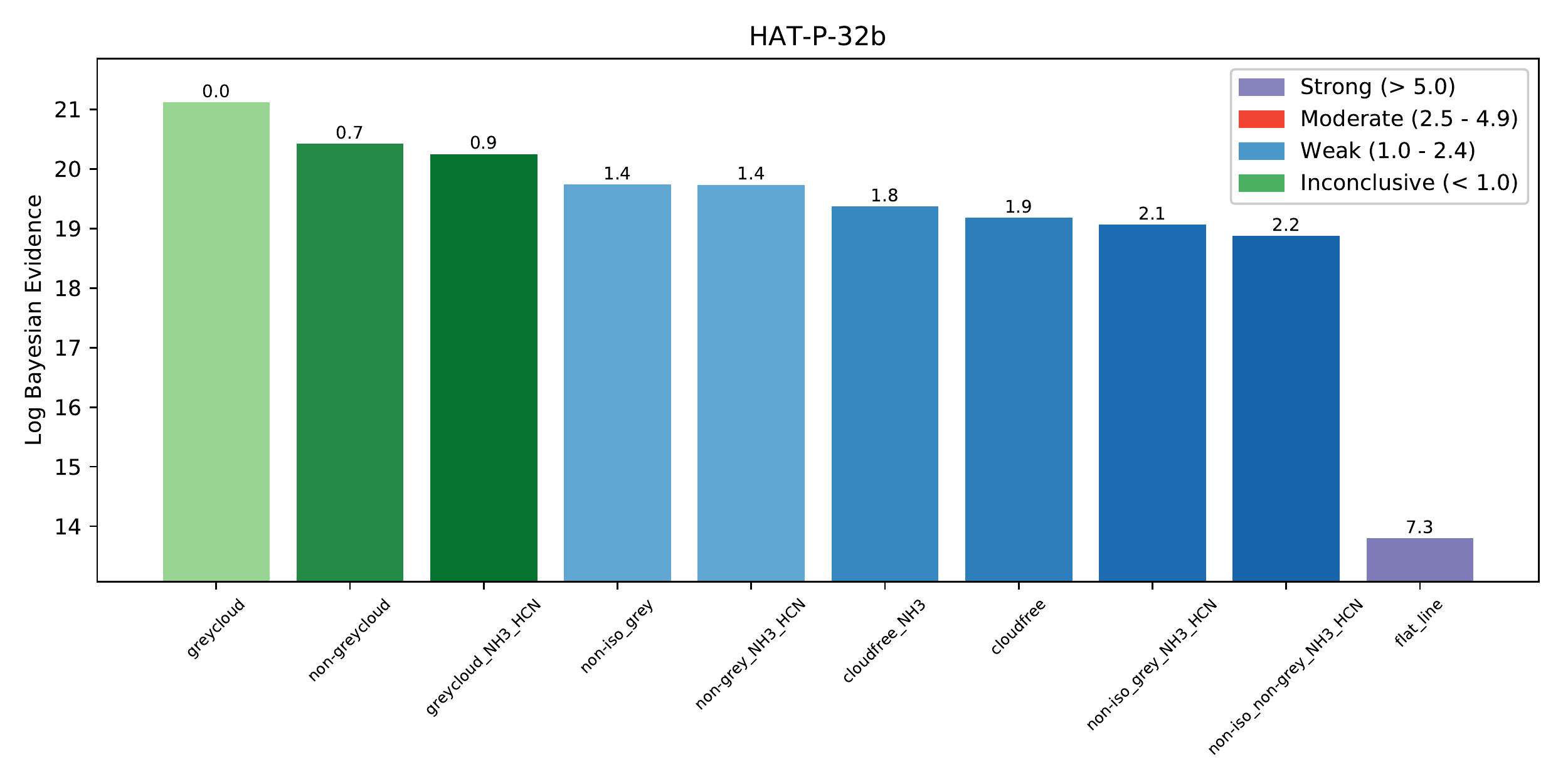}
\includegraphics[width=0.65\columnwidth]{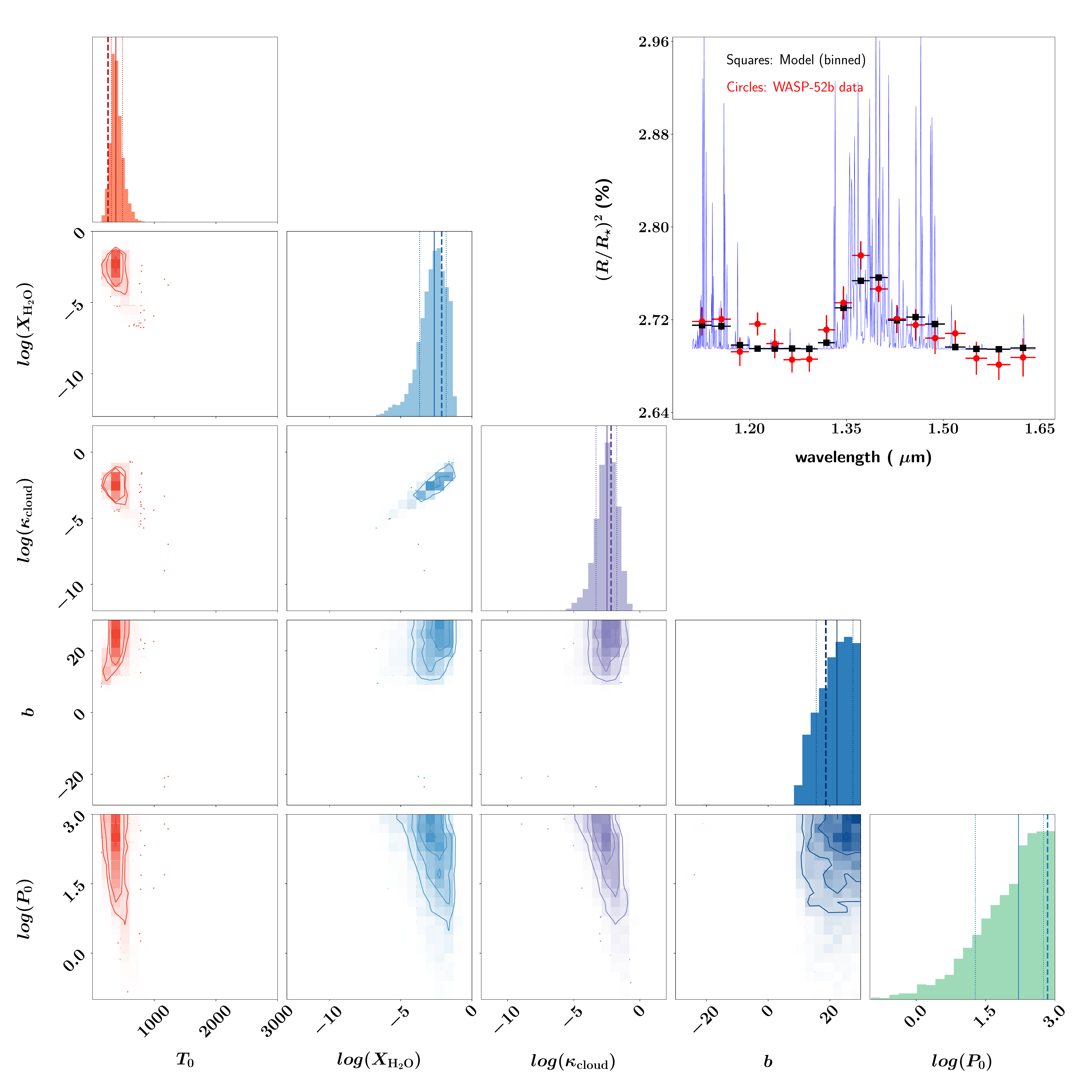}
\hspace{0.1in}
\includegraphics[width=1.2\columnwidth]{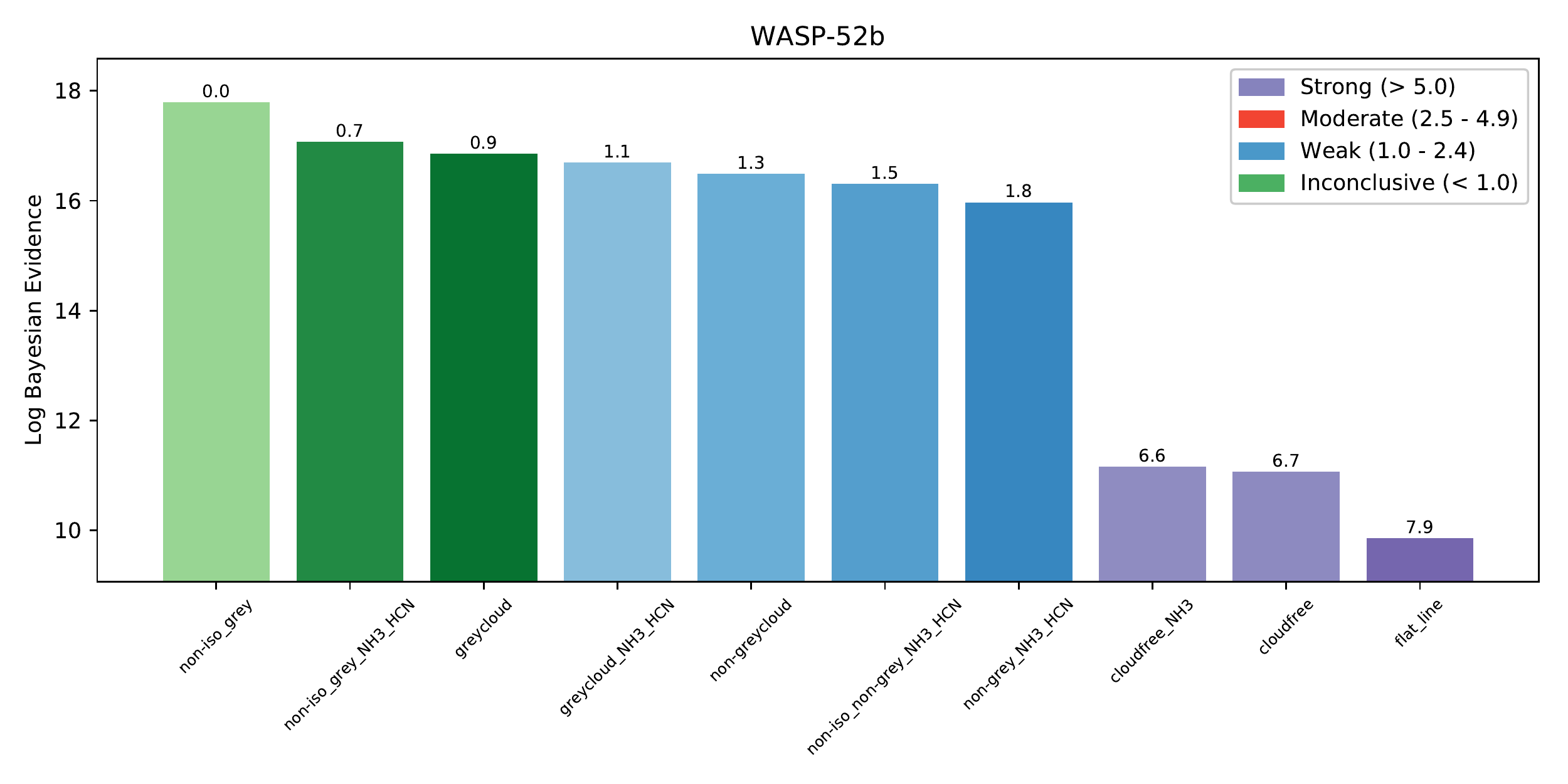}
\includegraphics[width=0.65\columnwidth]{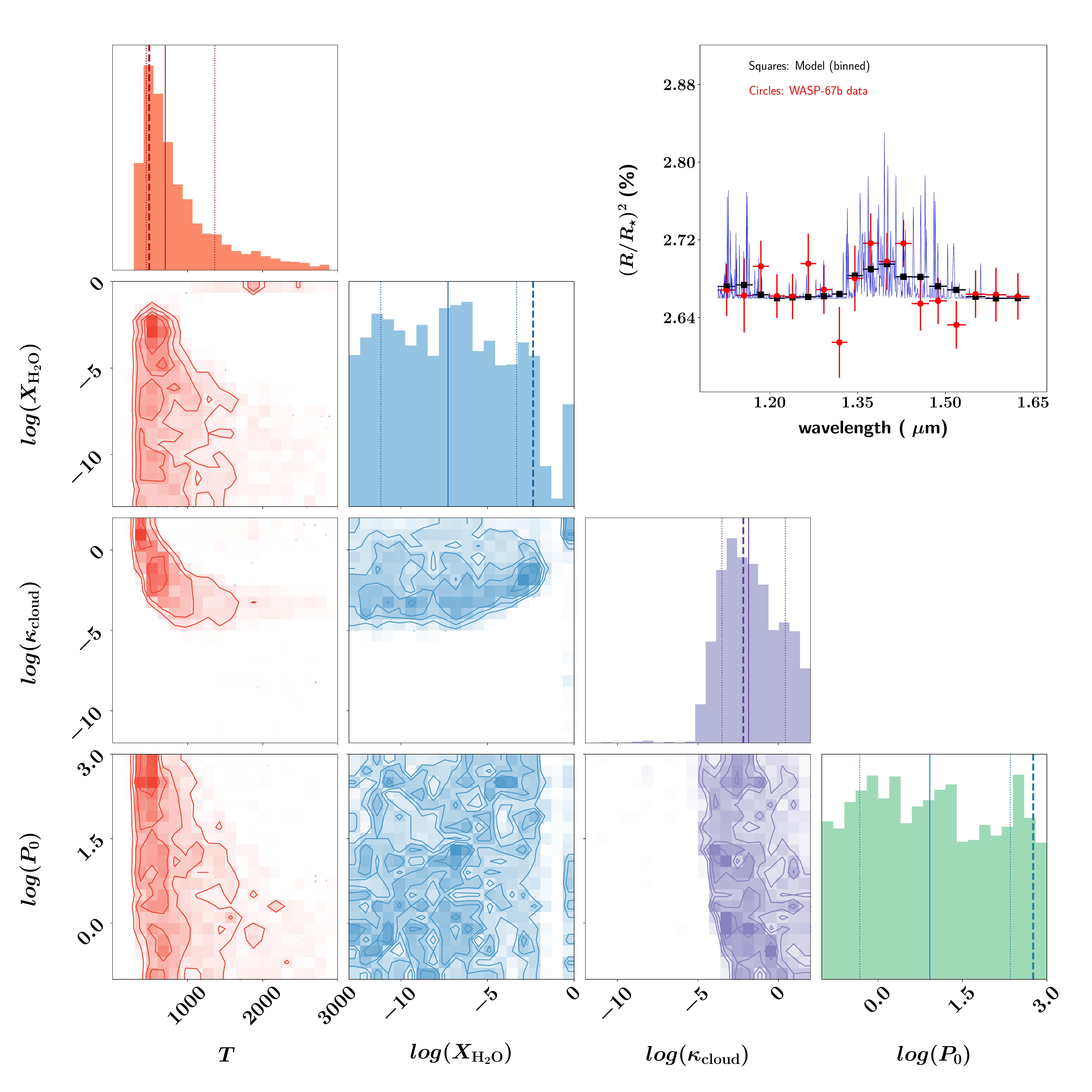}
\hspace{0.1in}
\includegraphics[width=1.2\columnwidth]{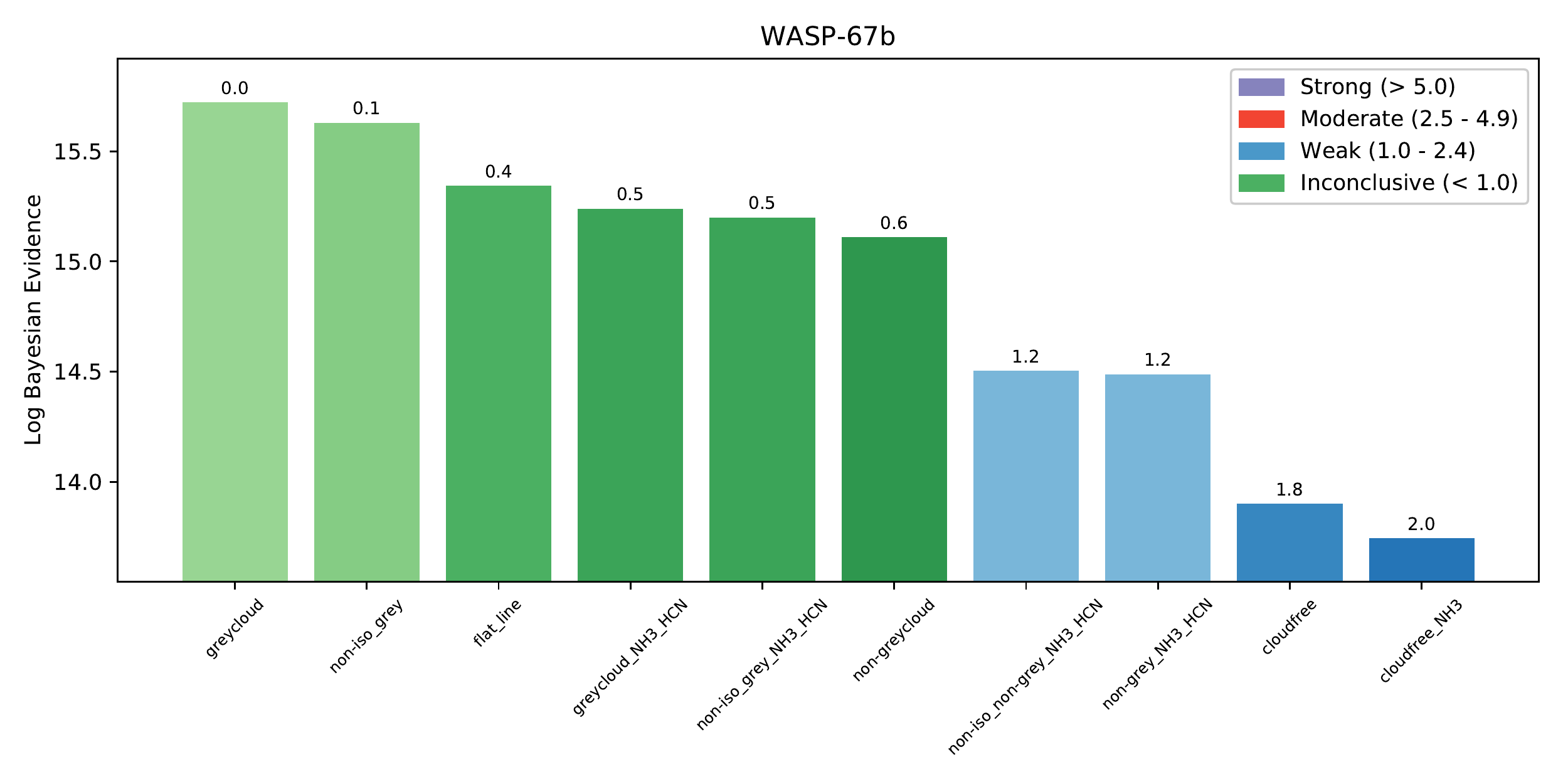}
\end{center}
\vspace{-0.1in}
\caption{Same as Figure \ref{fig:cloud-free_vs_nh3}, but for the rest of the hot Jupiters: HAT-P-17b, HAT-P-32b, WASP-52b and WASP-67b.}
\label{fig:others_1}
\end{figure*}

\begin{figure*}
\begin{center}
\includegraphics[width=0.65\columnwidth]{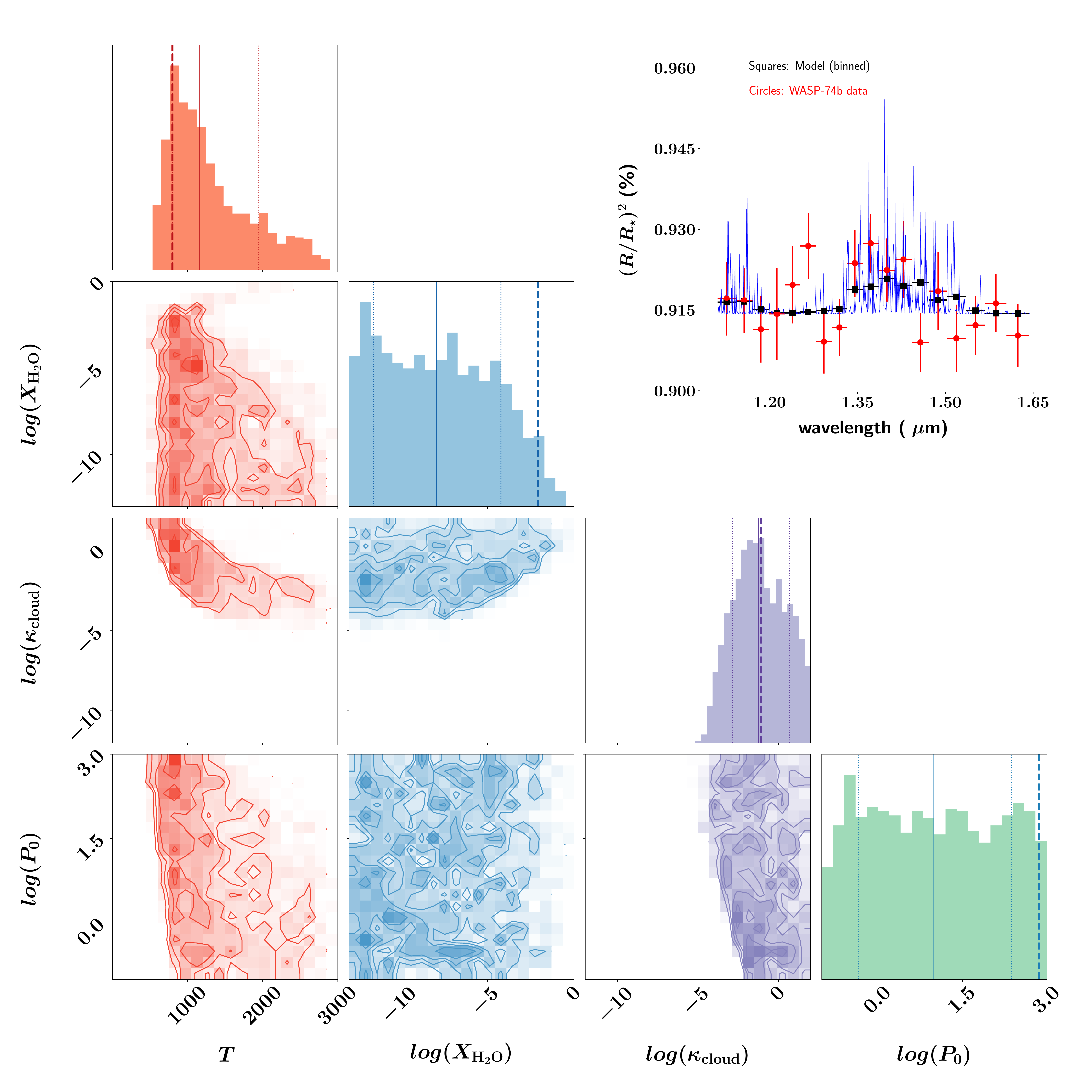}
\hspace{0.1in}
\includegraphics[width=1.2\columnwidth]{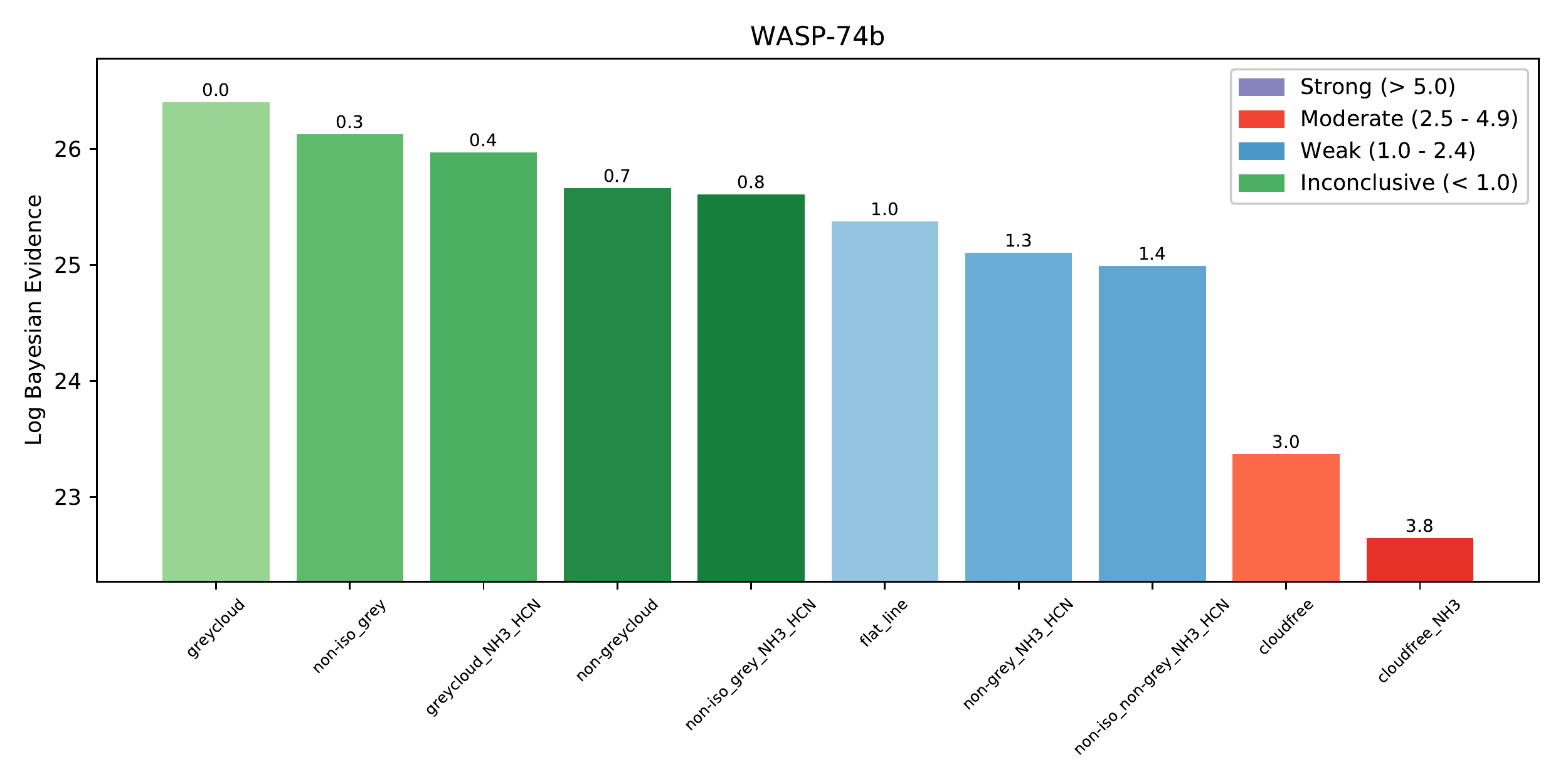}
\includegraphics[width=0.65\columnwidth]{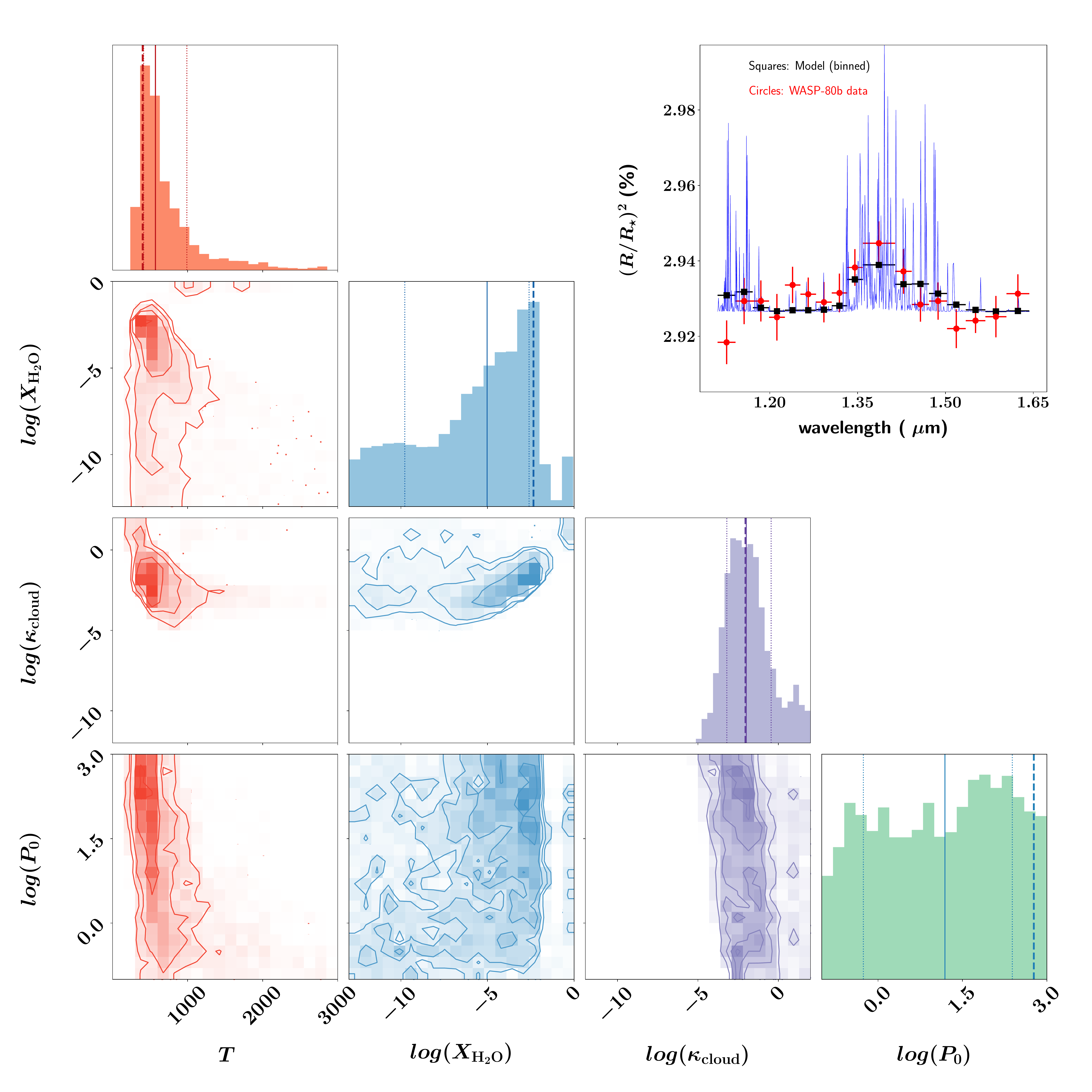}
\hspace{0.1in}
\includegraphics[width=1.2\columnwidth]{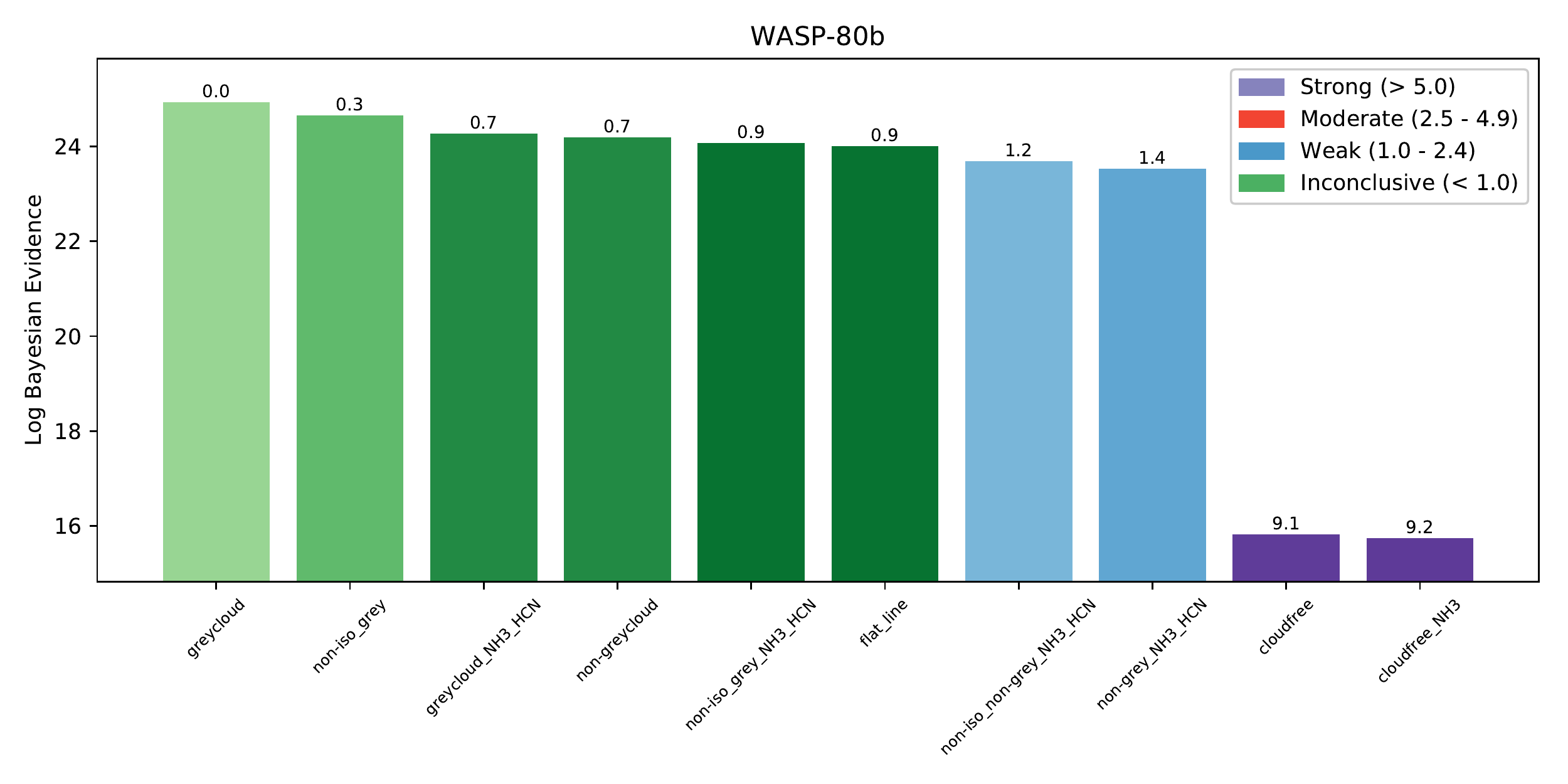}
\includegraphics[width=0.65\columnwidth]{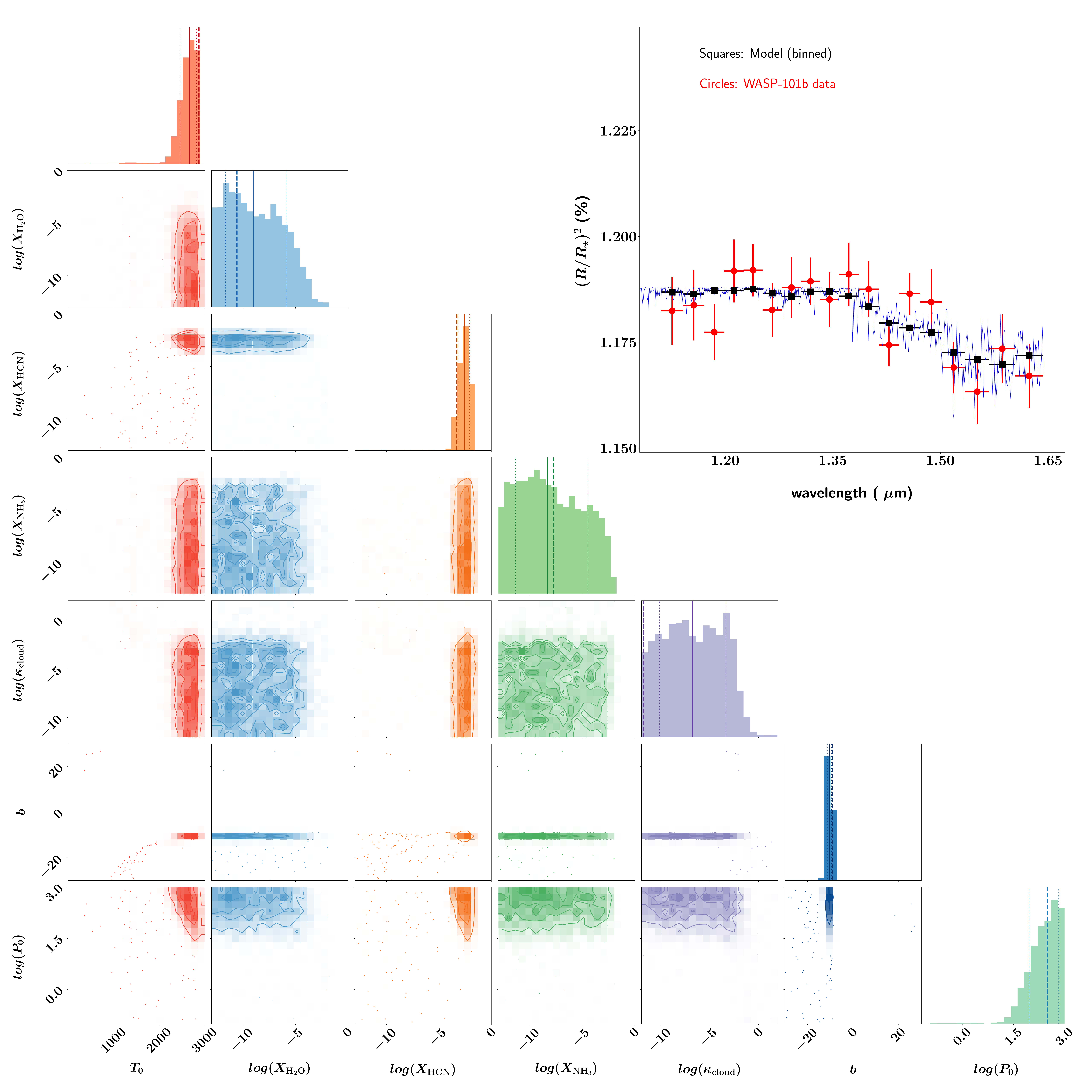}
\hspace{0.1in}
\includegraphics[width=1.2\columnwidth]{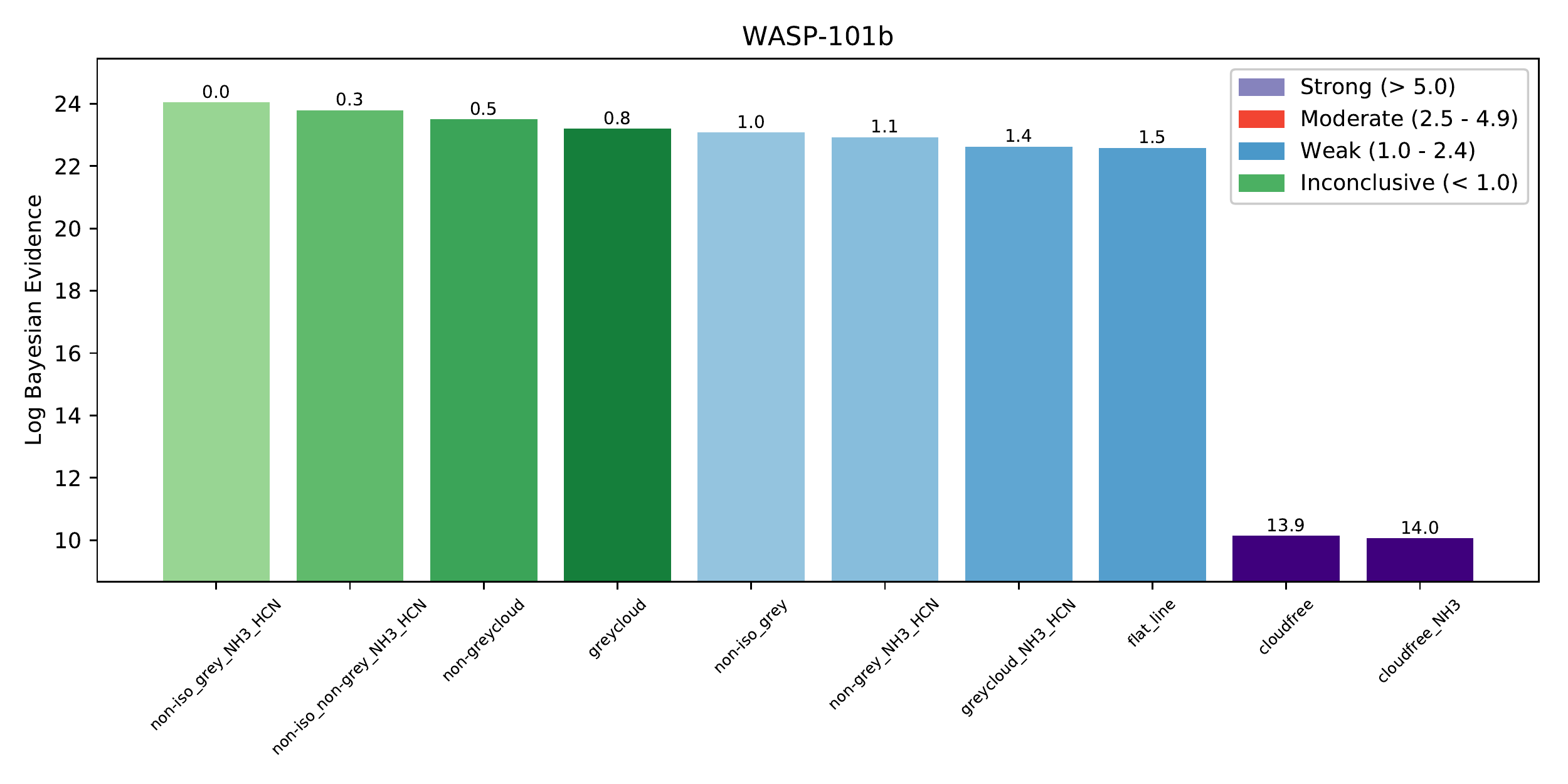}
\end{center}
\vspace{-0.1in}
\caption{Continuation of Figure \ref{fig:others_1} for the rest of the hot Jupiters: WASP-74b, WASP-80b and WASP-101b.  WASP-101b stands out as the only object for which HCN is significantly detected over water and ammonia.}
\label{fig:others_2}
\end{figure*}

\begin{figure*}
\begin{center}
\includegraphics[width=0.65\columnwidth]{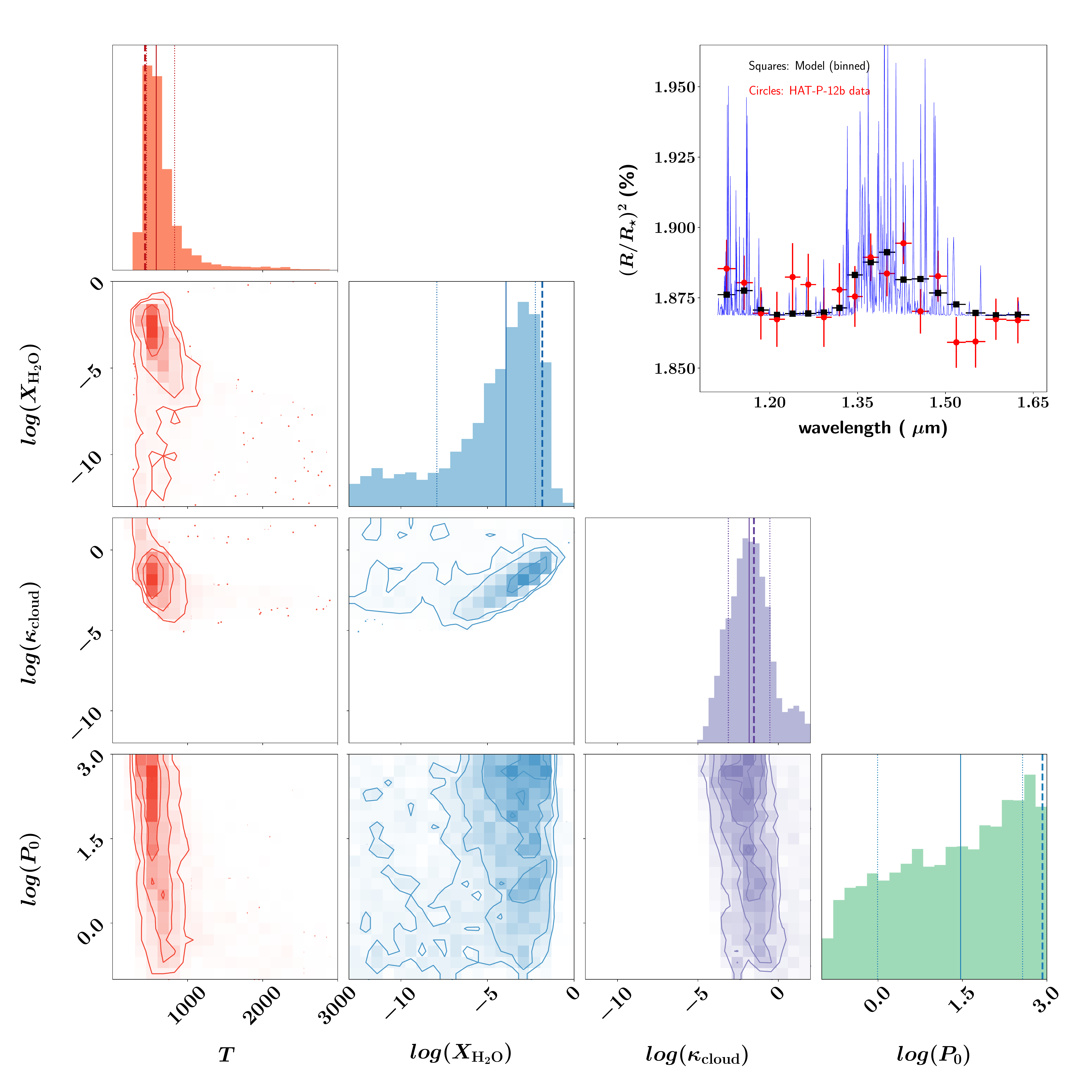}
\hspace{0.1in}
\includegraphics[width=1.2\columnwidth]{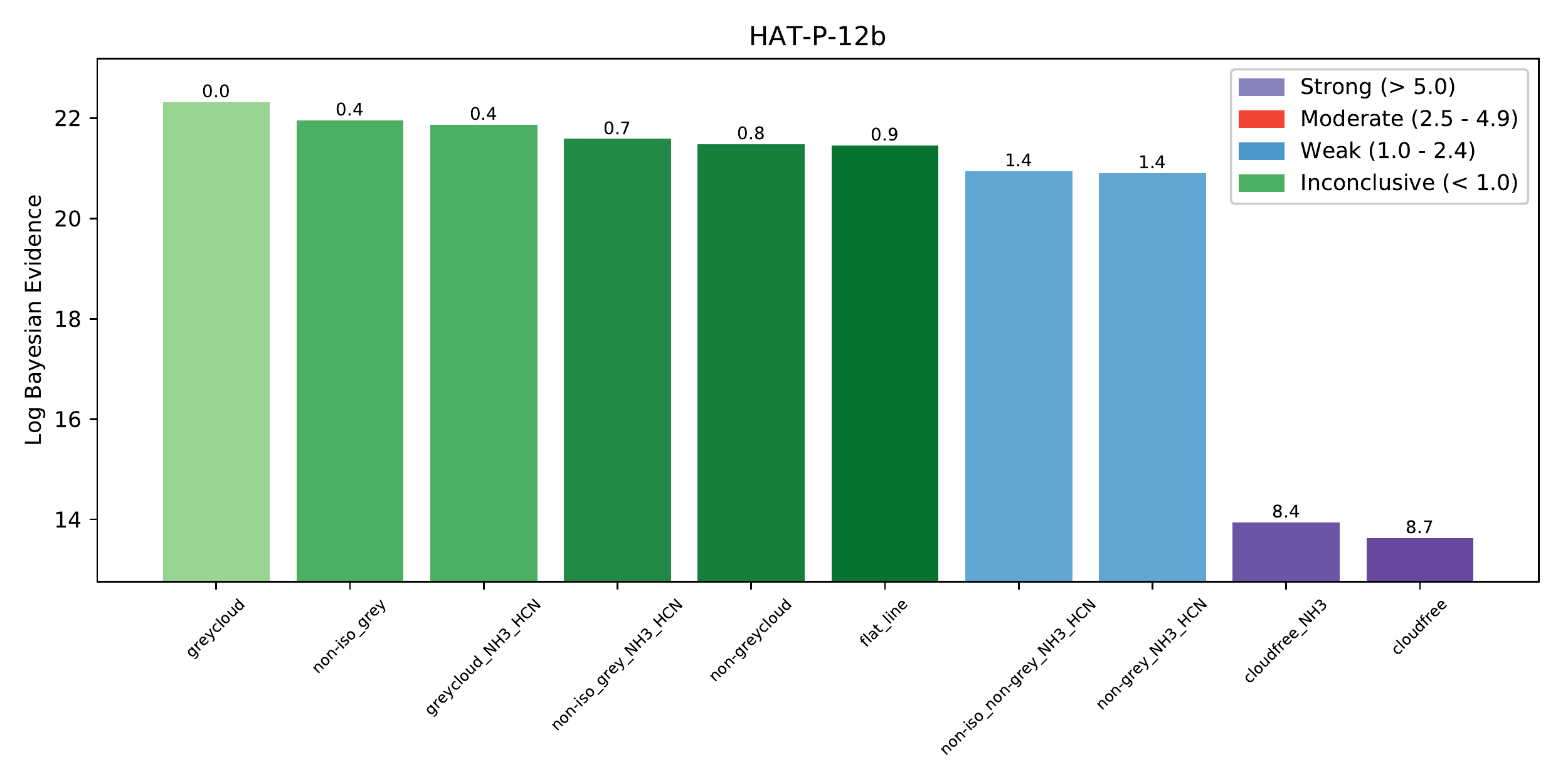}
\includegraphics[width=0.65\columnwidth]{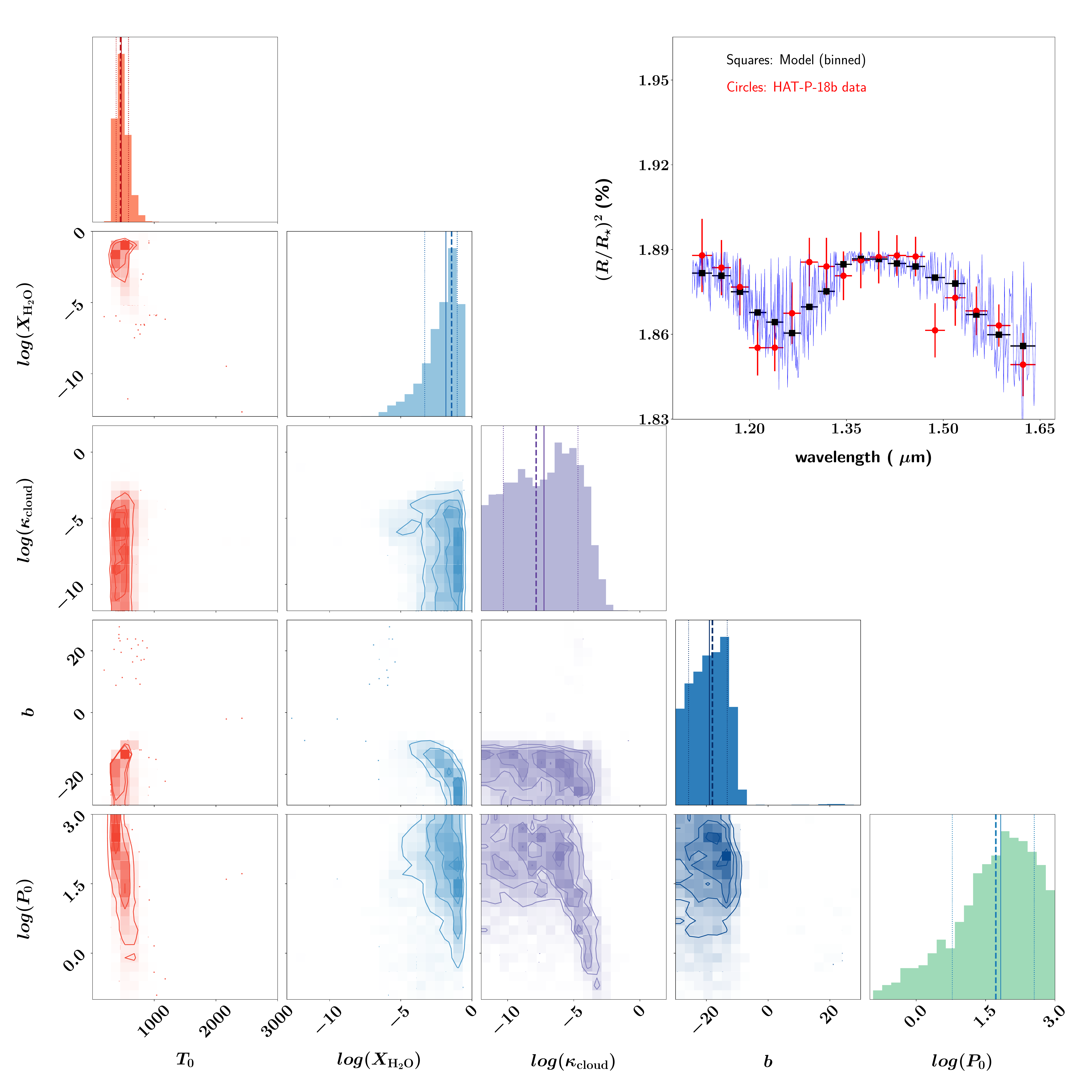}
\hspace{0.1in}
\includegraphics[width=1.2\columnwidth]{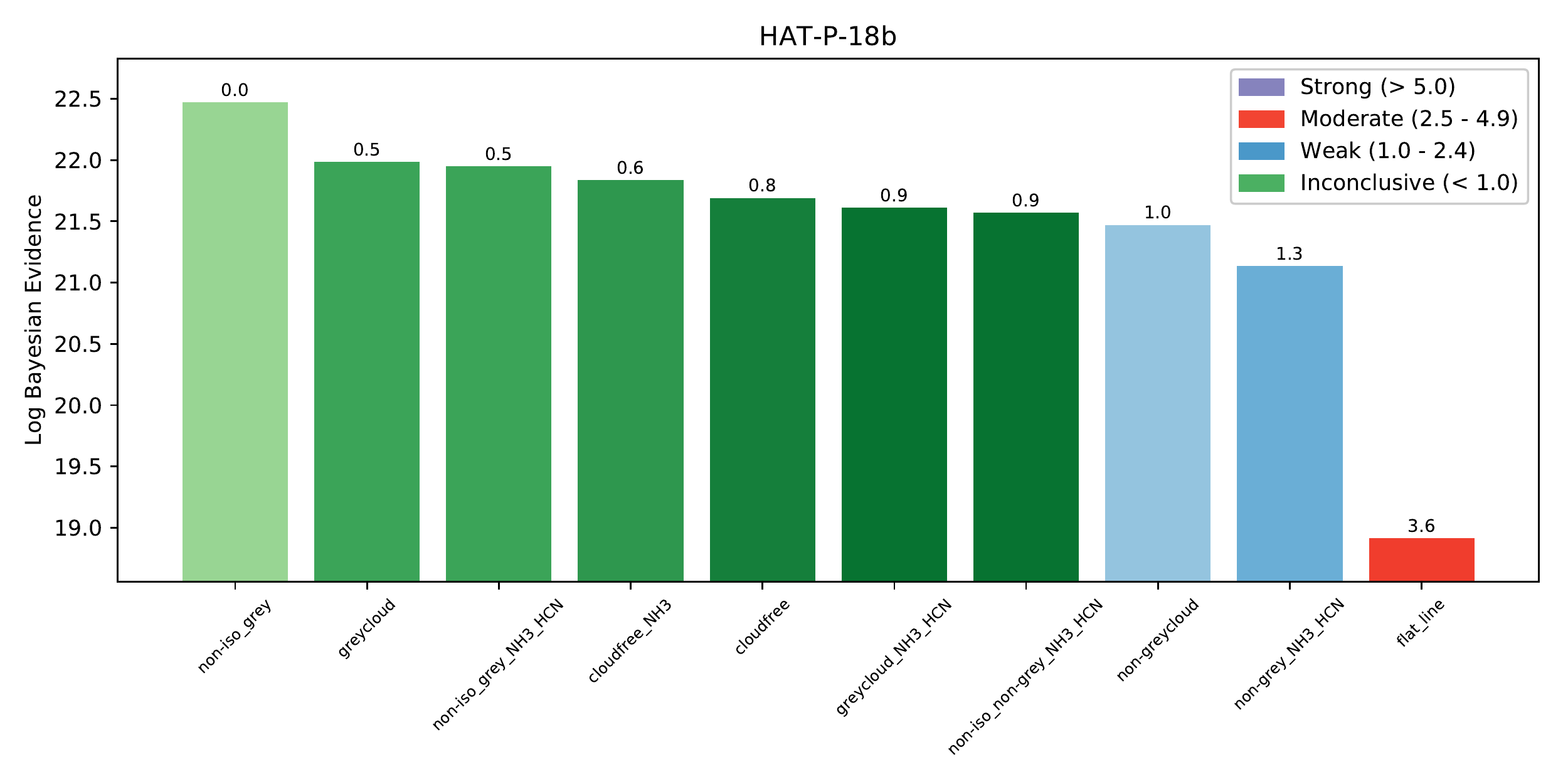}
\includegraphics[width=0.65\columnwidth]{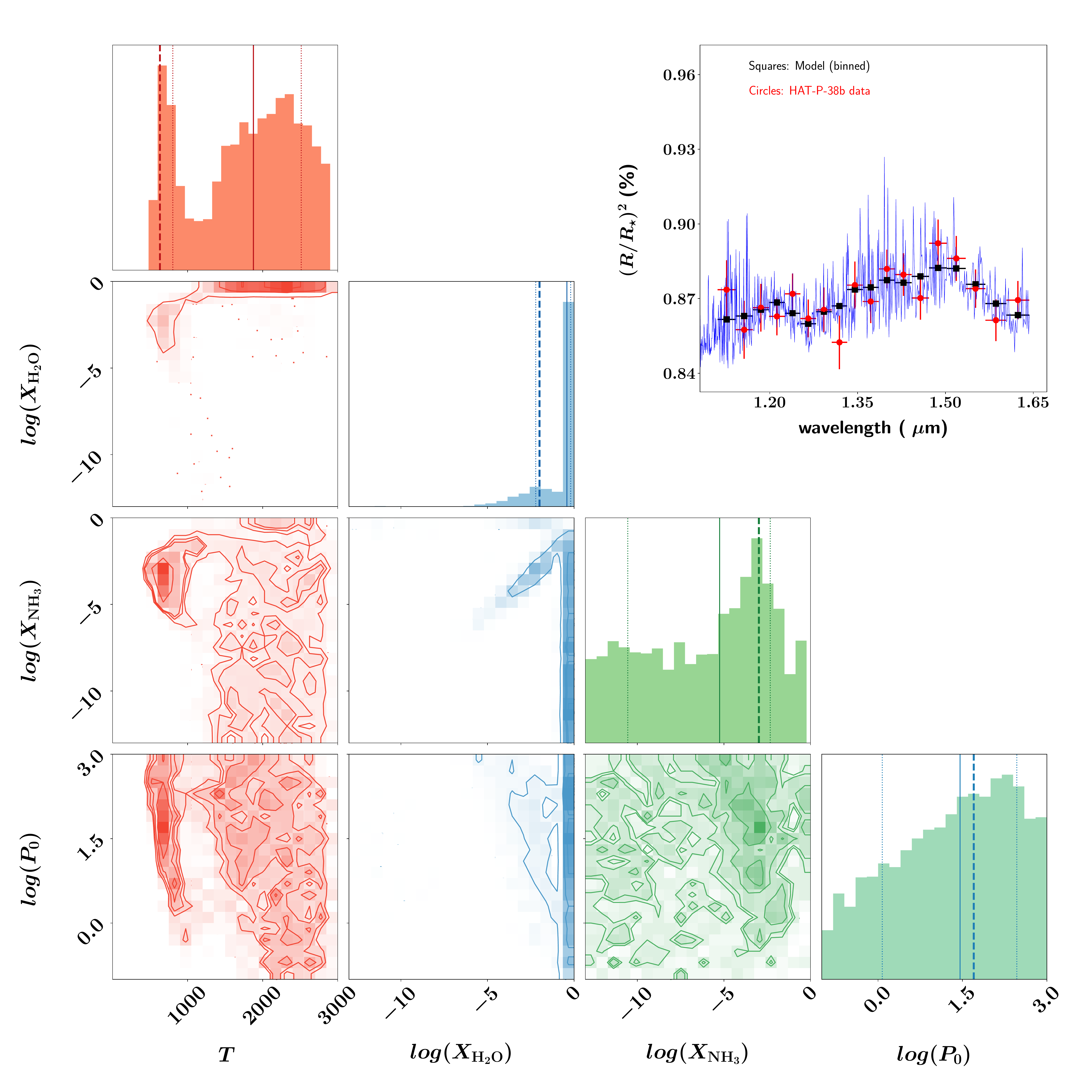}
\hspace{0.1in}
\includegraphics[width=1.2\columnwidth]{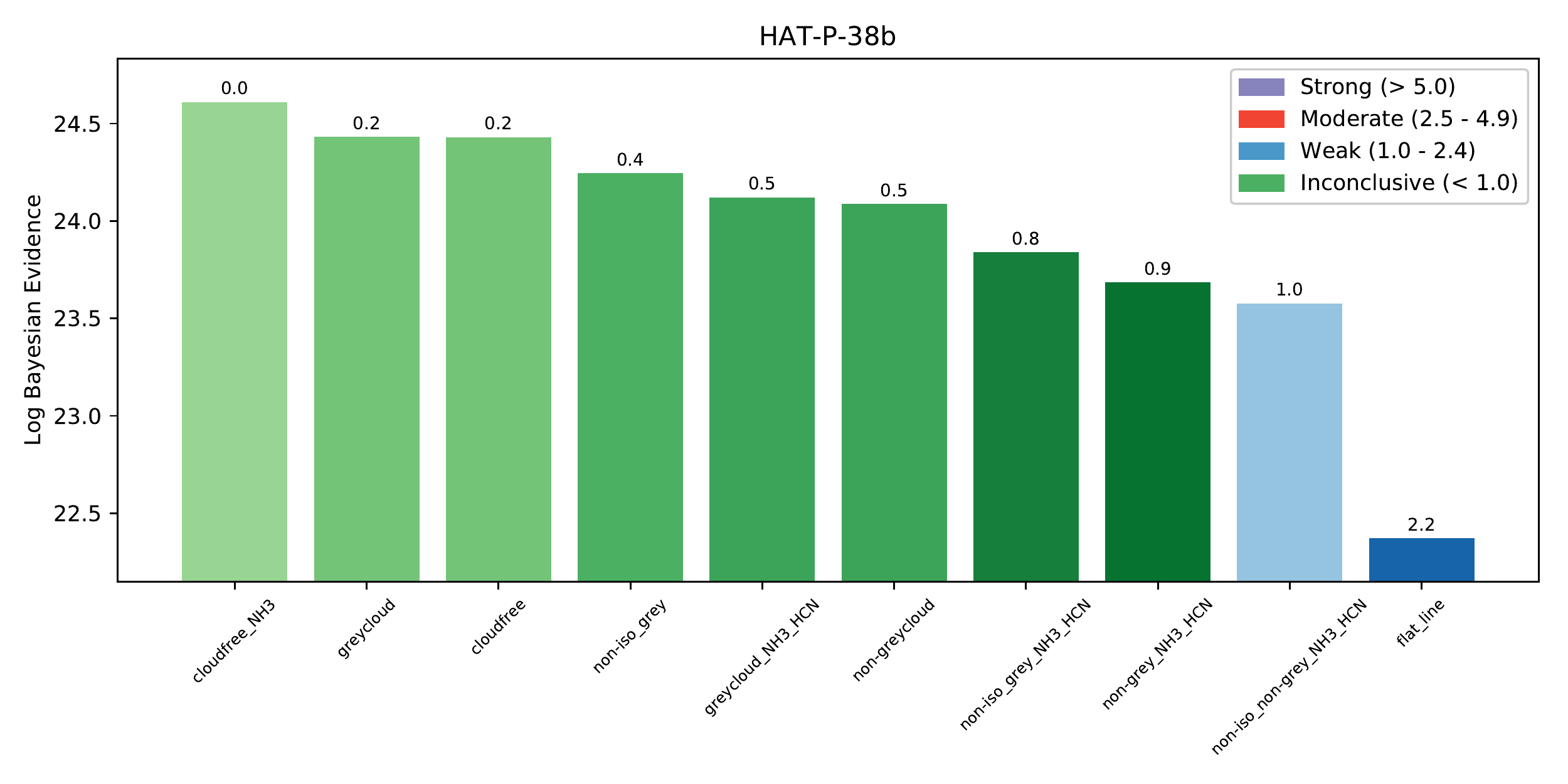}
\includegraphics[width=0.65\columnwidth]{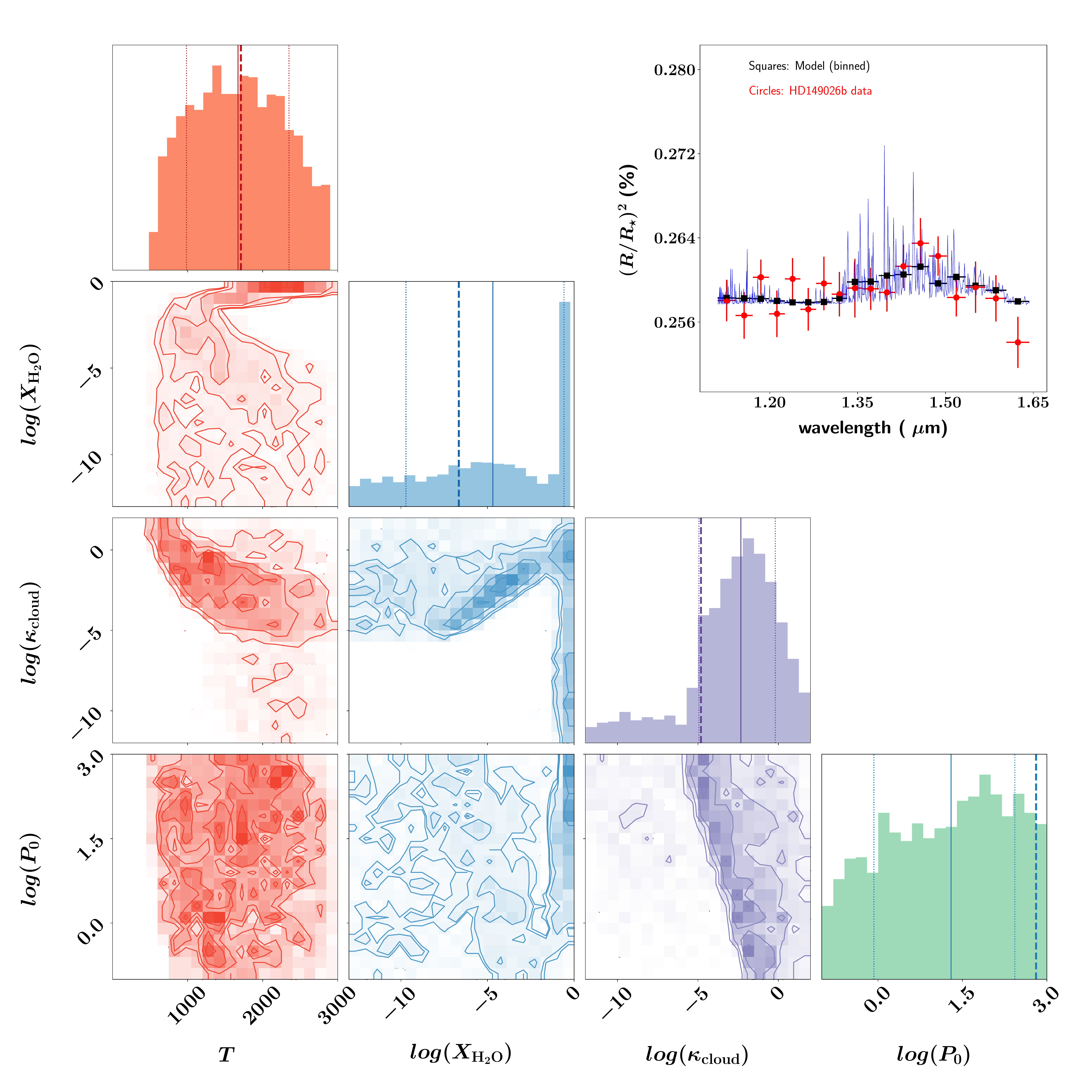}
\hspace{0.1in}
\includegraphics[width=1.2\columnwidth]{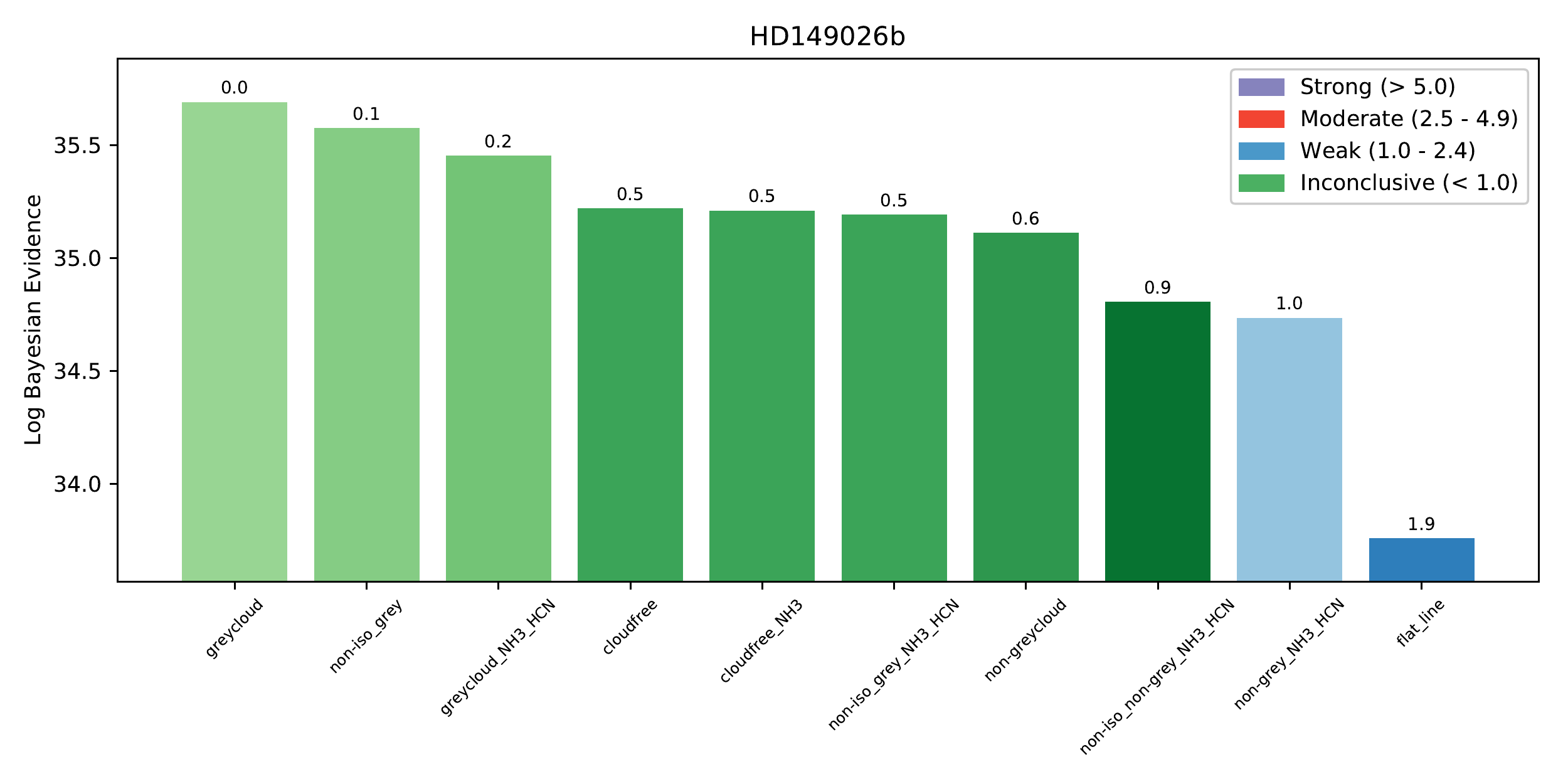}
\end{center}
\vspace{-0.1in}
\caption{Same as Figure \ref{fig:cloud-free_vs_nh3}, but for exo-Saturns (0.2--$0.4 M_{\rm J}$): HAT-P-12b, HAT-P-18b, HAT-P-38b and HD 149026b.}
\label{fig:saturns}
\end{figure*}

\begin{figure*}
\begin{center}
\includegraphics[width=0.65\columnwidth]{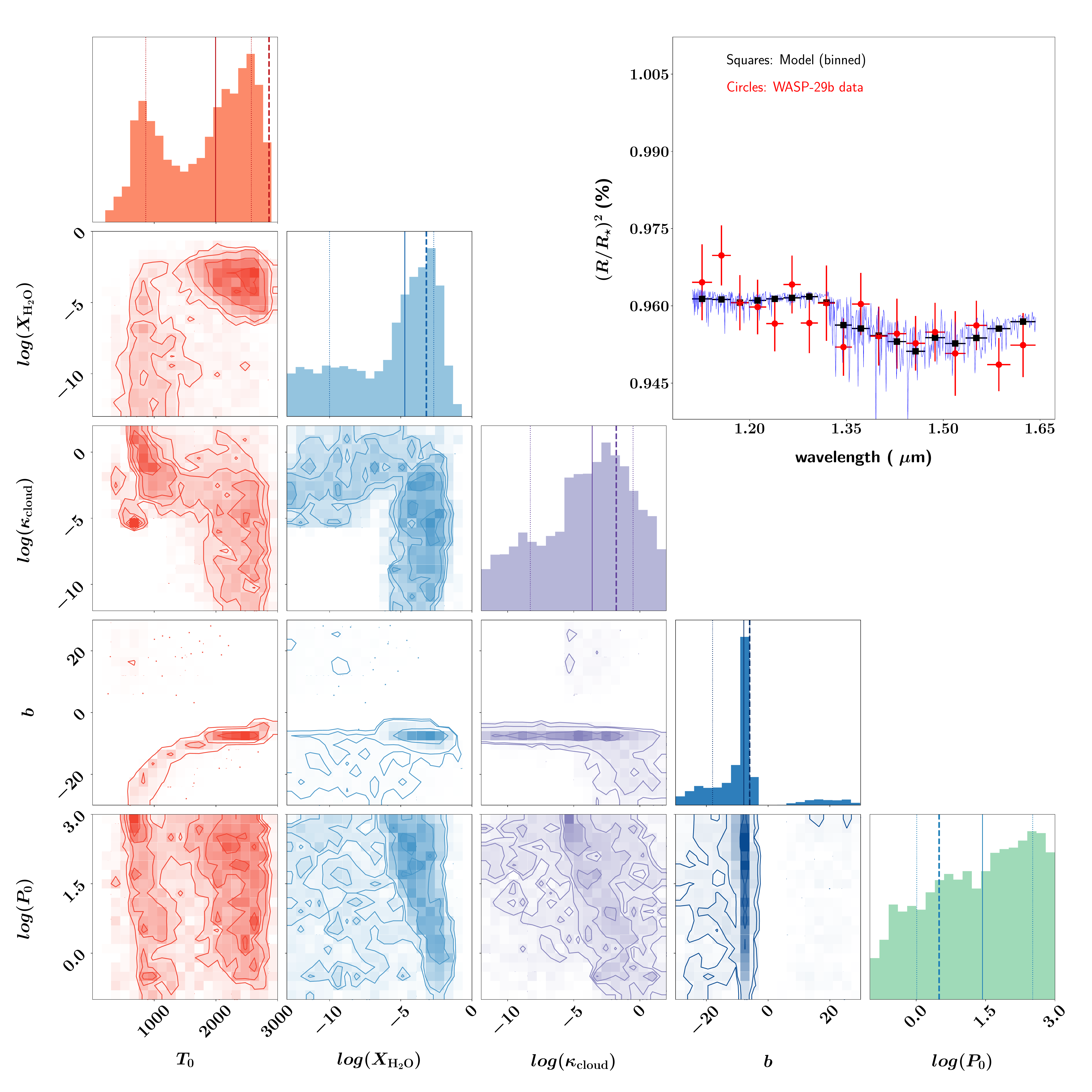}
\hspace{0.1in}
\includegraphics[width=1.2\columnwidth]{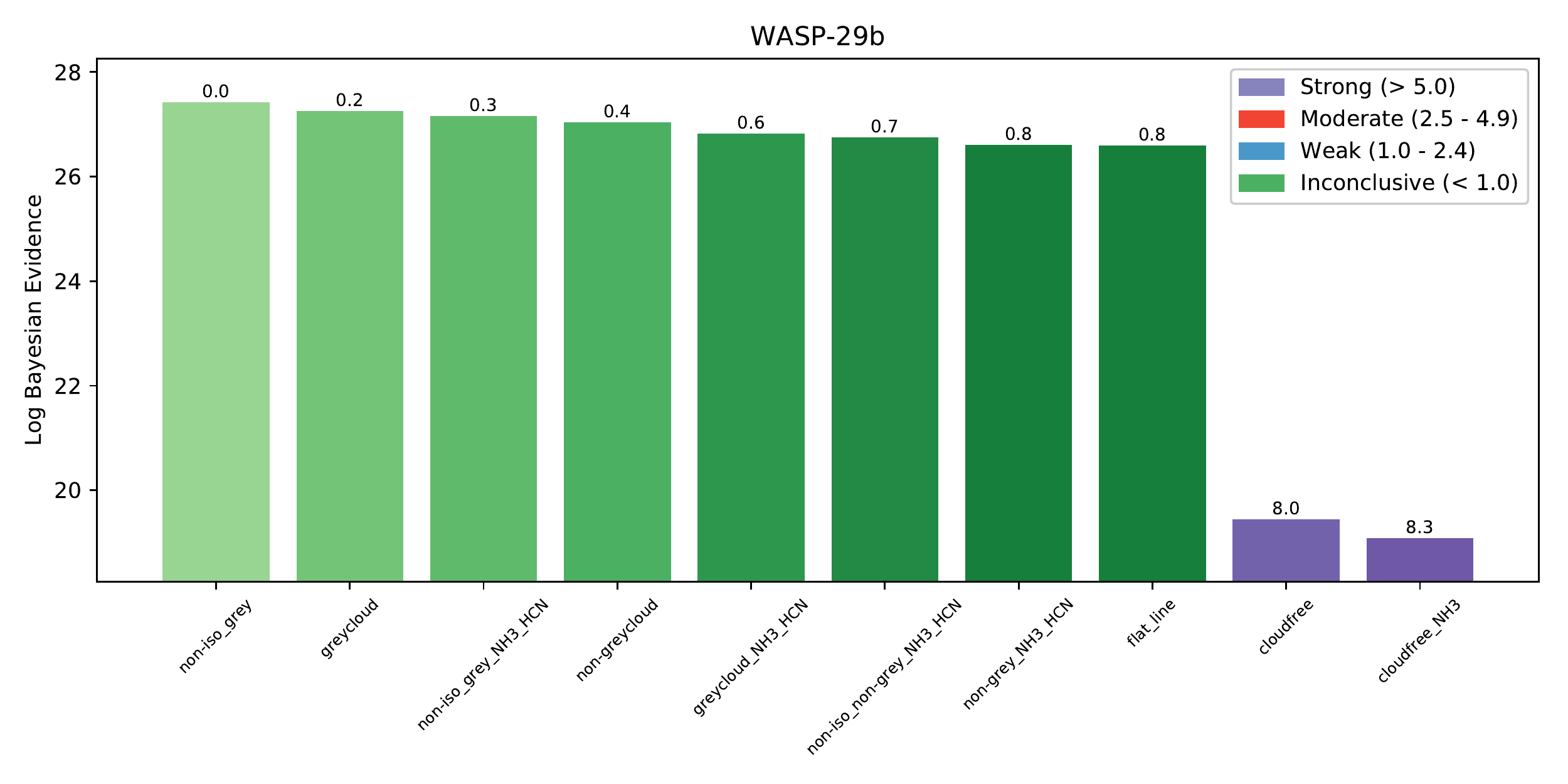}
\includegraphics[width=0.65\columnwidth]{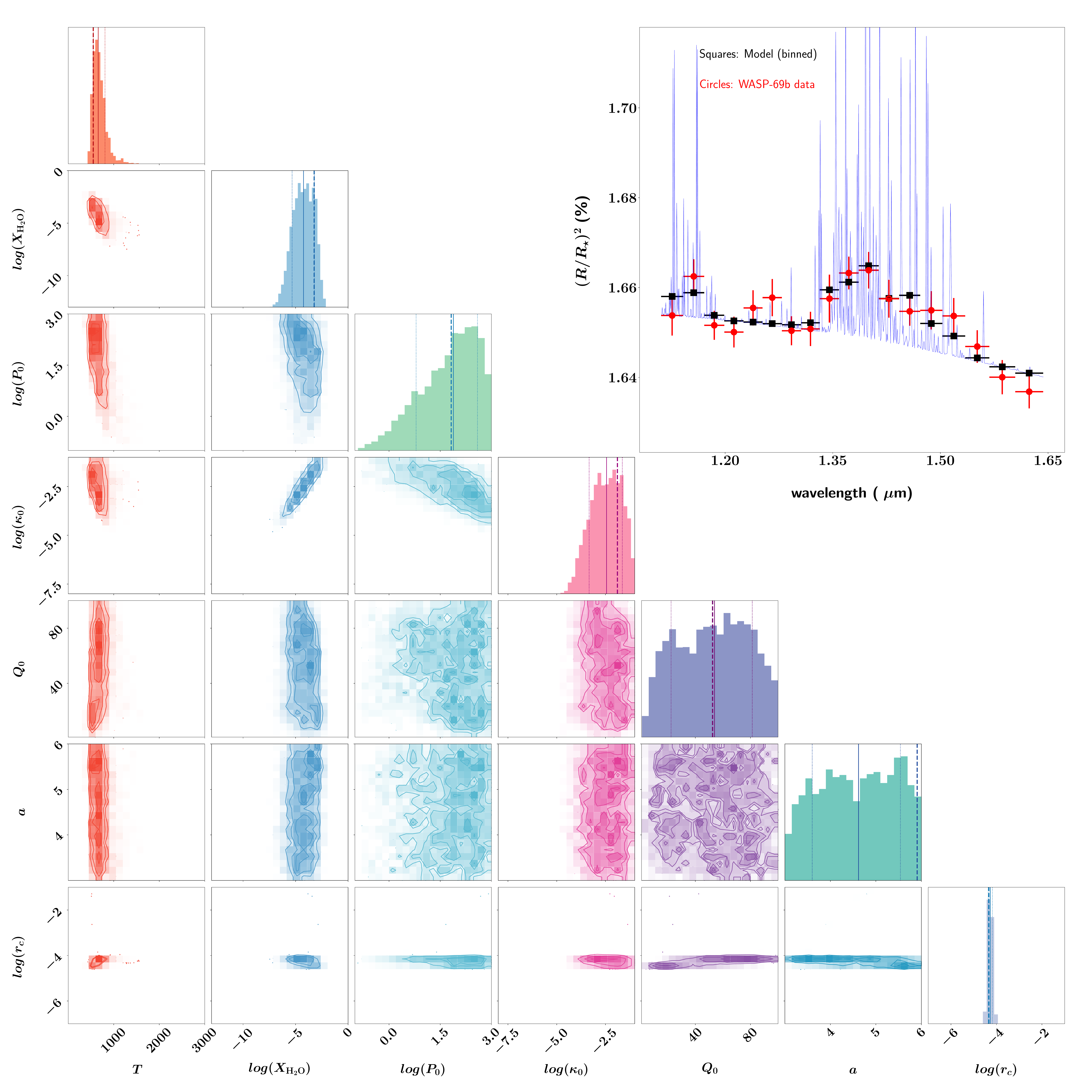}
\hspace{0.1in}
\includegraphics[width=1.2\columnwidth]{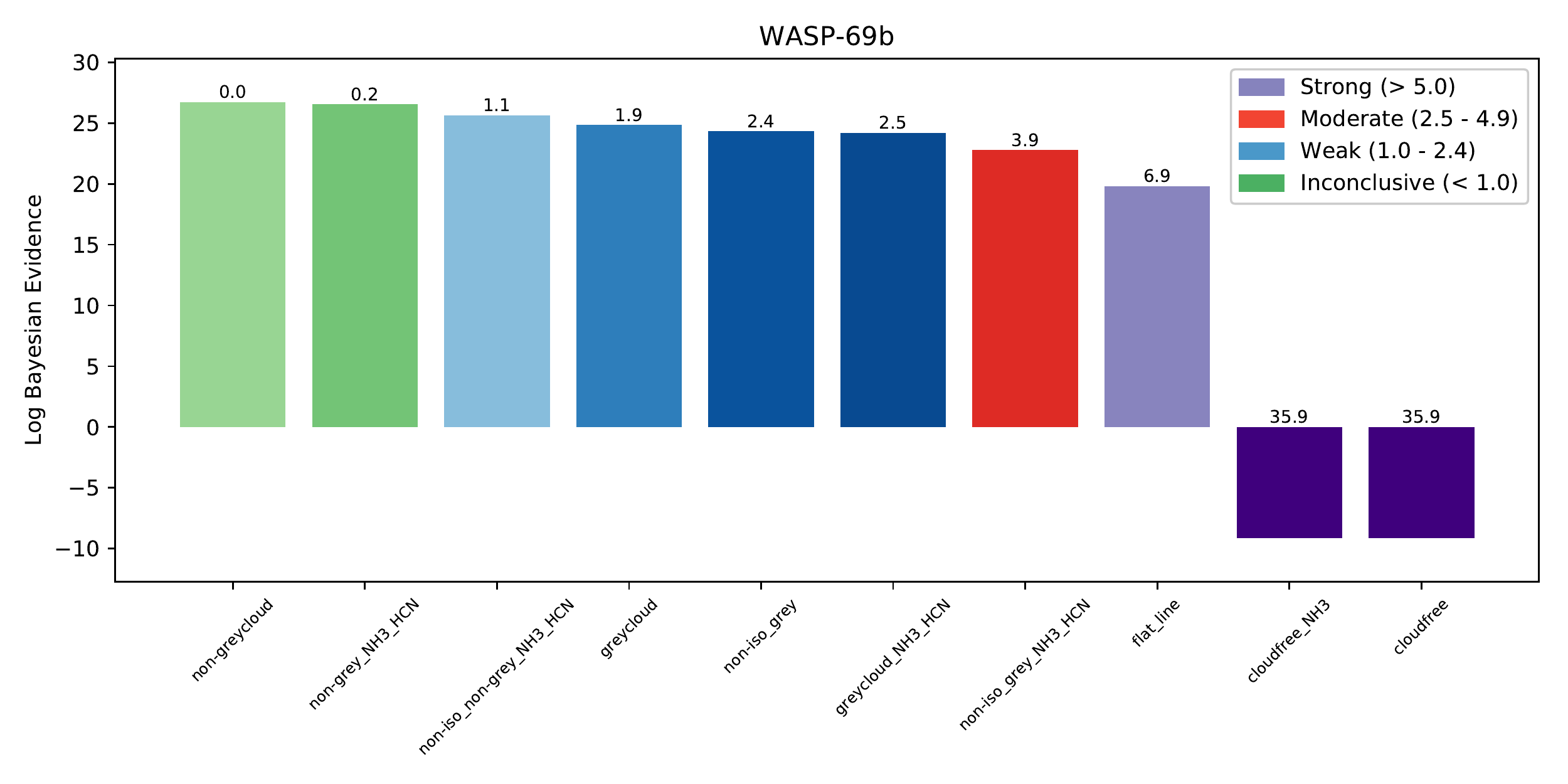}
\end{center}
\vspace{-0.1in}
\caption{Continuation of Figure \ref{fig:saturns} for exo-Saturns (0.2--$0.4 M_{\rm J}$): WASP-29b and WASP-69b.  Additionally, WASP-69b is one of two objects where non-grey clouds are needed to fit the data.}
\label{fig:saturns_2}
\end{figure*}

\begin{figure*}
\begin{center}
\includegraphics[width=0.65\columnwidth]{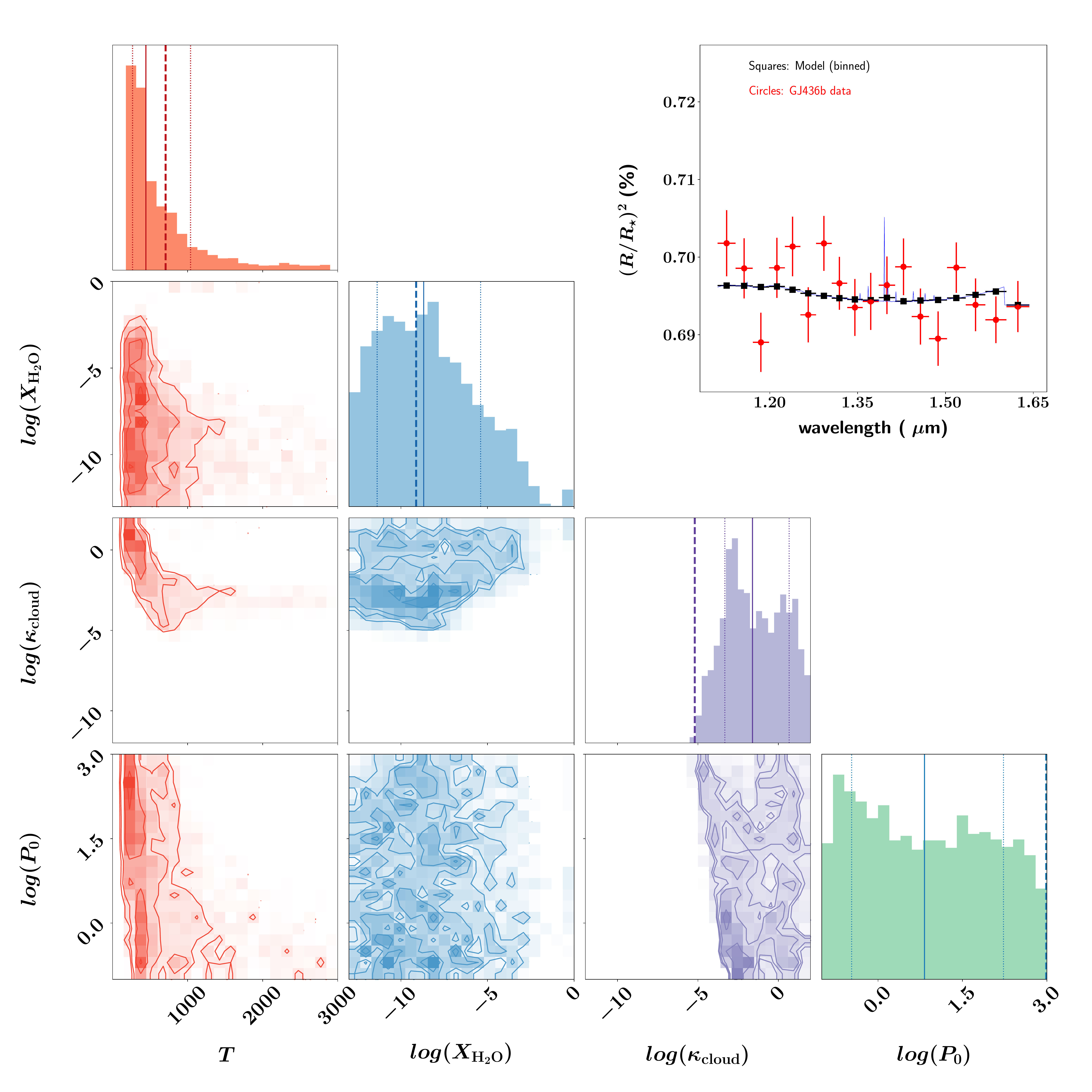}
\hspace{0.1in}
\includegraphics[width=1.2\columnwidth]{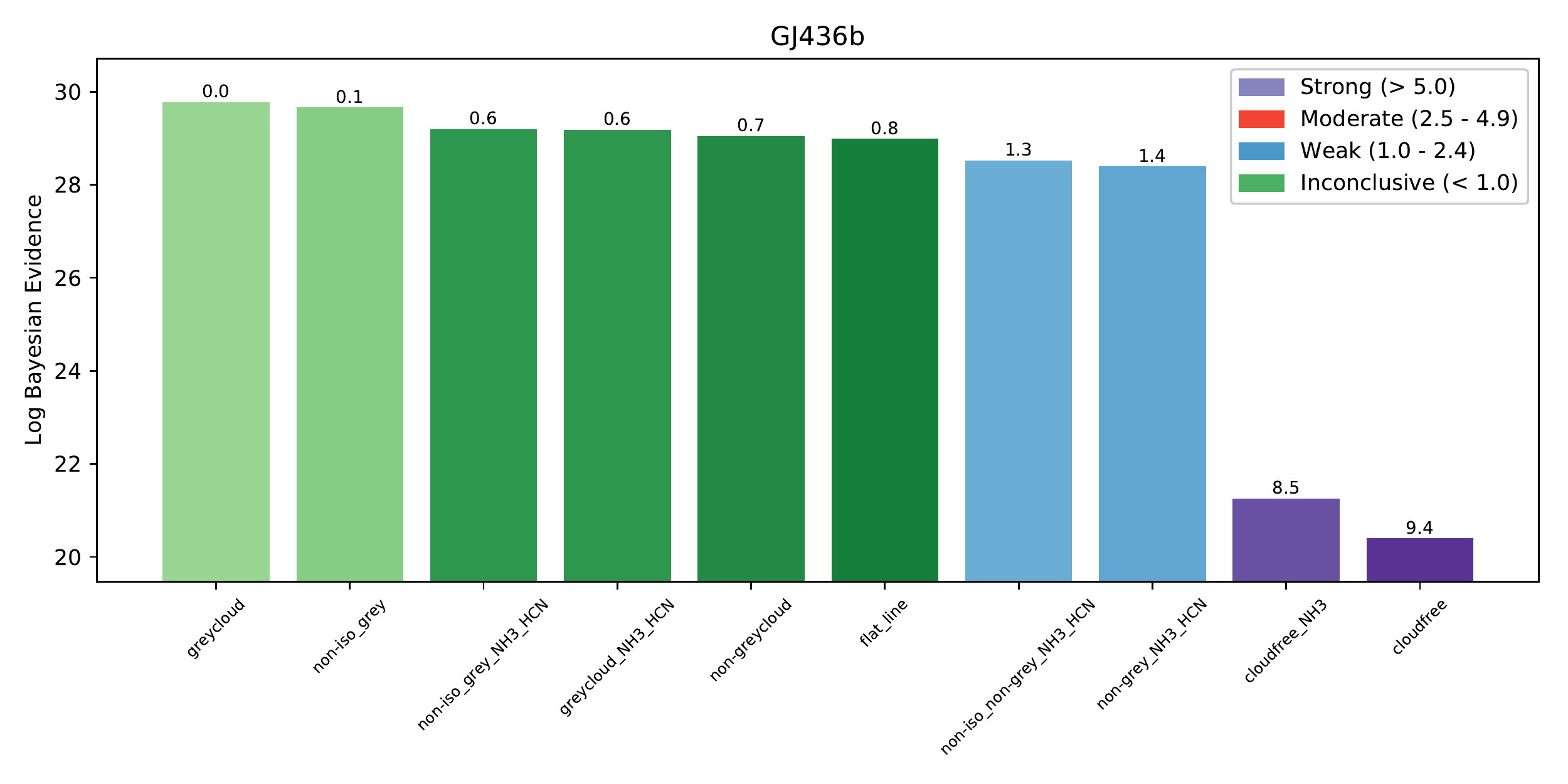}
\includegraphics[width=0.65\columnwidth]{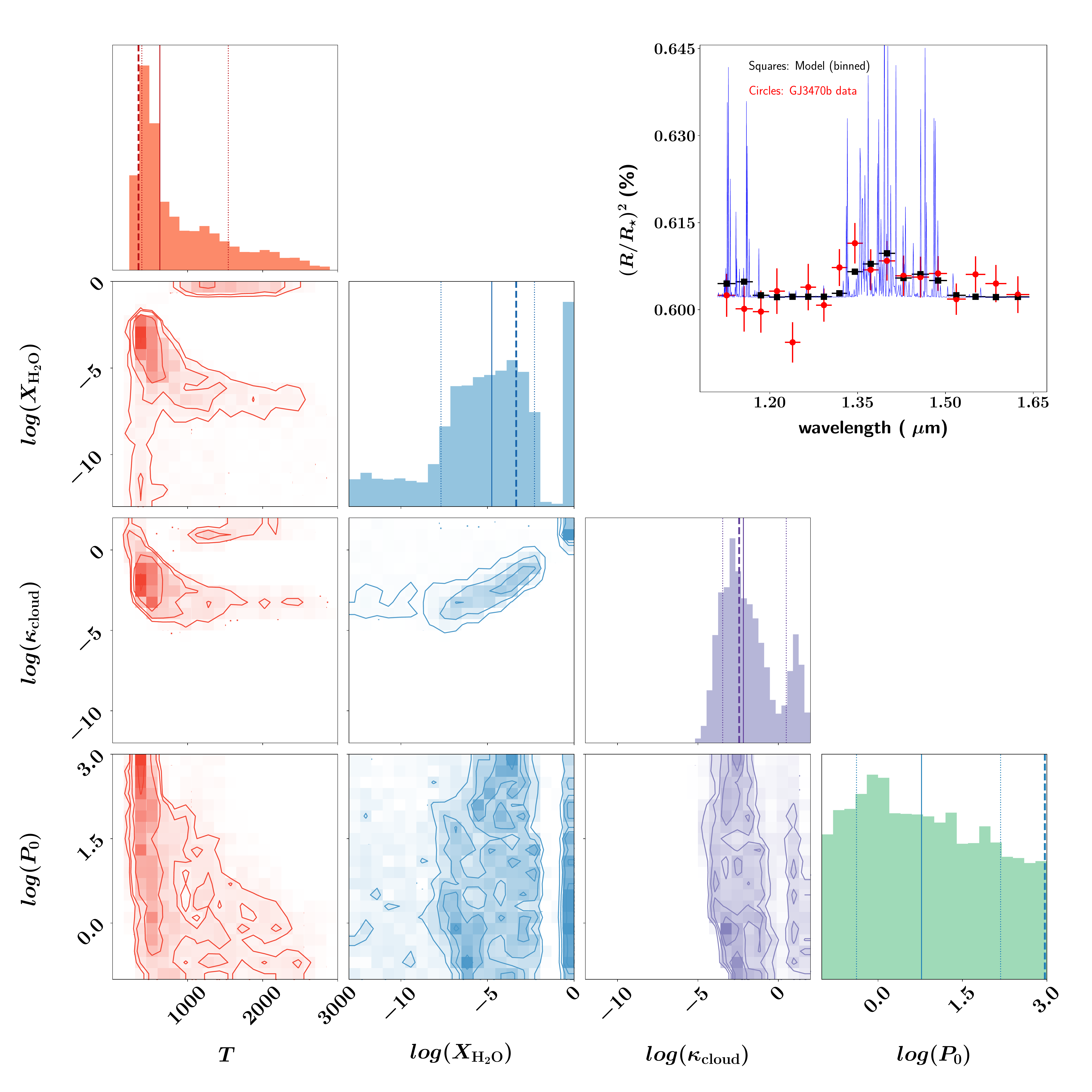}
\hspace{0.1in}
\includegraphics[width=1.2\columnwidth]{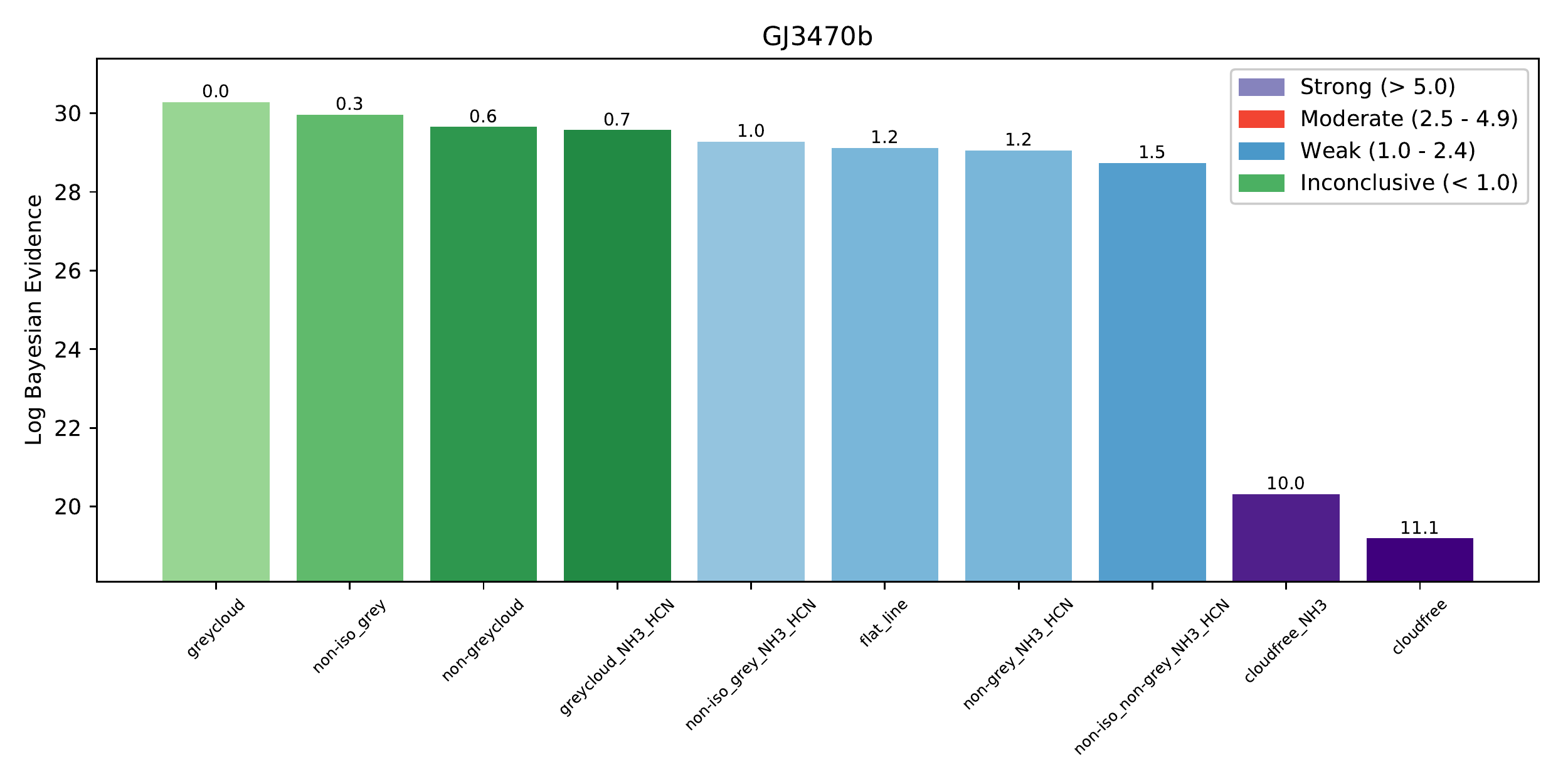}
\includegraphics[width=0.65\columnwidth]{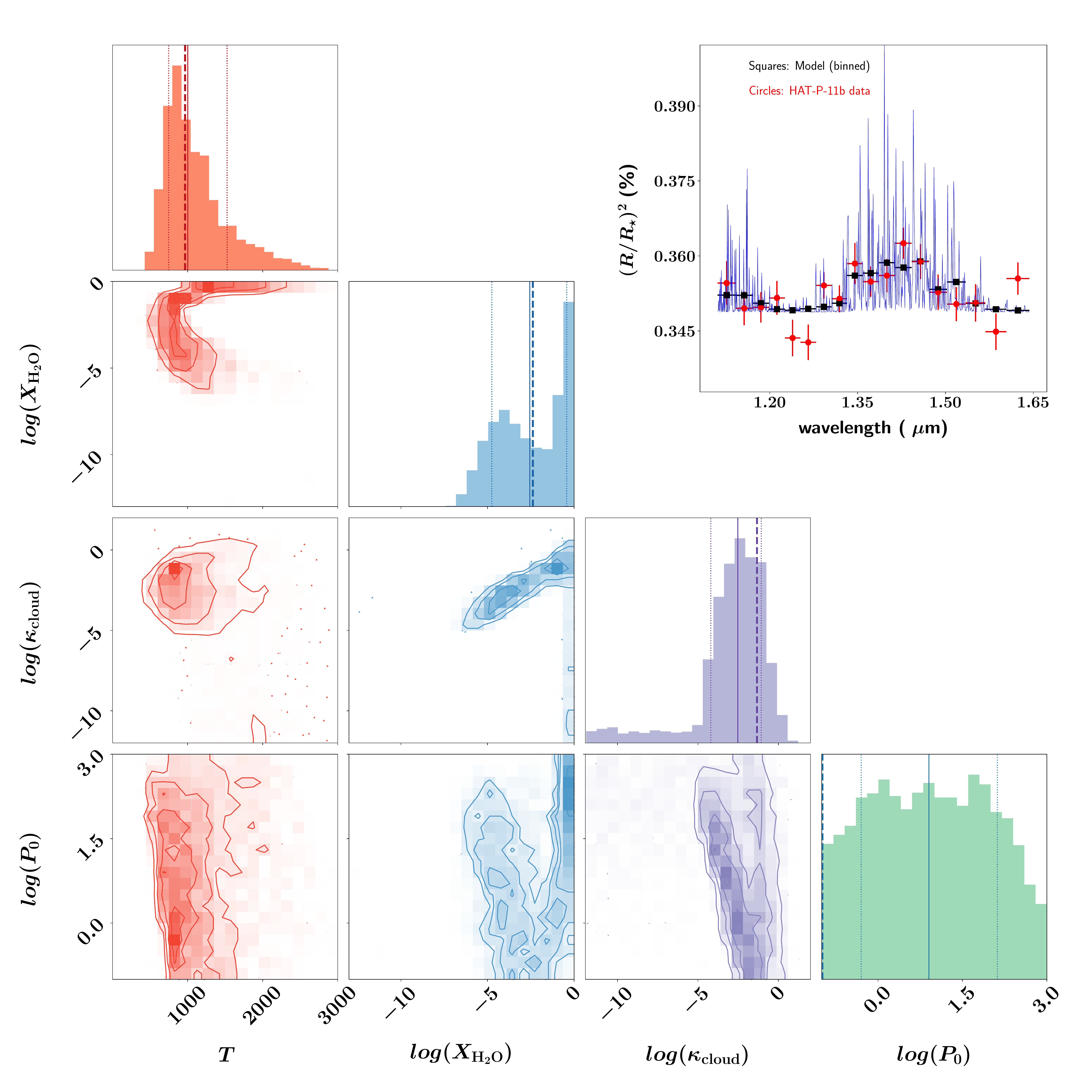}
\hspace{0.1in}
\includegraphics[width=1.2\columnwidth]{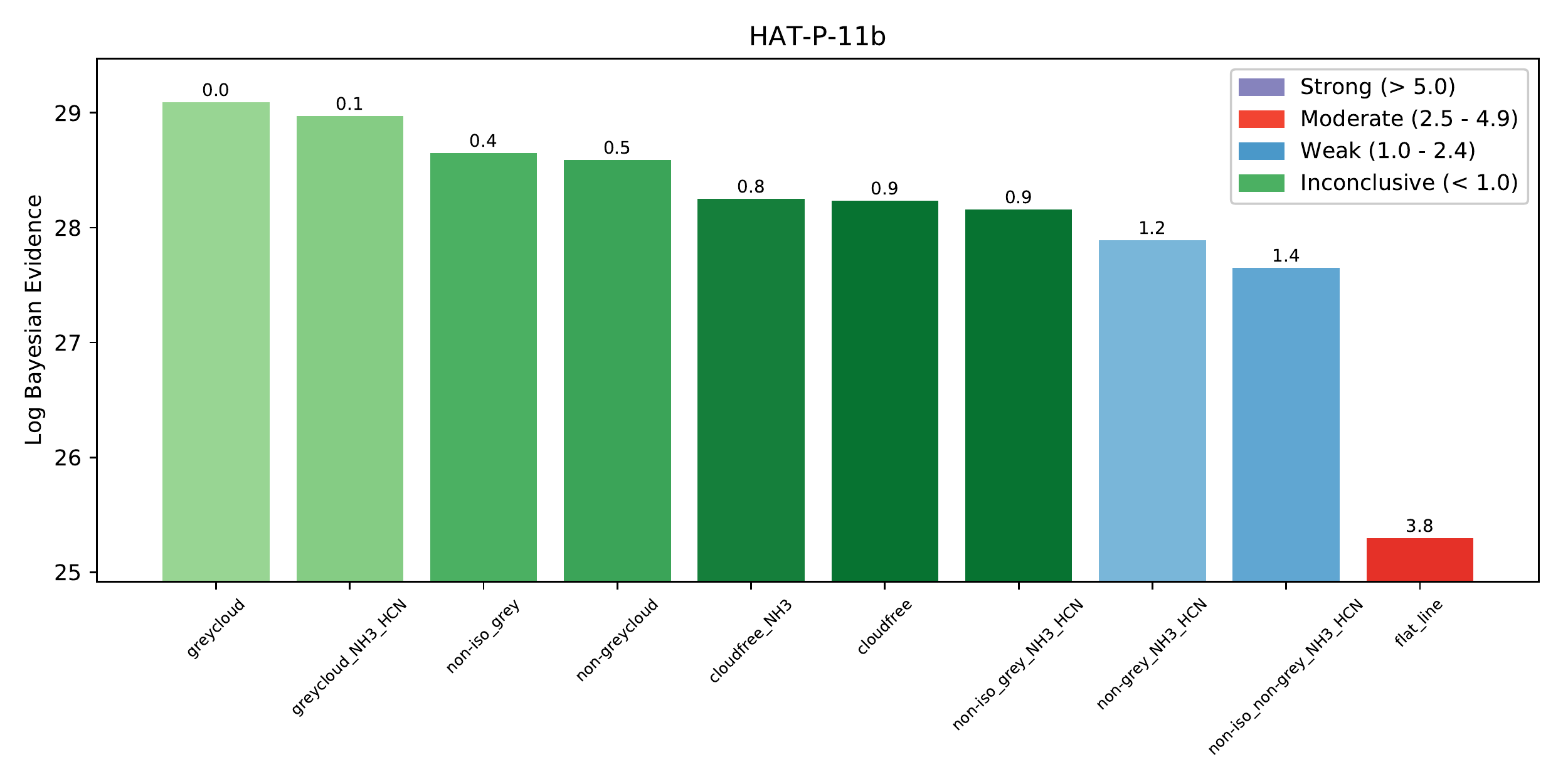}
\includegraphics[width=0.65\columnwidth]{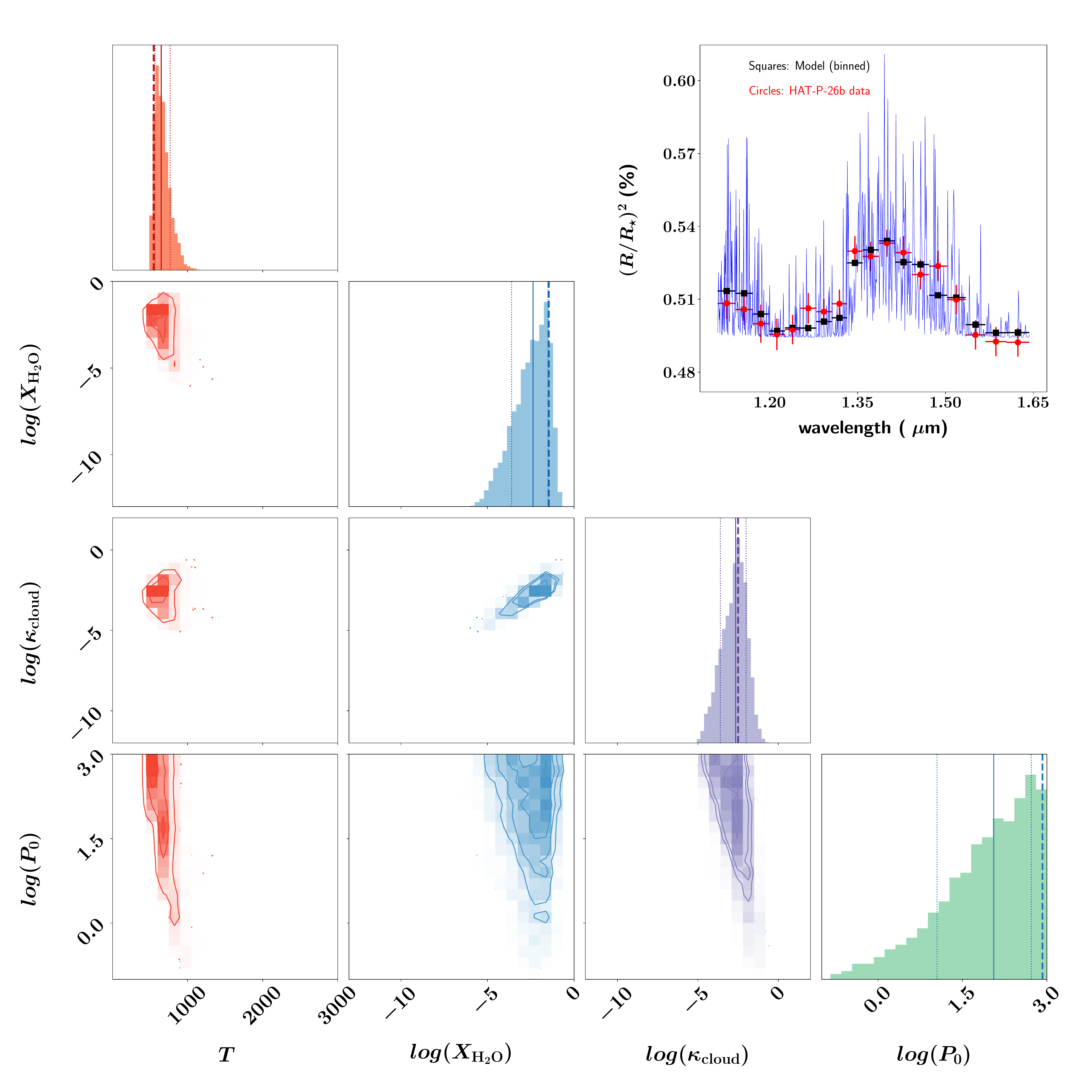}
\hspace{0.1in}
\includegraphics[width=1.2\columnwidth]{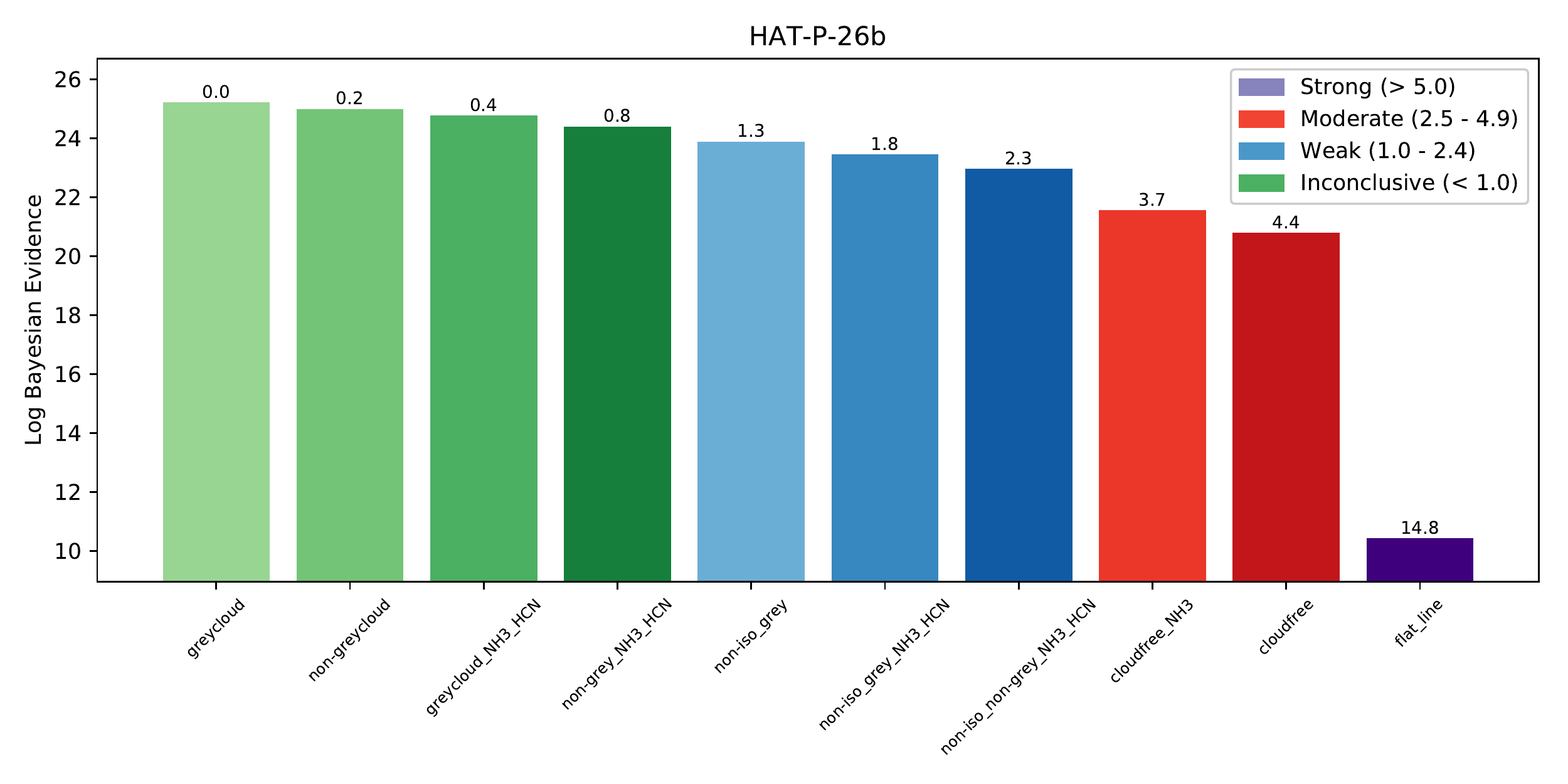}
\end{center}
\vspace{-0.1in}
\caption{Same as Figure \ref{fig:cloud-free_vs_nh3}, but for exo-Neptunes:  GJ 436b, GJ 3470b, HAT-P-11b and HAT-P-26b.}
\label{fig:neptunes}
\end{figure*}

\begin{figure*}
\begin{center}
\includegraphics[width=0.65\columnwidth]{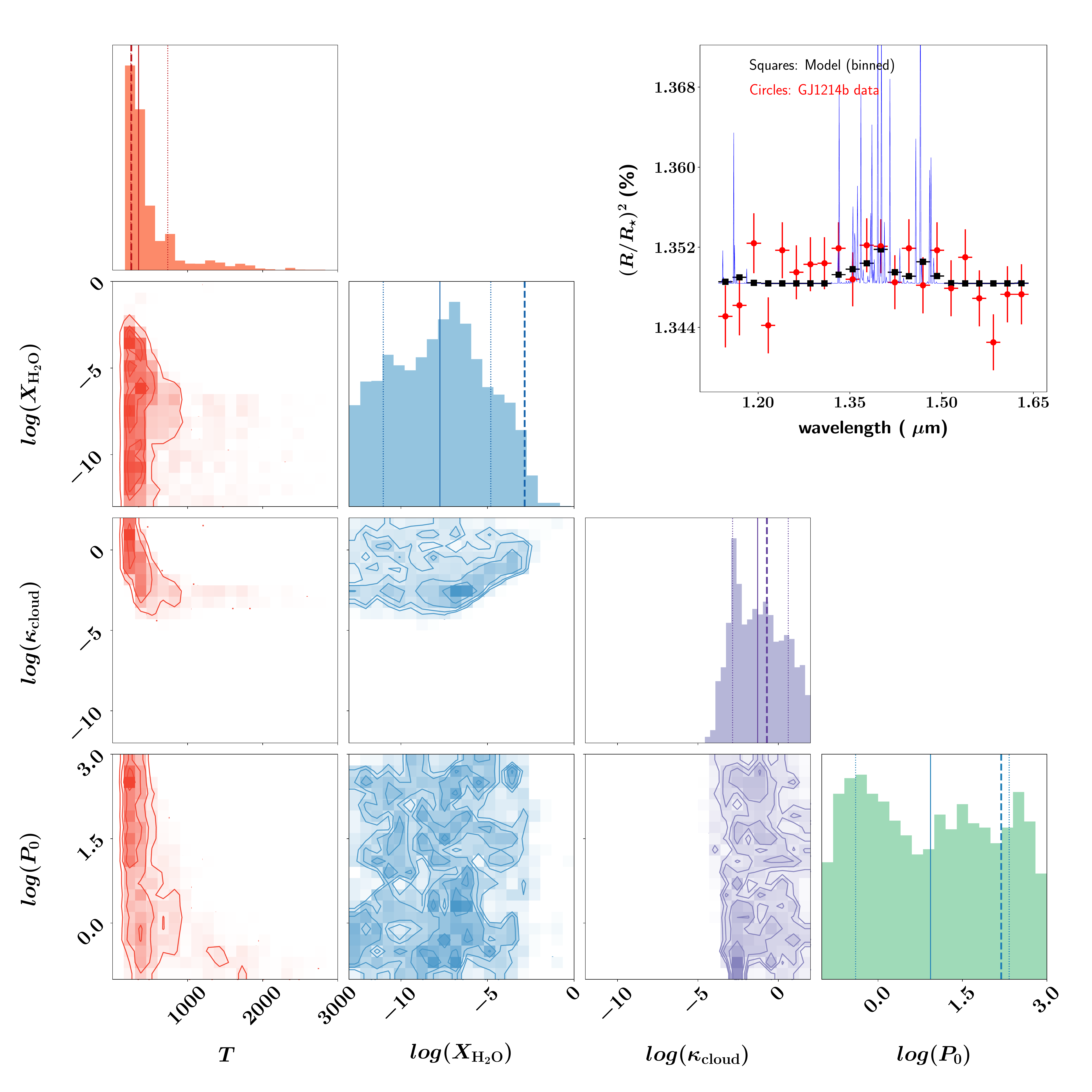}
\hspace{0.1in}
\includegraphics[width=1.2\columnwidth]{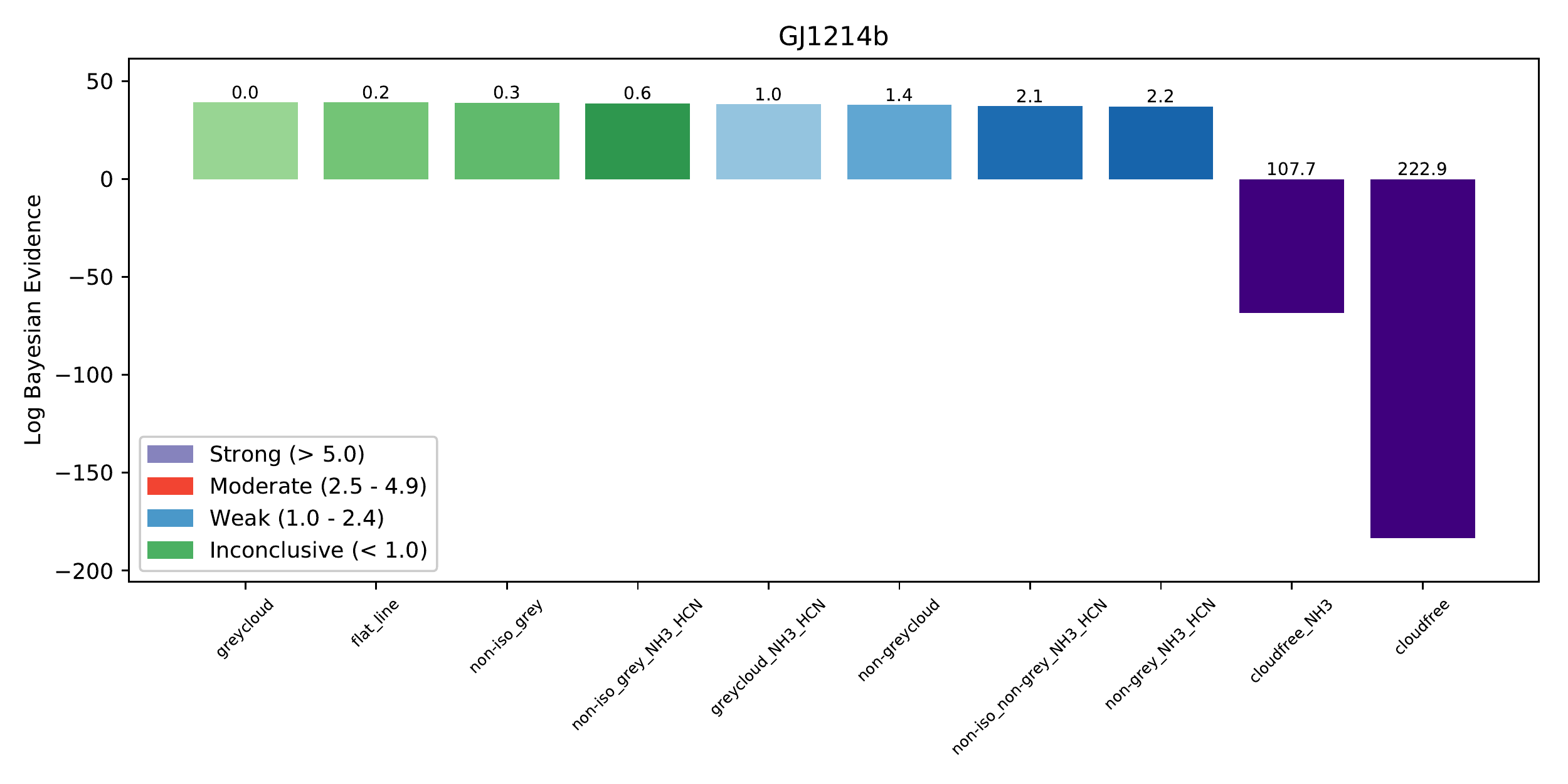}
\includegraphics[width=0.65\columnwidth]{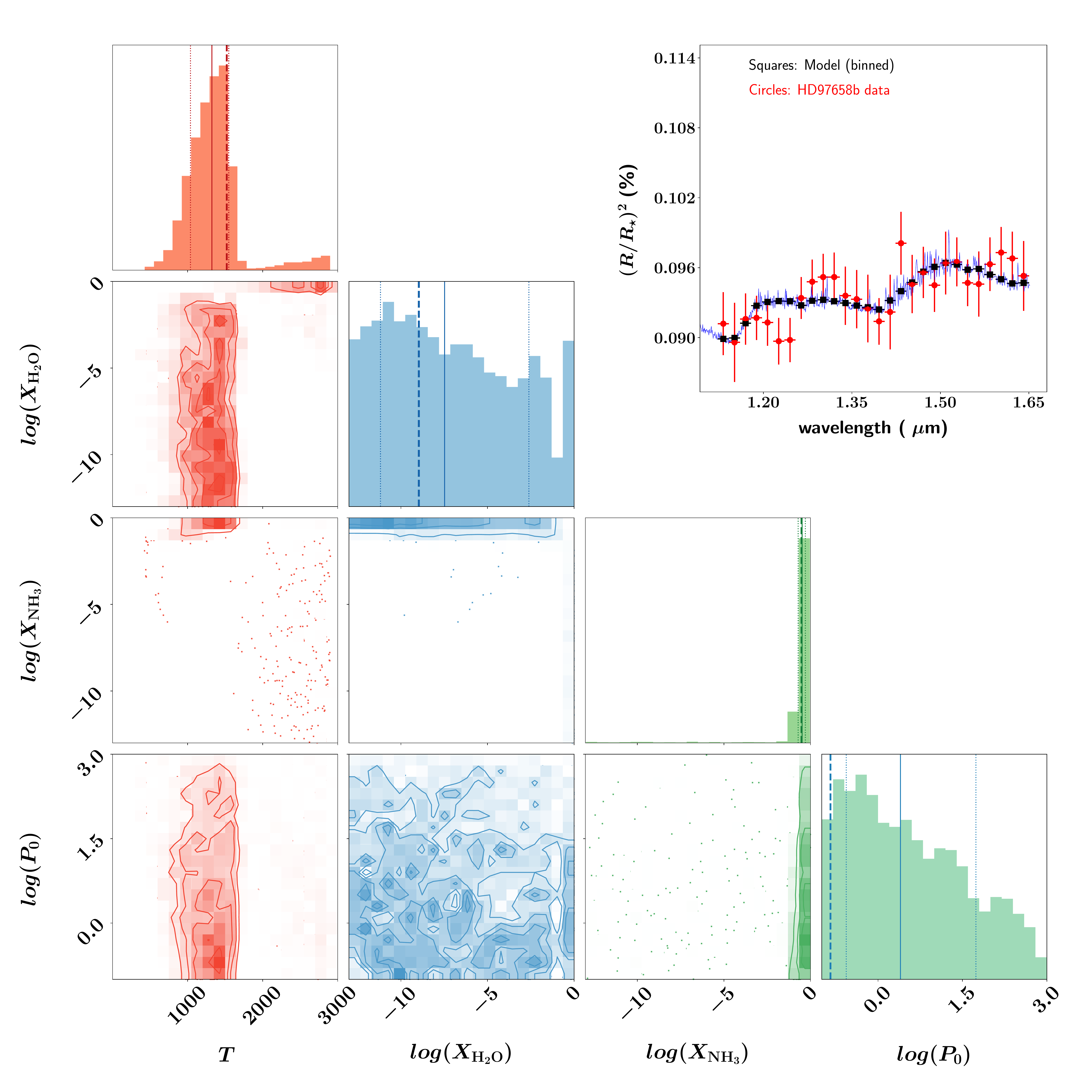}
\hspace{0.1in}
\includegraphics[width=1.2\columnwidth]{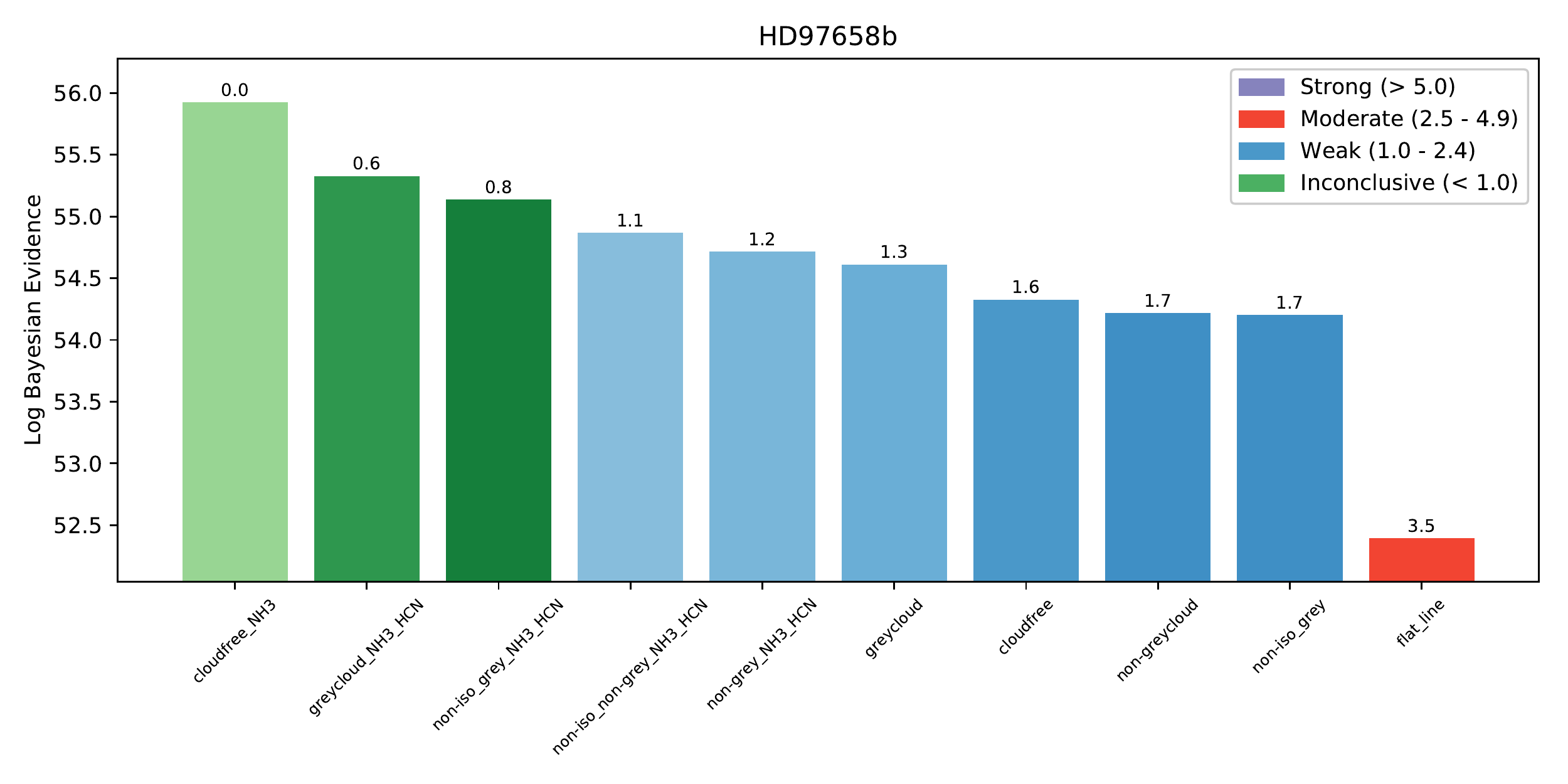}
\end{center}
\vspace{-0.1in}
\caption{Same as Figure \ref{fig:cloud-free_vs_nh3}, but for super Earths: GJ 1214b and HD 97658b.  HD 97658b stands out as an object where ammonia is significantly detected but the abundance of water is essentially unconstrained.}
\label{fig:searths}
\end{figure*}

\begin{figure*}
\begin{center}
\includegraphics[width=0.65\columnwidth]{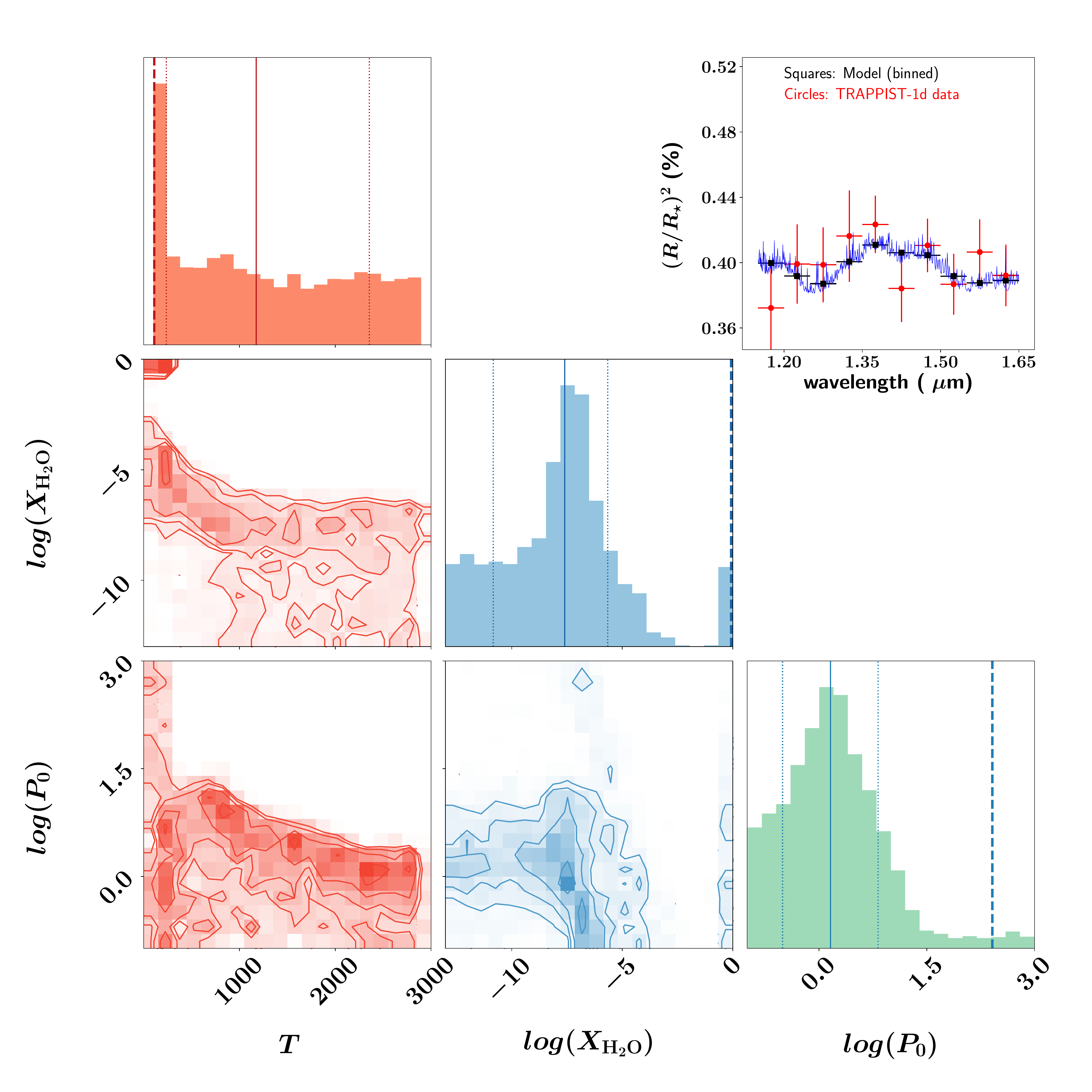}
\hspace{0.1in}
\includegraphics[width=1.2\columnwidth]{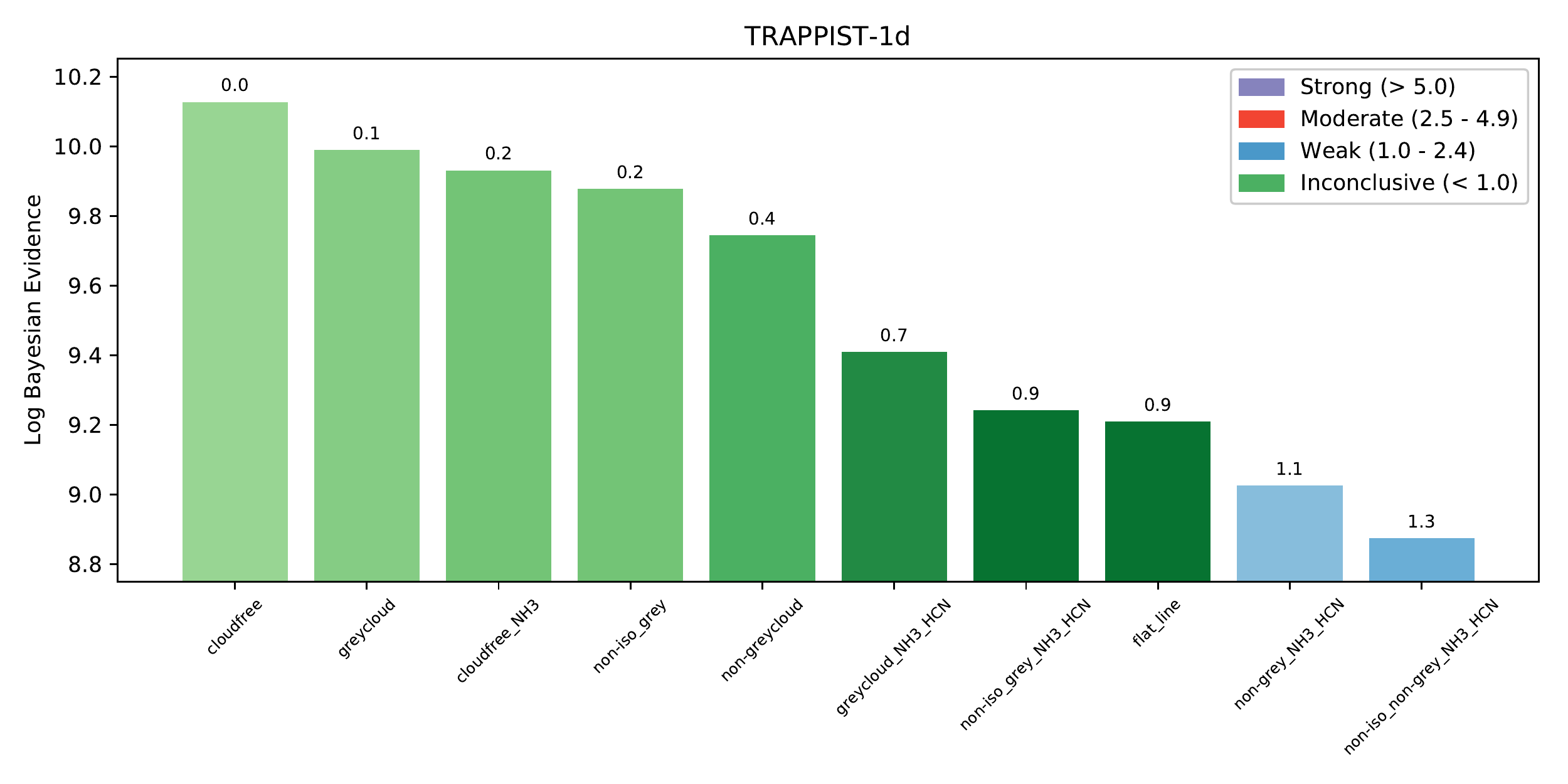}
\includegraphics[width=0.65\columnwidth]{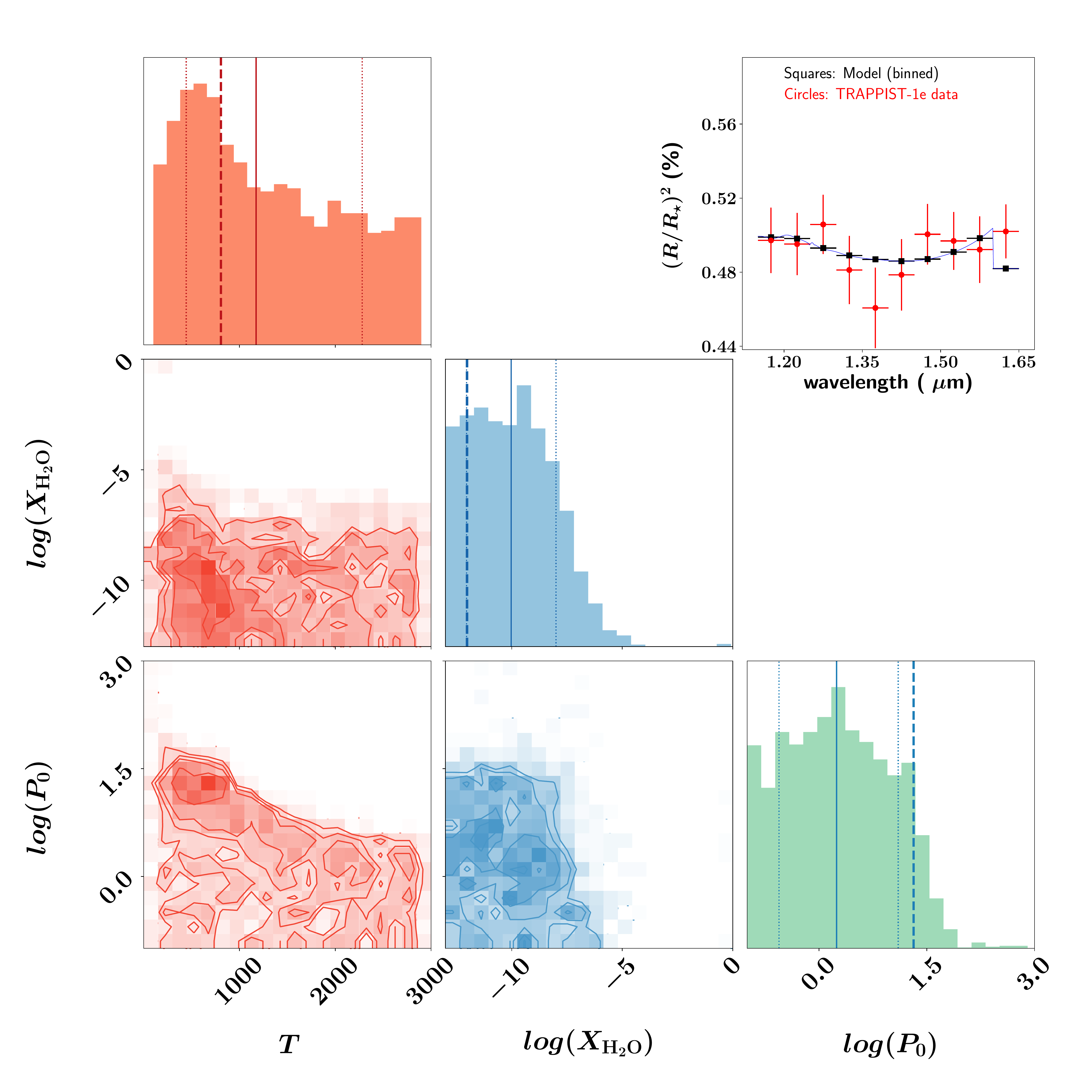}
\hspace{0.1in}
\includegraphics[width=1.2\columnwidth]{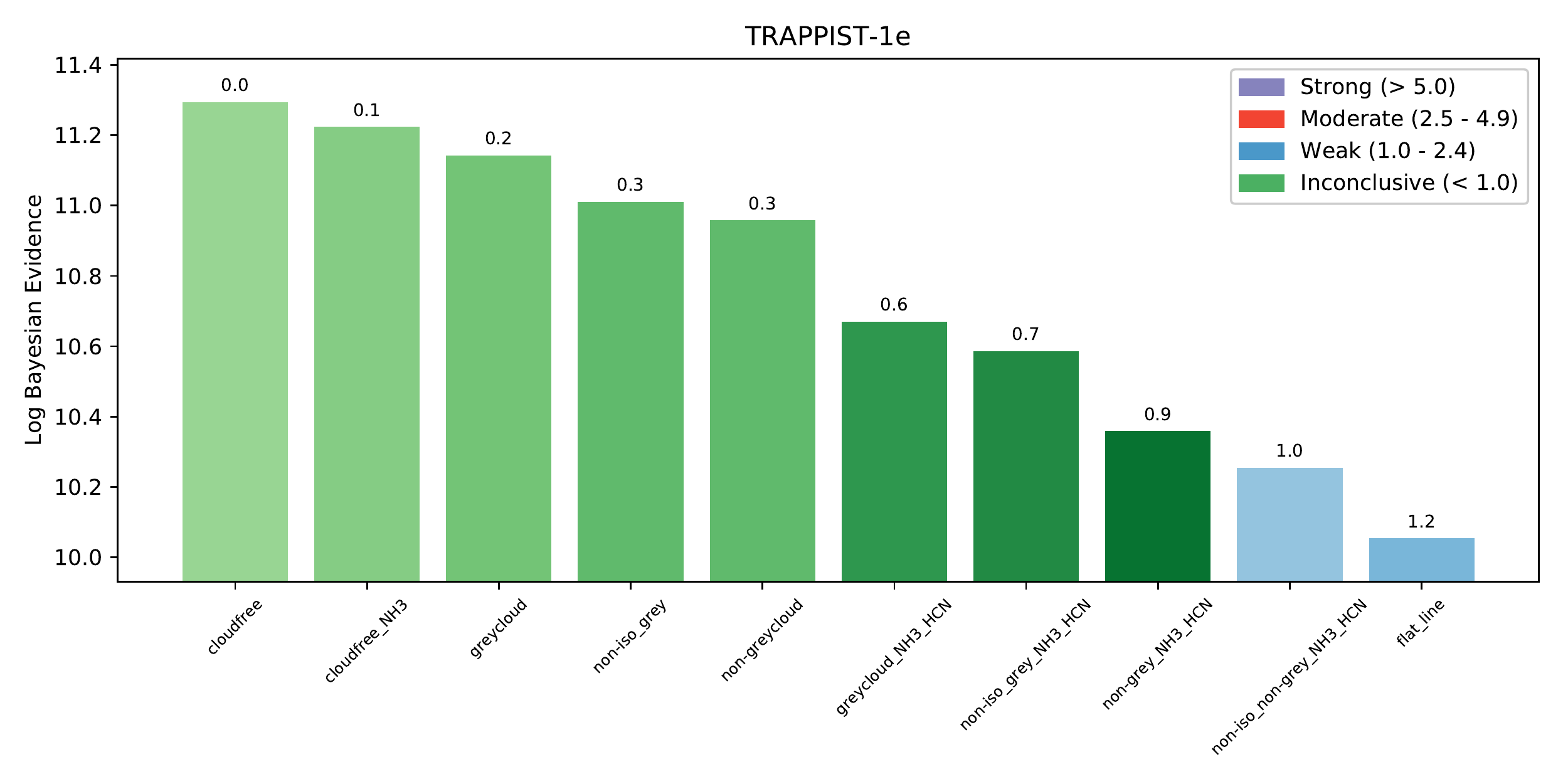}
\includegraphics[width=0.65\columnwidth]{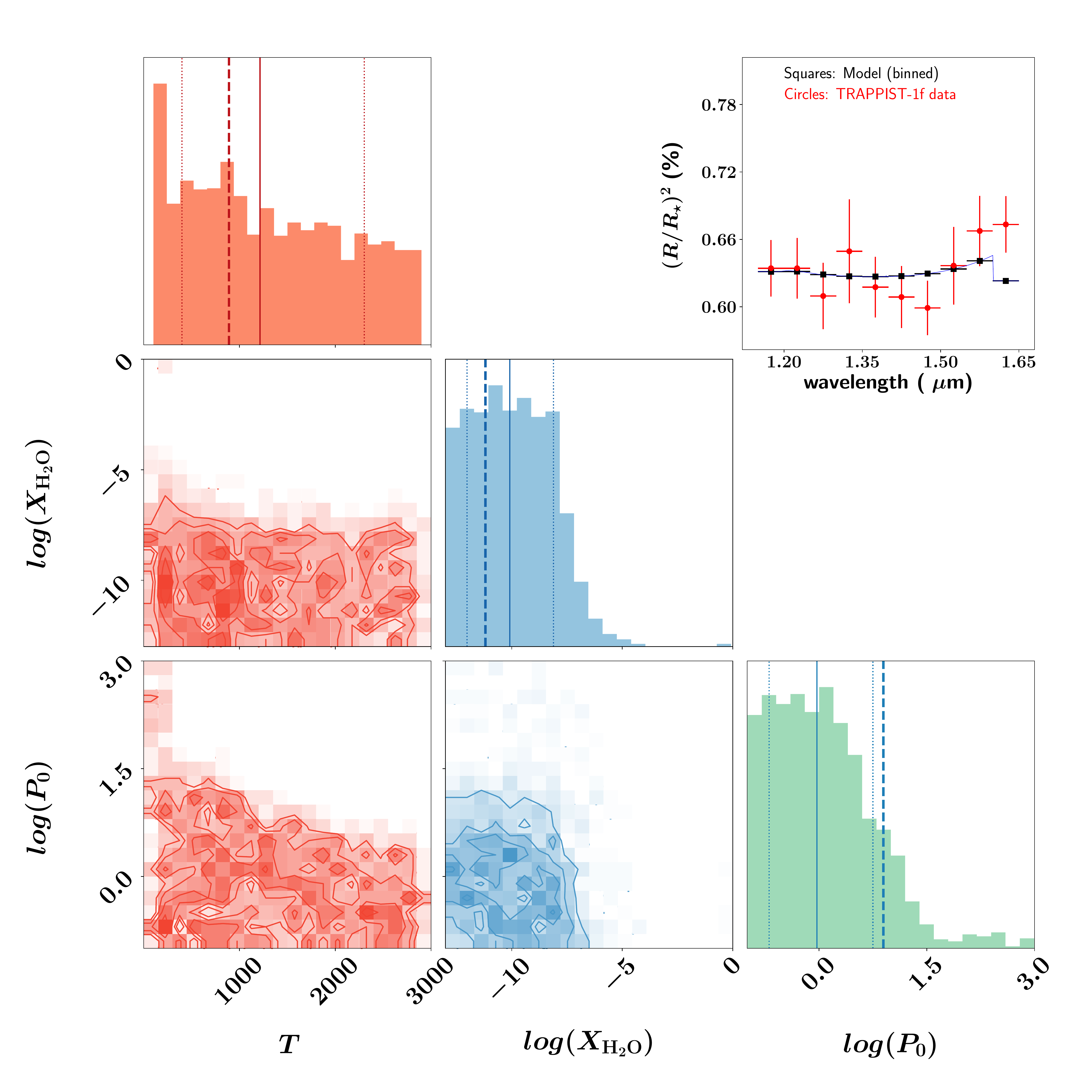}
\hspace{0.1in}
\includegraphics[width=1.2\columnwidth]{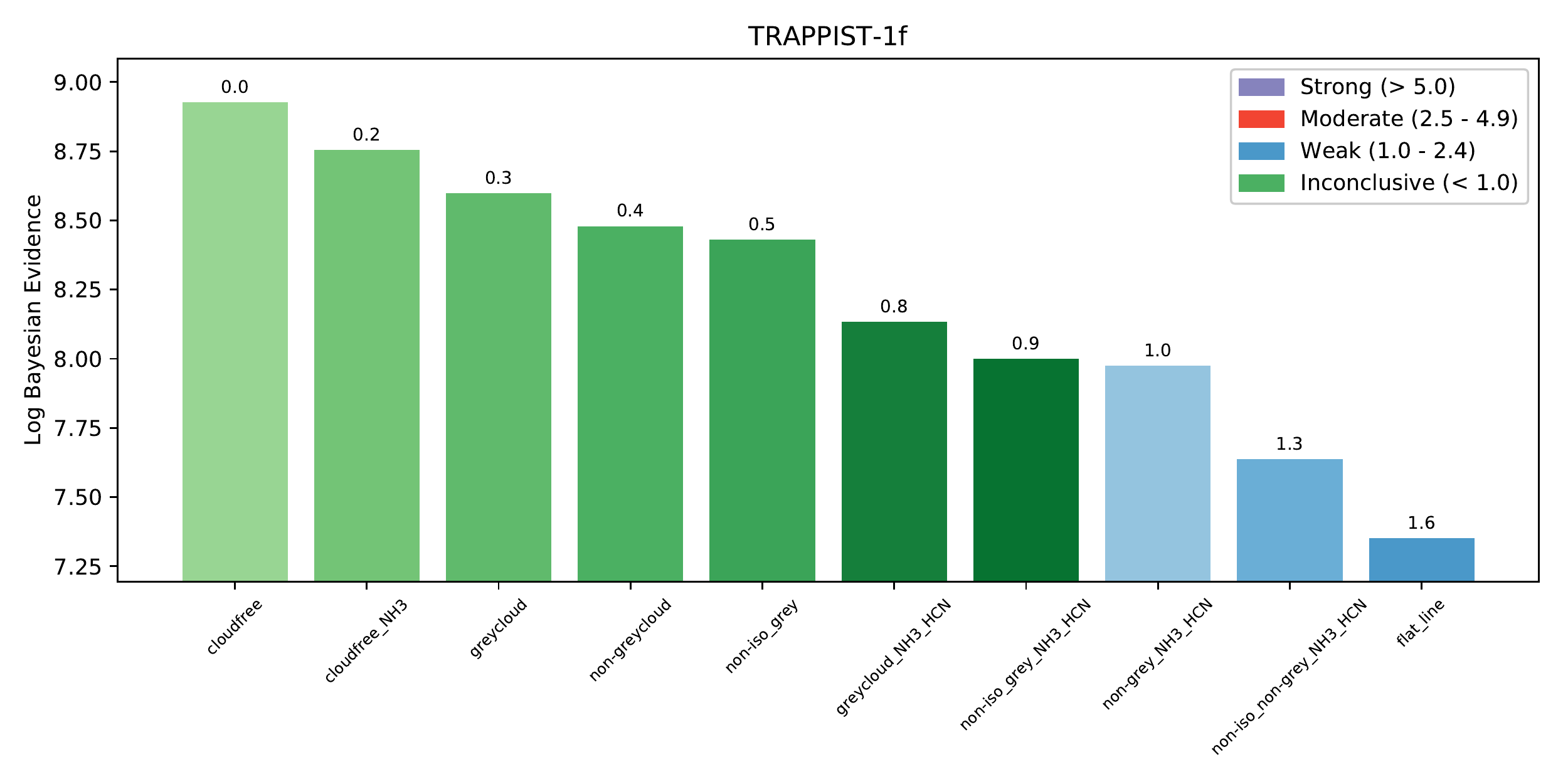}
\includegraphics[width=0.65\columnwidth]{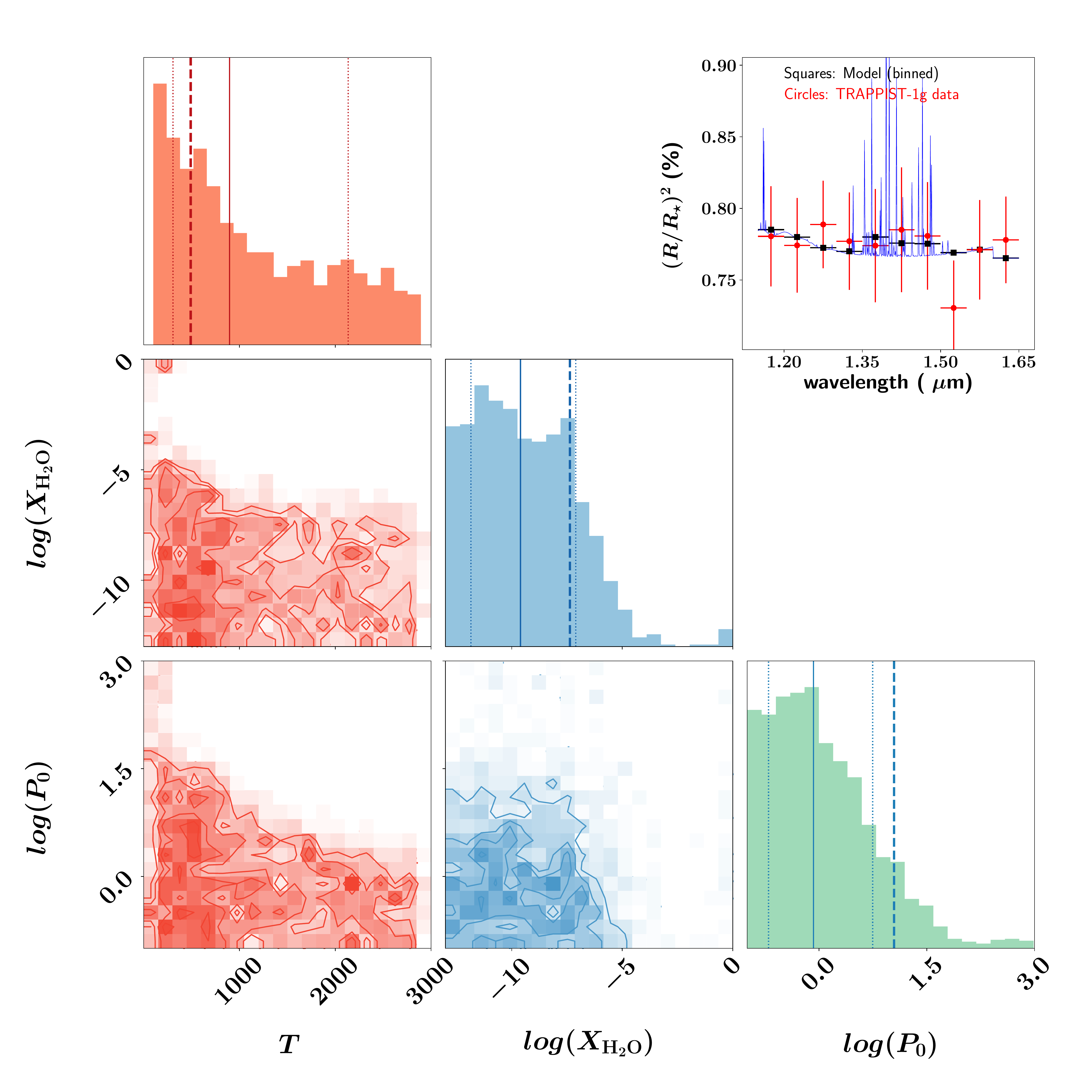}
\hspace{0.1in}
\includegraphics[width=1.2\columnwidth]{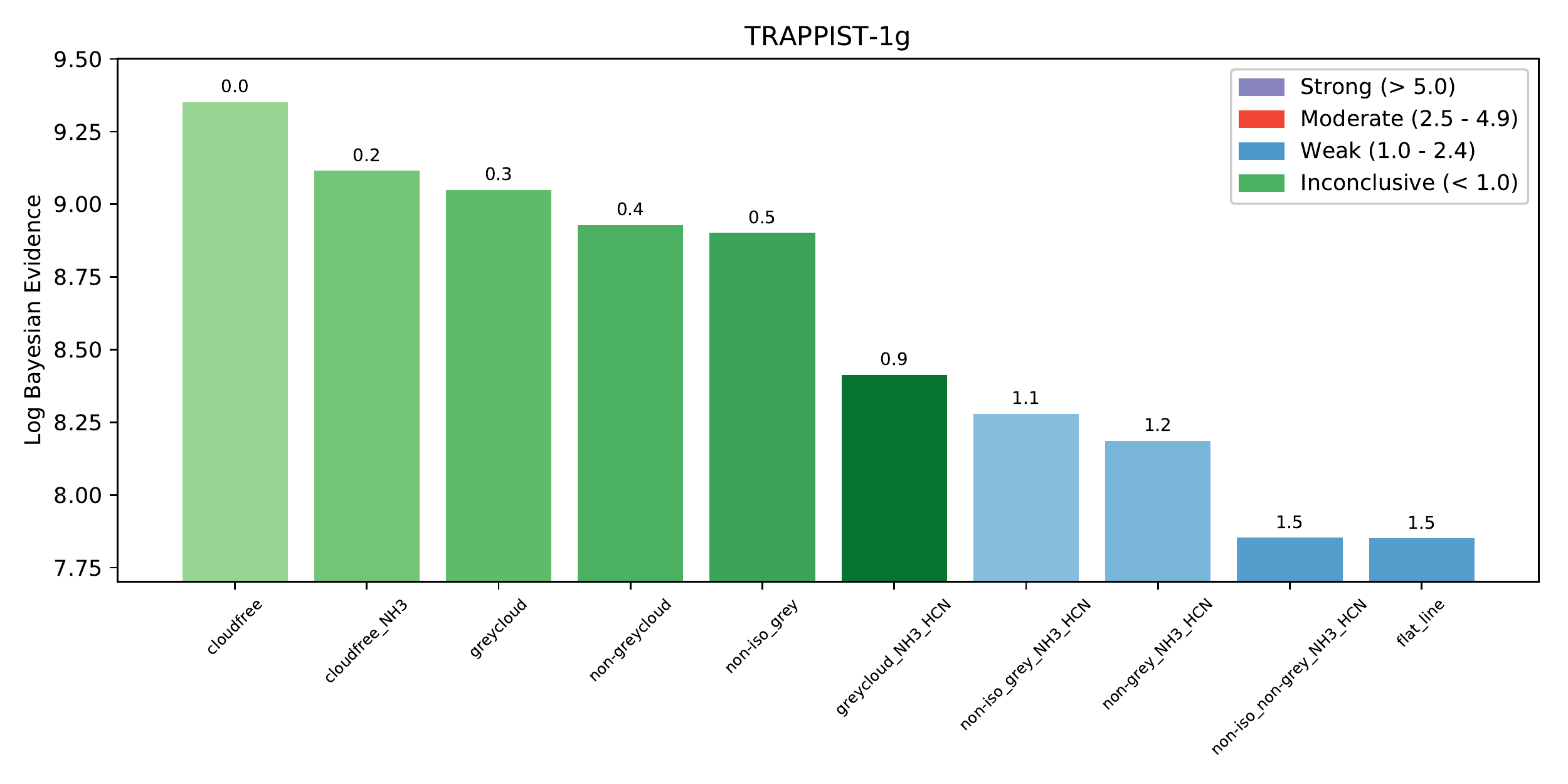}
\end{center}
\vspace{-0.1in}
\caption{Same as Figure \ref{fig:cloud-free_vs_nh3}, but for the TRAPPIST-1 exoplanets assuming Earth-like atmospheres ($m=29 ~m_{\rm H}$).}
\label{fig:trappist}
\end{figure*}

\begin{figure*}
\begin{center}
\includegraphics[width=0.65\columnwidth]{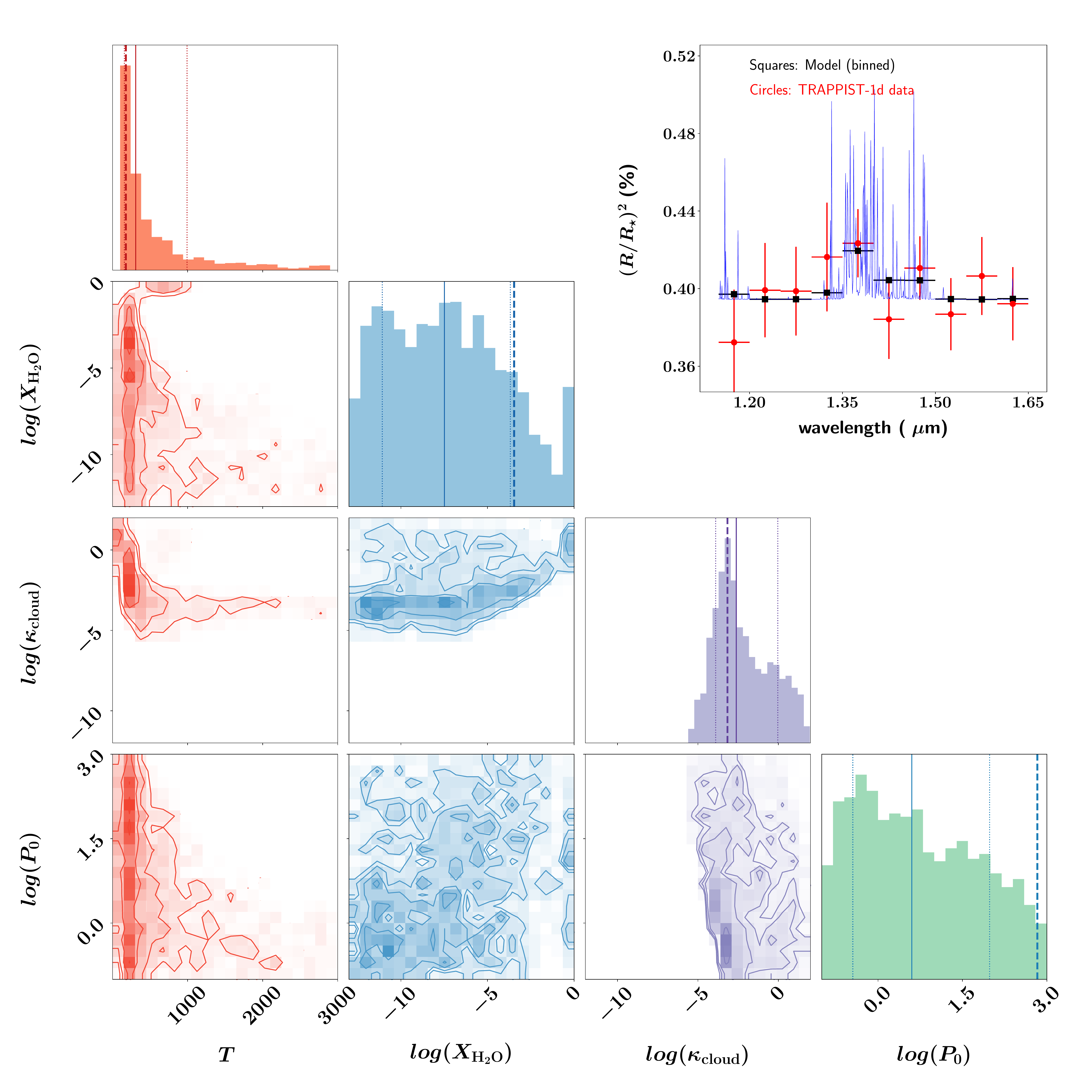}
\hspace{0.1in}
\includegraphics[width=1.2\columnwidth]{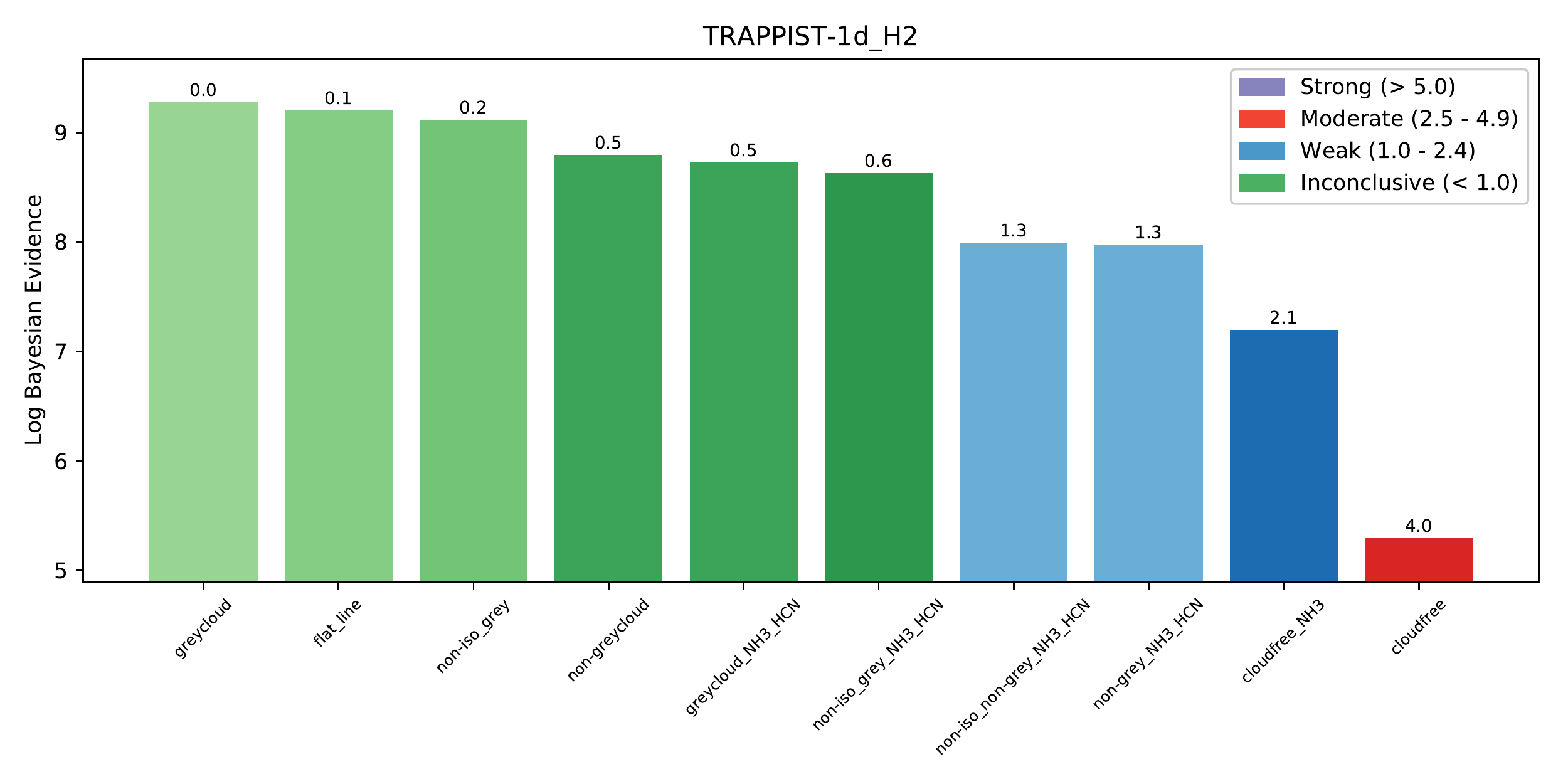}
\includegraphics[width=0.65\columnwidth]{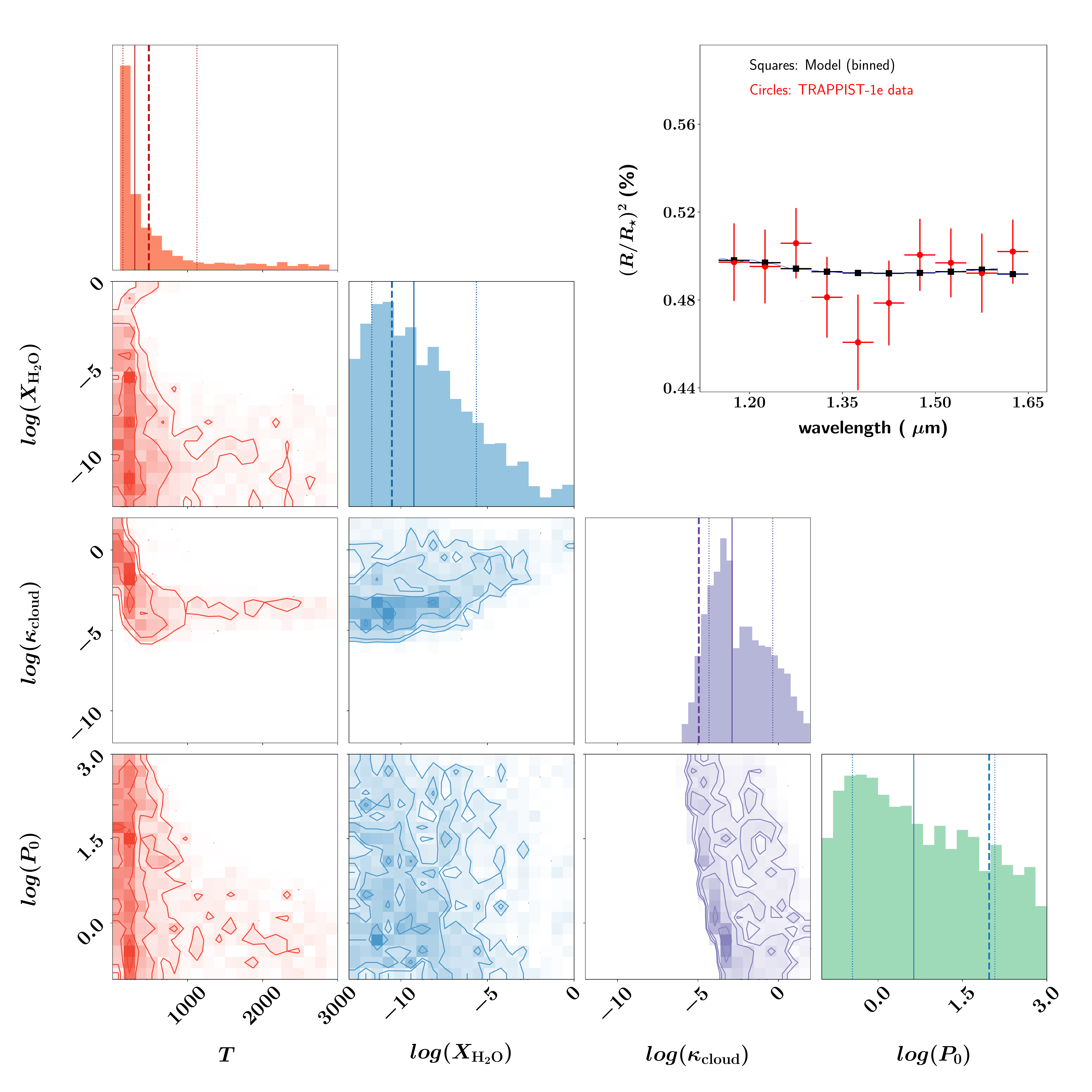}
\hspace{0.1in}
\includegraphics[width=1.2\columnwidth]{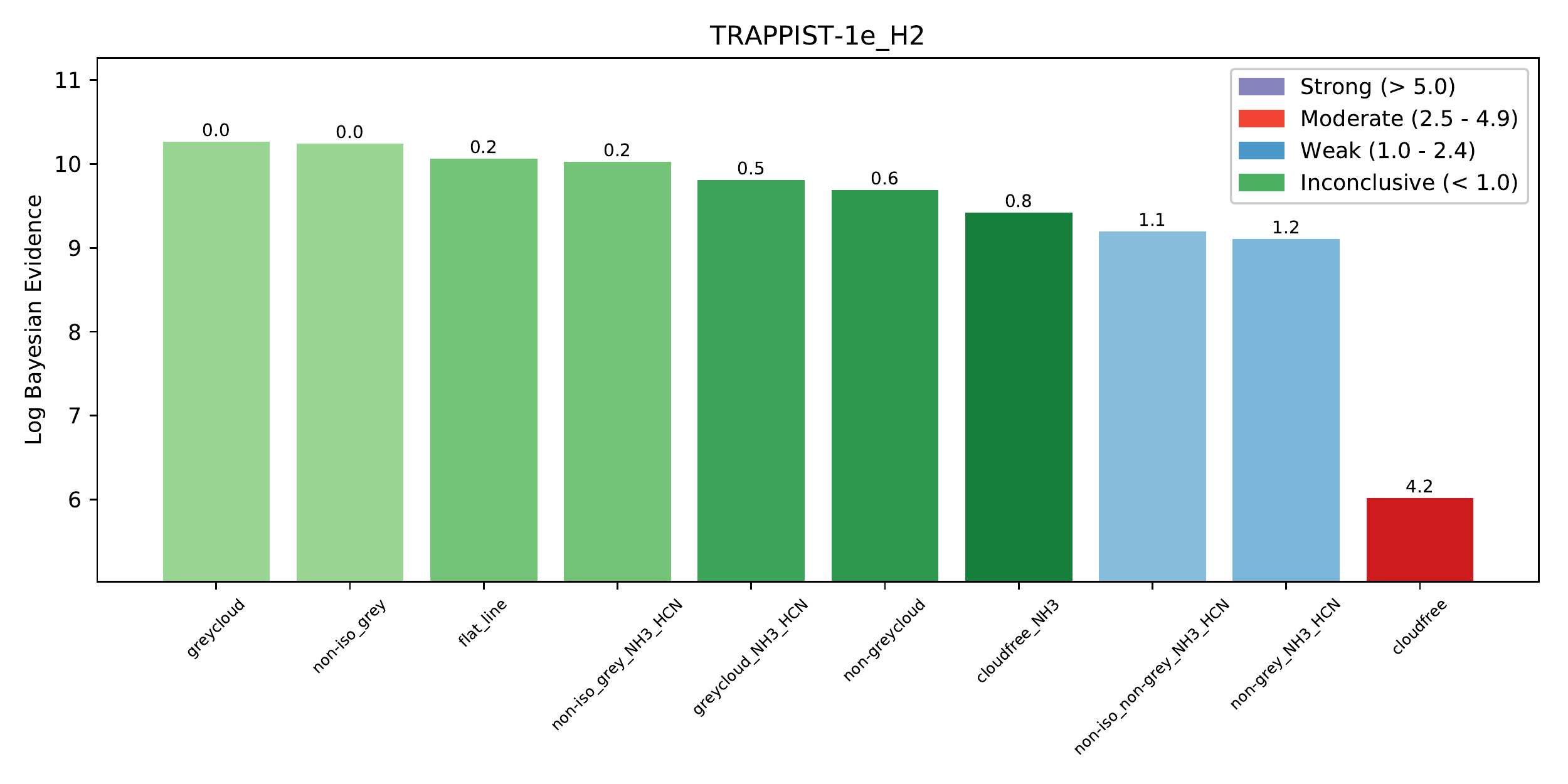}
\includegraphics[width=0.65\columnwidth]{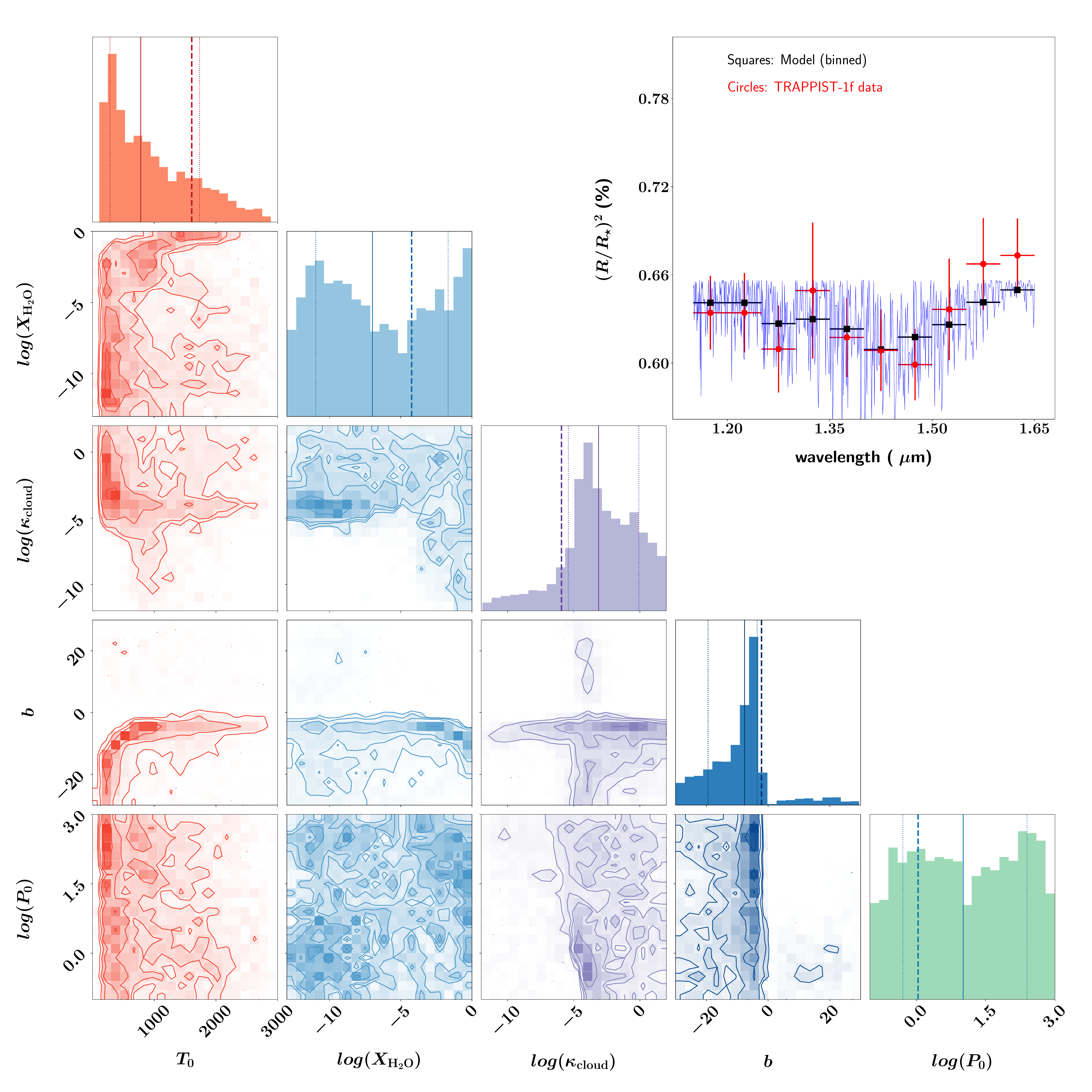}
\hspace{0.1in}
\includegraphics[width=1.2\columnwidth]{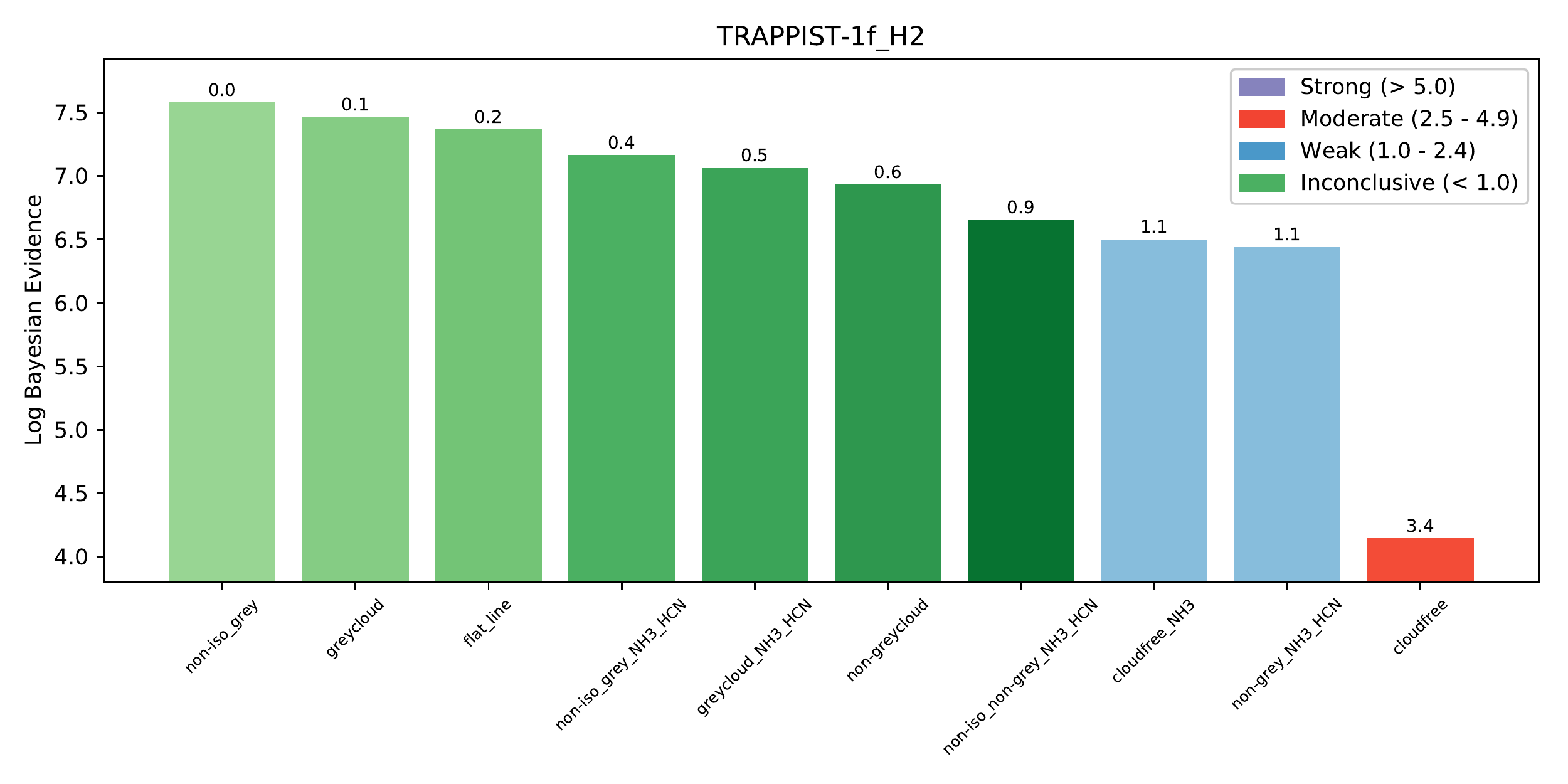}
\includegraphics[width=0.65\columnwidth]{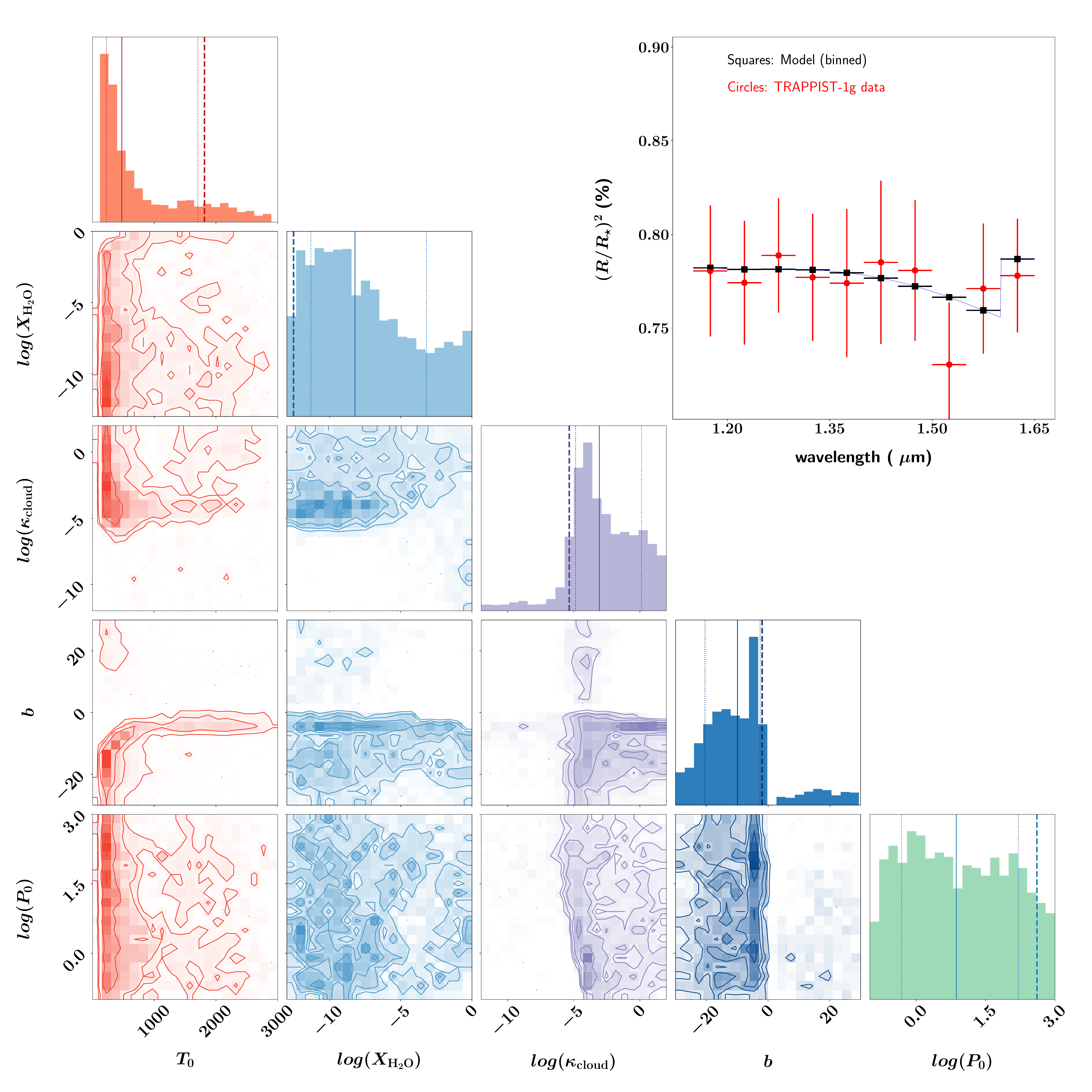}
\hspace{0.1in}
\includegraphics[width=1.2\columnwidth]{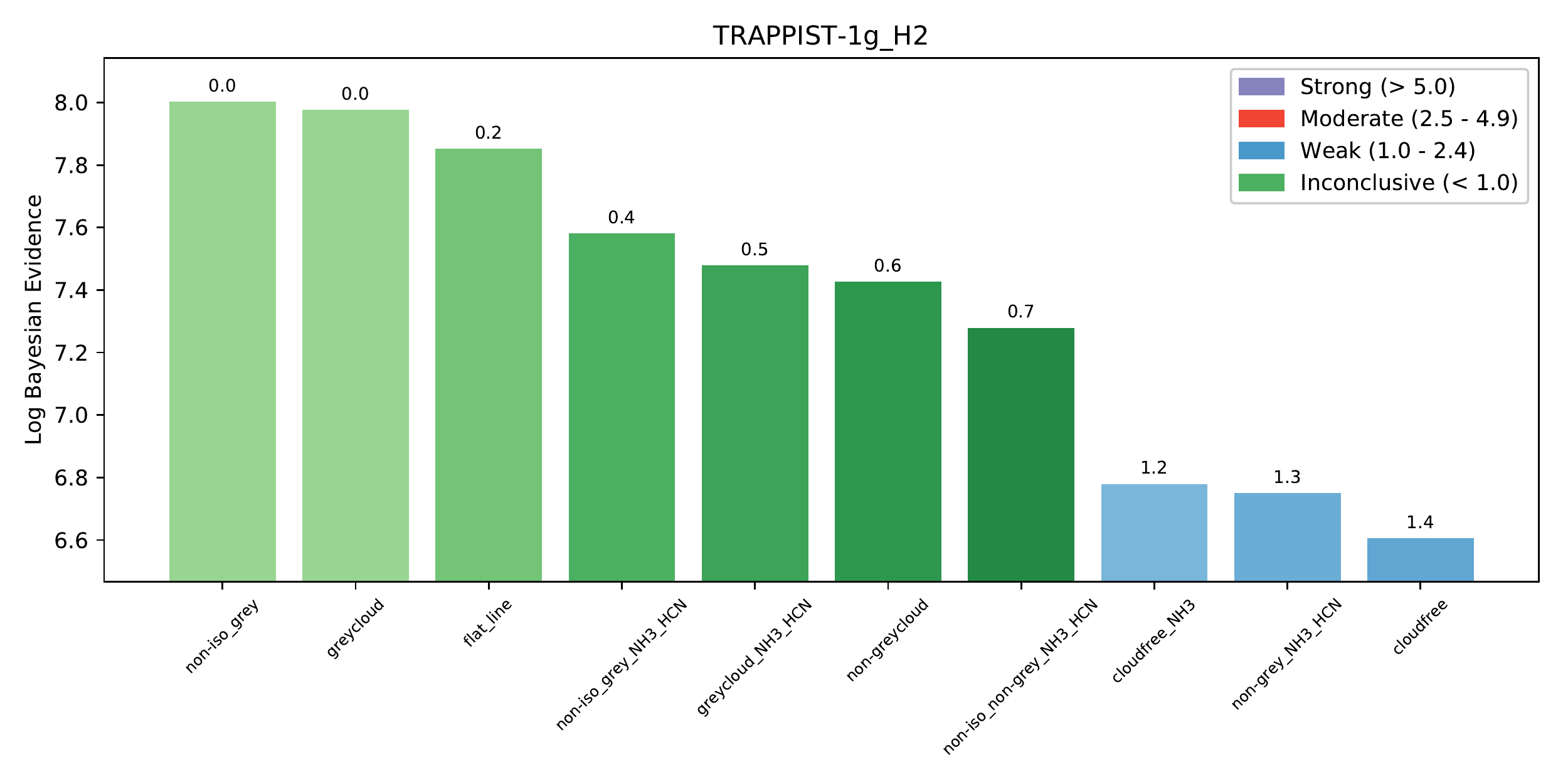}
\end{center}
\vspace{-0.1in}
\caption{Same as Figure \ref{fig:trappist}, but assuming atmospheres dominated by molecular hydrogen (variable $m$), where the pressure scale height is larger by about an order of magnitude.}
\label{fig:trappist_2}
\end{figure*}

\begin{figure*}
%\vspace{-0.1in}
\begin{center}
\includegraphics[width=\columnwidth]{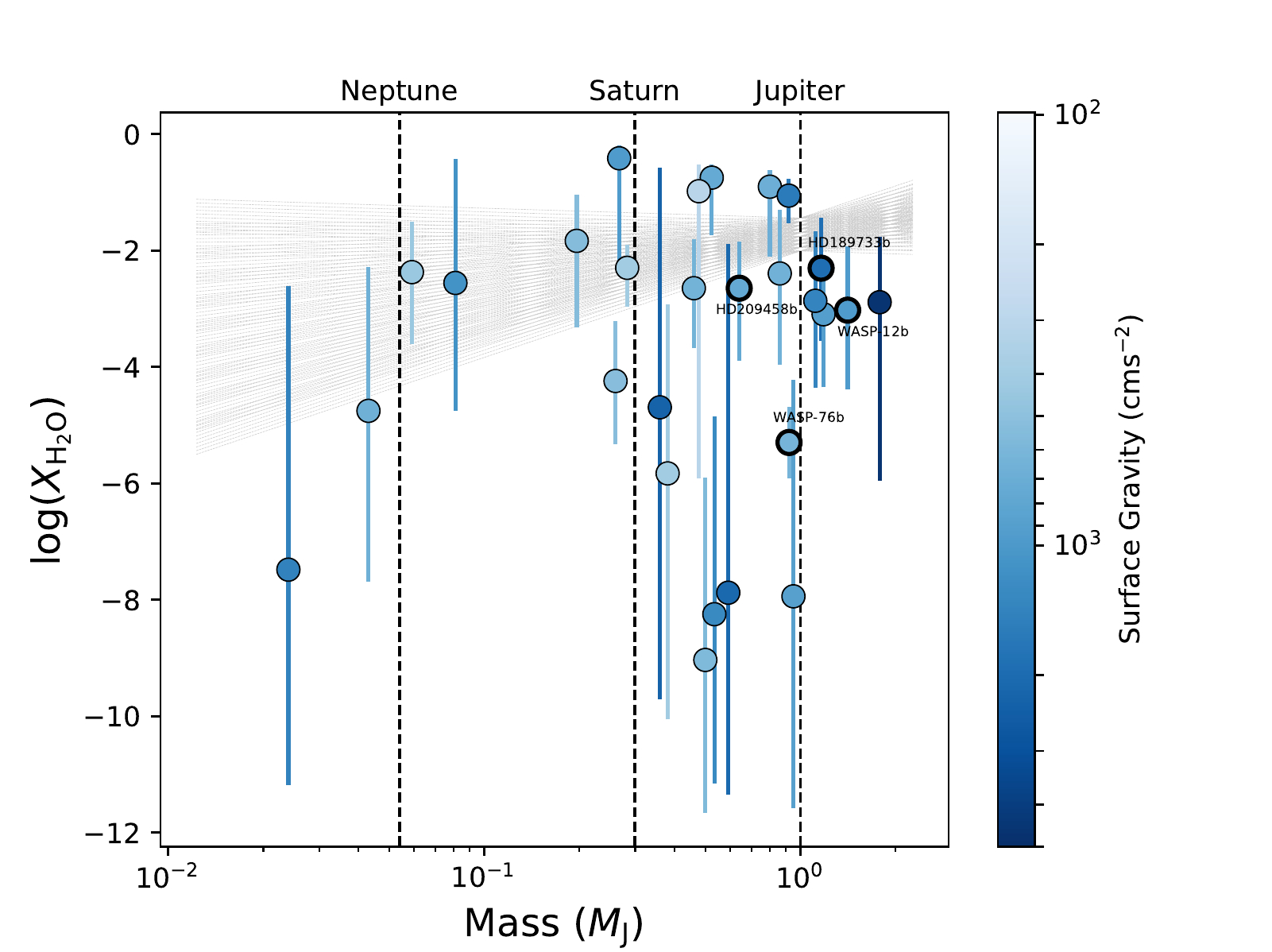}
\includegraphics[width=\columnwidth]{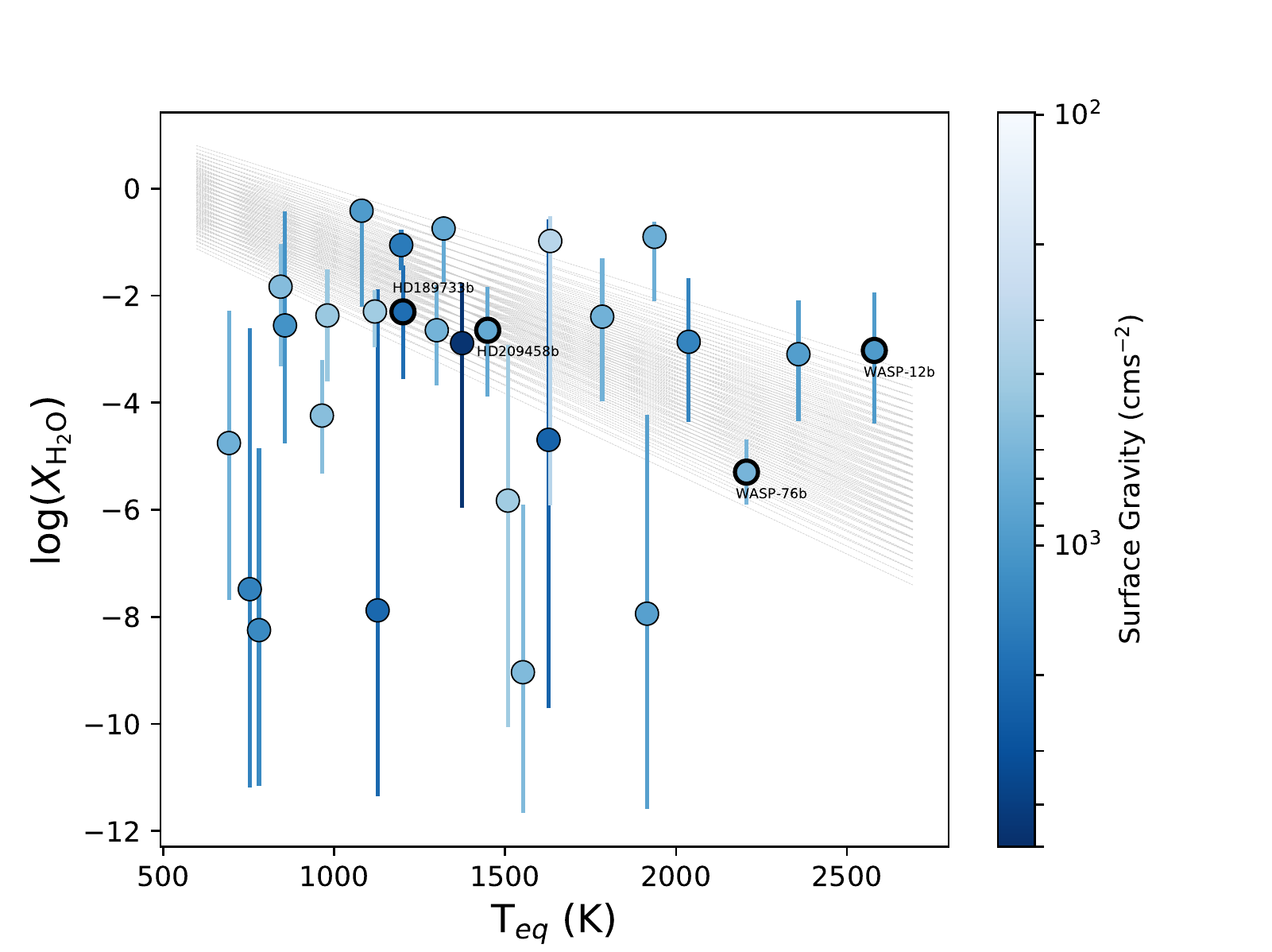}
\includegraphics[width=\columnwidth]{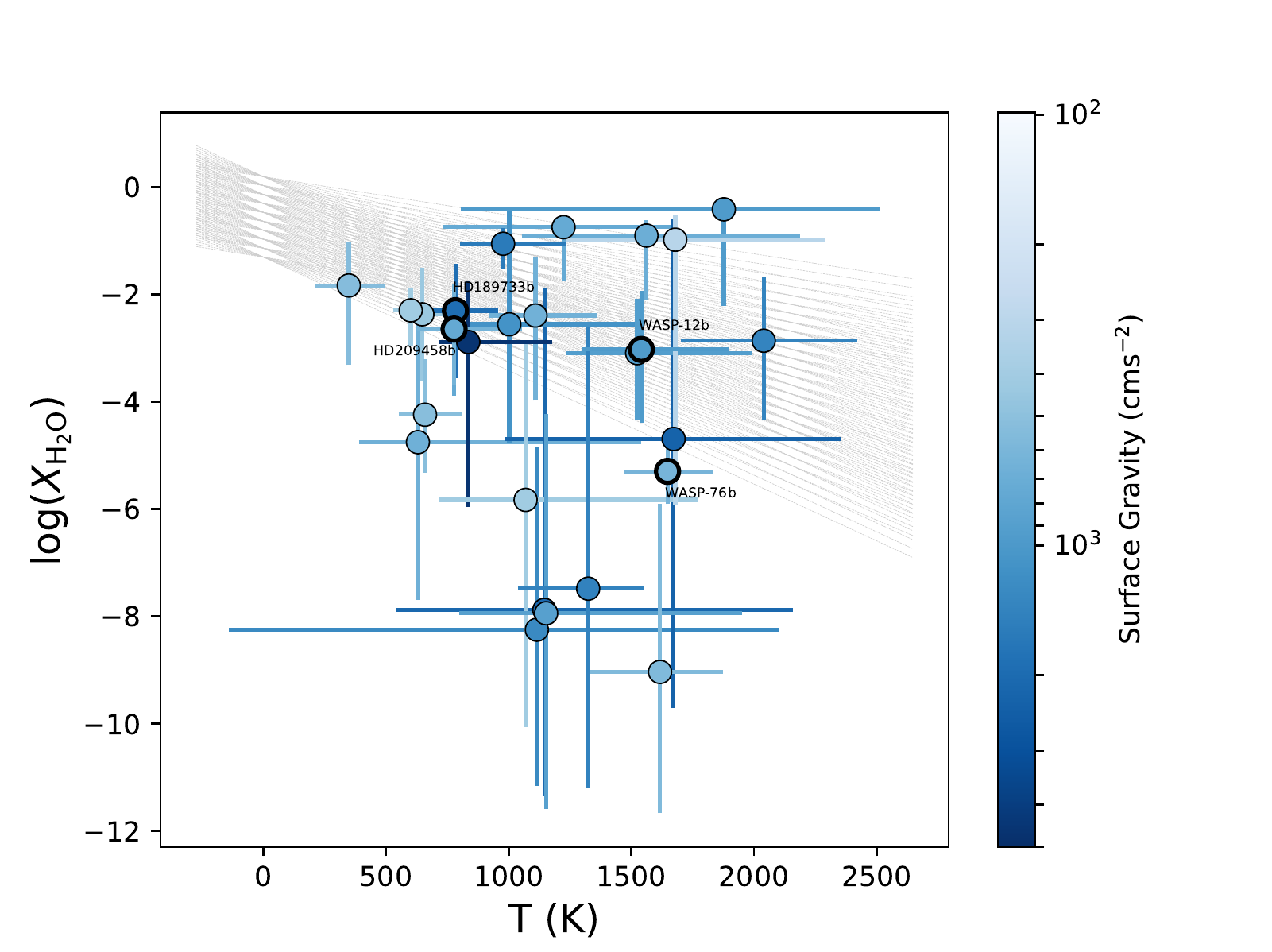}
\includegraphics[width=\columnwidth]{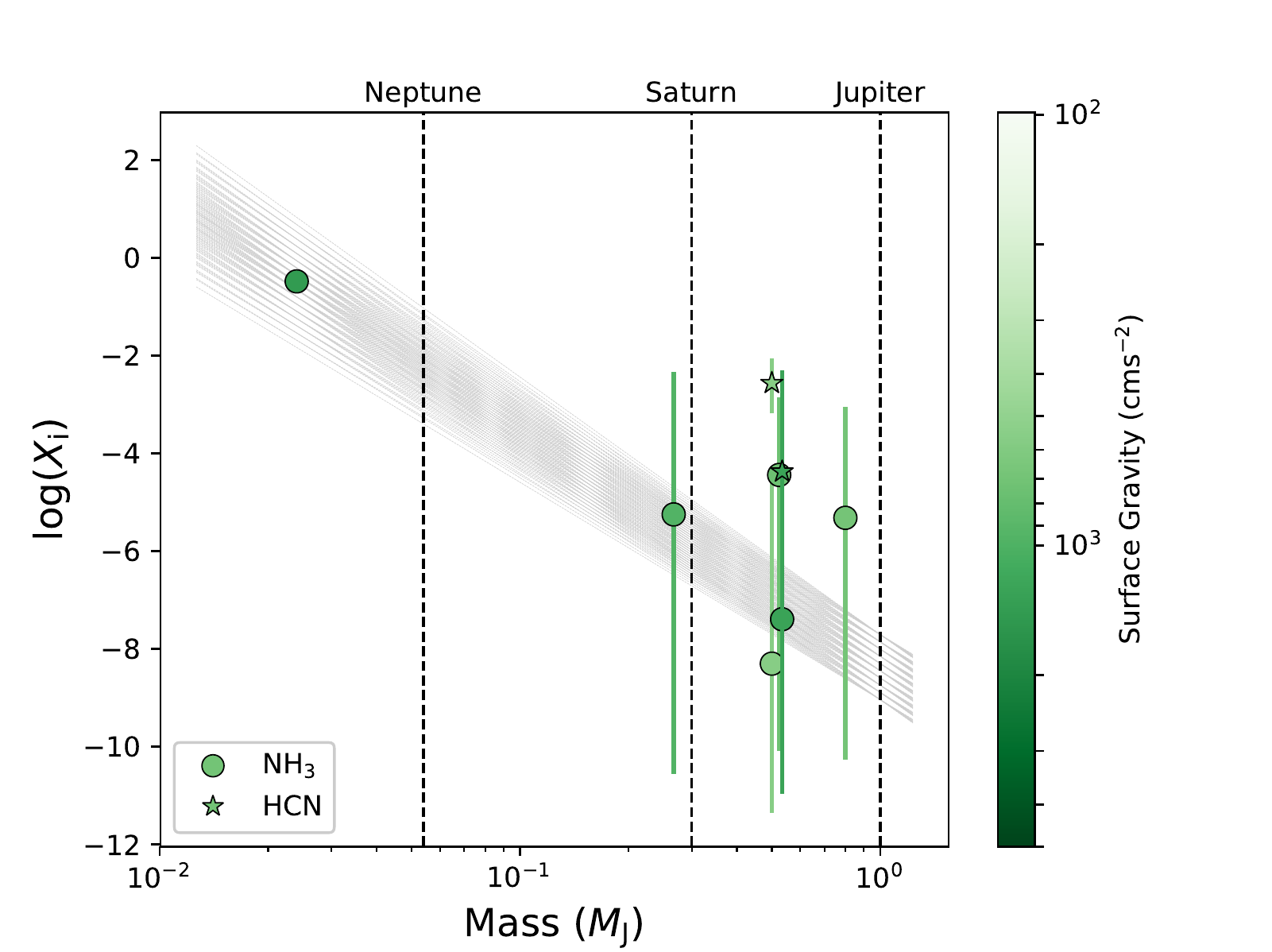}
\includegraphics[width=\columnwidth]{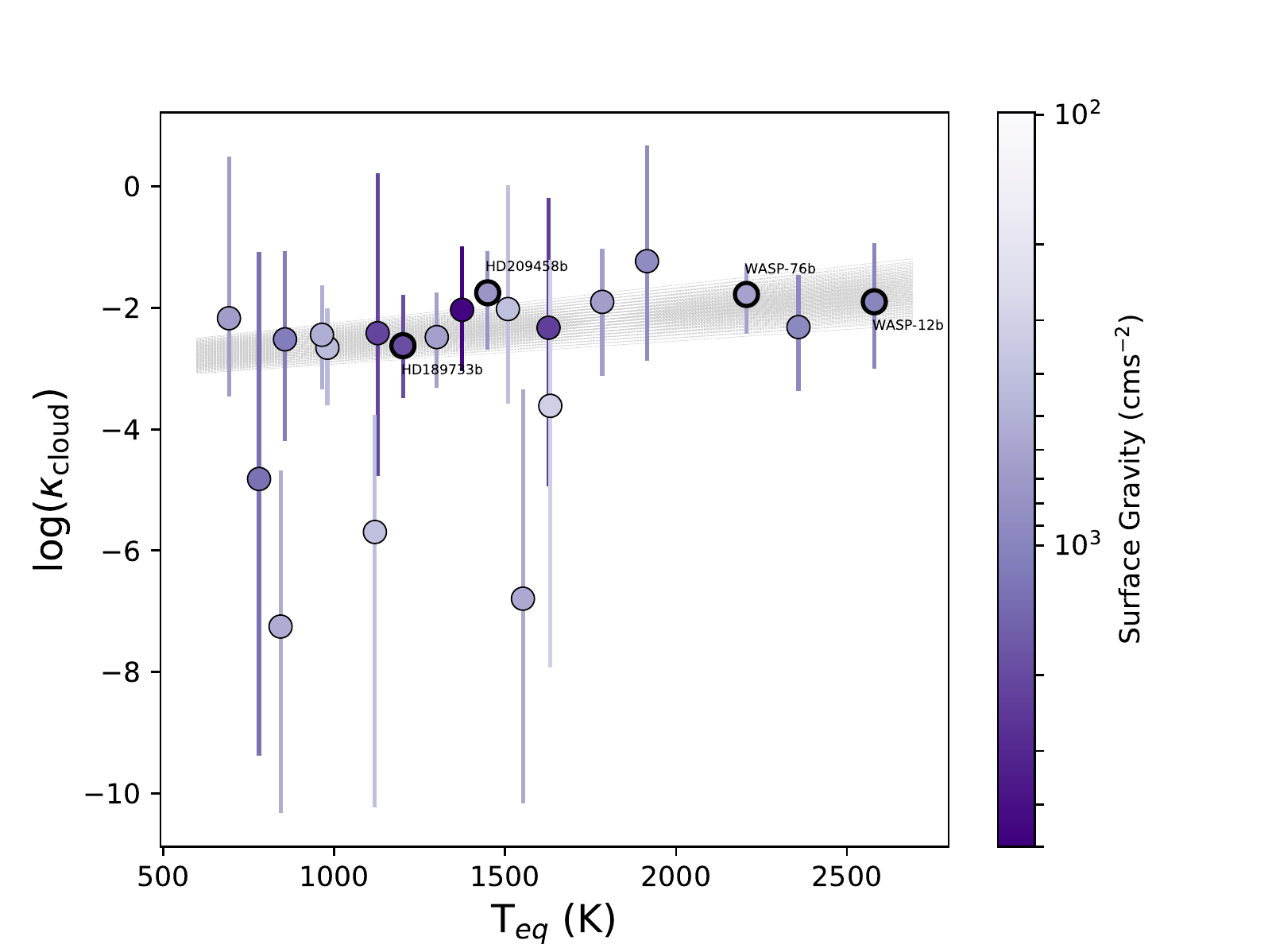}
\includegraphics[width=\columnwidth]{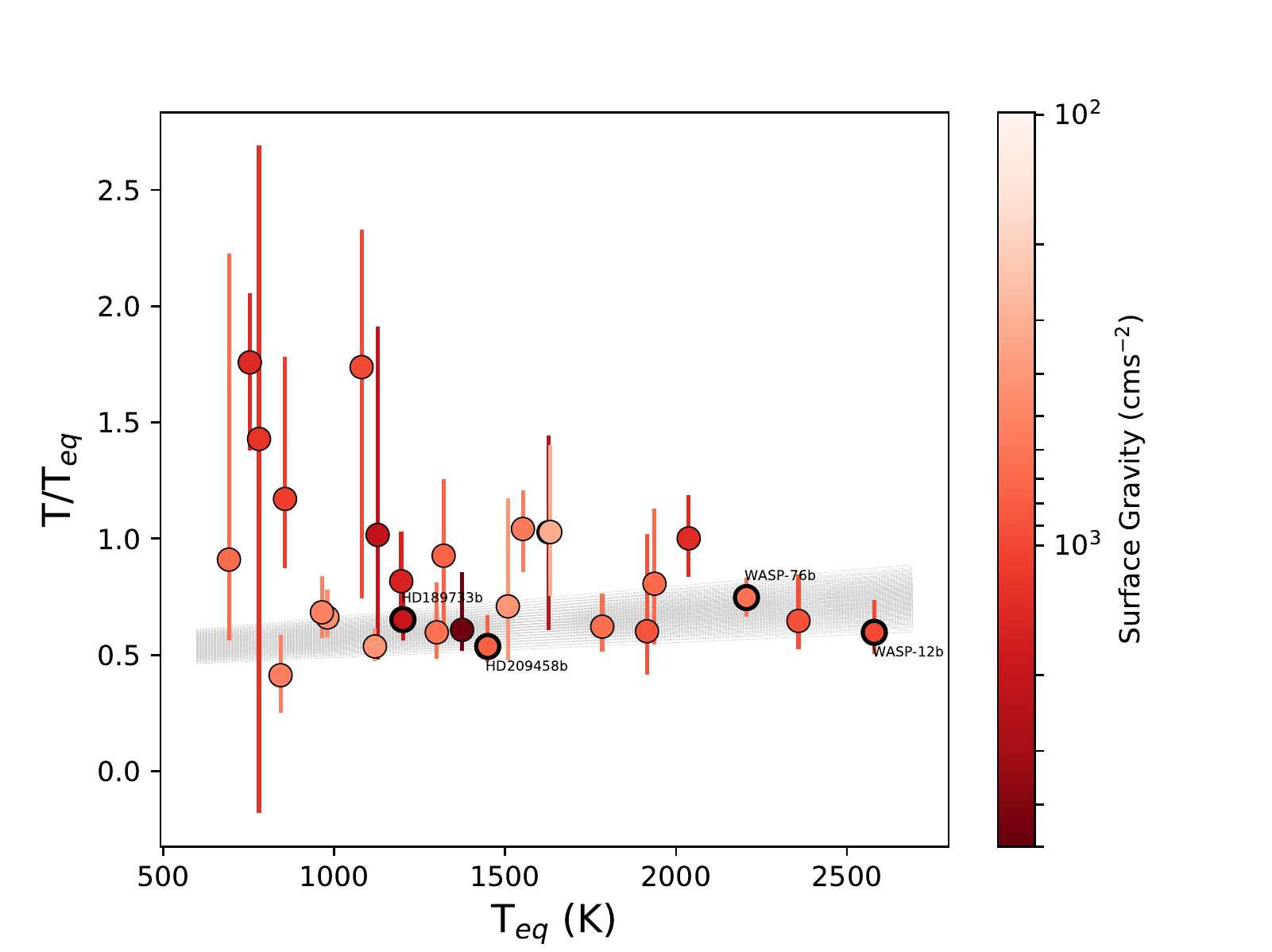}
\end{center}
\caption{A search for trends between the retrieved atmospheric properties based on the best model (highest Bayesian evidence) of each object.  We have excluded 7 objects that can be adequately fitted by a one-parameter flat line.  We have also excluded the 4 TRAPPIST-1 exoplanets.  The family of lines in each panel shows the Monte Carlo fits of a two-parameter straight line (slope and constant offset).  Top row: Water volume mixing ratio versus exoplanet mass (top left panel; slope of $0.94 \pm 1.11$) and equilibrium temperature (top right panel; slope of $-0.00245 \pm 0.00054$ K$^{-1}$).  Middle left panel: Water volume mixing ratio versus retrieved atmospheric temperature; slope of $-0.00134 \pm 0.00078$ K$^{-1}$.  Middle right panel: Ammonia and hydrogen cyanide volume mixing ratios versus exoplanet mass; slope of $-4.81 \pm 0.46$.  Bottom left panel: Grey cloud opacity versus equilibrium temperature; slope of $0.000517 \pm 0.000135$ K$^{-1}$.  Bottom right panel: Ratio of retrieved to equilibrium temperatures versus equilibrium temperature; slope of $0.000103 \pm 0.000038$ K$^{-1}$.}
\label{fig:trends}
\end{figure*}

\begin{table*}
\begin{center}
\caption{Summary of retrieval outcomes (38 objects, 42 sets of retrievals)}
\label{tab:results}
\begin{tabular}{lcccccccc}
\hline
\hline
Name & $T_{\rm eq}$ (K) & $T$ (K) & $log(X_{\rm H_2O})$ & $log(X_{\rm HCN})$ & $log(X_{\rm NH_3})$ & Cloudy? & Non-grey clouds? & $log(\kappa_{\rm cloud})$ (cm$^2$ g$^{-1}$) \\
\hline
\hline
GJ436b & 633 & $\dagger$ &$\dagger$ & $\dagger$ & $\dagger$ & $\dagger$ & $\dagger$ & $\dagger$ \\
GJ3470b & 692 & $629^{+911}_{-239}$ & $-4.75^{+2.47}_{-2.93}$ & $-$ & $-$ & Yes & No & $-2.17^{+2.67}_{-1.29}$ \\
HAT-P-1b & 1320 & $1223^{+435}_{-492}$ & $-0.75^{+0.23}_{-1.0}$ & $-$ & $-4.44^{+1.59}_{-5.65}$ & Maybe & $-$ & $-$ \\
HAT-P-3b & 1127 & $1145^{+1011}_{-604}$ & $-7.88^{+5.99}_{-3.47}$ & $-$ & $-$ & Maybe & $-$ & $-2.42^{+2.63}_{-2.35}$ \\
HAT-P-11b & 856 & $1002^{+524}_{-255}$ & $-2.56^{+2.13}_{-2.2}$ & $-$ & $-$ & Maybe & $-$ & $-2.52^{+1.45}_{-1.68}$ \\
HAT-P-12b & 958 & $\dagger$ &$\dagger$ & $\dagger$ & $\dagger$ & $\dagger$ & $\dagger$ & $\dagger$ \\
HAT-P-17b & 780 & $1114^{+984}_{-1255}$ & $-8.25^{+3.4}_{-2.91}$ & $-4.37^{+2.08}_{-5.17}$ & $-7.39^{+4.02}_{-3.58}$ & Yes & No & $-4.82^{+3.74}_{-4.56}$ \\
HAT-P-18b & 843 & $347^{+146}_{-137}$ & $-1.83^{+0.8}_{-1.48}$ & $-$ & $-$ & Maybe & $-$ & $-7.25^{+2.57}_{-3.07}$ \\
HAT-P-26b & 980 & $647^{+118}_{-82}$ & $-2.37^{+0.86}_{-1.24}$ & $-$ & $-$ & Yes & No & $-2.66^{+0.65}_{-0.94}$ \\
HAT-P-32b & 1784 & $1109^{+251}_{-190}$ & $-2.39^{+1.09}_{-1.57}$ & $-$ & $-$ & Yes & No & $-1.9^{+0.88}_{-1.22}$ \\
HAT-P-38b & 1080 & $1876^{+637}_{-1074}$ & $-0.41^{+0.22}_{-1.8}$ & $-$ & $-5.25^{+2.92}_{-5.31}$ & Maybe & $-$ & $-$ \\
HAT-P-41b & 1937 & $1561^{+624}_{-507}$ & $-0.9^{+0.28}_{-1.2}$ & $-$ & $-5.32^{+2.27}_{-4.94}$ & Maybe & $-$ & $-$ \\
HD149026b & 1627 & $1672^{+679}_{-687}$ & $-4.69^{+4.11}_{-5.02}$ & $-$ & $-$ & Maybe & $-$ & $-2.33^{+2.14}_{-2.61}$ \\
HD189733b & 1201 & $782^{+172}_{-107}$ & $-2.3^{+0.87}_{-1.26}$ & $-$ & $-$ & Yes & No & $-2.62^{+0.84}_{-0.86}$ \\
HD209458b & 1449 & $777^{+193}_{-95}$ & $-2.65^{+0.81}_{-1.24}$ & $-$ & $-$ & Yes & No & $-1.75^{+0.69}_{-0.94}$ \\
WASP-29b & 963 & $\dagger$ &$\dagger$ & $\dagger$ & $\dagger$ & $\dagger$ & $\dagger$ & $\dagger$ \\
WASP-31b & 1576 & $\dagger$ &$\dagger$ & $\dagger$ & $\dagger$ & $\dagger$ & $\dagger$ & $\dagger$ \\
WASP-39b & 1119 & $600^{+86}_{-72}$ & $-2.3^{+0.4}_{-0.67}$ & $-$ & $-$ & Yes & No & $-5.69^{+1.94}_{-4.54}$ \\
WASP-43b & 1374 & $835^{+340}_{-121}$ & $-2.89^{+1.13}_{-3.07}$ & $-$ & $-$ & Yes & No & $-2.03^{+1.04}_{-1.01}$ \\
WASP-52b & 1300 & $776^{+278}_{-149}$ & $-2.65^{+0.84}_{-1.03}$ & $-$ & $-$ & Yes & No & $-2.48^{+0.74}_{-0.83}$ \\
WASP-63b & 1508 & $1068^{+700}_{-352}$ & $-5.83^{+2.9}_{-4.23}$ & $-$ & $-$ & Yes & No & $-2.02^{+2.03}_{-1.56}$ \\
WASP-67b & 1026 & $\dagger$ &$\dagger$ & $\dagger$ & $\dagger$ & $\dagger$ & $\dagger$ & $\dagger$ \\
WASP-69b & 964 & $658^{+148}_{-107}$ & $-4.24^{+1.03}_{-1.09}$ & $-$ & $-$ & Yes & Yes & $-2.44^{+0.81}_{-0.89}$ \\
WASP-74b & 1915 & $1152^{+798}_{-354}$ & $-7.94^{+3.72}_{-3.64}$ & $-$ & $-$ & Yes & No & $-1.23^{+1.9}_{-1.64}$ \\
WASP-76b & 2206 & $1647^{+185}_{-178}$ & $-5.3^{+0.61}_{-0.61}$ & $-$ & $-$ & Yes & Yes & $-1.78^{+0.47}_{-0.65}$ \\
WASP-80b & 824 & $\dagger$ &$\dagger$ & $\dagger$ & $\dagger$ & $\dagger$ & $\dagger$ & $\dagger$ \\
WASP-101b & 1552 & $1616^{+256}_{-288}$ & $-9.03^{+3.13}_{-2.63}$ & $-2.56^{+0.5}_{-0.62}$ & $-8.3^{+3.84}_{-3.06}$ & Yes & No & $-6.79^{+3.45}_{-3.37}$ \\
WASP-121b & 2358 & $1523^{+468}_{-290}$ & $-3.09^{+1.01}_{-1.26}$ & $-$ & $-$ & Yes & No & $-2.32^{+0.86}_{-1.05}$ \\
XO-1b & 1196 & $977^{+254}_{-174}$ & $-1.06^{+0.29}_{-0.47}$ & $-$ & $-$ & Maybe & $-$ & $-$ \\
GJ1214b & 573 & $\dagger$ &$\dagger$ & $\dagger$ & $\dagger$ & $\dagger$ & $\dagger$ & $\dagger$ \\
HD97658b & 753 & $1323^{+224}_{-286}$ & $-7.48^{+4.88}_{-3.7}$ & $-$ & $-0.48^{+0.19}_{-0.23}$ & Maybe & $-$ & $-$ \\
WASP-17b & 1632 & $1678^{+610}_{-448}$ & $-0.98^{+0.46}_{-4.94}$ & $-$ & $-$ & Maybe & $-$ & $-3.61^{+2.4}_{-4.31}$ \\
WASP-19b & 2037 & $2039^{+381}_{-338}$ & $-2.86^{+1.2}_{-1.49}$ & $-$ & $-$ & Maybe & $-$ & $-$ \\
WASP-12b & 2580 & $1540^{+358}_{-242}$ & $-3.02^{+1.09}_{-1.36}$ & $-$ & $-$ & Yes & No & $-1.9^{+0.97}_{-1.11}$ \\
TRAPPIST-1d & 288 & $\dagger$ &$\dagger$ & $\dagger$ & $\dagger$ & $\dagger$ & $\dagger$ & $\dagger$ \\
 &  & $\dagger\clubsuit$ &$\dagger\clubsuit$ & $\dagger\clubsuit$ & $\dagger\clubsuit$ & $\dagger\clubsuit$ & $\dagger\clubsuit$ & $\dagger\clubsuit$ \\
TRAPPIST-1e & 251 & $\dagger$ &$\dagger$ & $\dagger$ & $\dagger$ & $\dagger$ & $\dagger$ & $\dagger$ \\
 &  & $1173^{+1108}_{-729}$$^\clubsuit$ & $-10.02^{+2.02}_{-2.02}$$^\clubsuit$ & $-$ & $-$ & Maybe$^\clubsuit$ & $-$ & $-$ \\
TRAPPIST-1f & 219 & $\dagger$ &$\dagger$ & $\dagger$ & $\dagger$ & $\dagger$ & $\dagger$ & $\dagger$ \\
&  & $1214^{+1089}_{-815}$$^\clubsuit$ & $-10.09^{+1.98}_{-1.94}$$^\clubsuit$ & $-$ & $-$ & Maybe$^\clubsuit$ & $-$ & $-$ \\
TRAPPIST-1g & 199 & $\dagger$ &$\dagger$ & $\dagger$ & $\dagger$ & $\dagger$ & $\dagger$ & $\dagger$ \\
 &  & $896^{+1238}_{-590}$$^\clubsuit$ & $-9.61^{+2.5}_{-2.24}$$^\clubsuit$ & $-$ & $-$ & Maybe$^\clubsuit$ & $-$ & $-$ \\
\hline
\hline
\end{tabular}\\
%\vspace{0.05in}
\end{center}
\textbf{For ``Cloudy?":} ``Yes" refers to cases where all of the cloud-free models have Bayes factors of unity or more.  ``No" means only cloud-free models have Bayes factor of less than unity.  ``Maybe" means a mixture of cloud-free and cloudy models have Bayes factor of less than unity. \textbf{For ``Non-grey clouds?":} ``Yes" refers to cases where only non-grey-cloud models have Bayes factors of less than unity.  ``No" means a mixture of non-grey-cloud and grey-cloud models have Bayes factors of less than unity.\\
%$\spadesuit$: XO-1b may also be explained by a cloud-free atmosphere with water only. \\
$\clubsuit$: For the TRAPPIST-1 exoplanets, we also examine Earth-like atmospheres ($m=29m_{\rm H}$).\\
$\dagger$: Flat-line fit has Bayes factor of less than unity and no atmospheric properties may be retrieved.
\end{table*}

\begin{table*}
\begin{center}
\caption{Summary of input parameters (38 objects, 42 sets of retrievals)}
\label{tab:parameters}
\begin{tabular}{lcccccc}
\hline
\hline
Name & $R_{\star} (R_{\odot})$ & $M$ ($M_{\rm J}$) & $R_0 (R_{\rm J})$ & $g$ (cm s$^{-2}$) & References & $\bar{R}_{\rm WFC3} (R_{\rm J})$ \\
\hline
\hline
GJ436b & 0.455 & $0.078^{+0.007}_{-0.008}$ & 0.3532 & 1318 & \cite{vb12} & 0.3693 \\
GJ3470b & 0.48 & $0.043 \pm 0.005$ & 0.3287 & 676 & \cite{biddle14} & 0.3630 \\
HAT-P-1b & 1.115 & $0.524 \pm 0.031$ & 1.213 & 750 & \cite{john08,sing16} & 1.272 \\
HAT-P-3b & 0.799 & $0.591 \pm 0.018$ & 0.8383 & 2138 & \cite{chan11} & 0.8559 \\
HAT-P-11b & 0.75 & $0.081 \pm 0.009$ & 0.4077 & 1122 & \cite{bakos10} & 0.4332 \\
HAT-P-12b & 0.701 & $0.211 \pm 0.012$ & 0.8770 & 562 & \cite{hartman09} & 0.9341 \\
HAT-P-17b & 0.838 & $0.534 \pm 0.018$ & 0.9677 & 1288 & \cite{howard12} & 0.9880 \\
HAT-P-18b & 0.717 & $0.196 \pm 0.008$ & 0.9349 & 542 & \cite{esp14} & 0.9552 \\
HAT-P-26b & 0.788 & $0.059 \pm 0.007$ & 0.4741 & 447 & \cite{hartman11} & 0.5475 \\
HAT-P-32b & 1.219 & $0.860 \pm 0.164$ & 1.714 & 661 & \cite{hartman11b} & 1.804 \\
HAT-P-38b & 0.923 & $0.267 \pm 0.020$ & 0.8010 & 977 & \cite{sato12} & 0.8380 \\
HAT-P-41b & 1.683 & $0.800 \pm 0.102$ & 1.568 & 692 & \cite{hartman12} & 1.662 \\
HD149026b & 1.368 & $0.359^{+0.022}_{-0.021}$ & 0.6536 & 2291 & \cite{torres08} & 0.6774 \\
HD189733b & 0.805 & $1.162 \pm 0.058$ & 1.200 & 1950 & \cite{boya15} & 1.221 \\
HD209458b & 1.203 & $0.64 \pm 0.09$ & 1.350 & 759 & \cite{boya15} & 1.414 \\
WASP-29b & 0.808 & $0.244 \pm 0.020$ & 0.7330 & 891 & \cite{hellier10} & 0.7692 \\
WASP-31b & 1.252 & $0.478 \pm 0.029$ & 1.379 & 456 & \cite{anderson11} & 1.535 \\
WASP-39b & 0.918 & $0.283 \pm 0.041$ & 1.207 & 414 & \cite{mac16} & 1.297 \\
WASP-43b & 0.67 & $1.78 \pm 0.10$ & 1.029 & 4699 & \cite{hellier11} & 1.039 \\
WASP-52b & 0.79 & $0.46 \pm 0.02$ & 1.199 & 646 & \cite{heb13} & 1.266 \\
WASP-63b & 1.88 & $0.38 \pm 0.03$ & 1.316 & 417 & \cite{hellier12} & 1.437 \\
WASP-67b & 0.87 & $0.42 \pm 0.04$ & 1.314 & 501 & \cite{hellier12} & 1.383 \\
WASP-69b & 0.813 & $0.260 \pm 0.017$ & 0.9563 & 532 & \cite{anderson14} & 1.017 \\
WASP-74b & 1.64 & $0.95 \pm 0.06$ & 1.456 & 891 & \cite{hellier15} & 1.528 \\
WASP-76b & 1.73 & $0.92 \pm 0.03$ & 1.635 & 631 & \cite{west16} & 1.752 \\
WASP-80b & 0.586 & $0.538^{+0.035}_{-0.036}$ & 0.9562 & 1396 & \cite{triaud15} & 0.9760 \\
WASP-101b & 1.29 & $0.50 \pm 0.04$ & 1.274 & 575 & \cite{hellier14} & 1.364 \\
WASP-121b & 1.458 & $1.183^{+0.064}_{-0.062}$ & 1.633 & 940 & \cite{delrez16} & 1.717 \\
XO-1b & 0.934 & $0.918^{+0.081}_{-0.078}$ & 1.172 & 1626 & \cite{torres08} & 1.197 \\
GJ1214b & 0.211 & $0.019 \pm 0.003$ & 0.2135 & 768 & \cite{ang13} & 0.2385 \\
HD97658b & 0.741 & $0.024^{+0.003}_{-0.002}$ & 0.2036 & 1466 & \cite{vg14} & 0.2208 \\
WASP-17b & 1.583 & $0.477 \pm 0.033$ & 1.709 & 316 & \cite{southworth12} & 1.897 \\
WASP-19b & 1.004 & $1.114 \pm 0.036$ & 1.311 & 1419 & \cite{tr13} & 1.378 \\
WASP-12b & 1.57 & $1.41 \pm 0.10$ & 1.748 & 977 & \cite{hebb09,k15} & 1.836 \\
TRAPPIST-1d & 0.121 & $9.34^{+1.10}_{-1.23} \times 10^{-4}$ & 0.05402 & 474 & \cite{grimm18,vg18} & 0.07436 \\
 & & & 0.07268$^\clubsuit$ & & \\
TRAPPIST-1e & 0.121 & $2.43^{+0.24}_{-0.25} \times 10^{-3}$ & 0.07329 & 912 & \cite{grimm18,vg18} & 0.08250 \\
 & & & 0.08174$^\clubsuit$ & & \\
TRAPPIST-1f & 0.121 & $2.94 \pm 0.25 \times 10^{-3}$ & 0.08490 & 837 & \cite{grimm18,vg18} & 0.09366 \\
 & & & 0.09294$^\clubsuit$ & & \\
TRAPPIST-1g & 0.121 & $3.61^{+0.30}_{-0.31} \times 10^{-3}$ & 0.09580 & 854 & \cite{grimm18,vg18} & 0.1036 \\
 & & & 0.1030$^\clubsuit$ & & \\
\hline
\hline
\end{tabular}\\
%\vspace{0.05in}
$\clubsuit$: For the TRAPPIST-1 exoplanets, we also examine Earth-like atmospheres ($m=29m_{\rm H}$).
\end{center}
\end{table*}

%%% REFERENCES %%%

% Don't change these lines
\bsp	% typesetting comment
\label{lastpage}

\begin{thebibliography}{99}

\bibitem[Abramowitz \& Stegun(1970)]{abram} Abramowitz, M., \& Stegun, I.A. \ 1970, Handbook of Mathematical Functions, 9th printing (New York: Dover Publications)

\bibitem[Anderson et al.(2011)]{anderson11} Anderson, D.R., Collier Cameron, A., Hellier, C., et al. \ 2011, A\&A, 531, A60

\bibitem[Anderson et al.(2014)]{anderson14} Anderson, D.R., Collier Cameron, A., Delrez, L., et al. \ 2014, MNRAS, 445, 1114

\bibitem[Anglada-Escud\'{e} et al.(2013)]{ang13} Anglada-Escud\'{e}, G., Rojas-Ayala, B., Boss, A.P., Weinberger, A.J., \& Lloyd, J.P. \ 2013, A\&A, 551, A48

\bibitem[Arcangeli et al.(2018)]{arc18} Arcangeli, J., D\'{e}sert, J.-M., Line, M.R., et al. \ 2018, ApJL, 855, L30

\bibitem[Arfken \& Weber(1995)]{arfken} Arfken, G.B., \& Weber, H.J. \ 1995, Mathematical Methods for Physicists, fourth edition (San Diego: Academic Press)

\bibitem[Barber et al.(2006)]{barber06} Barber, R.J., Tennyson, J., Harris, G.J., \& Tolchenov, R.N. \ 2006, MNRAS, 368, 1087

\bibitem[Barber et al.(2014)]{barber14} Barber, R.J., Strange, J.K., Hill, C., et al. \ 2014, MNRAS, 437, 1828

\bibitem[Bakos et al.(2010)]{bakos10} Bakos, G.A., Torres, G., P\'{a}l, A., et al. \ 2010, ApJ, 710, 1724

\bibitem[Batalha et al.(2017)]{batalha17} Batalha, N., Bean, J., Stevenson, K., et al. \ 2017, JWST Proposal ID 1366, Cycle 0 Early Release Science

\bibitem[Benneke \& Seager(2012)]{bs12} Benneke, B., \& Seager, S. \ 2012, ApJ, 753, 100

\bibitem[Benneke \& Seager(2013)]{bs13} Benneke, B., \& Seager, S. \ 2013, ApJ, 778, 153

\bibitem[B\'{e}tr\'{e}mieux \& Swain(2017)]{bs17} B\'{e}tr\'{e}mieux, Y., \& Swain, M.R. \ 2017, MNRAS, 467, 2834

\bibitem[Biddle et al.(2014)]{biddle14} Biddle, L.I., Pearson, K.A., Crossfield, I.J.M., et al. \ 2014, MNRAS, 443, 1810

\bibitem[Boyajian et al.(2015)]{boya15} Boyajian, T., von Braun, K., Feiden, G. A., et al. \ 2015, MNRAS, 447, 846

\bibitem[Brown(2001)]{brown01} Brown, T.M. \ 2001, ApJ, 553, 1006

\bibitem[Buchner et al.(2014)]{buchner14} Buchner, J., Georgakakis, A., Nandra, K., et al. \ 2014, A\&A, 564, A125

\bibitem[Chan et al.(2011)]{chan11} Chan, T., Ingemyr, M., Winn, J.N., et al. \ 2011, AJ, 141, 179

\bibitem[Cox(2000)]{cox} Cox, A.N. \ 2000, Allen's Astrophysical Quantities, 4th edition (New York: Springer-Verlag)

\bibitem[Delrez et al.(2016)]{delrez16} Delrez, L., Santerne, A., Almenara, J.-M., et al. \ 2016, MNRAS, 458, 4025

\bibitem[de Wit \& Seager(2013)]{ds13} de Wit, J., \& Seager, S. \ 2013, Science, 342, 1473

\bibitem[de Wit et al.(2018)]{dewit18} de Wit, J., Wakeford, H.R., Lewis, N., et al. \ 2018, Nature Astronomy, 2, 214

\bibitem[Ducrot et al.(2018)]{ducrot18} Ducrot, E., Sestovic, M., Morris, B.M., et al. \ 2018, arXiv:1807.01402

\bibitem[Esposito et al.(2014)]{esp14} Esposito, M., Covino, E., Mancini, L., et al. \ 2014, A\&A, 564, L13

\bibitem[Feroz \& Hobson(2008)]{feroz08} Feroz, F., \& Hobson, M.P. \ 2008, MNRAS, 384, 449

\bibitem[Feroz et al.(2009)]{feroz09} Feroz, F., Hobson, M.P., \& Bridges, M. \ 2009, MNRAS, 398, 1601

\bibitem[Feroz et al.(2013)]{feroz13} Feroz, F., Hobson, M.P., Cameron, E., Pettitt, A.N. \ 2013, arXiv:1306.2144

\bibitem[Foreman-Mackey et al.(2013)]{fm13} Foreman-Mackey, D., Hogg, D.W., Lang, D., \& Goodman, J. \ 2013, PASP, 125, 306

\bibitem[Fortney et al.(2016)]{fortney16} Fortney, J.J., et al. \ 2016, White Paper for NASA's Nexus for Exoplanet System Science (NExSS) (arXiv:1602.06305)

\bibitem[Freedman et al.(2014)]{f14} Freedman, R.S., Lustig-Yaeger, J., Fortney, J.J., et al. \ 2014, ApJS, 214, 25

\bibitem[Fu et al.(2017)]{fu17} Fu, G., Deming, D., Knutson, H., et al. \ 2017, ApJL, 847, L22

\bibitem[Griffith(2014)]{g14} Griffith, C.A. \ 2014, Philosophical Transactions of the Royal Society A, 372, 86

\bibitem[Grimm \& Heng(2015)]{gh15} Grimm, S.L., \& Heng, K. \ 2015, ApJ, 808, 182

\bibitem[Grimm et al.(2018)]{grimm18} Grimm, S.L., Demory, B.-O., Gillon, M., et al. \ 2018, A\&A, 613, A68

\bibitem[Hartman et al.(2009)]{hartman09} Hartman, J. D., Bakos, G. A., Torres, G., et al. \ 2009, ApJ, 706, 785

\bibitem[Hartman et al.(2011a)]{hartman11} Hartman, J.D., Bakos, G.A., Kipping, K.M., et al. \ 2011a, ApJ, 728, 138

\bibitem[Hartman et al.(2011b)]{hartman11b} Hartman, J.D., Bakos, G.A., Torres, G., et al. \ 2011b, ApJ, 742, 59

\bibitem[Hartman et al.(2012)]{hartman12} Hartman, J.D., Bakos, G.A., B\'{e}ky, B., et al. \ 2012, AJ, 144, 139

\bibitem[Hebb et al.(2009)]{hebb09} Hebb, L, Collier-Cameron, A., Loeillet, B., et al. \ 2009, ApJ, 693, 1920

\bibitem[H\'{e}brard et al.(2013)]{heb13} H\'{e}brard, G., Collier Cameron, A., Brown, D.J.A., et al. \ 2013, A\&A, 549, A134

\bibitem[Hellier et al.(2010)]{hellier10} Hellier, C., Anderson, D.R., Collier Cameron, A., et al. \ 2010, ApJL, 723, L60

\bibitem[Hellier et al.(2011)]{hellier11} Hellier, C., Anderson, D.R., Collier Cameron, A., et al. \ 2011, A\&A, 535, L7

\bibitem[Hellier et al.(2012)]{hellier12} Hellier, C., Anderson, D.R., Collier Cameron, A., et al. \ 2012, MNRAS, 426, 739

\bibitem[Hellier et al.(2014)]{hellier14} Hellier, C., Anderson, D.R., Collier Cameron, A., et al. \ 2014, MNRAS, 440, 1982

\bibitem[Hellier et al.(2015)]{hellier15} Hellier, C., Anderson, D.R., Collier Cameron, A., et al. \ 2015, AJ, 150, 18

\bibitem[Heng(2016)]{heng16} Heng, K. \ 2016, ApJL, 826, L16

\bibitem[Heng \& Tsai(2016)]{ht16} Heng, K., \& Tsai, S.-M. \ 2016, ApJ, 829, 104

\bibitem[Heng \& Kitzmann(2017)]{hk17} Heng, K., \& Kitzmann, D. \ 2017, MNRAS, 470, 2972

\bibitem[Heng(2017)]{heng17} Heng, K. \ 2017, Exoplanetary Atmospheres: Theoretical Concepts and Foundations (Oxford: Princeton University Press)

\bibitem[Heng(2018)]{heng18} Heng, K. \ 2018, RNAAS, 2, 128

\bibitem[Howard et al.(2012)]{howard12} Howard, A.W., Bakos, G.A., Hartman, J., et al. \ 2012, ApJ, 749, 134

\bibitem[Hubbard et al.(2001)]{h01} Hubbard, W.B., Fortney, J.J., Lunine, J.I., et al. \ 2001, ApJ, 560, 413

\bibitem[Huitson et al.(2013)]{huitson13} Huitson, C.M., Sing, D.K., Pont, F., et al. \ 2013, MNRAS, 434, 3252

\bibitem[Iyer et al.(2016)]{iyer16} Iyer, A.R., Swain, M.R., Zellem, R.T., et al. \ 2016, ApJ, 823, 109

\bibitem[Johnson et al.(2008)]{john08} Johnson, J. A., Winn, J. N., Narita, N., et al. \ 2008, ApJ, 686, 649

\bibitem[Kilpatrick et al.(2017)]{kil17} Kilpatrick, B.M., Cubillos, P.E., Stevenson, K.B., et al. \ 2018, AJ, 156, 103

\bibitem[Kitzmann \& Heng(2018)]{kh18} Kitzmann, D., \& Heng, K. \ 2018, MNRAS, 475, 94

\bibitem[Knutson et al.(2014)]{knutson14} Knutson, H.A., Dragomir, D., Kreidberg, L., et al. \ 2014, ApJ, 794, 155

\bibitem[Knutson et al.(2014b)]{knutson14b} Knutson, H.A., Benneke, B., Deming, D., et al. \ 2014, Nature, 505, 66

\bibitem[Kreidberg et al.(2014a)]{k14} Kreidberg, L., Bean, J.L., D\'{e}sert, J.-M., et al. \ 2014a, Nature, 505, 69

\bibitem[Kreidberg et al.(2014b)]{k14b} Kreidberg, L., Bean, J.L., D\'{e}sert, J.-M., et al. \ 2014b, ApJL, 793, L27

\bibitem[Kreidberg et al.(2015)]{k15} Kreidberg, L., Line, M.R., Bean, J.L., et al. \ 2015, ApJ, 814, 66

\bibitem[Lavie et al.(2017)]{lavie17} Lavie, B., Mendon\c{c}a, J.M., Mordasini, C., et al. \ 2017, AJ, 154, 91

\bibitem[Lecavelier des Etangs et al.(2008)]{lec08} Lecavelier des Etangs, A., Pont, F., Vidal-Madjar, A., \& Sing, D. \ 2008, A\&A, 481, L83

\bibitem[Lee et al.(2013)]{lee13} Lee, J.-M., Heng, K., \& Irwin, P.G.J. \ 2013, ApJ, 778, 97

%\bibitem[Line et al.(2012)]{line12} Line, M.R., Zhang, X., Vasisht, G., et al. \ 2012, ApJ, 749, 93

%\bibitem[Line et al.(2013a)]{line13a} Line, M.R., Wolf, A.S., Zhang, X., et al. \ 2013a, ApJ, 775, 137

\bibitem[Line et al.(2013)]{line13} Line, M.R., Knutson, H., Deming, D., Wilkins, A., \& Desert, J.-M. \ 2013, ApJ, 778, 183

%\bibitem[Line et al.(2014)]{line14} Line, M.R., Knutson, H., Wolf, A.S., \& Yung, Y.L. \ 2014, ApJ, 783, 70

\bibitem[Line \& Parmentier(2016)]{lp16} Line, M.R., \& Parmentier, V. \ 2016, ApJ, 820, 78

%\bibitem[Louden et al.(2017)]{louden17} Louden, T., Wheatley, P.J., Irwin, P.G.J., Kirk, J., \& Skillen, I. \ 2017, MNRAS, 470, 742

\bibitem[MacDonald \& Madhusudhan(2017)]{mm17a} MacDonald, R.J., \& Madhusudhan, N. \ 2017, MNRAS, 469, 1979

%\bibitem[MacDonald \& Madhusudhan(2017b)]{mm17b} MacDonald, R.J., \& Madhusudhan, N. \ 2017b, ApJL, 850, L15

\bibitem[Maciejewski et al.(2016)]{mac16} Maciejewski, G., Dimitrov, D., Mancini, L., et al. \ 2016, AcA, 66, 55

\bibitem[Madhusudhan \& Seager(2009)]{madhu09} Madhusudhan, N., \& Seager, S. \ 2009, ApJ, 707, 24

%\bibitem[Madhusudhan \& Seager(2010)]{madhu10} Madhusudhan, N., \& Seager, S. \ 2010, ApJ, 725, 261

%\bibitem[Madhusudhan \& Seager(2011)]{madhu11} Madhusudhan, N., \& Seager, S. \ 2009, ApJ, 729, 41

\bibitem[Madhusudhan et al.(2014)]{madhu14} Madhusudhan, N., Crouzet, N., McCullough, P.R., Deming, D., \& Hedges, C. \ 2014, ApJL, 791, L9

\bibitem[Mandell et al.(2013)]{mandell13} Mandell, A.M., Haynes, K., Sinukoff, E., et al. \ 213, ApJ, 779, 128

\bibitem[Mansfield et al.(2018)]{mansfield18} Mansfield, M., Bean, J.L., Line, M.R., et al. \ 2018, AJ, 156, 10

\bibitem[Miller \& Fortney(2011)]{mf11} Miller, N., \& Fortney, J.J. \ 2011, ApJL, 736, L29

\bibitem[Morris et al.(2018)]{morris18} Morris, B.M., Agol, E., Davenport, J.R.A., \& Hawley, S.L. \ 2018, ApJ, 857, 39

\bibitem[Nikolov et al.(2018)]{nikolov18} Nikolov, N., Sing, D.K., Fortney, J.J., et al. \ 2018, Nature, 557, 526

\bibitem[Rackham et al.(2018)]{rackham18} Rackham, B.V., Apai, D., \& Giampapa, M.S. \ 2018, ApJ, 853, 122

\bibitem[Rothman et al.(1987)]{rothman87} Rothman, L.S., Gamache, R.R., Goldman, A., et al. \ 1987, Applied Optics, 26, 4058

\bibitem[Rothman et al.(1992)]{rothman92} Rothman, L.S., Gamache, R.R., Tipping, R.H., et al. \ 1992, Journal of Quantitative Spectroscopy \& Radiative Transfer, 48, 469

\bibitem[Rothman et al.(1998)]{rothman98} Rothman, L.S., Rinsland, C.P., Goldman, A., et al. \ 1996, Journal of Quantitative Spectroscopy \& Radiative Transfer, 60, 665

\bibitem[Rothman et al.(2003)]{rothman03} Rothman, L.S., Barbe, A., Benner, D.C., et al. \ 2003, Journal of Quantitative Spectroscopy \& Radiative Transfer, 82, 5

\bibitem[Rothman et al.(2005)]{rothman05} Rothman, L.S., Jacquemar, D., Barbe, A., et al. \ 2005, Journal of Quantitative Spectroscopy \& Radiative Transfer, 96, 139

\bibitem[Rothman et al.(2009)]{rothman09} Rothman, L.S., Gordon, I.E., Barber, R.J., et al. \ 2010, Journal of Quantitative Spectroscopy \& Radiative Transfer, 111, 2139

\bibitem[Rothman et al.(2010)]{rothman10} Rothman, L.S., Gordon, I.E., Barbe, A., et al. \ 2009, Journal of Quantitative Spectroscopy \& Radiative Transfer, 110, 533

\bibitem[Rothman et al.(2013)]{rothman13} Rothman, L.S., Gordon, I.E., Babikov, Y., et al. \ 2013, Journal of Quantitative Spectroscopy \& Radiative Transfer, 130, 4

\bibitem[Sato et al.(2012)]{sato12} Sato, B., Hartman, J.D., Bakos, G.A., et al. \ 2012, PASJ, 64, 97

\bibitem[Sing et al.(2016)]{sing16} Sing, D.K., Fortney, J.J., Nikolov, N., et al. \ 2016, Nature, 529, 59

\bibitem[Skilling(2006)]{skilling06} Skilling, J. \ 2006, Bayesian Analysis, 1, 833

\bibitem[Sneep \& Ubachs(2005)]{su05} Sneep, M., \& Ubachs, W. \ 2005, Journal of Quantitative Spectroscopy \& Radiative Transfer, 92, 293

\bibitem[Southworth et al.(2012)]{southworth12} Southworth, J., Hinse, T.C., Dominik, M., et al. \ 2012, MNRAS, 426, 1338

\bibitem[Stevenson et al.(2014)]{stevenson14} Stevenson, K.B., D\'{e}sert, J.-M., Line, M.R., et al. \ 2014, Science, 346, 838

\bibitem[Tennyson \& Yurchenko(2017)]{ty17} Tennyson, J., \& Yurchenko, S.N. \ 2017, Molecular Astrophysics, 8, 1

\bibitem[Thorngren et al.(2016)]{thorngren16} Thorngren, D.P., Fortney, J.J., Murray-Clay, R., \& Lopez, E.D. \ 2016, ApJ, 831, 64

\bibitem[Torres et al.(2008)]{torres08} Torres, G., Winn, J.N., \& Holman, M.J. \ 2008, ApJ, 677, 1324

\bibitem[Tregloan-Reed et al.(2013)]{tr13} Tregloan-Reed, J., Southworth, J., \& Tappert, C. \ 2013, MNRAS, 428, 3671

\bibitem[Triaud et al.(2015)]{triaud15} Triaud, A.H.M.J., Gillon, M., Ehrenreich, D., et al. \ 2015, MNRAS, 450, 2279

\bibitem[Trotta(2008)]{trotta08} Trotta, R. \ 2008, Contemporary Physics, 49, 71

\bibitem[Tsiaras et al.(2018)]{tsi18} Tsiaras, A., Waldmann, I.P., Zingales, T., et al. \ 2018, AJ, 155, 156

\bibitem[van Grootel et al.(2014)]{vg14} van Grootel, V., Gillon, M., Valencia, D., et al. \ 2014, ApJ, 786, 2

\bibitem[van Grootel et al.(2018)]{vg18} van Grootel, V., Fernandes, C.S., Gillon, M., et al. \ 2018, ApJ, 853, 30

\bibitem[von Braun et al.(2012)]{vb12} von Braun, K., Boyajian, T.S., Kane, S.R., et al. \ 2012, ApJ, 753, 171

\bibitem[Waldmann et al.(2015)]{wald15} Waldmann, I.P., Tinetti, G., Rocchetto, M., et al. \ 2015, ApJ, 802, 107

%\bibitem[Waldmann et al.(2015b)]{wald15b} Waldmann, I.P., Rocchetto, M., Tinetti, G., et al. \ 2015a, ApJ, 813, 13

\bibitem[Wakeford et al.(2017)]{wakeford17} Wakeford, H.R., Sing, D.K., Kataria, T., et al. \ 2017, Science, 356, 628 

\bibitem[Wakeford et al.(2018)]{wakeford18} Wakeford, H.R., Sing, D.K., Deming, D., et al. \ 2018, AJ, 155, 29

\bibitem[West et al.(2016)]{west16} West, R.G., Hellier, C., Almenara, J.-M., et al. \ 2016, A\&A, 585, A126

\bibitem[Wyttenbach et al.(2015)]{wyttenbach15} Wyttenbach, A., Ehrenreich, D., Lovis, C., Udry, S., \& Pepe, F. \ 2015, A\&A, 577, A62

\bibitem[Yurchenko et al.(2011)]{y11} Yurchenko, S.N., Barber, R.J., \& Tennyson, J. \ 2011, MNRAS, 413, 1828

\bibitem[Yurchenko et al.(2013)]{y13} Yurchenko, S.N., Tennyson, J., Barber, R.J., \& Thiel, W. \ 2013, Journal of Molecular Spectroscopy, 291, 69

\bibitem[Yurchenko \& Tennyson(2014)]{yt14} Yurchenko, S.N., \& Tennyson, J. \ 2014, MNRAS, 440, 1649

\bibitem[Yurchenko et al.(2018)]{y18} Yurchenko, S.N., Al-Refaie, A.F., \& Tennyson, J. \ 2018, A\&A, 614, A131

\end{thebibliography}
\end{document}